\documentclass[preprint,12pt]{elsarticle}

\usepackage{amsmath}
\usepackage{mathtools}
\usepackage[utf8]{inputenc}
\usepackage[OT2,T2A,T1]{fontenc}
\usepackage[main=english,russian]{babel}
\usepackage{lmodern}
\usepackage{graphicx}
\usepackage{float}
\usepackage{subfloat}
\usepackage{caption}
\usepackage{subcaption}
\usepackage[export]{adjustbox} 
\usepackage{esvect}
\usepackage{amsfonts}
\usepackage[11pt]{moresize}
\usepackage{titlesec}
\usepackage{array}
\usepackage{multirow}
\usepackage{multicol}
\usepackage{arydshln}
\usepackage{placeins}
\usepackage{pdfpages}
\usepackage[a4paper,left=2cm,right=2cm,top=2cm,bottom=2cm]{geometry}
\usepackage{enumitem}
\usepackage{tocloft} 
\usepackage[titletoc,title]{appendix} 
\usepackage{url}
\usepackage{comment}
\usepackage{hyperref}

\biboptions{sort&compress}

\extrafloats{100}

\makeatletter
\def\ps@pprintTitle{%
 \let\@oddhead\@empty
 \let\@evenhead\@empty
 \def\@oddfoot{\centerline{\thepage}}%
 \let\@evenfoot\@oddfoot}
\makeatother

\makeatletter
\let\sv@endpart\@endpart
\def\@endpart{\thispagestyle{empty}\sv@endpart}
\makeatother

\newlength{\figurecustomwidth}
\begin{document}

\begin{frontmatter}
\phantomsection
\belowpdfbookmark{Abstract}{Abstract}

\title{Complex systems approach to natural language}

\author[ifj]{Tomasz Stanisz}
\author[ifj,pk]{Stanis{\l}aw Dro\.zd\.z\corref{cor2}}
\cortext[cor2]{stanislaw.drozdz@ifj.edu.pl}
\author[ifj]{Jaros{\l}aw Kwapie\'n}

\address[ifj]{Complex Systems Theory Department, Institute of Nuclear Physics, Polish Academy of Sciences, ul.~Radzikowskiego 152, 31-342 Krak\'ow, Poland}
\address[pk]{Faculty of Computer Science and Telecommunication, Cracow University of Technology, \\ ul.~Warszawska 24, 31-155 Krak\'ow, Poland}

\begin{abstract}
The science of complexity aims to answer the question of what rules nature chooses when assembling the basic constituents of matter and energy into structures and dynamical patterns that cascade through the entire hierarchy of scales in the Universe. A related phenomenon -- natural language -- can successfully mirror such structures as reflected by its ability to encode and transmit information about them and among them. It is thus legitimate to expect that natural language carries the essence of complexity. And indeed, in the human's speaking and writing it is particularly true that {\it more is different}. Natural language thus deserves a central place in the related quantitative study within the science of complexity. 

With this in mind the present review summarizes the main methodological concepts used in this domain and documents their applicability and utility in identifying universal as well as system-specific features of natural language in its written representation in several major Western languages. In particular, three main complexity-related current research trends in quantitative linguistics are exhaustively covered. The first part addresses the issue of word frequencies in texts and, in particular, demonstrates that taking punctuation into consideration largely restores scaling whose violation in the Zipf's law for the most frequent words is commonly modeled by the so-called Mandelbrot's correction. The second part introduces methods inspired by time series analysis, used in studying various kinds of long-range correlations in written texts. The related time series are generated on the basis of text partition into sentences or into phrases between consecutive punctuation marks. It turns out that these series develop features often found in signals generated by complex systems: the presence of long-range correlations along with fractal or even multifractal structures. Moreover, it appears that the distances between consecutive punctuation marks quite universally across languages comply with the discrete variant of the Weibull distribution, often appearing in survival analysis. In the third part, the application of the network formalism to natural language is reviewed, particularly in the context of word-adjacency networks whose structure reflects the word co-occurrence in texts. Various parameters characterizing topology of such networks can be used for classification of texts, for example, from a stylometric perspective. Network approach can also be applied in semantic analysis to represent a hierarchy of words and associations between them based on their meaning. Structure of such networks turns out to be significantly different from that observed in random networks, revealing genuine properties of language. Finally, punctuation appears to have a significant impact not only on the language's information-carrying ability but also on its key statistical properties, hence it seems recommended to consider punctuation marks on a par with words.

\end{abstract}

\begin{keyword}
natural language, complexity, power laws, fractals, complex networks, punctuation
\end{keyword}

\end{frontmatter}

\clearpage

\phantomsection
\belowpdfbookmark{\contentsname}{toc}
\tableofcontents

\selectlanguage{english}

\clearpage

\FloatBarrier


\section{Natural language}
\label{sect::natural_language}

One of the reasons for which humans can be considered extraordinary among all species present on Earth is the ability to think in abstract categories and to efficiently communicate the results of such a thinking process~\cite{CherryC-1980a,MildnerV-2007a}. Although research indicates that some animals are capable of solving tasks appearing to require abstract reasoning and are able to communicate with each other~\cite{FriendT-2005a} and even with humans~\cite{PepperbergIM-2017a}, the complexity and sophistication of human cognitive and communication abilities are enormously superior to those possessed by any other known organisms. The ability to use language is a key factor that allowed for the development of civilization and culture~\cite{IrikiA-2012a}, things that are considered unique for humans.

Language is such an important and multifaceted phenomenon that it draws the attention of a great variety of academic disciplines. To grasp the diverse properties of language and to be able to describe it possibly comprehensively, an interdisciplinary approach is required. Language is a set of symbols and rules, an organism's ability to generate sounds, a communication tool, a logical system of notions guiding the thinking process, as well as a social and cultural phenomenon. Therefore the fields actively studying subjects related to language range from humanities through social sciences to natural and formal sciences, each of them focusing on a~different~\mbox{perspective}.

Mathematics and physics offer tools that can be successfully applied to language study. A number of concepts originating in these sciences have found their use in the quantitative description of natural language. In physics, the complex-system approach seems to be particularly fruitful. Complex systems are a class of systems that typically consist of a large number of constituents, whose general properties usually cannot be deduced only from the properties of those constituents. Such systems can often be characterized by the phrase ``the whole is something beside the parts''. This is exactly the case of natural language, whose complicated multilevel structure cannot be simply reduced to a set of rules and laws. A number of traits that are shared among complex systems can be found here, like hierarchical structure, long-range correlations, fractality, and the presence of power laws. 

Description of the systems consisting of a large number of elements is naturally provided with the use of statistical mechanics. Statistical properties of such systems can often be grasped by models using stochastic processes, while the empirical and simulated data can be studied by means of tools that have their origin in mathematics and statistical physics, like multifractal analysis and network theory, among others. Quantitative research on natural language, which is concerned with language structure, dynamics, and evolution, often takes advantage of these tools and results can have practical application in automatic language processing, for example, which is going to have likely a growing impact on everyday life.

\subsection{Studying natural language from various perspectives}

Depending on a context and one's interest, language can be defined in multiple ways. Human language, which spontaneously evolved with the development of human communities, is often referred to as \textit{natural language}, as opposed to \textit{formal language}, which is a mathematical object (a set of sequences derived from a finite set of symbols). Another opposing term is  \textit{constructed language} (sometimes called \textit{conlang}), which shares many properties of a natural language, but whose structure comes from a planned activity: artificially designed vocabulary, grammar, and phonology. Among the examples one can list Esperanto, Interlingua, Interslavic, and the languages created by a fantasy writer J.R.R. Tolkien.

From the most natural perspective, language is a structured system of communication. The ability to communicate is not an extraordinary phenomenon among animals. Among well-known examples one may list birds singing to attract mates and to repel rivals~\cite{Catchpole2012}, bees dancing to inform their nestmates about the distance and direction to food sources~\cite{Frisch1974,Kirchner1988}, or dolphins whistling to recognize each other~\cite{Janik2000,Janik2006,Lopez2011}. There are plenty of forms of communication between animals, with various types of signals: visual, auditory, olfactory, tactile, etc.~\cite{Domb2001,Gosling2001,Slabbekoorn2002,Braeuer2005,Schloegl2007,Paulos2007,Salazar2008,Slobodchikoff2008}. However, human language is unique compared to communication systems of other animals. In 1960s, Charles Hockett defined a collection of essential characteristics of language~\cite{Hockett1960}, the so-called \textit{design features}, potentially useful in setting language apart from animal communication. The original list evolved over time and has been modified. Although its practical use is in some cases limited~\cite{Wacewicz2014}, the idea of studying the proposed features of language strongly influenced linguistics. Among those features, it is worth to mention \textit{displacement}, \textit{productivity}, \textit{cultural transmission}, \textit{duality of patterning}, \textit{learnability}, and \textit{reflexiveness}.

Displacement is the possibility of referring to events remote in space and time, to objects that are not present in the immediate environment, or even do not exist. Productivity is the ability of language users to create and understand new expressions that can convey any message; the number of possible utterances in every human language is infinite. Cultural transmission is constituted by the fact that language is learned by interactions with individuals already capable of using~it. Although predisposition to use language may be innate, the key factor in language acquisition is social setting (it determines, for example, which particular language is acquired as the first). Duality of patterning refers to the organization of language simultaneously on two levels: meaningless constituents (sounds, letters) are combined into units that have particular meaning (words); these units can be further combined into a complete message. Learnability means that a speaker of some language can learn other languages. Reflexiveness is the ability of language to describe itself. Humans can use language to define what language is, to discuss its structure, or to talk about its usage. Human language is the only known system in nature that exhibits all the aforementioned features; animal communication systems either possess only some of them in a limited form or do not have them at~all~\cite{Grady1997,Yule2010}.

The uniqueness of human language poses a question about its origin. However, the development of the study on when and how human language came into existence has been severely limited by the lack of empirical evidence that could prove or disprove numerous hypotheses~\cite{Gibson2011,Hauser2014}. The direct evidence of the existence of language can be obtained by discovering the earliest traces of writing (which are dated to about 3000 years BC), but speech is much older than written language~\cite{Yule2010,JeanAitchison2012}. Contemporary research on this subject relies on indirect information supplied by paleontology, archaeology, biology, linguistics, and cognitive science. Studying fossil record may reveal human ancestors' anatomical traits of potential relevance to language, for example brain size. However, this line of reasoning faces certain limitations, like the lack of possibility of reconstructing brain internal structure, or the absence of data regarding evolution of the vocal tract~\cite{Gibson2011}. A related approach concentrates on the artefacts left by early humans: depending on the level of sophistication, advanced tools or art might indicate the capability of abstract and symbolic thinking~\cite{Johansson2005}. This can confirm certain cognitive skills at a given stage of human development, but it is rather useless in determining how and when exactly the language appeared as it could precede fossilizable art and advanced tools~\cite{Mcbrearty2000}.

A different line of enquiry employs genetics and genomics to study the origin and migration of human populations, as well as to identify the points in time when the lineage of modern humans diverged from other species~\cite{Paeaebo2003,Gibson2011,Fisher2016}. This allows one, for example, to put constraints on the time period when language was born. Studies of human genome are also aimed at determining, which genes are relevant for language capacity. However, due to complexity of this problem, a precise and consistent view on how genes are related to the emergence of language has not yet been reached~\cite{Hauser2014, Mountford2017, DeSalle2017}. Another area of investigation is related to the research on animal cognition and communication. Its major direction is the study of language-related traits in non-human primates~\cite{Gibson2011, Fitch2015}, in particular chimpanzees, which are the closest living relatives to humans (the last common ancestor of the two species is estimated to have lived between 4 and 8 million years ago~\cite{Ruvolo1997,Bernard2000,Chen2001,Paeaebo2003,Bradley2008,Wood2011}). Assuming that a trait that is present in all the species sharing a certain common ancestor was probably also present in that ancestor, one may attempt to determine which species preceding humans exhibited particular traits necessary for the development of language.

Research on other, more distantly related species may also be informative. An important concept here is convergent evolution, i.e.,  a process of independent development of a similar feature in a few different species whose last common ancestor did not have that feature at all (e.g., the streamlined body shape shared by penguins, fish, and some aquatic mammals~\cite{Urry2017}). Convergent evolution is a result of adaptation to a similar environment in a similar way. Therefore, studying selected animal traits at least partially related to language (like vocal learning, which occurs in whales, dolphins, bats and some birds~\cite{Jarvis2000}) might be helpful in explaining the mechanisms that have driven the emergence of language~\cite{Fitch2015}. The problem with the language-related research is that it is hard to find and choose animal features that can be unequivocally linked to language since human language has no comparably complex counterpart among animals. Also, a number of results in this field are disputable and difficult to~interpret (for example, whether the great ape gestural communication possesses any characteristics of human language)~\cite{Gibson2011,Scott-Phillips2015,Moore2015,Scott-Phillips2015a,Tomasello2018a}.

With all the methods of investigating the early history of natural language having their issues and limitations, the question about the language origin remains unanswered~\cite{LockeJL-2006a,Yule2010}. It is not known whether human language features have developed as an adaptation for some early forms of communication or whether they are a result or a byproduct of an adaptation to other tasks, like tool-making or numerical reasoning~\cite{Hauser2002}. It is not known when language started to emerge, how long this process took, and whether the spoken form of language was preceded by gestural communication~\cite{Johansson2005}. The answers provided by contemporary science are still to a large extent speculative. It has been stated that, within this research field, ``the richness of ideas is accompanied by a poverty of evidence''~\cite{Hauser2014}. However, a constant improvement of research methods gives hope that knowledge on the language origin will get more and more complete in future.

\begin{figure}[ht]
\centering
\begin{minipage}{\figurecustomwidth}
\centering
\includegraphics[width=0.98\textwidth]{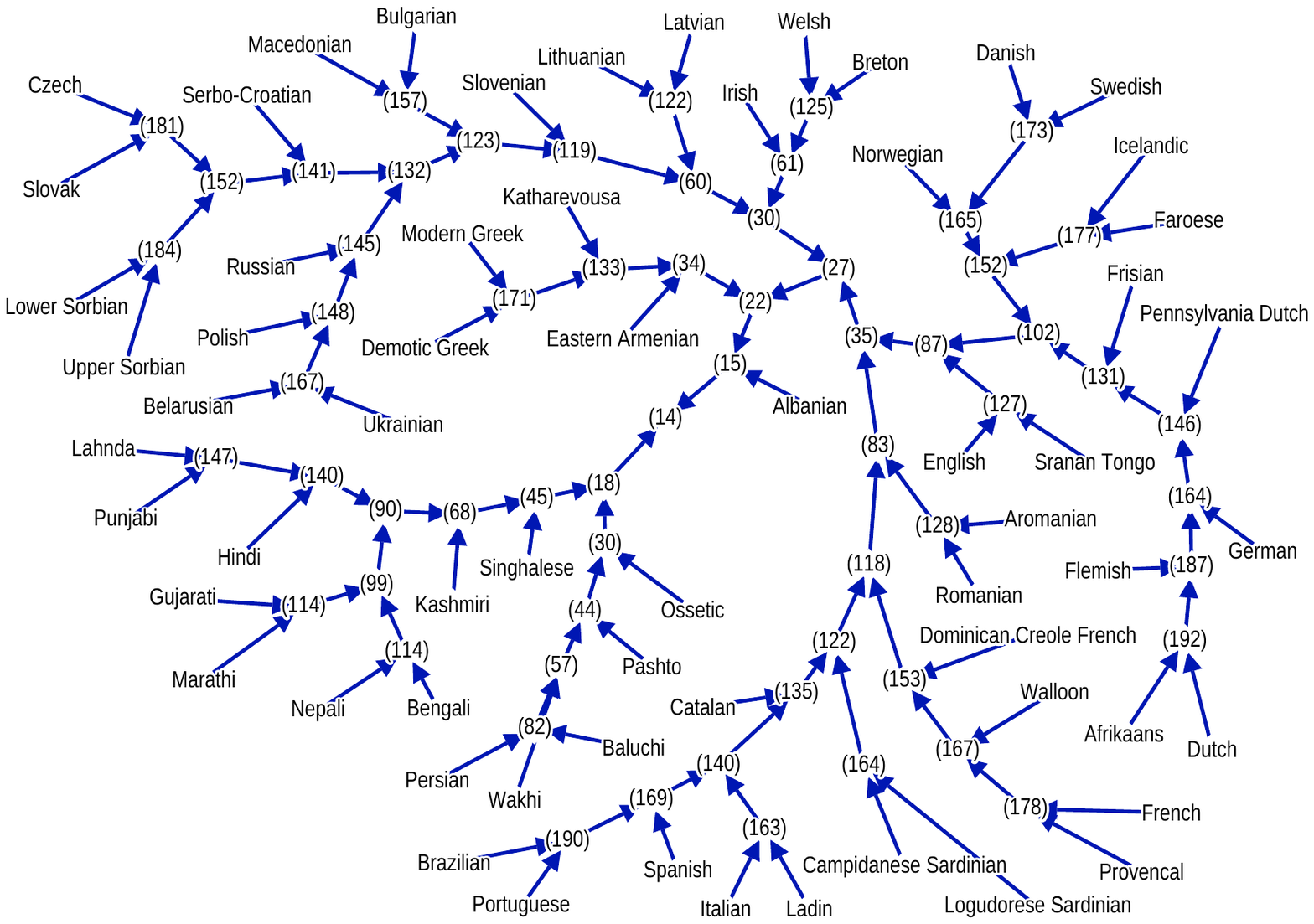}
\caption{A network representing lexical relationships between 63 selected languages from the Indo-European family based on a subset of data used in~\cite{Dyen1992}. The data is a multilingual Swadesh list with $N=200$ entries. Each entry corresponds to one meaning and consists of the words representing that meaning in different languages. These words divided into groups reflect their possible common origin. As a result, each pair of words under a given entry is judged as ``cognate'', ``doubtfully cognate'', or ``not cognate''. One can define the proximity $n_c(l_1,l_2)$ between two languages $l_1,l_2$ as the total number of word pairs judged as ``cognate'' among all entries. Consequently, the distance between $l_1$ and $l_2$ can be expressed as $d(l_1,l_2)=N-n_c(l_1,l_2)$. The network presented above is a directed tree representing hierarchical clustering of the studied languages using the so-defined distance. Each leaf (a node with no incident edges) corresponds to one language, and each internal node (a node with at least one incident edge) is a cluster of languages. Consecutive groupings into bigger and bigger clusters are represented by arrows (directed edges). Each cluster is labeled with its internal minimum proximity: if $k$ is the number labeling the cluster, then the proximity $n_c$ (i.e., the number of shared cognates) between any two languages belonging to that cluster is not smaller than~$k$. More advanced methods of analyzing the distances between language lexicons can be useful in reconstruction of the language evolutionary trees~\cite{Gray2003}.}
\label{fig::language_relationships}
\end{minipage}
\end{figure}

Language is not a static entity. Languages continually undergo gradual changes of their lexical, phonological, syntactic, and semantic nature. These changes are driven by a number of factors, like migration and language contact, the development of technology, or people's willingness to use the language that they associate with a certain degree of social prestige~\cite{Yule2010,Hock2009}. Studying how language changes over time allows one to get an insight into certain cultural and social processes~\cite{Lansdall-Welfare2017,Lansdall-Welfare2014,Michel2010}. Languages influence each other, some languages die out, and new languages can be born. Therefore, the number of living, actively used languages also changes over time. Currently, this number is estimated to be around 7000~\cite{Glottolog2020,Ethnologue2020}, but the exact number depends, for example, on whether some varieties are classified as separate languages or dialects. Research on language history aims to find laws that govern the process of language change and to answer a question how particular languages are mutually related. Often, it employs a method of comparing phonological, morphological, and syntactic features of different languages, as well as their lexicons. A noteworthy example of such a method is the so-called Swadesh list~\cite{swadesh1971}. It consists of 100 or 200 words (depending on the version; other numbers are also in use)~\cite{Mcmahon2006}, which are assumed to represent basic vocabulary (some English examples are: \textit{water}, \textit{hand}, \textit{tree}, \textit{tooth}, \textit{rain}, \textit{moon}, \textit{long}, \textit{cold}, \textit{give}, and \textit{sleep}). A similar list can be constructed for each language by identifying words with the same meaning. By creating the Swadesh lists for two languages and determining how many words are cognates (words of common etymological origin), one can analyze the lexical relationships between these languages. A quantitative description of such relationships (Fig.~\ref{fig::language_relationships}) may be useful in answering a question if two given languages have derived from a common parent language and, if so, when their divergence took place~\cite{Gray2003,Petroni2008}.

One may ask not only how language occurred and evolved in human species, but also about the mechanisms driving the development of language in every individual. Two major schools of thought regarding language acquisition may be distinguished: the first one states that language is an ability that is learned in the way similar to other cognitive skills~\cite{Skinner1992,Tomasello2003}, the other one (established primarily by Noam Chomsky) says that some features of language are innate~\cite{Chomsky1965,Pinker1994}. The proponents of the latter argue that the amount and diversity of the information that children are exposed to is too small to properly acquire language skills from the basics. This argument, called the \textit{poverty of the stimulus}, was the key reason for introducing the concepts of \textit{language acquisition device} -- a hypothetical innate component of human brain responsible for certain linguistic skills and \textit{universal grammar} -- a set of highly abstract rules and characteristics shared among all the world languages, which are encoded in each human's brain. According to the theories using these concepts, children acquire language relatively quickly regarding the amount of the available ``linguistic data'', because the core cognitive features required for that task are known in advance; they only have to adjust the parameters that can vary among languages. However, the assumptions behind this point of view have been questioned and debated~\cite{Pullum2002,Legate2002,Fernald2006,Bot2015}. Current research emphasizes the role of learning in the process of language acquisition and, by using the results from neuroscience, aims to discover the mechanisms driving this process~\cite{Kuhl2010}.

One of important perspectives of natural language research is a relationship between language and brain function. It is focused on brain specialised language-related areas of the cortex and the internal mechanisms responsible for learning and language processing. This research field benefits much from the development of neuroimaging techniques~\cite{Price2012, Zubicaray2019}, which allow one to conduct respective experiments and measurements. For a long time a prevalent view on that matter was that language comprehension and production is, to a large extent, contained within two regions of the cerebral cortex: Broca's and Wernicke's. These regions were identified as crucial for the ability to use language by the 19th-century physicians who worked with patients suffering from an impairment of language abilities caused by brain damage and who were able to link the symptoms with a damage to specific parts of the brain~\cite{Geschwind1970}. However, modern research has found that treating the language-related processes as dependent on two brain regions only is an oversimplification~\cite{Tremblay2016}. It turns out that the system of language comprehension and production constitutes a distributed network involving multiple brain regions. Moreover, this network cannot be easily divided into separate modules responsible for different tasks, because a single task meant to interfere with one particular linguistic ability can actually activate multiple parts of the whole system~\cite{Hickok2007,Saur2008,Hickok2009,Kemmerer2015,Zubicaray2019,Debowski2021}. This shows that it has a complicated structure and exhibits complex patterns of activity. Studying the way in which language is processed in the brain can potentially lead both to better understanding of natural language and to practical contributions to other disciplines; for example, it can be helpful in treating people suffering from language disorders~\cite{Nasios2019}.

Another interesting area of investigation is a relationship between language and thinking. There are a number of theories claiming that language affects at least some of the other aspects of cognition. There exists an idea that the process of thinking has structure similar to the structure of language: it combines simple concepts into complex thoughts in the way analogous to the way that syntax combines words into sentences. This hypothetical structure has been named the language of thought, sometimes also called ``mentalese''~\cite{Fodor1979,Pinker1994,Tillas2015}. This hypothesis has been a subject of debate. There has been, for example, a contrary line of argument stating that thinking requires explicit usage of natural (and native) language~\cite{Kaye1995,Garfield1997,Carruthers1998,Viger2005}. Another influential concept, the so-called \textit{linguistic relativity} (also known as \textit{the Sapir-Whorf hypothesis}) states that a particular language used by an individual influences their perception and way of thinking~\cite{Hill1992,Lucy2001}. Whorf's idea was based on a study of the Hopi language, in which the conceptualization of time is different to the one usually appearing in the other languages. Whorf pointed out that the Hopi language lacks a word referring to time and its verbs do not distinguish between present, past, and future. Therefore, a view on the surrounding reality possessed by a Hopi-language native speaker may be different from a view typical for, for instance, a user of English~\cite{Whorf1956a,Whorf1956b}. Although later studies showed that Whorf's conclusions about an inability to refer to time in the Hopi language could be exaggerated~\cite{Malotki1983}, his work initiated the research on how particular languages influence thinking. A language also investigated in this context is the Pirahã language, which has no words for numerals and lacks a notion of counting. Its native users seem to have extreme difficulty with acquiring even the most basic numeracy skills~\cite{Everett2005,Frank2008}. Another example of research are the experiments suggesting that presence of the grammatical gender affects a way that its native speakers describe particular objects and also influences their ability to memorize names given to those objects~\cite{Boroditsky2003}. Research on the relationships between language and cognitive skills like color perception, spatial orientation, and numerical reasoning show that language, understood as a mental ability universal for humans, serves as a powerful tool in cognitive processes and it is particularly useful in transforming various pieces of information into a convenient representation~\cite{Uenal2016}.

Since the process of thinking often employs combining simpler concepts into more sophisticated ones, it benefits much from a system capable of representing virtually any concept in a standardized, usable manner. Language is such a system and, therefore, it can be considered as a human mind's resource of substantial importance, indispensable in complex intellectual activities.

\subsection{Computational and quantitative approach to language}

The origins of the usage of quantitative and computational methods to study language can be dated back to 1940s and 1950s. The development of this research field was related to a desire to create systems capable of automatic processing of natural language. One of the first related tasks that gained much attention was machine translation~\cite{Jones1994}. After a period of optimism and enthusiasm, it became clear that not only this particular task was difficult, but also modeling language in general was much more challenging than it used to appear initially. Language has a multilevel, complex structure, and its processing requires typically multiple steps which utilize various tools. A language user needs to be able to extract and recognize a sequence of words from an audio signal and to transform a sequence of words into an audio signal. These actions require knowledge of phonology and phonetics, which describe which sounds are needed to pronounce each word and how these sounds are physically realized. Words may have various forms and to use them correctly it requires knowledge about morphology, which specifies how words can be divided into components and what information is carried by these components (for instance, a distinction between a singular and a plural form of a noun). Maintaining relationships between individual words (e.g., their order) requires knowledge of syntactic rules. Knowledge of meaning of words and their allowed combinations -- the semantics -- is needed both to understand an utterance and to generate one, because appropriate words have to be found to express a thought. In a conversation, one needs to be aware of a context and a situation, in which the conversation takes place. In other words, one has to keep track of the discourse. This requires knowledge about the linguistic units larger than a single utterance. Of course, language users do not necessarily have to be able to verbalize what kind of knowledge they are using. Nevertheless, comprehending and generating utterances in natural language is inseparably connected to the activities mentioned above.

Each of these activities can be considered a separate task. Such tasks need to be performed by a human or a machine that uses or processes language. Describing objectives of such tasks by using mathematics and developing algorithms that carry out them is the subject of computational linguistics and natural language processing. Currently, due to a recent rapid increase in computing power availability and development of appropriate methods, natural language processing usually employs machine learning, especially deep learning~\cite{Young2018, Deng2018}. Before the epoch of machine learning, the language processing systems were usually ``rule-based'' -- they used to operate on a predefined set of rules designed to capture structure and relationships in a modeled aspect of language. Since such rules are explicit, they are interpretable, in the sense that one can track how a system using them produces a given result. This is unlike many machine learning methods whose intermediate stages of operation are often unreadable to a human. However, the systems of rules can be very complicated and might not be easily comprehensible. Nevertheless, construction of rules describing even a small part of the language structure can be useful from a theoretical point of view, because it helps in grasping the mechanisms driving various linguistic phenomena.

An influential concept in linguistic research is an idea of characterizing language structures by using formal languages. It was proposed by Noam Chomsky in 1950s and 1960s~\cite{Chomsky1957,Chomsky1965}. It treats the ability to use natural language as a certain sort of computational system, which uses a set of rules defined by an appropriate grammar to organize individual components into complete utterances. Formal language is a mathematical concept; it is a set whose elements are in a sense constructed from elements of some other set. Let $\Sigma$ be a finite set. From the elements of $\Sigma$ one can construct strings -- sequences of arbitrary (but finite) length consisting of elements of $\Sigma$. Let $\Sigma^{*}$ be a set of all such sequences. A formal language is a subset of $\Sigma^{*}$. In other words, it is a subset of the set of all finite strings that can be generated from the elements of $\Sigma$. Specification which strings in $\Sigma^{*}$ belong to a formal language can be done in a few ways, but a way that is particularly advantageous is formal grammar. Formal grammars are mathematical objects providing sets of rules able to generate strings from other strings by a selected production rule. See Appendix~\ref{appendix::formal.grammars} for more details and examples.

Formal languages are quite general and abstract mathematical objects. Therefore a number of concepts derived from formal language theory have found applications outside mathematics and linguistics. An interesting example are the so-called Lindenmayer systems (L-systems in short), i.e., the string-rewriting systems originally invented to model development and growth of some organisms (one of the first organisms studied in that context were algae~\cite{Lindenmayer1968,Lindenmayer1968a}). This theory was then applied more generally to various kinds of branching systems. L-systems are able to generate strings with recursively nested patterns, so they are well suited to model self-similar structures. Therefore, they are used to generate fractals (discussed in Sect.~\ref{sect::fractal_structure_of_language}) and other objects exhibiting similar properties. Examples of images created with the use of L-systems as well as their formal definition are presented in Appendix~\ref{appendix::lindenmayer.systems}. The fact that similar mathematical objects like formal grammars and L-systems are used to model self-similar, hierarchical systems as well as natural language is more than a mere coincidence. Multiple aspects of language organization are of strongly hierarchical nature. In fact, the ability to generate multilevel, recursively nested structures is sometimes considered as one of the defining features of natural language~\cite{Hauser2002}. This is one of the reasons why the analytical tools designed to study such structures (like fractal geometry) are useful in research on natural language.

An important subfield of language research is called quantitative linguistics. It investigates language using statistical methods. It concentrates on the properties of language that can be described in terms of probability distributions, statistical models, time series and related tools, and attempts to formulate linguistic laws pertaining to those properties. Studying such laws and an origin of their presence allows one to formulate hypotheses about cognitive processes behind language and about language origin, evolution, and learning. It has a practical purpose also since knowledge about statistical patterns in language can be applied to design natural language processing tools and methods. In order to propose linguistic laws and to verify them empirically, quantitative linguistics uses appropriately large samples of language. In the case of written language, such a sample called a corpus is a collection of texts written in a given language. Diversity and the number of used texts depends on a particular application. A large enough corpus can be treated as representative for the studied language. Therefore, observations based on such a corpus can be generalized to particular language. By analyzing multiple corpora in various languages it is possible to draw conclusions about laws universal across languages.

Among the most fundamental linguistic laws is Zipf's law~\cite{ZIPF_1936,ZIPF_1949,PIANTADOSI_2014} that specifies a distribution of word frequencies in texts. Heaps' law (also known as Herdan's law)~\cite{Heaps1978,Egghe2007,Chacoma2020} describes how the number of different words in a text sample varies with the text length. The Menzerath-Altmann law~\cite{Altmann1980,Milicka2014} states that the size of a linguistic construct is negatively correlated with the size of its constituents -- for example, the size of sentences measured by the number of clauses is negatively correlated with the size of clauses themselves measured by the average number of words in a clause. These examples are, among others~\cite{Torre2017,Corral2020,Altmann2016}, a part of the field of active research, which attempts to observe general statistical patterns in language, to describe them with appropriate formalism, and to explain their origin. Two examples of the widely known linguistic laws -- Zipf's law and Heaps' law -- are discussed in more detail in Sect.~\ref{sect::word_statistics}.

It is worth to note in this context that there is a distinction between spoken and written language. Natural languages are primarily spoken; writing is a complement to speech -- an invented system representing spoken language visually. Spoken language is, in some sense, more ``natural'': it existed before writing was invented and it is acquired without specific instructions in childhood. In contrast, learning to write requires significant deliberate effort. Also, some languages do not have a written form at all~\cite{Grady1997,Yule2010}. Speech and writing differ in a number of traits. When writing down a spoken utterance, some information -- conveyed by voice modulation, for instance -- might be lost. Written language is often more formal than speech: the things like grammatical errors, hesitations, and repetitions are typically present in spoken, but not in written language. An analysis of one selected representation of language (spoken or written) is treated in many situations as sufficient to draw conclusions about language in general. However, it is important to be aware of that some of the observed effects may be specific to a chosen representation.

To investigate the properties of natural language using statistical methods, language samples of considerable size are needed. Type of the samples varies depending on specific area of study: different tools are needed to study spoken language and written language. For the latter, a convenient source of linguistic data are literary texts written in prose. Obviously, it has to be remembered that narrative texts constitute a specific type of data and do not account for the whole language. However, among the literary forms, prose is the one that to the greatest extent mimics a natural flow of speech and uses grammatical structures typical for the everyday  language. A fact that the language used in literary texts usually adheres more strictly to grammatical rules and uses more refined vocabulary than the colloquial language can be considered as an advantage: it allows one to study structures of certain degree of sophistication and complexity, which often lack in the everyday language. Many narrative texts have a form of book and books are typically large enough to be subject to a statistical analysis on their own. These properties of narrative texts make them highly useful in quantitative study of language.


\section{Complexity and complex systems}
\label{sect::complexity_and_complex_systems}

Among the most important goals of science, especially physics, is to understand natural phenomena, to explain them using models, and to make predictions based on these models. The models are typically designed in such a way that they can grasp the relevant information about the studied systems, but they also remain as simple as possible, avoiding unnecessary intricacies. Therefore, an essential method of explaining and modeling natural phenomena is treating them as effects of other, more fundamental phenomena. In this view, characteristics of a system are the result of interactions among the elements of this system. This line of reasoning is known as \textit{reductionism}~\cite{WeinbergS-1995a,KaluszynskaE-1998a}. All the phenomena observed in the Universe, regardless of how complicated they are, are a direct effect of interactions whose complete description can be given by a set of ``fundamental'' physical laws. Such an approach, applied to describe various processes in nature, led to a huge number of achievements, both enhancing humanity's understanding of the Universe and allowing for the development of technology solving many practical problems. This is one of the reasons why it has become the dominant paradigm in scientific activity over the last few centuries, in which modern science has developed.

However, for complex systems, the description in purely reductionist manner poses significant difficulties~\cite{DysonF-1996a,LaughlinRB-1999a}. These are the systems in which the relationship between macroscopic and microscopic properties might not be straightforward and direct. A complex system consists of a large number of nonlinearly interacting elements that are able to exhibit a mixture of disorder and collective behaviour, and, by interacting with its surroundings, it is able to modify its internal structure and patterns of activity~\cite{Kwapien2012}. A common trait of the complex systems is emergence: a phenomenon that cannot be reliably deduced or predicted based solely on the knowledge about the properties of the constituents and their interactions. Complexity is often summarized with phrases such as ``more is different''~\cite{Anderson1972} or ``the whole is something beside the parts''~\cite{Aristotle2005}. Emergence occurs when interactions between elements on a microscopic scale give rise to a spontaneous appearance of a macroscopic order. This may happen when the interactions inside a system can be propagated over long distances. In this case, local effects (occurring, for example, due to fluctuations) can be transformed into a collective behaviour, depending on how the system interacts with the surroundings. This is possible under specific circumstances, particularly when the system is near a critical state~\cite{Kwapien2012,Roli2017}. When a system approaches a critical point, one can observe divergence of correlation length (a quantity representing typical, characteristic range of correlations between the states of individual elements interacting within the system). This leads to a situation where the maximal range of correlations is limited only by the size of the system. Fluctuations might then propagate over arbitrarily long distances and a collective behaviour can occur on all possible scales. Keeping a system in the vicinity of a critical state often requires some effort to maintain delicate control over the external parameters characterizing environment (e.g., temperature). However, many complex systems seem to be able to evolve spontaneously towards a critical state. This is why it is frequently stated that complex systems operate ``on the edge between order and chaos''~\cite{Sole1996,Roli2017,Langton1986,Langton1990,Melby2000,Mitchell2009,Baym2006,Landa2011}.

The nontrivial relationship between the global characteristics of a complex system and the properties of its constituents gives rise to the development of analytical tools designed specifically to study systems belonging to this class. A research field concentrated on such systems is sometimes considered a separate scientific discipline -- complexity science. Distinguishing this field is justified by the universality of the observed features. Examples of phenomena related to complexity are: convection~\cite{Getling1998}, phase transitions~\cite{Roli2017,Sole1996,Stanley1987}, formation of landforms and coastlines~\cite{Mandelbrot1967,Werner1999,Wiggs2001,Rak2019}, organization of the Internet~\cite{Park2005,Willinger2002,Barabasi1999,Albert2002}, population dynamics in ecosystems~\cite{Turchin2003,Hofbauer1998}, brain activity~\cite{Bassett2011,Beggs2003,Beggs2004,Chialvo2004,Friedrich1991,Sporns2002,Telesford2011,Bullmore2009,Chialvo2010,Chialvo2018}, speculative bubbles and functioning of financial markets~\cite{Sornette2003,Gerlach2019,Bouchaud2003,Watorek2021,Giardina2003,Drozdz2020,Mantegna2000,Johnson2003,Oswiecimka2005,Kwapien2005}, climate~\cite{Fan2021,Rind1999,Lovejoy2017,Weber2001,Boers2021,Zhou2015}, epidemics~\cite{Wang2015,Rhodes1997,Ferrari2006,Cai2015,Rhodes1996,PastorSatorras2015,Newman2002,PastorSatorras2001}, and organization of social systems~\cite{Palla2007,Guimera2003,Raafat2009,Zhao2011,Drozdz2017,Liljeros2001,Schelling1971}. These and many other phenomena share some aspects of complexity although not all the properties typical for complex systems have to be present in each particular system. Among such properties are power laws, self-organization, criticality, long-range correlations, fractality, multilevel hierarchical structure, and nontrivial organization of a network representation.

To be able to assess the complexity of various systems in one unified manner, it would be beneficial to use some kind of quantity able to measure the degree of complexity of an arbitrary system. There have been multiple attempts to construct such a quantity; each of proposed measures has its own rationale, but also has significant drawbacks, limiting its use. An important concept in this context is algorithmic complexity, introduced independently by Solomonoff, Kolmogorov and Chaitin~\cite{Kolmogorov1963,Kolmogorov1968,Solomonoff1964,Solomonoff1964a,Chaitin1969}. Algorithmic complexity of a string (a sequence of symbols) can be viewed as the length of the description (formulated in some computationally universal language, that is, a language in which any Turing machine can be implemented) of the shortest possible algorithm generating that string. This quantity, although important in the fields of information theory and computability theory, has limited practical use due to its uncomputability for arbitrary sequences~\cite{Vitanyi2020} and due to the fact that it loses its functionality when dealing with random data. The latter stems from the observation that to specify a truly random sequence one needs to give it explicitly -- as there are no regularities to exploit -- and therefore the description of the relevant algorithm has the length comparable to the sequence itself. A quantity related to algorithmic complexity is the so-called effective complexity~\cite{GellMann1996,GellMann2004}. To avoid treating random sequences as complex, it is designed to measure only the complexity of the non-random contribution to a sequence. However, determining the extent to which a string is random involves some degree of arbitrariness~\cite{McAllister2003,Ay2010}.

Another approach to quantifying complexity utilizes the notion of logical depth -- which is also related to algorithmic complexity. Logical depth of a string can be interpreted as the time needed by a universal computer (a device capable of computing what can be computed by a Turing machine) to execute the algorithm which generates the string and has description as short as possible~\cite{Bennett1988,Ay2010}. According to this idea, complex objects (logically deep objects) are the objects which require large computational effort to be generated. It reflects the intuition that complex structures are often created by complicated processes; it also treats random data as relatively ``shallow''.

A concept originating in physics and often employed in quantifying complexity, is information entropy (Shannon entropy~\cite{Shannon1948a,Shannon1948b}). Information entropy is a quantity measuring the level of uncertainty of a random variable. For a discrete random variable which has $n$ possible values $x_i$ occurring with probabilities $p_i$ ($i=1,2,...,n$), information entropy is defined as:
\begin{equation}
H = - \sum_{i=1}^{n} p_i \log_2 p_i.
\label{def_entropy}
\end{equation}
Depending of the choice of logarithm base in the above definition, entropy can be expressed in various units; if the logarithm base is 2, then entropy is given in bits; choosing the base $e$ gives entropy in nats (``natural units''). Information entropy can be considered a generalization of the notion of entropy known from statistical physics. If in Eq.~(\ref{def_entropy}) all probabilities $p_i$ are equal ($p_i=1/n$), then the equation simplifies to  $H = \log_2 n$. If $n$ denotes the number of possible microstates which can yield a given macrostate of some system, then after changing the logarithm base and introducing a multiplicative constant $k_B$ (the Boltzmann constant), the formula can be recognized as Boltzmann's definition of entropy: 
\begin{equation}
H = k_B \ln n.
\label{entropy_Boltzmann}
\end{equation}
For a discrete random variable, the lowest possible value of information entropy is~0; it is attained when the variable has only one possible value (other values either do not exist or have zero probability; for values with zero probability the product $p_i \log_2 p_i$ is assigned the value 0, in accordance with the limit: $\lim_{x \to 0^+} x \log_2 x = 0$). Information entropy of a random variable with a fixed number of possible values is maximized when probability is uniformly distributed over those values. Therefore entropy is interpreted as the degree of uncertainty or randomness inherent in a random variable. It can also be treated as an average amount of information contained in a single measurement of a quantity described by the considered variable. This view can be presented as follows: if some system can be in one of $n$ states, and the probability is distributed approximately uniformly among the states, which means that the entropy of the system is high, then a single measurement revealing the system's state gives much information, because it would be difficult to make a correct assumption about the state before the measurement. Conversely, when the distribution of probability among states is highly nonuniform and the entropy is low, a measurement is not very informative -- as it typically leads to an expected result -- that the system is in one of the states of high probability.

The usefulness of information entropy as a direct measure of complexity is limited due to the fact that it attains the highest values for systems with the highest degree of randomness -- and systems organized in a purely random fashion (random strings of symbols, for instance) cannot be considered complex. However, many methods aiming to quantify complexity employ entropy and related concepts. An example of a quantity rooted in statistical physics and intended to measure complexity is thermodynamic depth~\cite{Lloyd1988}, which can be considered a physical counterpart of logical depth. It is based on the assumption that complex systems are the systems which are difficult to assemble or create. In that view, complexity is measured by the amount of information required to specify the trajectory (a history of system's past states) that the system followed to arrive at its present state. It can be expressed as the entropy of the distribution of trajectories leading to system's current state. Although it is in agreement with the intuitive comprehension of complexity (being a product of a complicated process), thermodynamic depth encounters serious problems with its practical application. One limitation is the fact that the knowledge about the whole history of a system is usually unavailable; another difficulty is arbitrariness involved in determining the trajectory followed by the system~\cite{Crutchfield1999, Mitchell2009}.

Another measure of complexity, designed to study symbolic sequences and objects that can be described by such sequences, based on identifying repeating patterns, is Lempel-Ziv complexity~\cite{Lempel1976}. There exist several definitions of Lempel-Ziv complexity, but all rely on the same idea -- iterative processing of the string and identifying patterns which are copies of patterns encountered at earlier stages. This became a backbone of the Lempel-Ziv algorithm -- a lossless data compression algorithm, existing in multiple variants (LZ77, LZ78, LZW, and others~\cite{Ziv1977,Ziv1978,Welch1984}), and being of huge importance for computer science and practical applications of information theory~\cite{LZ77_milestone}. The key part of the method proposed by Lempel and Ziv can be briefly presented as an appropriate string partition procedure. One of its variants relies on dividing a string $S$ into substrings $S_1, S_2, ..., S_N$, called \textit{phrases}, such that their concatenation is equal to $S$ and that each consecutive phrase $S_i$ is the shortest possible phrase different from each of the phrases $S_1, S_2, ..., S_{i-1}$ (except for the last one, $S_N$, which might not be unique). For example, according to that procedure, the string $AABABBBABAABABBBABBABBA$ is divided into $A|AB|ABB|B|ABA|ABAB|BB|ABBA|BB|A$ (vertical lines separate consecutive phrases $S_1, S_2, ..., S_N$). For a string containing many repeating substrings, the number of unique phrases grows with string's length more slowly than in case of a string in which symbol sequences are rarely repeated. Hence, string complexity can be measured in terms of the number of unique phrases. Using the presented scheme to compress the string relies on the observation that each of the consecutive phrases of length greater than 1 is a copy of some of the previously encountered phrases, concatenated with a single symbol (the symbol determining that the phrase is distinct from each of the previously encountered phrases). Therefore, instead of specifying the phrase explicitly, one can specify how it can be constructed from previous phrases: if $S_j$ is a phrase which can be obtained by appending a single symbol to a phrase $S_i$ already encountered in the string $S$, then $S_j$ can be described by 3 parameters: the position in $S$ at which $S_i$ starts, the length of $S_i$, and the symbol that needs to be appended to $S_i$ to get $S_j$. For a string of sufficient length, containing many repeated substrings, such an approach allows to represent the string in a significantly more compact form.

Measuring complexity using the idea proposed by Lempel and Ziv has certain advantages -- like the fact that Lempel-Ziv complexity can be relatively easily computed for arbitrary strings -- but, in the context of complex systems, it suffers from the same problem as information entropy: it assigns high complexity to random sequences (as they have no systematically repeating patterns) and randomness is different from complexity. In fact, Lempel-Ziv complexity is related to entropy -- for example, procedures of identifying repeating patterns in strings, similar to the one presented above, are used in methods of estimating information entropy of symbolic~sequences~\cite{Kontoyiannis1998,Gao2008}.

Complexity can also be understood in relation to how system's parts interact with each other. In this view, the more relationships and connections are present between the individual elements of the system, the higher the system complexity. The occurrence of such connections can manifest itself by statistical dependence of variables representing the states of system's constituents. There exist a number of tools designed to measure dependence of this kind. A basic example is correlation function -- which expresses how, on average, the values of variables pertaining to individual states of system's constituents are related to each other, depending on the distance (in space or time) between those constituents. Another example is mutual information, which is the difference between the sum of entropies $H(X) + H(Y)$ and the joint entropy $H(X,Y)$ of two variables $X$ and $Y$. If the mutual information $I(X,Y) = H(X) + H(Y) - H(X,Y)$ is greater than zero, it means that observing the value of one variable allows to reduce the ``uncertainty'' of the other; in other words, some information is shared by $X$ and $Y$, and they are mutually dependent. Statistical dependence and correlations, especially those of long-range and nonlinear character, are common traits of complex systems; however, in some cases, their presence might be due to reasons more obvious and straightforward than complexity~\cite{Kwapien2012, Bennett1990}. For example, a multi-component system in which all components are in the same state and evolve in the same way, does exhibit very strong internal correlations, but such a system is not considered complex. Therefore, measuring complexity solely on the basis of correlations' strength is not a method that could be reliably applied to all types of systems.

Finally, complexity can be related to fractals and multifractals. Many systems in nature have a hierarchical, multilevel organization that is precisely self-similar or statistically self-similar~\cite{Mandelbrot1982,Feder1988,Tel1988,Stanley1988}. A structure exhibiting such behaviour is sometimes called \textit{scale-free} due to a fact that it has similar characteristics regardless of the scale at which it is inspected. Certain properties of fractals, like ``rough'', irregular shape, which is difficult to describe with the standard approach based on Euclidean geometry, and recursively nested patterns, can be intuitively interpreted as the signs of complexity. Fractal analysis allows to identify those properties and to characterize them quantitatively. From that perspective, the most complex objects among the fractals are \textit{multifractals}, which can be thought of as the systems consisting of many different convoluted fractals. However, despite the fact that fractality is abundant in nature, it is not necessarily present in all complex systems. There also exist systems, in which fractality, although present, can be difficult to detect. Therefore, a possibility of quantifying complexity by identifying structural fractals or multifractals is restricted only to a certain subclass of complex systems.

The methods of quantifying complexity presented above certainly do not exhaust all the ways, in which complexity can be expressed and measured~\cite{Kinsner2010}. Instead, they show that virtually any kind of approach has its specific limitations, either conceptual or practical. Due to the huge diversity of complex systems, each proposed method is usually designed to deal with a specific class of systems or signals, but it might be insufficient to characterize systems in which complexity is understood differently. Therefore, when studying complex systems, one usually investigates a number of characteristics related to complexity rather than applying one unified approach. It is not necessary that all such characteristics occur in one system, though identifying only some of them is typically sufficient to identify a given system as complex.

\subsection{Language complexity}

Natural language is a clear example of a complex system. Properties of its multilevel structure can be considered as displaying emergence in multiple aspects. Higher levels of its organization usually cannot be reduced to a sum of the elements involved. For example, phonemes and letters basically do not have any meaning, but the words consisting of them refer to specific objects and concepts. Likewise, knowing the meaning of separate words does not necessarily provide the understanding of a sentence composed of them, as this sentence can carry additional information, like an emotional load or a metaphorical message. And the meaning of a sentence can be fully understood when analyzed in an appropriate context constituted by other sentences. The presence of many different types of relationships between various structures in language, each typical for a specific level of language organization, extends to even higher levels: for written language, the examples of such levels are paragraphs, chapters, and whole books. Complexity of language structure is reflected in the number of academic disciplines involved in language research. The lowest levels of language organization are studied by biology and physiology, the higher ones -- by linguistics and its various sub-fields, and the highest ones -- by sociology, psychology, and literary studies.

Language reveals its complexity also when a system of rules and laws governing the relationship between various elements of language is considered. On the one hand, the rules of grammar have to be precise enough to allow one for generating utterances, which can be understood by other language users. On the other hand, there is some degree of freedom in constructing of an utterance: the rules allow one for new forms and can also evolve over time. A large part of linguistic structures cannot be characterized by simple rules that are not subject to exceptions. Therefore, language can be considered as a system that displays a certain form of balance between regularity and irregularity. This is one of the reasons, which render description of natural language difficult and which make studying language from various perspectives particularly valuable.

Another perspective on language complexity is related to how language evolves over time. The conditions that language has to satisfy to remain an effective communication tool and the way, in which it changes indicate that language is subject to self-organization. Language self-organization is usually studied from two perspectives~\cite{Boer2011}. The first one considers language as a system of thought expression that is individual for each human and which optimizes with respect to the ease of its acquisition and use~\cite{Liljencrants1972,Lindbloma,Ke2003}. The other approach treats language as a system of communication between individuals belonging to some population. In this view, self-organization is a process driven by mutual interactions between language users that leads to continual language change and adaptation~\cite{Kvasnicka1999,Berrah1999,STEELS_2000,Schwaemmle2009}. It is important to point out that language evolution is hard to being described quantitatively, because it is itself a complex process, driven by many factors, like the evolution of language users (humans), the influence that language users have on each other, and the interactions they have with the environment.

Among the traits typical for complex systems which are also observed in natural language, one can mention the presence of power laws including Zipf's law. If language sample is treated as a signal, one can usually observe long-range correlations and scale-free fluctuations, which are also described by power laws. This corresponds to a fractal or multifractal structure, which is yet another sign of complexity. By exploiting a network approach, language reveals complicated patterns of organization, some of them being typical for networks representing other complex systems.


\section{Word statistics}
\label{sect::word_statistics}


\subsection{Basic definitions and relations}

\subsubsection{Probability distribution functions}

In many complex systems the distributions of certain quantities describing system's structure or behaviour are given by power laws. This property is exhibited by a great variety of systems, including physical, biological, economic, and social ones~\cite{Newman2005,Markovic2014,Sornette2012}. A quantity $x$ is distributed according to a \textit{power-law distribution} (also called shortly just a \textit{power law}, when the context is clear), if its probability density function (for a continuous variable) or its probability mass function (for a discrete variable) is of the form:
\begin{equation}
p(x) = Cx^{-\beta},
\end{equation} 
where $C$ is a normalization constant and $\beta>0$. It is assumed that $x$ is bounded from below by some positive constant $x_{\rm min}$, being the lowest possible value of $x$, as for $x \to 0$ the function $x^{-\beta}$ diverges. It is common that the distribution of the quantity of interest adheres to a power law only in the tail -- in such case $x_{\rm min}$ is a threshold above which the analysis of power-law behaviour of $x$ is relevant. The power law given above does not necessarily have to be obeyed exactly -- asymptotic agreement is usually considered sufficient. If the support of the distribution is bounded from above, then $\beta$ can be any positive number. However, in many typically encountered situations, the support is right-unbounded ($x$ can take arbitrarily large values), and this is the case considered here. Then $\beta$ has to be greater than 1, to allow for proper normalization. When the support is right-unbounded and $x$ is discrete, that is, when $x \in \left\lbrace x_{\rm min}, \, x_{\rm min} \! + \! 1, \, x_{\rm min} \! + \! 2, ... \right\rbrace $, the normalization is given by:
\begin{equation}
1 = \sum_{k = x_{\rm min}}^{+\infty} \!\! Ck^{-\beta}.
\label{eq_powerlaw_norm_discr}
\end{equation} 
For continuous $x$, the normalization is:
\begin{equation}
1 = \int \limits_{x_{\rm min}}^{+\infty} \!\! Cx^{-\beta}dx = \frac{C}{1-\beta} \Big[ x^{-\beta+1}  \Big]_{x=x_{\rm min}}^{x=+\infty}
\label{eq_powerlaw_norm_cont}
\end{equation}
Both the series in Eq.~(\ref{eq_powerlaw_norm_discr}) and the integral in Eq.~(\ref{eq_powerlaw_norm_cont})` are convergent only when $\beta>1$, hence the restriction of the possible values of $\beta$.

\subsubsection{Survival function}

Since it is often the tail of the distribution that is under consideration when studying power laws, it is convenient to express a power-law distribution in terms of its \textit{complementary cumulative distribution function} $\overline{F}$ (also called \textit{survival function} or \textit{tail distribution}). For a random variable $X$ the survival function $\overline{F}$ can be defined~as:
\begin{equation}
\overline{F}(x) = P(X \geq x),
\label{eq_powerlaw_CCDF_def_1}
\end{equation}
where $P(X \geq x)$ denotes the probability that $X$ takes on a value greater than or equal to $x$. Equivalently, $\overline{F}$ can be expressed as:
\begin{equation}
\overline{F}(x) = 1-F(x),
\end{equation}
where $F$ is the cumulative distribution function. Depending on how exactly $F$ is defined (as right-continuous or as left-continuous), the inequality in Eq.~(\ref{eq_powerlaw_CCDF_def_1}) can be strict or not (this distinction is important only for discrete distributions); here it is assumed that $\overline{F}(x) = P(X \geq x)$. Like cumulative distribution function, survival function fully specifies the studied distribution.

The survival function of a power-law distribution is a power function. For a continuous variable it is given by:
\begin{equation}
\overline{F}(x) = \int \limits_{x}^{+\infty} \!\! Ct^{-\beta}dt = \frac{C}{1-\beta} \; x^{-\beta + 1}.
\end{equation}
For a discrete variable, the survival function is:
\begin{equation}
\overline{F}(x) = \sum_{k = x}^{+\infty} Ck^{-\beta}.
\end{equation} 
Although the sum above does not follow a power law exactly, it can be approximated for large $x$ (using, for example, Euler's summation formula~\cite{Apostol1974,Stutz1968}) by an integral:
\begin{equation}
\sum_{k = x}^{+\infty} Ck^{-\beta} \approx \int \limits_{x}^{+\infty} \!\! Ct^{-\beta}dt.
\label{eq_powerlaw_CCDF_approx}
\end{equation} 
Therefore, it can be stated that for sufficiently large $x$ both continuous and discrete power-law distributions have survival functions behaving like power functions:
\begin{equation}
\overline{F}(x) \propto x^{-\beta+1}.
\end{equation}
One can introduce the notation: $\overline{F}(x) \propto x^{-\alpha}$, where $\alpha = \beta-1$; both $\alpha$ and $\beta$ can be called the \textit{exponents} of a power law, depending on the context. Since power-law-like behaviour of probability density function or probability mass function is closely related to the same type of behaviour of survival function, the identification of a power-law distribution can be performed by observing that any of the mentioned functions is a power function. And due to the fact that certain sums and integrals can be asymptotically approximated by one another (like in Eq.~(\ref{eq_powerlaw_CCDF_approx})), many characteristics of continuous power-law distributions are valid also for their discrete counterparts (calculations for one variant of the distribution might be much more tractable than for the other, however). For that reason, from now on, the presented properties of power-law distributions are given for their continuous variants.

Power laws belong to the class of the so-called \textit{heavy-tailed distributions}. A~distribution with survival function $\overline{F}$ has a (right) heavy tail~\cite{Foss2013}, when for any $\lambda>0$:
\begin{equation}
\limsup_{x\to \infty} \frac{\overline{F}(x)}{e^{-\lambda x}} = \infty,
\end{equation}
that is, a distribution has a heavy tail when for $x \to \infty$ its survival function decays slower than any decreasing exponential function. There exists an important subclass of heavy-tailed distributions -- the so-called \textit{subexponential distributions}; most of typically encountered heavy-tailed distributions belong to the class of subexponential distributions. All power-law distributions are subexponential. A distribution is subexponential~\cite{Foss2013,Embrechts1997} when
\begin{equation}
\lim_{x\to\infty} \frac{\overline{F \! \ast \! F}(x)}{\overline{F}(x)} = 2.
\end{equation}
In the above formula, $F \! \ast \! F$ is the convolution of the cumulative distribution function $F$ with itself, which corresponds to the cumulative distribution function of the sum of two independent random variables distributed according to $F$. Consequently, $\overline{F \! \ast \! F}$ is the survival function of such a sum. Subexponentiality is equivalent to the following property. Let $X_1, X_2, ..., X_n$ be independent, identically distributed random variables with a subexponential distribution. Then
\begin{equation}
\lim_{x\to\infty} \frac{ P\left( X_1 + X_2 + ... + X_n \geq x \right) }{ P \left( \max(X_1, X_2, ..., X_n) \geq x \right) } = 1,
\label{eq_powerlaw_bigjump}
\end{equation}
where $P\left( X_1 + X_2 + ... + X_n \geq x \right)$ denotes the probability that the value of the sum $X_1 + X_2 + ... + X_n$ is greater or equal to $x$ and $P \left( \max(X_1, X_2, ..., X_n) \geq x \right)$ denotes the probability that the largest value among $X_1, X_2, ..., X_n$ is greater or equal to~$x$. Eq.~\ref{eq_powerlaw_bigjump} expresses the fact that for large enough $x$, the sum of values drawn independently from a subexponential distribution exceeds $x$ with practically the same probability as the largest of those values does. In that sense, the behaviour of the sum is to a large extent ``determined'' by the behaviour of the largest value; this is known as the \textit{single big jump principle}, and is a substantial characteristic of processes described by subexponential distributions~\cite{Vezzani2019,Wang2019}.

Power laws are often characterized as distributions which can span over several orders of magnitude, in contrast to, for example, normal distribution or exponential distribution, for which one usually can identify a typical range of values or a characteristic scale. Whether a distribution covers multiple orders of magnitude depends on the distribution's parameters and on the units in which the studied quantity is measured, but power laws found in many systems in nature are indeed associated with quantities considered to have a wide range of possible values. This is often related to the fact that the $m$-th raw moment of a power-law distribution:
\begin{equation}
\left\langle x^{m} \right\rangle = \int \limits_{x_{min}}^{+\infty} \!\! Cx^{-\beta + m}dx  = \frac{C}{m + 1 - \beta} \Big[ x^{-\beta+m+1}  \Big]_{x=x_{min}}^{x=+\infty}
\end{equation}
is finite only for $\beta > m+1$. This implies that the expected value $\left\langle x \right\rangle$ exists only for $\beta > 2$ and the variance $\left\langle x^2 \right\rangle \! - \! \left\langle x \right\rangle^2$ only for $\beta > 3$. For that reason, for distributions with $\beta \leq 3$, which are quite common in nature, there is no finite expected value, or -- if it exists -- the average squared deviation from the expected value is infinite. A related effect, attributed to power laws with appropriately low exponents, is the high degree of ``non-uniformity''. When some quantity, which can be interpreted as a certain kind of resource, is distributed over some population according to such a power law, then a large fraction of the overall amount of the resource is concentrated within a small fraction of the population. In some areas this phenomenon has been called \textit{Pareto principle} or \textit{80-20 rule}; the former name refers to Vilfredo Pareto -- an economist who pioneered in using power laws to represent wealth distribution in society~\cite{Pareto1964}, the latter describes situations where 20\% of some population holds 80\% of some resource~\cite{Hardy2010}. It should be noted, however, that exact numbers expressing that effect may vary; the relationship 80\%-20\% is obtained for a specific value of power law exponent $\beta$. In a continuous distribution with probability density function $p(x)$ of the form
\begin{equation}
p(x) = (\beta-1) x_{\rm min}^{\beta-1} \, x^{-\beta},
\end{equation}
which is known as the Pareto distribution, $\beta$ has to be equal to $1 + \log_4 5 \approx 2.16$ to comply to 80-20 rule precisely.

\subsubsection{Scale invariance}

An important property of power laws is scale invariance. For a power function
\begin{equation}
f(x) = Cx^{-\beta}
\end{equation}
and a positive constant $\lambda$, the following condition is satisfied:
\begin{equation}
f(\lambda x) = C(\lambda x)^{-\beta} = \lambda^{-\beta}f(x),
\end{equation}
which means that scaling the argument of the function  by a constant $\lambda$ results in scaling the value of the function by the constant $\lambda^{-\beta}$. Therefore, a function of that type does not have any characteristic scale -- its properties are qualitatively the same in all possible scales. For that reason, all power functions with a particular exponent are in a sense equivalent, since they differ from each other only by a multiplicative constant. Scale invariance of power-law probability distributions can be interpreted as the presence of a certain kind of hierarchy -- for a power-law distribution with probability density function $p(x)=Cx^{-\beta}$ and any $x_1$, $x_2$ contained in the interval in which the power-law relationship is valid, the densities $p(x_1)$, $p(x_2)$ are bound by:
\begin{equation}
\frac{p(x_2)}{p(x_1)} = \left( \frac{x_2}{x_1} \right) ^{-\beta} \!.
\end{equation}

Identifying power-law distributions in empirical data usually employs the fact that a relationship described by a power function $f(x) = Cx^{-\beta}$ can be transformed into a linear relationship, by taking the logarithm of both sides:
\begin{equation}
\log (f(x)) = \log C -\beta \log x.
\end{equation}
Therefore, when $f(x)$ is presented on a log-log plot (which might be a plot in log-log scale or a graph of $\log(f(x))$ vs. $\log x$), observing a linear relationship allows to conclude that $f$ is a power function of $x$. The exponent can be determined from the slope of the line.

To investigate if a sample comes from a power-law distribution, one can compute the empirical survival function $\widetilde{\overline{F}}$, defined as:
\begin{equation}
\widetilde{\overline{F}}(x) = \frac{N_{[x;\infty)}}{N},
\end{equation}
where $N_{[x;\infty)}$ is the number of observations greater or equal to $x$ in the sample, and $N$ is the total number of observations; $\widetilde{\overline{F}}$ is a step function, with steps at points corresponding to unique values in the sample, therefore it is usually computed only for $x$ being such values. After $\widetilde{\overline{F}}$ is determined, one can plot the set of points $(\log x, \log \widetilde{\overline{F}}(x))$. If the points lie on a straight line for some range of $x$, then within that range $\widetilde{\overline{F}}$ is a power function of $x$, and since $\widetilde{\overline{F}}$ approximates the survival function of the underlying distribution, the distribution can be recognized as a power law. The exponent $\alpha$, describing the behaviour of the survival function ($\overline{F} \propto x^{-\alpha}$), which is usually of primary interest, can be obtained by determining the slope of the observed line. The slope is very often computed by using the least squares method to fit a linear relationship to  $\widetilde{\overline{F}}(x)$. The advantage of this approach is simplicity; maximum likelihood estimation of $\alpha$ is superior in terms of accuracy and error estimation, but requires choosing carefully the range in which the power-law relationship holds and, depending on the type of the data, it might involve solving a transcendental~equation~\cite{Clauset2009}.

\subsubsection{Rank-size distributions}

A tool similar to survival function plots and often used in identifying power-law distributions in empirical data, is the so-called rank-size distribution. If some collection of values is presented in the form of ranking -- that is, a list in which the first element is the largest observation, the second element is the second largest observation, and so on, then the rank-size distribution is the function which relates the value with its position in the ranking. If this function is a power function, the underlying probability distribution is a power-law distribution. This can be understood with the following line of reasoning. Let the values in a $N$-element sample drawn from some distribution be sorted into a non-increasing sequence $(x_1,x_2,...,x_N)$. The rank $R(x_k)$ of an observation $x_k$ ($ k=1,2,...,N$) can be defined in a few ways, two possibilities are considered here:
\begin{eqnarray}
(1)& R(x_k) & \!\!\! = k {\label{eq_powerlaws_rank_def_usual}}\\
(2)& R(x_k) & \!\!\! = \max\{j : x_j = x_k\} {\label{eq_powerlaws_rank_def_invertible}}
\end{eqnarray}
In the first variant, the rank $R$ of the observation $x_k$ is the position of $x_k$ in the ranking; in the second one, $R$ is the number of observations greater or equal to $x_k$ in the sample. They differ only when the values in the sample might repeat (which often happens for data coming from a discrete distribution); however, even when they do, the differences typically appear for large $R$, and it is the range of small $R$ that is usually of interest, as it corresponds to the tail of the probability distribution. Therefore the distinction between definitions in Eq.~(\ref{eq_powerlaws_rank_def_usual}) and Eq.~(\ref{eq_powerlaws_rank_def_invertible}) is rarely significant in practical calculations, but while the first is often used in the literature, the second one is more suited to the derivation of the formulas given here. With rank defined as in Eq.~(\ref{eq_powerlaws_rank_def_invertible}), one can relate each unique value $x$ in the sample with its rank $R$; the function $x(R)$ is called the rank-size distribution (sometimes also the rank-frequency distribution, if the data represents the frequencies or counts). Stating that $x$ has rank $R$ is equivalent to stating that exactly $R$ observations in the sample are larger or equal to $x$. It means that $R/N$ is an estimate of $P(X \geq x)$ -- the probability that a random variable $X$ with the considered probability distribution takes on a value greater or equal to $x$. That is, $R/N$ is equal to the value of the empirical survival function at point~$x$:
\begin{equation}
\frac{R}{N} = \widetilde{\overline{F}}(x).
\label{eq_powerlaws_rank_and_survival}
\end{equation}
Therefore, rank-size distribution contains information sufficient to fully characterize a sample, same as empirical survival function. If the rank-size distribution $x(R)$ is a power law:
\begin{equation}
x \propto R^{-\gamma} \text{ \quad for some } \gamma > 0,
\end{equation}
then by raising both sides of that relationship to the power $-1/\gamma$, the inverse function $R(x)$ is obtained:
\begin{equation}
R \propto x^{-1/\gamma}.
\end{equation}
Since $N$ is a constant for a given sample, $R/N$ behaves in the same way as $R$ with respect to $x$, that is: $R/N \propto x^{-1/\gamma}$; using Eq.~(\ref{eq_powerlaws_rank_and_survival}), one gets:
\begin{equation}
\widetilde{\overline{F}}(x) \propto x^{-1/\gamma}.
\end{equation}
This shows that a power law in the rank-size distribution $x(R) \propto R^{-\gamma}$ corresponds to a power-law form of the empirical survival function $\widetilde{\overline{F}}(x) \propto x^{-\alpha}$, and the exponents $\alpha$ and $\gamma$ are related by:
\begin{equation}
\alpha = \frac{1}{\gamma}.
\label{eq_powerlaws_rank_size_and_survival_exponents}
\end{equation}
Therefore, a power law in the rank-size distribution of some sample indicates that the sample is drawn from a power-law probability distribution.

It is important to note that the above-presented methods of identifying power laws based on studying the behaviour of survival function or of rank-size distribution pertain to situations where the probability distribution function (or probability mass function) is a power function with the exponent $\beta$ greater than 1. The formulas describing the functional form of the survival function and of the rank-size distribution cease to be valid for $\beta \leq 1$. For distributions with $\beta$ below or close to~1, other methods of detecting the power-law relationship should be used; it should be pointed out that such distributions have to have a bounded support or a cutoff at some point to be normalizable. For example, one can use a histogram as a piecewise constant approximation of probability density function or of probability mass function, and study its behaviour in log-log scale. Since power-law distributions, especially those with low exponents, typically span a wide range of values, it is often beneficial to construct the histograms of power laws with the use of bins of varying length.


\subsection{Zipf's law}

A fundamental statistical property of natural language, first observed by J.B.~Estoup~\cite{MANNING_1999}, later systematically studied and popularized by G.K.~Zipf~\cite{ZIPF_1936,ZIPF_1949,PIANTADOSI_2014} -- and therefore called Zipf's law -- is the power-law distribution of word frequencies in texts, or more generally, in linguistic corpora (a corpus is a text or a set of texts put together one after another). Assuming that a ``word'' is a sequence of letters between whitespace characters, Zipf's law can be summarized by a statement that the frequency (the number of occurrences) of a word in a text is inversely proportional to the rank of that word, where the rank is the position on the list of all different words appearing in the text, sorted by decreasing frequency. More precisely, if for some text $R$ denotes the rank of a word, and $\omega$ is the number of times that word appears in the text, then
\begin{equation}
\omega \propto R^{\,-\alpha},
\label{eq::Zipfs_law_statement}
\end{equation}
with $\alpha \approx 1$. Mathematically, Eq.~(\ref{eq::Zipfs_law_statement}) expresses a rank-size distribution; since the ``size'' here pertains to frequency, it is also called a rank-frequency distribution. The corresponding probability distribution can be characterized as follows. If $V$ is the vocabulary of a text, that is, the set of all distinct words appearing in the text, then the probability $p_{\omega}(\omega)$ that a word randomly chosen from $V$ has the frequency $\omega$ in the text is expressed by
\begin{equation}
p_\omega(\omega) \propto \omega^{-\beta},
\label{eq::Zipfs_inverse_law}
\end{equation}
where $\beta = 1/\alpha+1 \approx 2$. Both Eq.~(\ref{eq::Zipfs_law_statement}) and Eq.~(\ref{eq::Zipfs_inverse_law}) are referred to as Zipf's law; the latter is sometimes called the \textit{inverse Zipf's law}~\cite{CANCHO2002}.

\begin{figure}
\centering
\begin{minipage}{\figurecustomwidth}
\centering
\subfloat[English]{\includegraphics[width=0.48\textwidth]{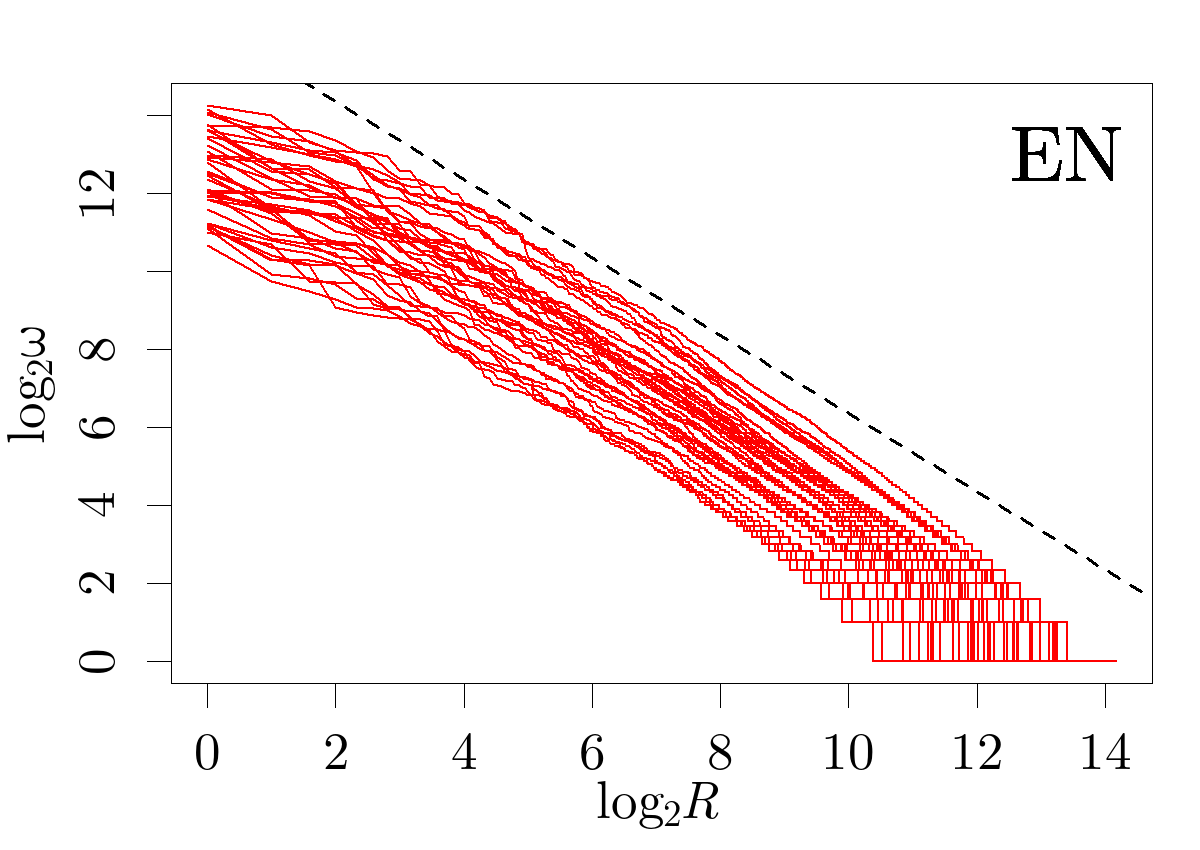}}
\hfill
\subfloat[German]{\includegraphics[width=0.48\textwidth]{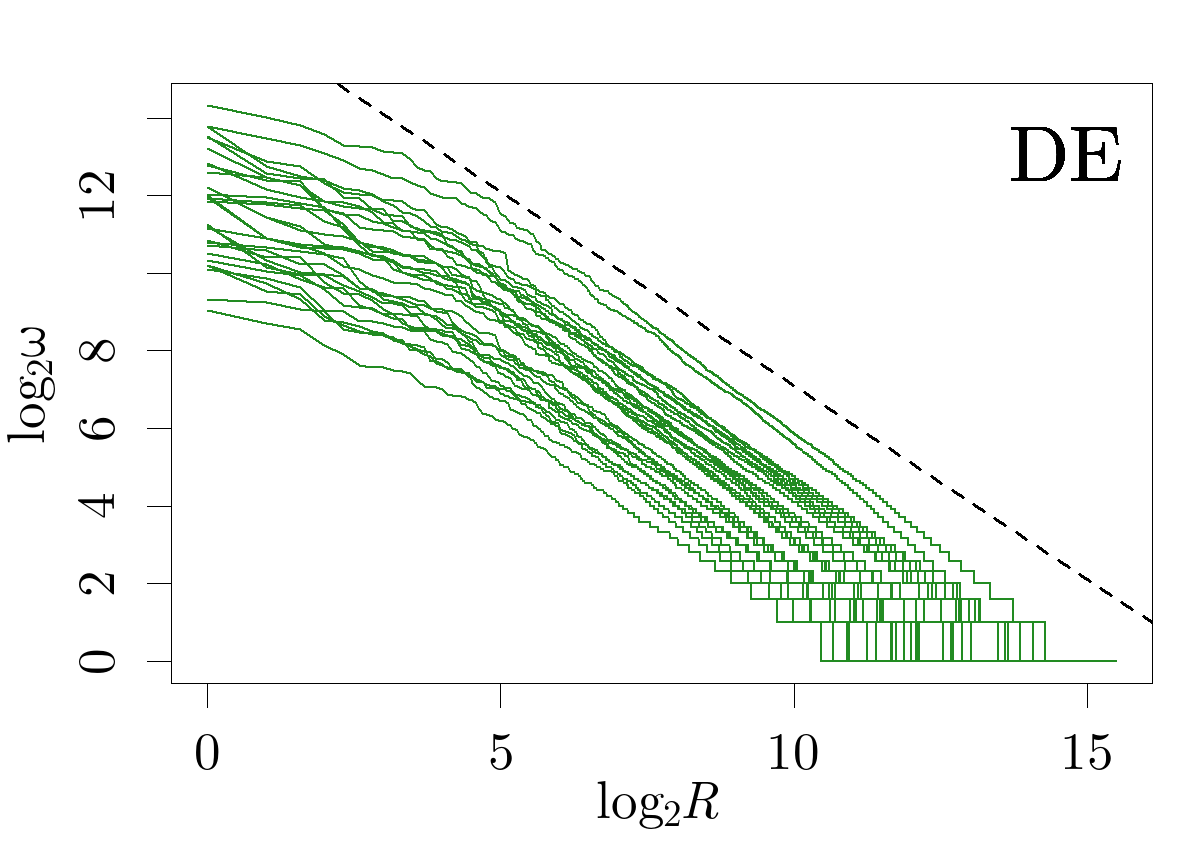}}
\hfill
\subfloat[French]{\includegraphics[width=0.48\textwidth]{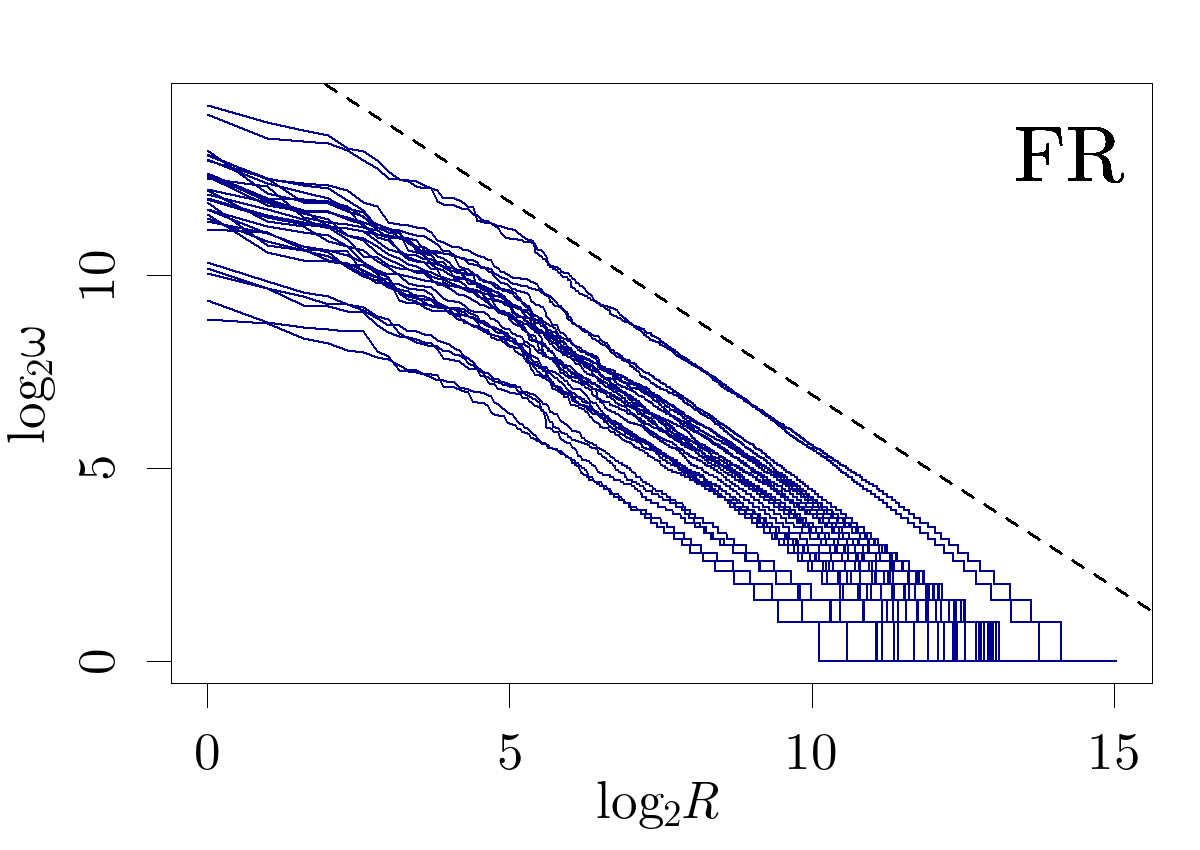}}
\hfill
\subfloat[Italian]{\includegraphics[width=0.48\textwidth]{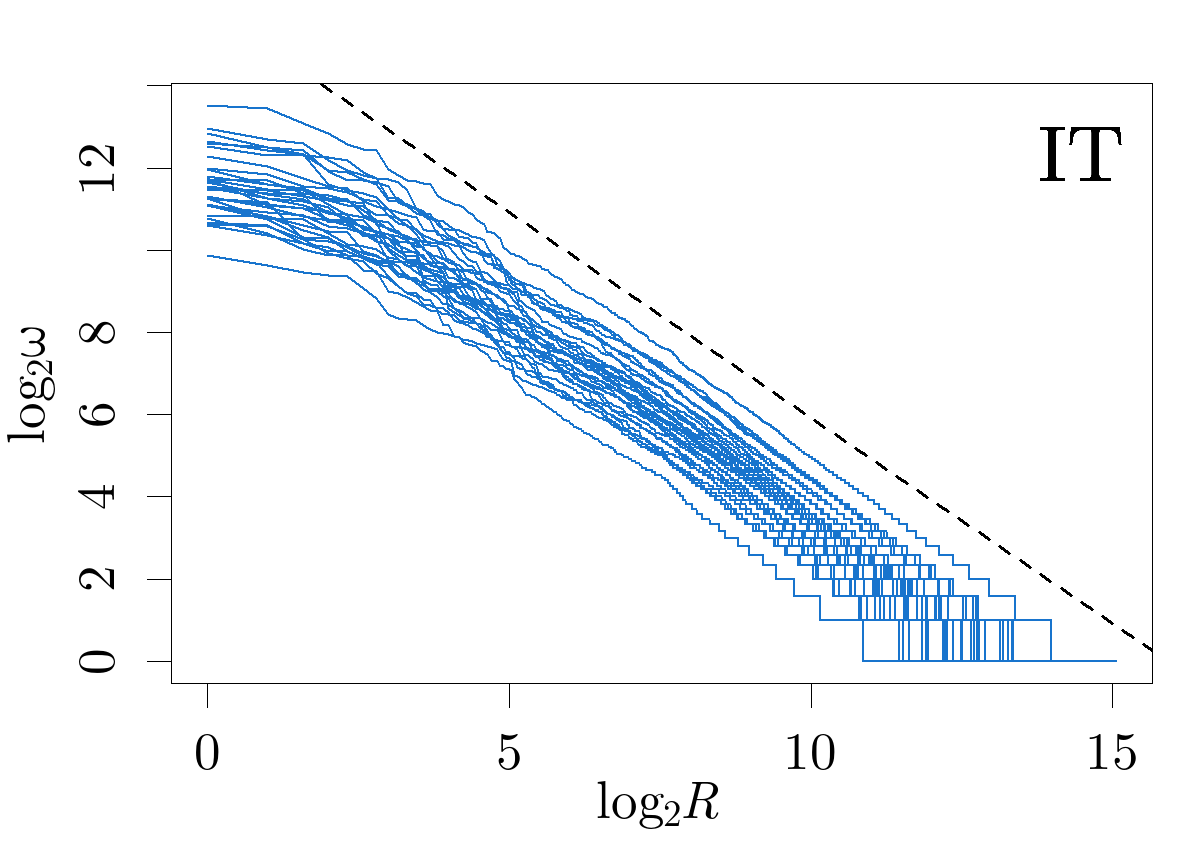}}
\hfill
\subfloat[Spanish]{\includegraphics[width=0.48\textwidth]{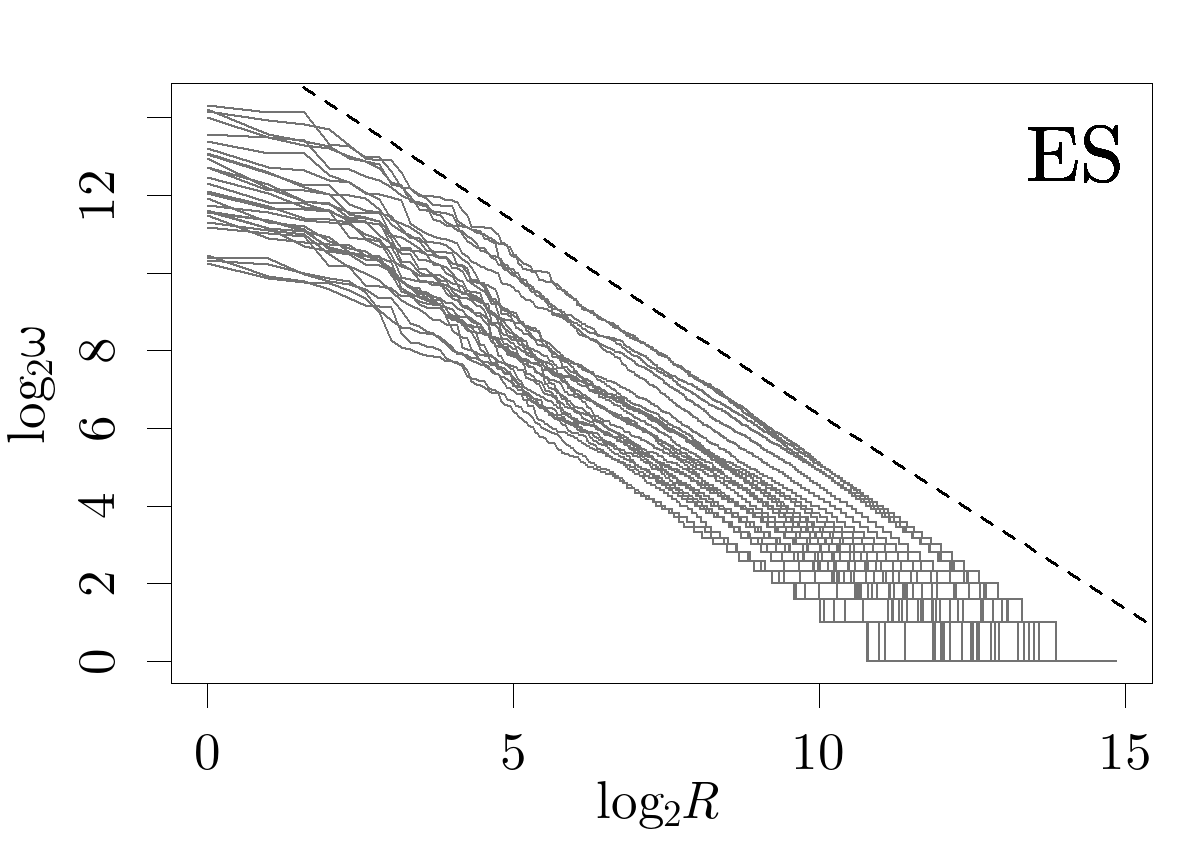}}
\hfill
\subfloat[Polish]{\includegraphics[width=0.48\textwidth]{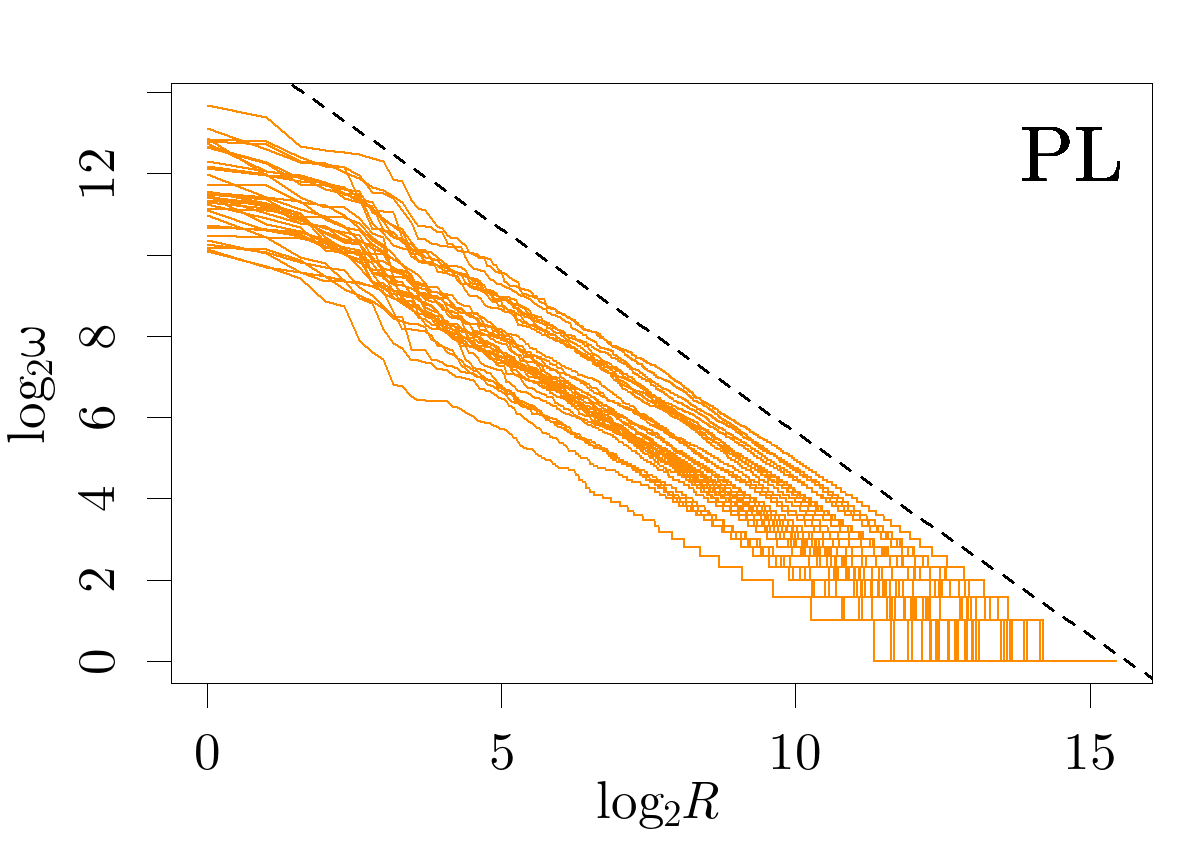}}
\hfill
\subfloat[Russian]{\includegraphics[width=0.48\textwidth]{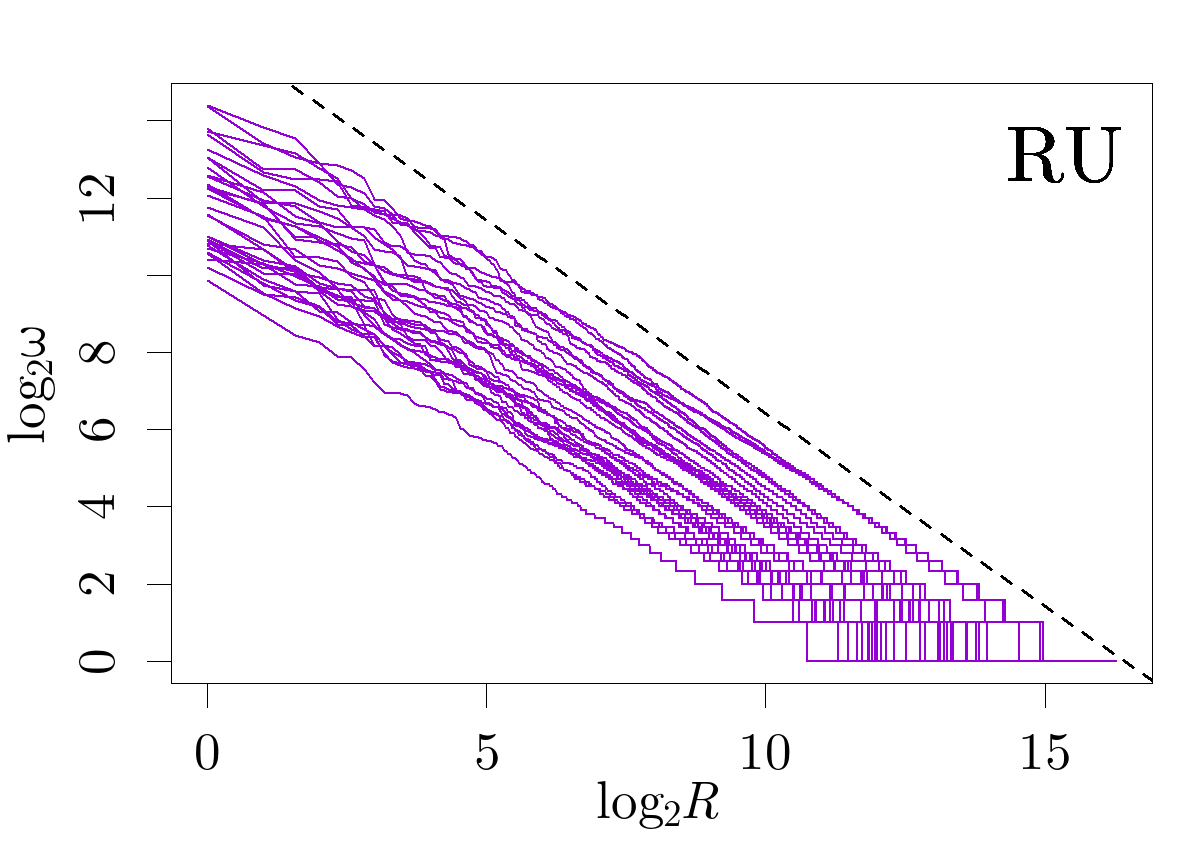}}
\hfill
\subfloat[Corpora]{\includegraphics[width=0.48\textwidth]{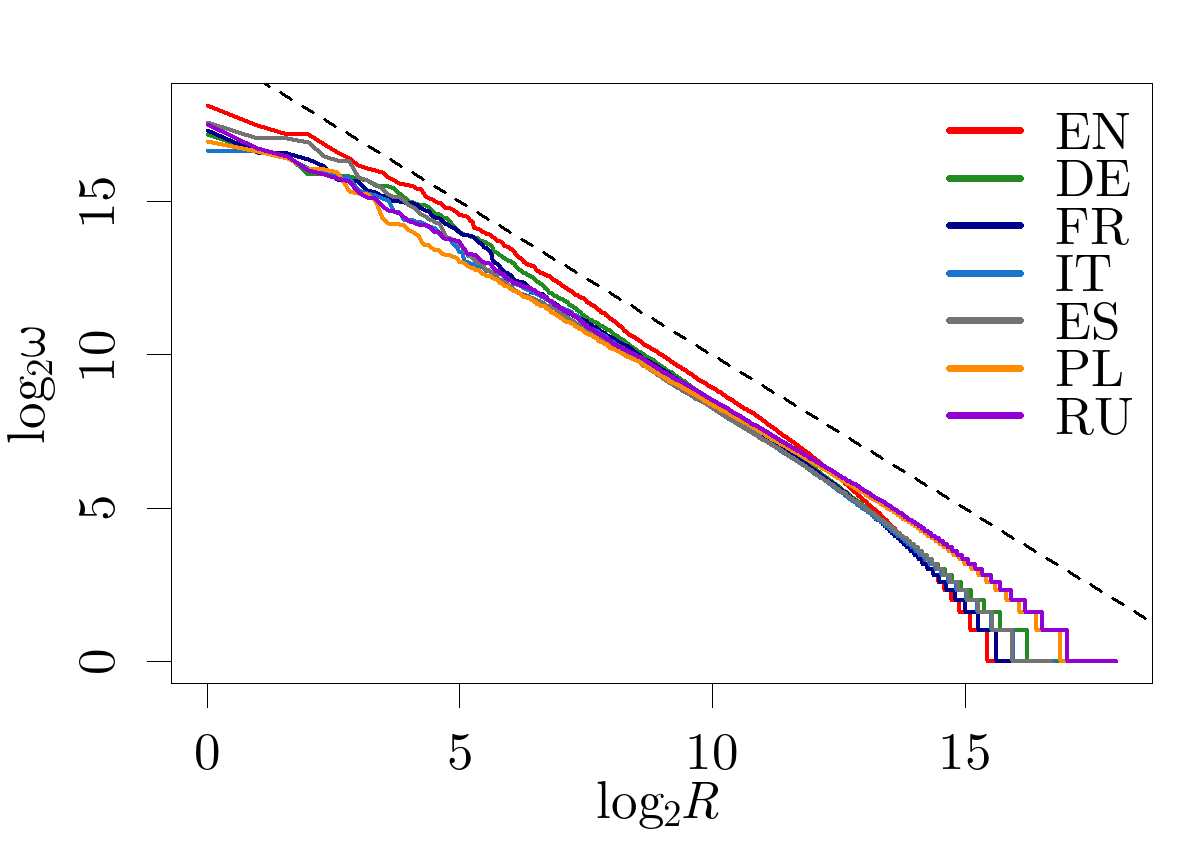}}
\hfill
\caption{(a)-(g) Zipf's plots for books written in various European languages. Each line represents a log-log plot of the rank-frequency distribution (continuous to make graphs more legible) for a single book. Words were not lemmatized. (h) Zipf's law for corpora constructed from a set of books in each language. In all panels the dashed line corresponds to a slope index equal to $-1$.}
\label{fig::Zipfs_law_in_various_languages}
\end{minipage}
\end{figure}

The Zipf's law is observed in the majority of languages studied with regard to this aspect, including artificial languages~\cite{MANARIS_2006}, and extinct languages~\cite{SMITH_2007}. Exceptions are languages using logographic writing systems (such as Chinese), but it has been shown that although Zipf's law might not hold for logograms, it might be exhibited in some other way, for example by combinations of logograms~\cite{CLARK_1990,SHTRIKMAN_1994}. An illustration of Zipf's law in 7 different languages is presented in Fig.~\ref{fig::Zipfs_law_in_various_languages}, where rank-frequency distributions of sample books and of corpora constructed from those books are shown.

Although the values of the exponents $\alpha$ and $\beta$ are to a large degree universal, it is possible to observe deviations from Zipf's law with $\alpha \approx 1$ and $\beta \approx 2$~\cite{Cancho2005a}. $\beta$ greater than 2 is a sign that a text contains lots of rare words (words with low frequencies); this may be a result of covering a wide range of topics (each with specific vocabulary), or a consequence of the richness of vocabulary of particular author. Conversely, $\beta<2$ indicates poor vocabulary, which may be specific to particular language user (for example $\beta<2$ is observed in schizophrenia and in language used by very young children) or due to specific circumstances (for example, in military communication, where non-essential words tend to be avoided). It is also worth mentioning that even when Zipf's law holds in its basic form (with $\beta \approx 2$), linguistic analysis relying on Zipf's law may reveal properties specific to individual texts. For example, one can define a distance between Zipf plots of two texts (based on the ordering of the words in rank-frequency relationship), and it seems that such a distance is lower for pairs of texts being similar in some sense (belonging to the same author or genre, for instance) than for pairs of ``unrelated'' texts~\cite{Havlin1995}. Another interesting observation derived from word frequency analysis pertains to how statistical regularities like Zipf's law are manifested in different parts of a text. For example, it has been observed that if a text is cut into two halves, there are statistically significant differences  between some statistical properties of the first and of the second half~\cite{Deng2021}.

Unlike words, other statistics related to written language do not show such a Zipfian behaviour. For example, sentence lengths studied in~\cite{AusloosM-2008a} showed a limited power-law dependence for the lowest 100-200 ranks only, while the remaining parts of the plots were decaying much faster. That study was based on only two books, but nevertheless this is a typical result for the sentence data.


\subsection{Heaps' law}

\begin{figure}
\centering
\begin{minipage}{\figurecustomwidth}
\centering
\includegraphics[width=0.8\textwidth]{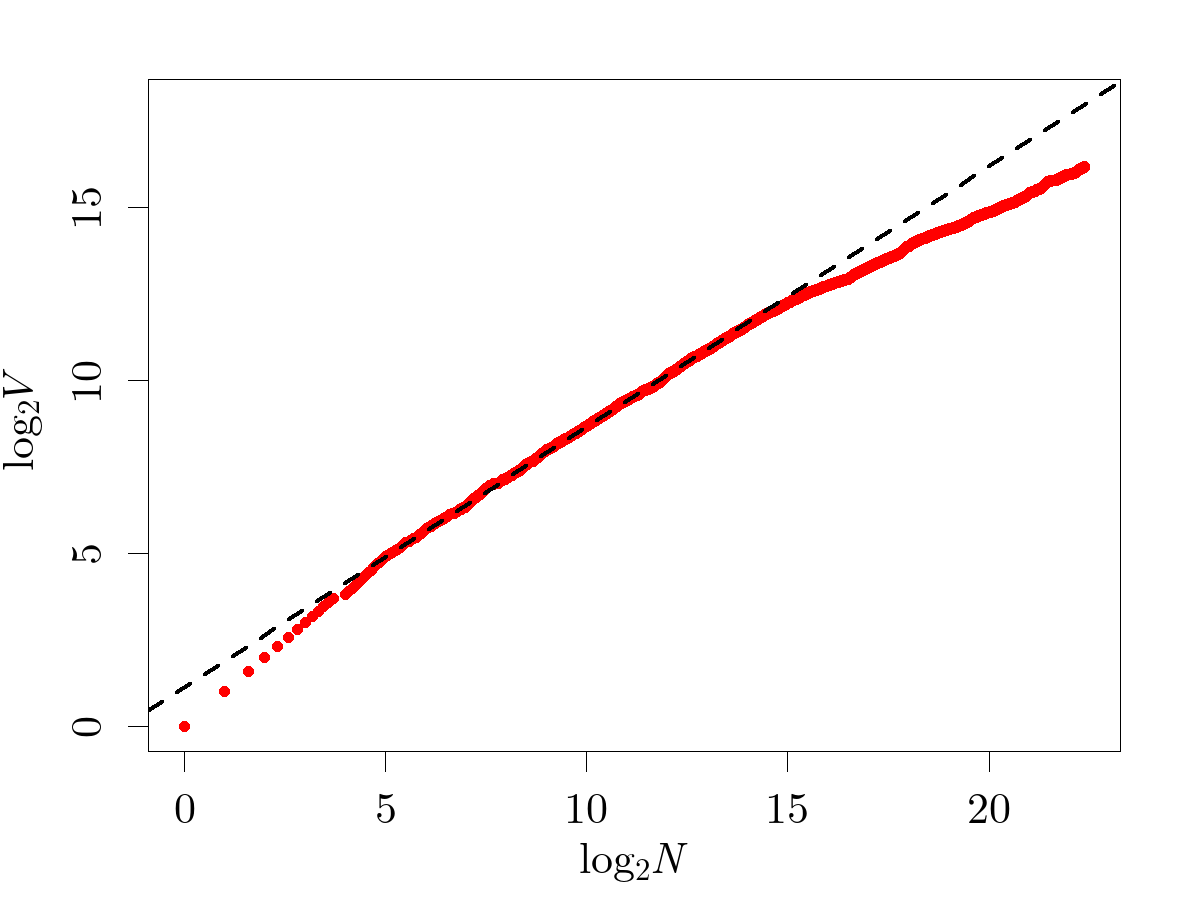}
\caption{An illustration of the Heaps' law created by using a corpus constructed from sample English books. The dots represent $V(N)$ -- the size of vocabulary as a function of text length. Slope of the dashed line is equal to $\eta = 0.75$. A power-law regime holds for a few orders of magnitude. For small $N$, the relationship $V(N)$ is practically linear as almost every consecutive word in the text expands vocabulary. For large $N$, however, the lack of new yet unencountered words makes $V(N)$ grow more slowly.}
\label{fig::Heaps_law}
\end{minipage}
\end{figure}

Another linguistic law involving power-law relationships is Heaps' law (also called Herdan's law)~\cite{Heaps1978,Egghe2007,Chacoma2020}. It describes how the number of distinct words increases with the increasing size of a text. If $N$ denotes the number of all words encountered up to some point in the text and $V(N)$ is the number of distinct words (the vocabulary size) up to that point, then Heaps' law can be formulated as:
\begin{equation}
V(N) \approx C_H N^\eta,
\label{eq::Heaps_law}
\end{equation}
where $\eta$ is a real number between 0 and 1, and $C_H > 0$ is a constant with respect to~$N$; it might depend on language and on the specific text. The relationship between $N$ and $V(N)$ given by Eq.~(\ref{eq::Heaps_law}) typically holds for a few orders of magnitude of $N$; for very long texts ($N \to \infty$), the increase of $V(N)$ becomes slower and slower, as there are less and less commonly used words in the set of words yet unused. For some time, Heaps' law was treated as a trait of language separate from Zipf's law, but it has been shown that it can be considered as related to Zipf's law. That is, it is possible to show that assuming that language is subject to Zipf's law leads to Heaps' law, under some (mild) additional assumptions~\cite{Lue2010,VANLEIJENHORST2005,Kornai1999}. Heaps' law is illustrated in Fig.~\ref{fig::Heaps_law}; the figure presents the log-log plots of $V(N)$ for the corpus constructed from sample English books.

\subsection{Possible explanations of the Zipf's law origin}

There are multiple mechanisms which might serve as explanations of the presence of power laws in various systems. This applies also to natural language. There have been many attempts to explain the origin of Zipf's law, some of them contradictory to each other, but no universally agreed theory has been proposed. Zipf's original explanation was the \textit{principle of least effort}. According to this principle, the language optimizes the information transfer between the speaker (information source) and the listener (information receiver). The messages need to be as short as possible, but at the same time they have to contain enough information to be understandable. The idea of the principle can be roughly presented with the following line of reasoning~\cite{Manin2009}. Let words be sequences of symbols, taken from an $n$-element alphabet. The cost of using a word is equal to its length (the number of symbols it contains). Therefore the most frequently used words are the shortest ones. Assuming that all possible sequences of symbols are used to form words, it can be stated that the cost of the word of rank $R$ can be approximated as $c_R \approx \log_{n}R$. This can be understood as follows: there are $n^l$ words of length $l$, so for an arbitrary word of length $l$ there is $m = \sum_{j=1}^{l} n^j$ words with lengths less or equal to $l$ and therefore occurring with frequencies higher or equal to the frequency of the considered word. Hence, $m$ can be interpreted as the rank of a word consisting of $l$ symbols, $m=R$. Then to approximate $R$ one can notice that
\begin{equation}
R = m = \sum_{j=1}^{l} n^j = n^l \sum_{j=0}^{l-1} \left(\frac{1}{n}\right)^j.
\end{equation}
Since the sum on the right-hand side of the above equation is always smaller than the sum of the geometric series $\sum_{j=0}^{\infty} (1/n)^j = n/(n-1)$, $R$ satisfies
\begin{equation}
n^l < R < n^l \frac{n}{n-1}.
\end{equation}
The larger the $n$, the better the approximation of $R$ by $R \approx n^l$. From that approximation one gets $l \approx \log_{n}R$. The cost of using a word is expressed by the length of that word, so the cost $c_R$ of using the word of rank $R$ can be expressed as: $c_R = l \approx \log_{n}R$. If $p_R$ is the normalized frequency of the word with rank $R$ (in other words, it is the probability that a word randomly chosen from the text is the word with rank~$R$), then the average cost per word $\left< c \right>$ is:
\begin{equation}
\left< c \right> = \sum_{R=1}^{\max\lbrace R \rbrace} \!\! p_R c_R.
\label{eq_least_effort_avg_cost}
\end{equation}
The average amount of information per word can be expressed by information entropy $H$ (here $\log(\boldsymbol{\cdot})$ is the natural logarithm and the entropy is given in nats):
\begin{equation}
H = - \!\! \sum_{R=1}^{\max\lbrace R \rbrace} \!\! p_R\log p_R.
\label{eq_least_effort_avg_information}
\end{equation}
According to the principle of least effort, word frequency distribution in language is such that the transmission of information is cost-efficient, that is, it minimizes the quantity $\left< c \right> \!/H$. The set of numbers $p_R$, which constitute the rank-frequency distribution, can be found by minimizing $\left< c \right> \! /H$ with the normalization constraint imposed on $p_R$: $\sum_{R=1}^{\max\lbrace R \rbrace} \! p_R = 1$. This can be done by Lagrange multipliers method -- treating $\left< c \right>$ and $H$ as functions of $p_R$ ($R=1,2,,..., \max\lbrace R \rbrace$) specified by Eq.~(\ref{eq_least_effort_avg_cost}) and Eq.~(\ref{eq_least_effort_avg_information}), one minimizes $\left< c \right>\!/H$ by solving for each $p_R$:
\begin{equation}
\frac{\partial}{\partial \, p_R} \left( \frac{\left< c \right>}{H} - \lambda \! \sum\limits_{r=1}^{\max\lbrace R \rbrace} \!\! p_r \right) = 0,
\end{equation}
where $\lambda$ is the Lagrange multiplier. Calculating the derivative transforms the above equation into:
\begin{equation}
\frac{c_R}{H} + \frac{\left< c \right> \! \left( \log p_R + 1 \right)}{H^2} - \lambda = 0 \quad \text{ for each } R,
\end{equation}
from which one gets
\begin{equation}
p_R \; = \; \exp \left(\frac{\lambda H^2}{\left< c \right>}-1 \right) \exp \left( - \frac{c_R H}{\left< c \right>} \right) \; = \; A_\lambda R^{ \, -H / \left( \left< c \right> \log n \right) },
\end{equation}
where $\exp(\boldsymbol{\cdot})$ is the exponential function, $A_\lambda = \exp \left( \lambda H^2/ \! \left< c \right> - 1 \right)$ serves as a normalization constant (which can be set by setting $\lambda$ appropriately), and the last equality follows from expressing the word usage cost $c_R$ in the form $c_R = \log_n R$. The above formula implies that minimizing $\left< c \right> \!/H$ leads to a power-law rank-frequency distribution $p_R \propto R^{-\alpha}$, with exponent $\alpha = H / \! \left( \left< c \right> \log n \right)$ (it is worth noting that this result does not give $p_R$ explicitly; to obtain a closed-form solution one needs to explicitly determine $\left< c \right>$ and $H$).

A model of text generation able to generate power-law rank-frequency distributions, based on a line of reasoning different than the one presented above, is the so-called model of \textit{intermittent silence} (also referred to as \textit{typewriting monkey}~\cite{Newman2005}), introduced by Miller~\cite{Miller1957}. It can be shown that under some general assumptions the basic idea of that model is in fact mathematically equivalent to the idea of the least effort principle~\cite{Simkin2010}, but the intuition behind it is slightly different. Let a text be generated by adding one symbol at a time, each symbol being either a letter from an $n$-element alphabet or a space. The symbol to append at each step is chosen randomly, the space is chosen with probability $p_s$; if the chosen symbol is not the space, then it is a letter picked randomly from an uniform distribution, so the probability of each letter is equal to $(1-p_s)/n$. The choices of symbols are independent of each other. To generate a particular word, a specific sequence of symbols followed by a space must occur. The probability $p_l$ of generating a specific word of length $l$ is therefore given by:
\begin{equation}
p_l = \left( \frac{1-p_s}{n} \right)^l \! p_s = p_s \exp \left( l \, \log \frac{1-p_s}{n} \right)
\end{equation}
where $\exp(\boldsymbol{\cdot})$ and $\log(\boldsymbol{\cdot})$ denote the exponential function and natural logarithm, respectively. By using the same approximate relation between the length $l$ and the rank $R$ of a word as before: $l \approx \log_n R = \log R / \log n$, one obtains the normalized frequency $p_R$ of the word with rank $R$ in the form:
\begin{equation}
p_R = p_l = p_s \exp \left( \frac{\log R}{\log n} \, \log \frac{1-p_s}{n} \right) = p_s R^{\,-\alpha} \quad \text{with } \alpha = 1 - \frac{\log \left( 1-p_s \right) }{\log n},
\label{eq::intermittent_silence_rank_frequency}
\end{equation}
where $\exp(\boldsymbol{\cdot})$ and $\log(\boldsymbol{\cdot})$ denote the exponential function and natural logarithm, respectively. Setting $n=26$ and $p_s = 0.18$, which are values used originally by Miller, taken from English language, one gets $\alpha \approx 1.06$, which is close to the exponent of the Zipf's law (Eq.~(\ref{eq::Zipfs_law_statement})).

Other mechanisms generating power laws, like Yule (also called Yule-Simon or preferential attachment) processes and their modified variants, have also been used as possible explanations of Zipf's law. Originally conceived for biological systems to model the emergence of power-law distributions describing the number of species in genera, or more generally, the number of subtaxa in taxa, they have found application in many other areas~\cite{Willis1922,Yule1925,Simon1955,Price1976,Newman2005}. A Yule process models the behaviour of a system composed of a collection of objects which have a certain positive quantity assigned to them, when both the number of objects and the total sum of the studied quantity in the system grow in consecutive time steps in a specific way. For illustrative purposes it is convenient to imagine the considered system as a collection of boxes, with balls inside them. Then one of the forms of the Yule process can be described as follows.

At each point in time, the system consists of a certain number of boxes: the $i$-th box has $k_i$ balls in it. A single time step of the process starts from adding $m>0$ new balls to the system and distributing them among boxes in the following way: $m$ boxes are chosen randomly from the system and one ball is added to each of them. The probability $P_i$ of choosing a particular box is an increasing linear function of the number $k_i$ of balls already present in that box:
\begin{equation}
P_i \propto (k_i + c),
\label{eq::powerlaws_Yule_pref_att_rule}
\end{equation}
where $c$ is a real constant. After inserting balls into the boxes, one new box with $K_0 \geq 0$ balls inside is added to the system. As a consequence, the number of boxes present in the system increases by 1. The time step ends here and, in the next step, the presented procedure is repeated. The constants $K_0$ and $c$ have to satisfy the condition $K_0 + c >0$, which ensures that $k_i + c$ is positive for any possible $k_i$, because $k_i \geq K_0$ for all $i$.

One of the roles of the constant $c$ in Eq.~(\ref{eq::powerlaws_Yule_pref_att_rule}) is to allow the boxes added to the system to participate in the process of distributing new balls among boxes when $K_0 = 0$. More generally, it allows to make the relationship between $P_i$ and $k_i$ more flexible. If $c=0$, then the probability of choosing a box is just directly proportional to the number of balls in that box: $P_i \propto k_i$.

Having defined what happens at each time step, the only thing that remains to be specified is the initial state of the system (i.e., the number of boxes and the number of balls in each of them at the beginning of the process). The influence of the initial state on the characteristics of the process becomes negligible in the limit of large number of time steps. Therefore, the initial state is to some degree arbitrary. Here it is assumed that in the initial state there are at least $m$ boxes and each of them has at least $K_0$ balls inside.

Based on the characterization given above, it can be concluded that a Yule process in the presented form is controlled by 3 parameters: a positive integer $m$, a non-negative integer $K_0$, and a real number $c$; the parameters $c$ and $K_0$ must satisfy: $c + K_0 > 0$. Power laws are generated in Yule processes in the limit of large number of steps.

\begin{figure}
\centering
\begin{minipage}{\figurecustomwidth}
\centering
\includegraphics[width=0.8\textwidth]{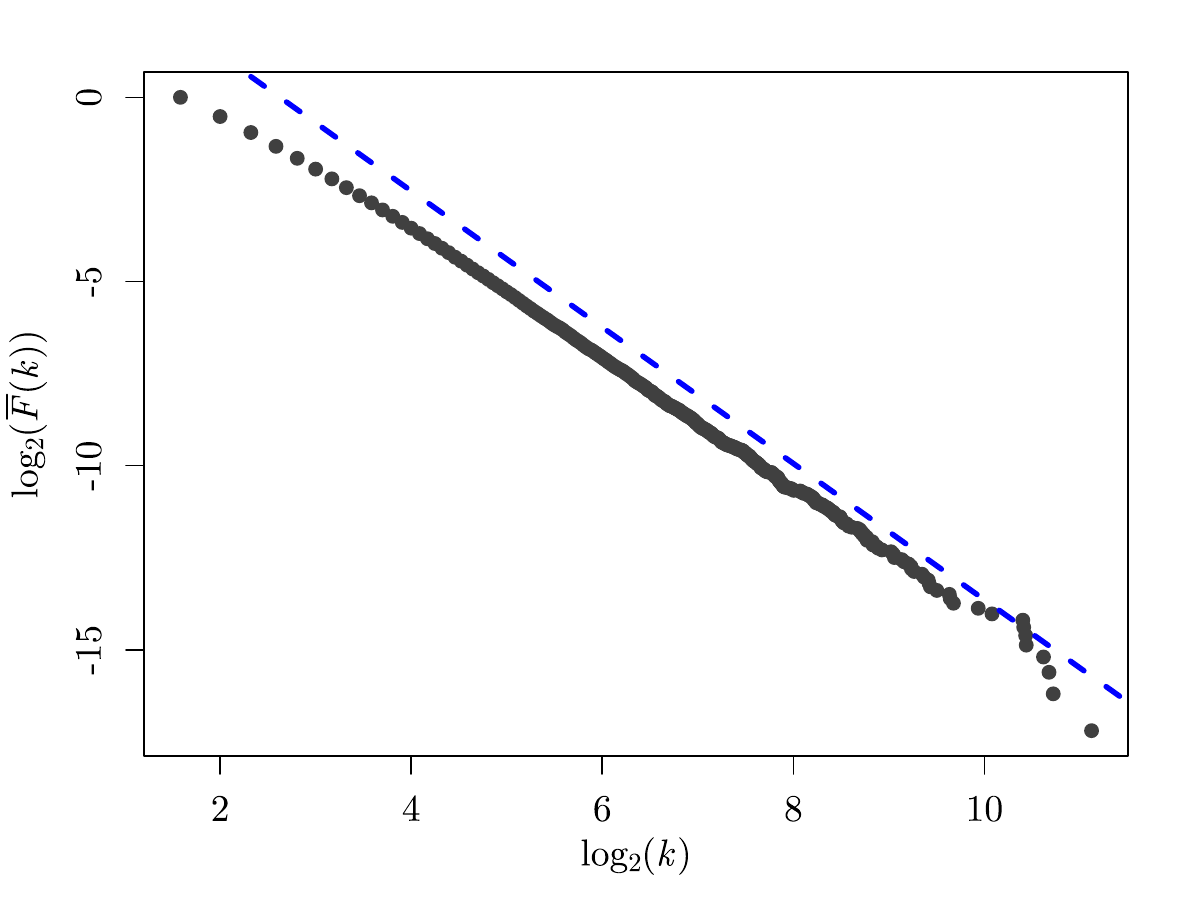}
\caption{Survival function $\overline{F}(k)$ of the distribution of the number of balls in a box generated by a realization of the Yule process with $1.5 \cdot 10^5$ time steps. The parameters of the process are: $m=5$, $K_0=3$, $c=1.25$. Since both the argument of the function and its value are under logarithm, a~straight line shape indicates the  presence of a power law. The dashed line has the slope $-\alpha=-1.85$, corresponding to the limiting distribution.}
\label{fig::Yule_process_distribution_example}
\end{minipage}
\end{figure}

If $n$ denotes the number of performed steps and $p(k)$ denotes the probability mass function of the distribution of the number of balls in a box, that is, the probability that a randomly chosen box has exactly $k$ balls inside, then after large number of steps ($n \to \infty$), the distribution $p(k)$ is a power-law distribution~\cite{Newman2005}:
\begin{equation}
p(k) \sim C\,k^{-\xi} \quad \text{ for } k \to +\infty \text{ and some constant } C.
\end{equation}
An example of such a distribution is presented in Fig.~\ref{fig::Yule_process_distribution_example}. The value of the exponent~$\xi$ is given by\cite{Newman2005}:
\begin{equation}
\xi = 2 + \frac{(K_0+c)}{m}.
\label{eq_powerlaws_Yule_exponent}
\end{equation}
Therefore, by tuning the values of the parameters $m$, $K_0$, and $c$, an arbitrary exponent greater than 2 can be obtained.

The key property of Yule processes, allowing to generate power laws, is expressed by Eq.~(\ref{eq::powerlaws_Yule_pref_att_rule}). It is the tendency to put new balls into boxes which already have many balls inside. It can be said that newly added items have the \textit{preference} to be placed where their concentration is already high, hence the name preferential attachment. Other names and phrases used to describe this effect are ``cumulative advantage'', ``rich get richer'' or ``Matthew effect''~\cite{Merton1968,Perc2014}. The last name refers to a verse in the biblical Gospel of Matthew: ``\textit{For to every one who has will more be given, and he will have abundance; but from him who has not, even what he has will be taken away}''. It should be noted, however, that only the first part of the verse correctly describes the considered effect as Yule processes do not contain mechanisms removing or relocating items already present in a system.

In a linguistic context, an example of this mechanism is the model of random text production studied by Simon~\cite{Simon1955}. According to that model, a text initially consists of a single word and then words are added in consecutive steps (one word in one step) in the following way. With probability $q$ a new word (a word not yet present in the vocabulary) is appended and with probability $1-q$ the appended word is randomly chosen from the words already present in the text. The probability of choosing a specific word is proportional to its frequency (the number of times it has already occurred in the text). This determines that the model can be considered a variant of the Yule process. The model gives the word frequency distribution of the form $p_\omega(\omega) \propto \omega^{-\beta}$, with $\beta = 1 + 1/(1-q)$ in the limit of large number of steps. With $q$ close to 0, $\beta$ is close to 2, which corresponds to Zipf's law in the form given by Eq.~(\ref{eq::Zipfs_inverse_law}).

Knowing that power-law distributions can be obtained in simple stochastic models, it may seem doubtful whether the fact that word frequencies in texts are described by power laws gives any significant information about language. However, the studied models are often clearly unrealistic and do not account for many essential traits of language. For example, the intermittent silence model does not take into account that words in natural language do not consist of random letters -- only some letter sequences are allowed as others might even not be pronounceable. Also, the distribution of word lengths does not correspond to what is observed in natural language~\cite{CANCHO2002}. Nevertheless, the presented models and other procedures of similar type remain an important class of models showing it is possible to obtain a power-law distribution as a result of a rather simple process~\cite{Debowski2020a}. 

A different view on the origin of the Zipf-type relations in empirical data was presented in~\cite{CristelliM-2012a}, where the authors pointed out that Zipf's law is coherent, i.e., neither the aggregation of two sets that are in agreement with this law individually nor division of the whole Zipfian set into two smaller subsets leads to the law's inheritance. Typically, the resulting sets do not show Zipf's property. It seems that it is required for a system to evolve organically to show a Zipfian form. This refers not only to language data sets, but also to all the other systems that can be described by the Zipf's law like city size or income distributions. This requirement poses a crucial limitation on considering, e.g., specific part-of-speech tokens as a separate data set. A process that allows one to create the Zipf-type rank-frequency distribution must have such a property that, after each draw from the distribution, it has to reshape itself. This reshaping is called \emph{conditioned sampling} and consists of ``blocking'' this part of the underlying probability distribution function, from which the recent value has been drawn.


\subsection{Generalized Zipf's law}

In language samples taken from real-world texts, some deviations from Zipf's law in its original form can be observed. One of such deviations is particularly typical for very big corpora usually consisting of large numbers of texts. For large samples, the rank-frequency distribution $\omega(R)$ with exponent $\alpha \approx 1$ holds up to some rank $R_c$ and, for ranks $R>R_c$, it breaks down and transforms into another power law with exponent $\alpha^{\prime} $ larger than $\alpha$~\cite{Cancho2001, Montemurro2001}. This is often explained by the existence of two types of vocabulary: one being a kind of core vocabulary consisting of a few thousand words the most frequently used, while the other one being more specialized and consisting of less common words, which are specific to particular topics or circumstances.

\begin{figure}
\centering
\begin{minipage}{\figurecustomwidth}
\centering
\includegraphics[width=0.8\textwidth]{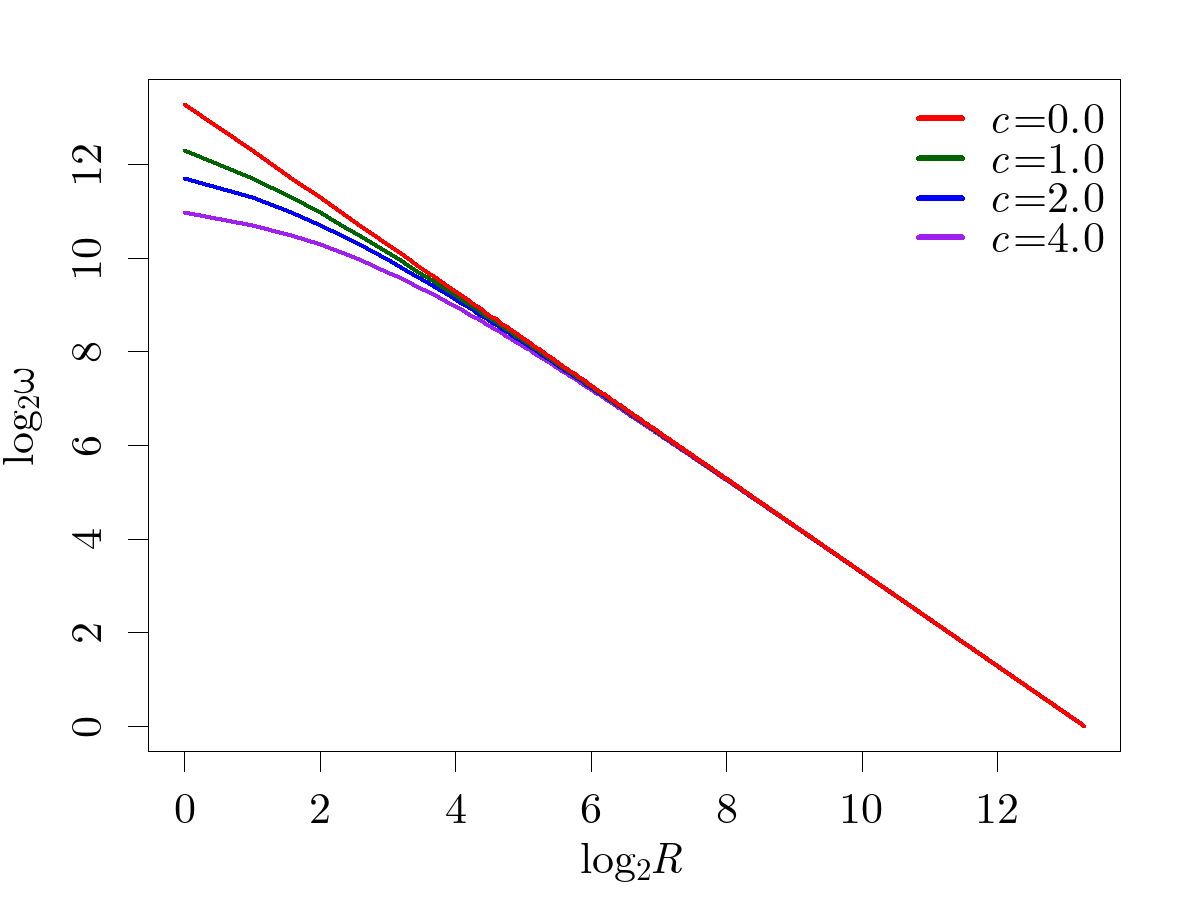}
\caption{Log-log plots of exemplary functions $\omega(R)$ given by Zipf-Mandelbrot law (Eq.~(\ref{eq::Zipf_Mandelbrot_law})) with $\alpha = 1$ and different values of $c$.}
\label{fig::Zipf_Mandelbrot_law_examples}
\end{minipage}
\end{figure}

Another frequently observed form of discrepancy between Zipf's law in its basic form and empirical data is the fact that usually words with lowest ranks have frequencies slightly lower than predicted by Zipf's law. Accordingly, Zipf's law holds for ranks above some rank $R_Z$, where usually $R_Z$ is on the order of 5 or 10. For $R<R_Z$, the frequencies $\omega(R)$ are below the frequencies given by exact power-law relationship between $\omega$ and $R$. To account for this effect, Mandelbrot introduced a correction to Zipf's rank-frequency relationship and the resulting formula, known as \textit{Zipf-Mandelbrot law}, can be written as:
\begin{equation}
\omega(R) \propto \left( R + c \right)^{\, -\alpha},
\label{eq::Zipf_Mandelbrot_law}
\end{equation}
where $R$ is the rank of a word, $\omega(R)$ is the frequency of the word of rank $R$ in the text, and $c$ is a non-negative constant. For $c=0$, Zipf-Mandelbrot law reduces to Zipf's law. Nonzero values of $c$ in the equation describing the rank-frequency relationship introduce the flattening of $\omega(R)$ for small values of $R$ and therefore allow for more accurate description of empirical data. An illustration of how the shape of $\omega(R)$ given by Zipf-Mandelbrot law depends on the value of $c$ is shown in~Fig.~\ref{fig::Zipf_Mandelbrot_law_examples}.

Zipf's law applies not only to the distribution of words -- which in written language can be understood as sequences of non-whitespace characters surrounded by whitespaces from both ends -- but also to some other distributions characterizing language. One can study, for example, the rank-frequency distribution of words after lemmatization, that is after reducing each word in the text to its basic, dictionary form called lemma. It turns out that lemmatized corpora also conform to Zipf's law~\cite{Corral2015}. In some situations, this allows for more meaningful comparison between languages, especially between languages with different degrees of inflection usage. Without lemmatization, the size of vocabulary can be artificially inflated in inflected languages (languages utilizing inflection to specify words' grammatical features), because various inflected forms of the same lemma are then counted as separate words.

\begin{figure}
\centering
\begin{minipage}{\figurecustomwidth}
\centering
\includegraphics[width=0.85\textwidth]{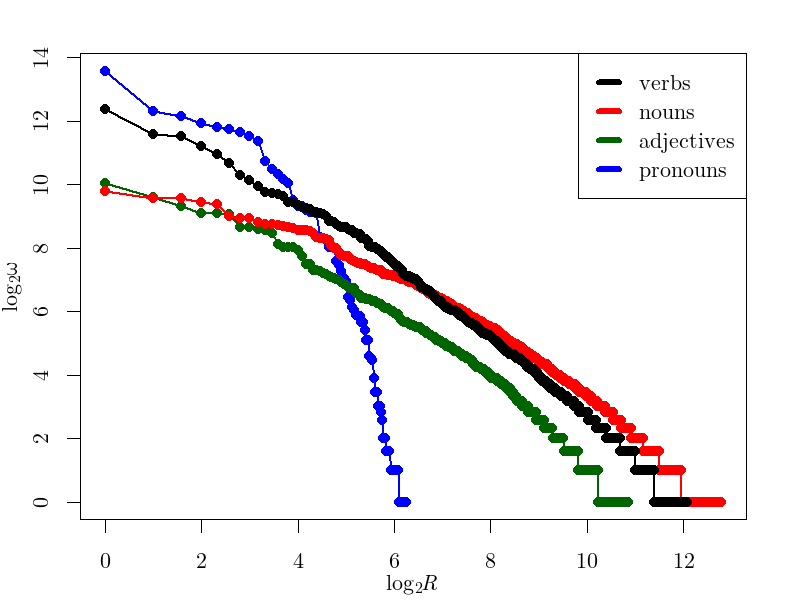}
\caption{Log-log plots of rank-frequency distributions $\omega(R)$ determined for selected parts of speech separately in a book \textit{David Copperfield} by Charles Dickens. Words are not lemmatized. A power-law relationship is the most closely followed by verbs.}
\label{fig::Zipfs_law_individual_parts_of_speech}
\end{minipage}
\end{figure}

Another possible type of word frequency analysis is investigating how subsets of words behave in terms of frequency distributions. For example, the set of all words in a corpus can be partitioned into subsets corresponding to different parts of speech and the frequency distribution within each subset can be studied~\cite{Kwapien2012,PIANTADOSI_2014}. An example of results of word frequency analysis utilizing such an approach is presented in Fig.~\ref{fig::Zipfs_law_individual_parts_of_speech}. When performed for corpora in English, the analysis of this type reveals various types of rank-frequency distributions, depending on part of speech under consideration. Some parts of speech are subject to power-law rank-frequency distributions, but some exhibit different type of behaviour. This is related to  the role of individual word classes in language. Words whose role is mostly grammatical (like conjunctions, prepositions, pronouns, and articles in English) are the ones used the most frequently and their rank-frequency distribution can be considered a power law for some range of low ranks, above which their frequencies quickly decay to zero. For words being references to specific objects and notions (nouns, for example), the agreement with power-law relationship is typically better outside the range of low ranks. The distinctive behaviour of verbs, which seem to conform to a power law in general (in a wide range of ranks), might be related to the fact that verbs play in a sense a dual role in language: most verbs are associated with some kind of action or state, but a group of verbs in English have special grammatical uses (like \textit{be, have} or \textit{will}).

\begin{figure}
\centering
\begin{minipage}{\figurecustomwidth}
\centering
\subfloat[English]{\includegraphics[width=0.48\textwidth]{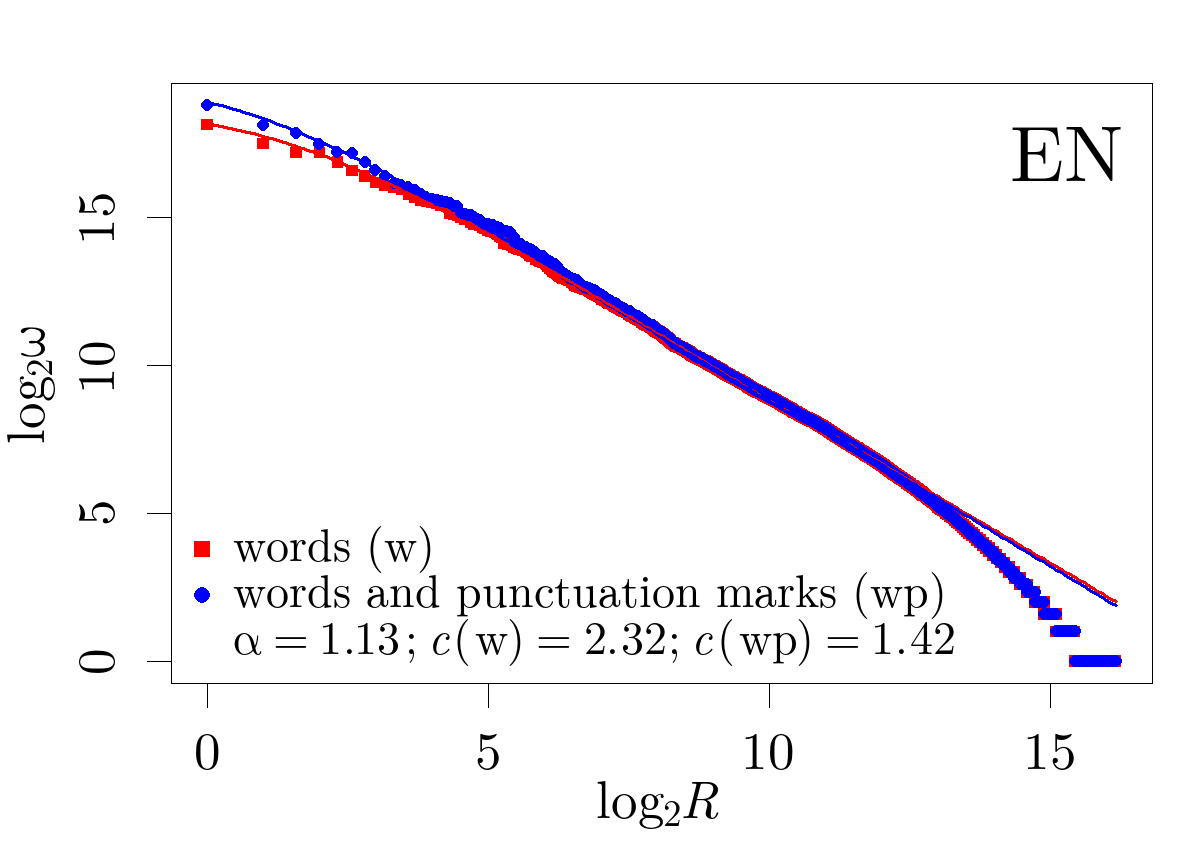}}
\hfill
\subfloat[German]{\includegraphics[width=0.48\textwidth]{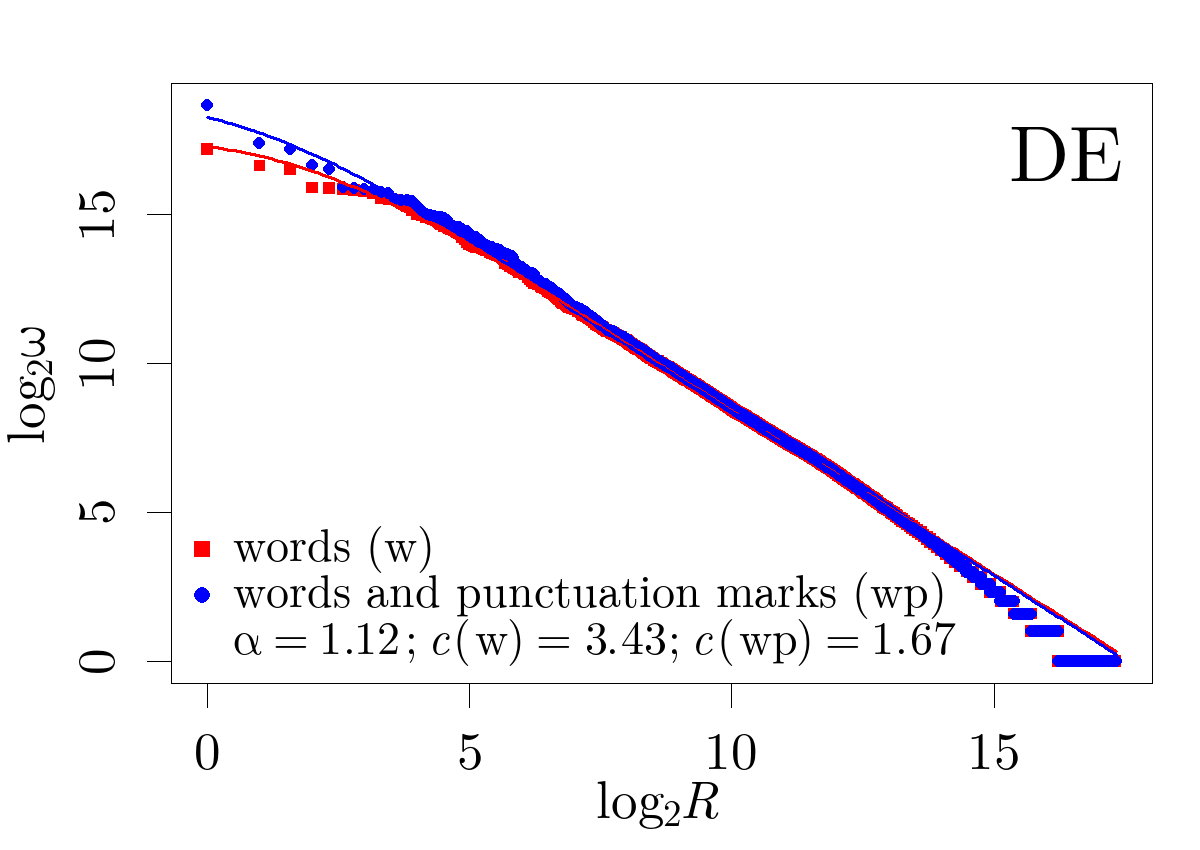}}
\hfill
\subfloat[French]{\includegraphics[width=0.48\textwidth]{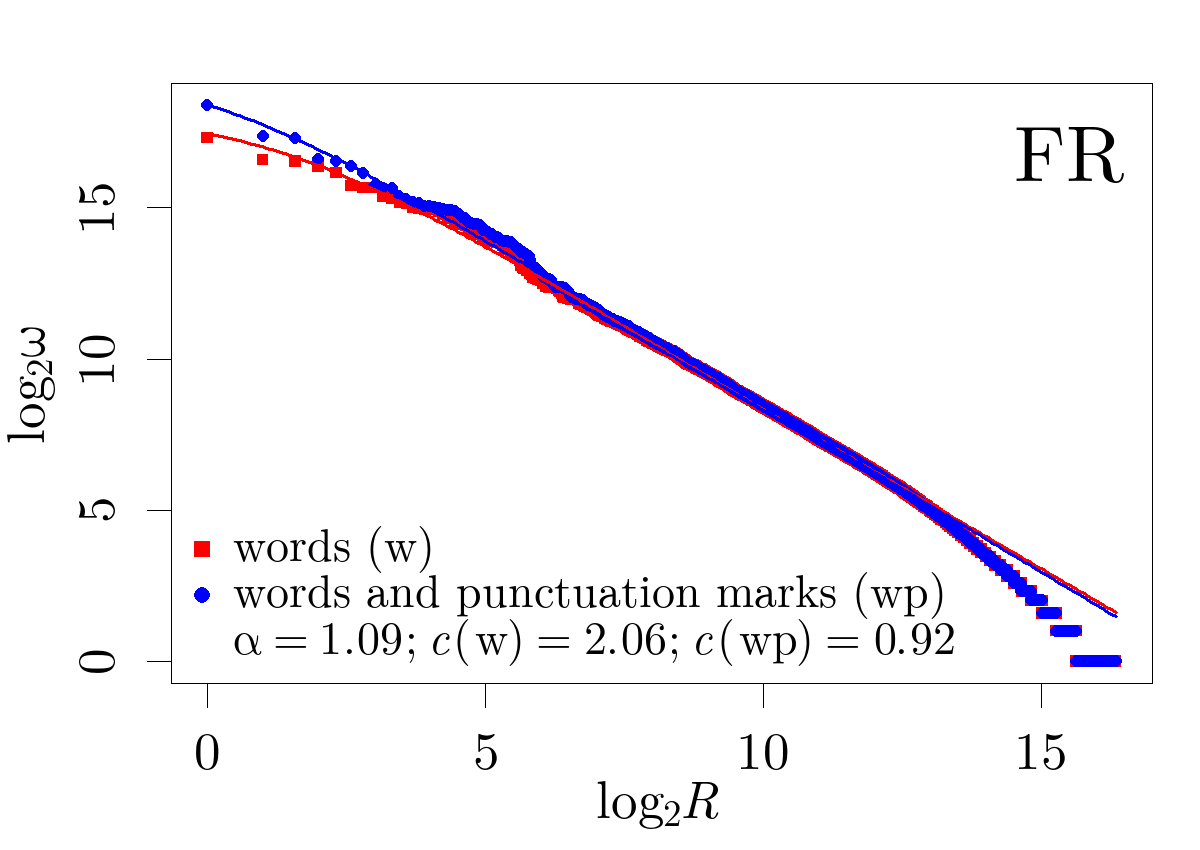}}
\hfill
\subfloat[Italian]{\includegraphics[width=0.48\textwidth]{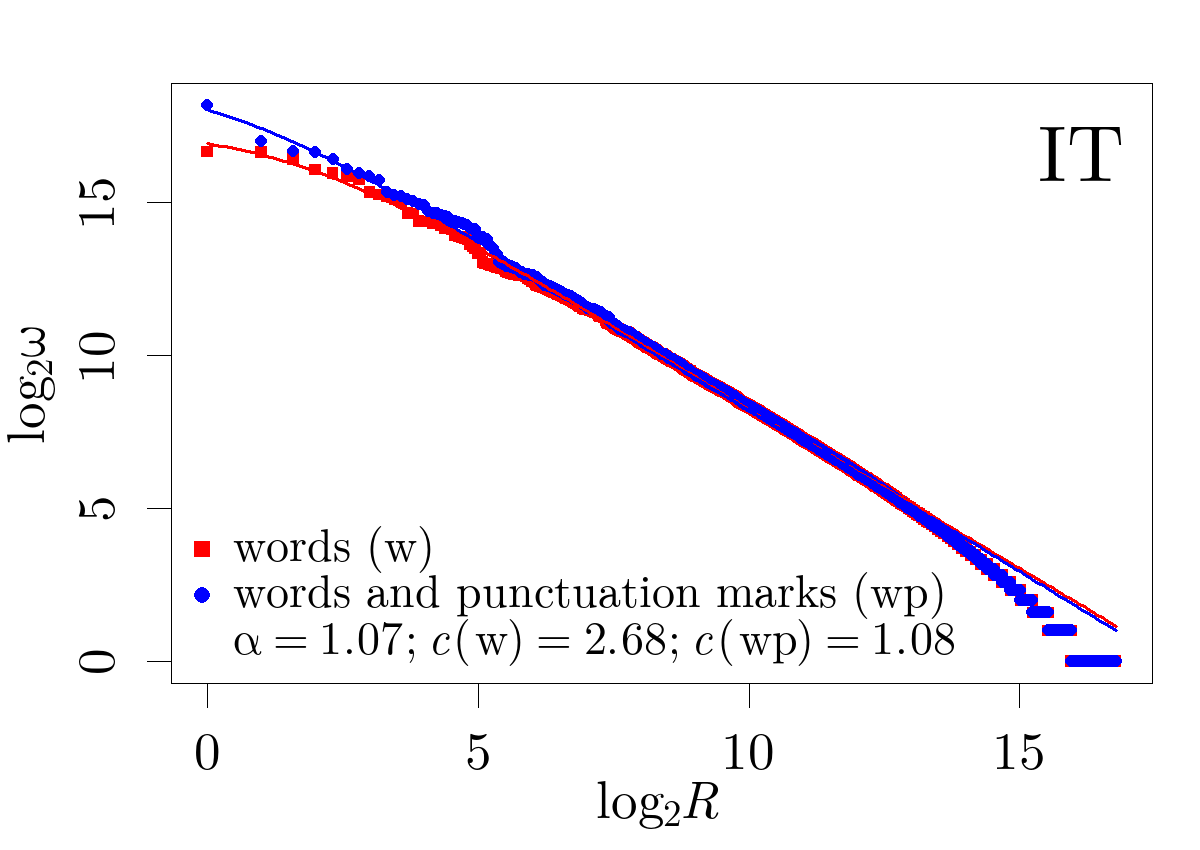}}
\hfill
\subfloat[Spanish]{\includegraphics[width=0.48\textwidth]{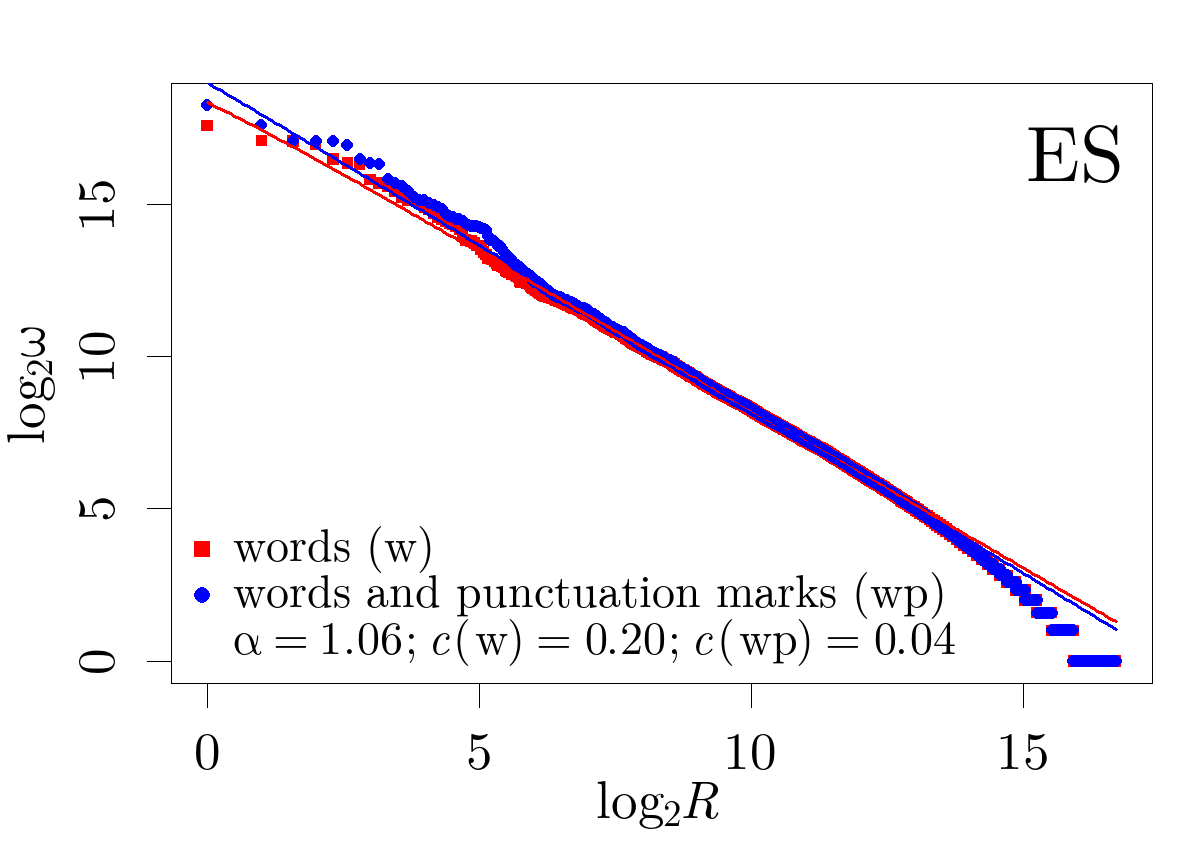}}
\hfill
\subfloat[Polish]{\includegraphics[width=0.48\textwidth]{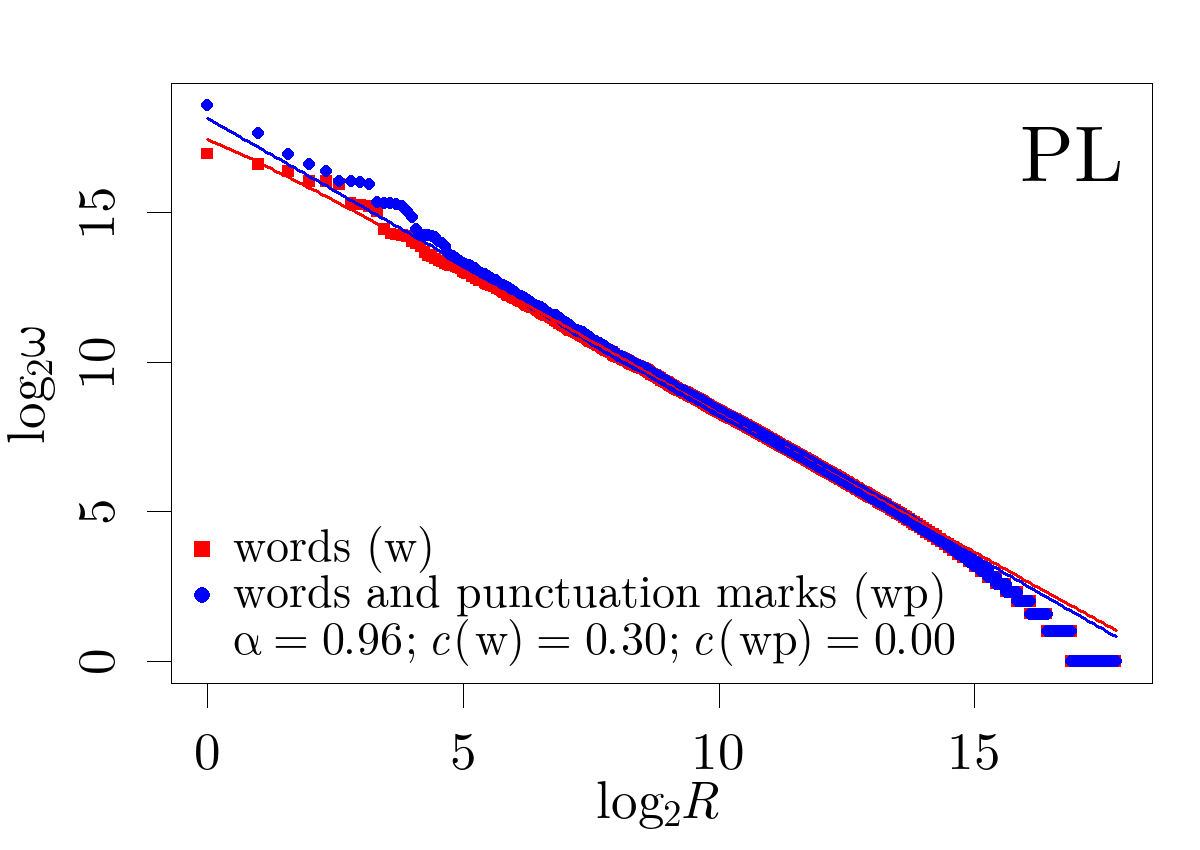}}
\hfill
\subfloat[Russian]{\includegraphics[width=0.48\textwidth]{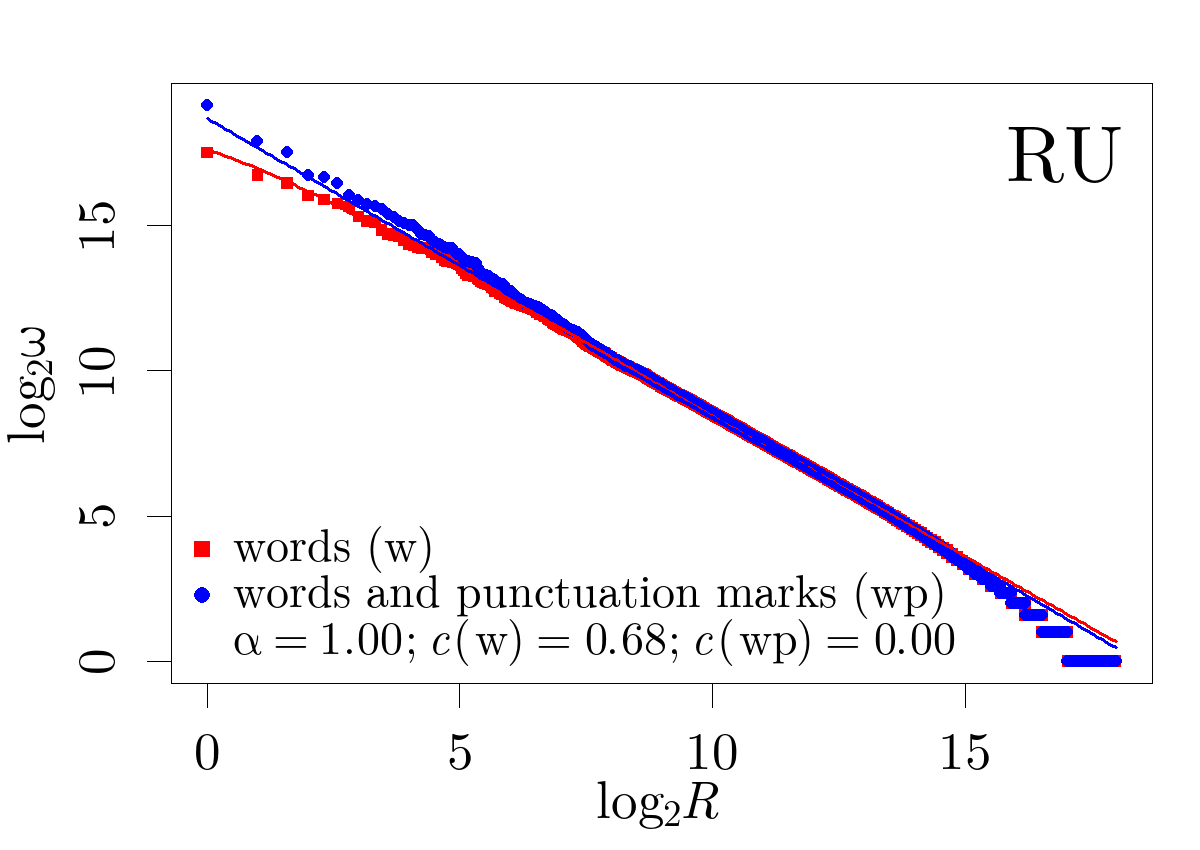}}
\hfill
\caption{Rank-frequency distributions for corpora constructed from sample books in selected European languages. Red squares represent the distribution for words, and blue dots -- the distribution for words and punctuation marks treated as words; in both cases words are not lemmatized. Zipf-Mandelbrot law (Eq.~(\ref{eq::Zipf_Mandelbrot_law})) fitted to the data is denoted by solid lines. The power law exponents $\alpha$, relevant for both distributions (words and words with punctuation marks), are given under each graph along with the values of $c$.}
\label{fig::Zipf_Mandelbrot_with_punctuation}
\end{minipage}
\end{figure}

Word frequency analysis can also be generalized to other entities occurring in written language. An interesting result has been obtained for punctuation marks treated as words and included into rank-frequency distributions of literary texts~\cite{Kulig2017}. Some of punctuation marks have frequencies comparable with the frequencies of the most frequent words. In fact, a comma and a period often occupy ranks between 1 and 3. It turns out that while rank-frequency distributions determined for words only are typically described by Zipf-Mandelbrot law, the distributions for words together with punctuation marks are closer to the regime given by ``pure'' Zipf's law. This effect is demonstrated in Figure~\ref{fig::Zipf_Mandelbrot_with_punctuation}. It can be stated that treating punctuation marks as words decreases the flattening of $\omega(R)$ for small $R$. A few European languages have been studied in that context and all of them have been identified as displaying the presented effect, but with varying intensity. Quantitatively, the decrease in the rank-frequency distribution's flattening for small $R$ corresponds to the decrease of the value of the constant $c$ in Eq.~(\ref{eq::Zipf_Mandelbrot_law}) expressing the Zipf-Mandelbrot law. Among the studied languages, the Germanic languages (English and German) have the weakest tendency to restore Zipf's law when punctuation is included ($c$ decreases, but remains significant), while for the Slavic languages (Polish and Russian) including punctuation into analysis results in $c$ dropping close to 0. Therefore, punctuation marks and words can be considered to fit into the same Zipfian regime of frequency distribution. Together with results from other methods of text analysis~\cite{Kulig2017}, this fact allows to state that, although punctuation marks and words are clearly different objects, their statistical properties are, in many respects, comparable. This conclusion leads to a hypothesis that, at least in some aspects, punctuation carries information in a way similar to words. It also justifies taking punctuation into account when computing word frequency characteristics in statistical analysis of language and its practical applications.


\section{Entropy and long-range correlations}
\label{sect::entropy_and_long-range_correlations}


\subsection{Basic definitions and relations}

\subsubsection{Entropy of symbolic sequences}

In this Section some selected methods of time series analysis that are useful in dealing with complex system data, including the natural language data, are briefly presented. We start from information entropy of symbolic sequences quantifying the extent to which variables in a sequence are dependent on each other. Let $\lbrace X(t) \rbrace$ be a stationary sequence of categorical random variables indexed by time $t \in \lbrace 0, \pm 1, \pm 2, ... \rbrace$ and let us assume that the possible values of each of the variables are symbols from some fixed set. The stationarity condition means that the joint cumulative distribution function $F[X(t),X(t+1),...,X(t+s)]=const$ for any $t$ and $s$. The bitwise entropy rate $H_{\rm X}$ of a stationary process $\lbrace X(t) \rbrace$ can be defined as
\begin{equation}
\begin{gathered}
H_X = \lim_{n \to \infty} \textbf{E} \bigg[ -\log_2 P \bigg( X(t) \big| X(t-1), X(t-2), ..., X(t-n) \bigg) \bigg],
\end{gathered}
\end{equation}
where $P \big( X(t) \big| X(t-1), ..., X(t-n) \big)$ is the conditional probability distribution of $X(t)$ given $X(t-1), X(t-2), ..., X(t-n)$, and $\textbf{E}\left[ \cdot \right]$ is the expectation operator. $H_{\rm X}$ measures the average degree of uncertainty about the value at some point $t$, when the prior values are known.

Practically, the entropy of a symbolic sequence can be calculated, for example, by using an estimator based on the Lempel-Ziv algorithm, which identifies repeated sequences in time series. If a sequence of symbols $x_0, x_1, x_2, ...$ is a realization of a stochastic process $\lbrace X(t) \rbrace$ and $L_i$ is the length of its shortest subsequence starting at position $i$, which does not appear anywhere up to position $i-1$, then
\begin{equation}
H_{\rm X} = \lim_{n \to \infty} \hat{H}_n = \lim_{n \to \infty} \left( \frac{1}{n} \sum_{i=2}^{n} \frac{L_i}{\log_2 i} \right)^{-1}
\label{eq::entropy_estimator_LZ}
\end{equation}
under some mild assumptions regarding the behaviour of the underlying process in large time scales~\cite{Kontoyiannis1998,Gao2008} (the convergence is relatively slow, however~\cite{Gao2008}).

\subsubsection{Long-range correlations}

If a time series is a sequence of numbers, basic methods allowing for the identification of the statistical dependencies often use the frequency domain analysis. Let $\{x_n\}$ be a finite ($n=0,...,N-1$) sequence of real numbers representing measurements of some quantity at equally spaced points in time. By using the discrete Fourier transform (DFT), the spectral density (periodogram) $S(k)$ of the sequence can be defined as
\begin{equation}
S(k) = C \left| \sum_{n=0}^{N-1} x_n \, e^{- i (2 \pi k/N) n} \right|^2,
\label{eq::periodogram}
\end{equation}
where $k=0,...,N-1$ indexes the harmonic frequencies of DFT's fundamental frequency corresponding to one cycle per whole sequence, while $C$ is a normalization constant. For infinite signals, the spectral density with an appropriate $C$ becomes the power spectrum $S(f)$. If $\{x_n\}$ is considered as a signal with finite energy, one can put $C=1$. From a technical point of view, it should be pointed out that sometimes determining a reliable estimation of the spectrum from finite-size data requires additional steps, like dividing the time series into windows, computing the periodogram for each window, and then averaging the results~\cite{Stoica2005}.

Spectral density of a signal describes how much a given frequency or frequency band contributes to the signal's total variability. It also gives insight into temporal correlations. For a stochastic process $\{X(t)\}$ with real values indexed by $t$, the autocorrelation function is defined as
\begin{equation}
\begin{gathered}
R_{\rm XX}(t,t+\tau) = \textbf{E} \bigg[ \bigg( X(t) - \textbf{E}\left[ X(t) \right] \bigg)   \bigg( X(t+\tau) - \textbf{E}\left[ X(t+\tau) \right] \bigg) \bigg] = \\
= \textbf{E} \bigg[ X(t)X(t+\tau) - \textbf{E}\left[ X(t) \right] \textbf{E}\left[ X(t+\tau) \right] \bigg],
\end{gathered}
\label{eq::autocovariance}
\end{equation}
where $\tau>0$. For a process that is weakly stationary, i.e., when (1) $\textbf{E}\left[ X(t) \right] = \textbf{E}\left[ X(t+\tau) \right]$ for all $t, \tau$, (2) $\textbf{E}\left[ X(t)^{2} \right] = \sigma^{2} < \infty$ for all $t$, and (3) $R_{\rm XX}(t,t+\tau)$ depends only on $\tau$ without dependence on $t$, the autocorrelation function has a simplified form of $R_{\rm X}(\tau)$. For a weakly stationary process with $\textbf{E}\left[X(t)\right] = 0$, the spectral density $S(k)$ is equal to the Fourier transform of the autocorrelation function (the Wiener-Khinchin theorem). A weakly stationary, zero-mean process has long-range correlations (or long memory) if $R_{\rm X}(\tau)$ decays so slowly with $\tau$ that its sum or integral is divergent~\cite{Beran1994,Gao2006}. A signal whose power spectrum satisfies
\begin{equation}
S(f) \propto \frac{1}{f^{\beta}}
\label{eq_powerlaw_1_over_f_noise_def}
\end{equation}
for sufficiently wide range of frequencies $f$ is called a $1/f^\beta$ noise. While typically $0 < \beta \le 2$, the special case of $\beta=1$ is sometimes used to refer to $1/f^{\beta}$ noise in general. A power spectrum of the form~\ref{eq_powerlaw_1_over_f_noise_def} implies the presence of a specific structure of correlations. For example, it can be shown~\cite{Rangarajan2000} that a weakly stationary, zero-mean, discrete-time process with long memory whose autocorrelation function for large $\tau$ has the form: 
\begin{equation}
R_{\rm X}(\tau) \propto \tau^{-\alpha}, \quad \alpha \in \left( 0; 1 \right),
\label{eq::autocorrelation.function.power.law}
\end{equation}
has the power spectrum also behaving like a power function for small $f$:
\begin{equation}
S(f) \propto \frac{1}{f^{1-\alpha}}.
\end{equation}
Therefore, $1/f^{\beta}$ spectrum for $\beta$ between 0 and 1 can be directly related to the long-range correlations. An example of a stochastic process whose autocorrelation function satisfies Eq.~(\ref{eq::autocorrelation.function.power.law}) is the discrete fractional Gaussian noise $\{B_H^{'}(t)\}$, in which all the variables $B_H'(t)$ have the same normal distribution with zero mean and standard deviation $\sigma$~\cite{Mandelbrot1968,Molz1997}. Its autocorrelation function is given by~\cite{Beran1994}:
\begin{equation}
R_{B_H'}\!(\tau) = \frac{\sigma^2}{2} \big( \left|\tau + 1\right|^{2H} - 2 \left| \tau \right| ^{2H} +  \left|\tau - 1\right|^{2H} \big) \propto \tau^{2H-2}.
\label{eq::autocorrelation.function.gaussian.noise}
\end{equation}
and depends on the Hurst exponent $H \in [0,1]$. The character of the correlations is determined by the value of $H$: for $H=\tfrac{1}{2}$ the variables $\lbrace B_H'(t) \rbrace$ are independent, for $H<\tfrac{1}{2}$ the process is antipersistent (its consecutive values are negatively correlated), and for $H>\tfrac{1}{2}$ it is persistent (its consecutive values are positively correlated). The corresponding spectral density $S(f)$ of $\{B_H'(t)\}$ for small $f$ satisfies~\cite{Beran1994,Mandelbrot1968,Molz1997}:
\begin{equation}
S(f) \propto \frac{1}{f\!\:^{2H-1}},
\label{eq_powerlaw_fGn_spectrum}
\end{equation}
which shows that the fractional Gaussian noise is long-range correlated for $H > \frac{1}{2}$.

The process $\{B_H'(t)\}$ can be viewed as an increment process of the fractional Brownian motion $\{B_H(t)\}$ with the increment correlations determined by the Hurst exponent $H$. $\{B_H(t)\}$ is a non-stationary process whose autocorrelation function depends on time $t$. Its power spectrum reads
\begin{equation}
S(f) \propto \frac{1}{f^{2H+1}}.
\label{eq_powerlaw_fBm_spectrum}
\end{equation}
Eqs.~(\ref{eq_powerlaw_fGn_spectrum}) and~(\ref{eq_powerlaw_fBm_spectrum}) show that one can generate $1/f^{\beta}$ noises with $\beta$ both below and above 1. A crucial property of the fractional Brownian motion is the behaviour of its variance, which for any $t_0$ and $t>0$ is given by~\cite{Beran1994,Delignieres2015}:
\begin{equation}
\textbf{E}\left[ \big( B_H(t_0+t) - B_H(t_0) \big)^2 \right] = \sigma^2 t^{2H},
\end{equation}
where $\sigma^2$ is the variance of $\{B_H'(t)\}$. In a more general sense, if $\lbrace Y(t) \rbrace$ is a process whose increments are given by a weakly stationary process $\lbrace X(t) \rbrace$ and $\lbrace Y(t) \rbrace$ satisfies
\begin{equation}
\textbf{E} \left[ \big( Y(t_0+t) - Y(t_0) \big)^2 \right] \propto t^{2H}
\label{eq::Hurst.exponent}
\end{equation}
for any $t_0$ and $t > 0$, then $H$ can be called the Hurst exponent of $\lbrace X(t) \rbrace$. 

A $1/f^{\beta}$ process with $\beta = 1$ is of special interest. It functions at the interface between two distinct regimes: the one corresponding to a correlated random-walk-like process ($1 < \beta \leq 2$) and the other one corresponding to its increment process ($0 \leq \beta < 1$). For that reason such a process is considered to reveal maximum complexity~\cite{Fossion2010,Voss1989}.


\subsection{Entropy in written language}

Samples of written language can be naturally represented in terms of symbolic sequences. In the most straightforward approach, a text can be treated as a sequence of characters. Each character can either be a unique symbol or a transformation can map these characters to a smaller set of symbols. For example, a ternary symbol space may be considered where all the characters used in a text sample are distributed into three sets: (1) consonants, (2) vowels and digits, and (3) punctuation marks. Then the elements of each set were denoted by the same symbol. Another approach is to consider words as individual symbols -- a text becomes then a sequence of words. Determining the entropy of such a sequence gives an insight into how much, on average, the occurrence of a symbol (word) in a text is determined by the specific symbol (word) sequence preceding the considered symbol (word). However, instead of using the entropy itself, one can also study the difference between the entropy $H_{\rm orig}$ of the original text and the entropy $H_{\rm rand}$ of the same text, but with words shuffled randomly. In a symbolic sequence in which the order of the symbols is random, the entropy is determined purely by the distribution of symbol frequencies. Hence, the difference $H_{\rm rand}-H_{\rm orig}$ provides information about the decrease in text's entropy caused by the specific order of words, compared to the entropy which would be observed if words were placed at random in the text. The usefulness of the quantity $H_{\rm rand}-H_{\rm orig}$, which in the considered context is referred to as \textit{relative entropy}, is due to the fact that it allows to remove the influence of purely frequency-based effects. For example, if words are not lemmatized (as is the case here), texts in the languages with extensive use of inflection typically have more unique words (symbols) than texts in the languages, in which inflection is less developed. This influences the frequency distribution -- and, consequently, entropy -- but since the distribution is the same in the original and in the randomized text, the relative entropy can be anticipated to capture only the effects related to word ordering. Relative entropy has been reported in the literature~\cite{Montemurro2011,Montemurro2014} to be approximately independent of language (this was tested on corpora in a number of languages), with values ranging from about 3 bits per word to about 4 bits per word. This suggests that despite the differences between the grammars and the vocabularies of individual languages, the amount of ``order'' contained in how words are placed with respect to each other is to a certain degree universal across languages.

\begin{figure}
\centering
\begin{minipage}{\figurecustomwidth}
\centering
\includegraphics[width=0.8\textwidth]{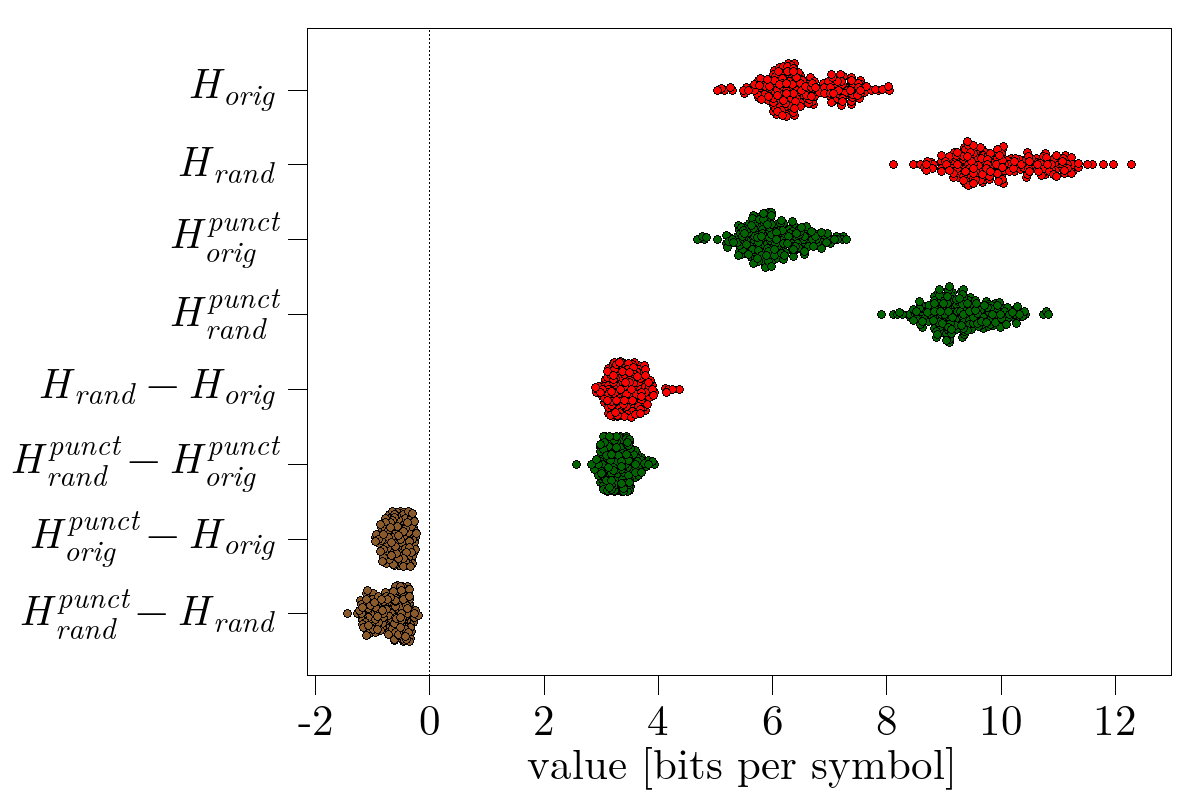}
\caption{Distributions of the values of several quantities constructed from entropy rates computed for sample texts. The values of entropy rates are obtained with the use of the estimator given in Eq.~(\ref{eq::entropy_estimator_LZ}). Each point on the plot corresponds to one text. Different positions along the vertical axis correspond to different quantities. The studied quantities are: entropy rate of a text treated as a sequence of words $H_{\rm orig}$, entropy rate of a text treated as a sequence of words and punctuation marks $H_{\rm orig}^{\rm punct}$, entropy rate of a randomly shuffled text treated as a sequence of words $H_{\rm rand}$, entropy rate of a randomly shuffled text treated as a sequence of words and punctuation marks $H_{\rm rand}^{\rm punct}$. Four additional quantities are constructed from the listed ones and presented in the plot:  $H_{\rm rand}-H_{\rm orig}$, $H_{\rm rand}^{\rm punct} - H_{\rm orig}^{\rm punct}$, $H_{\rm orig}^{\rm punct}-H_{\rm orig}$, $H_{\rm rand}^{\rm punct}-H_{\rm rand}$.}
\label{fig::entropy_rate_of_texts}
\end{minipage}
\end{figure}

Fig.~\ref{fig::entropy_rate_of_texts} shows the entropy (estimated using the estimator given by Eq.~(\ref{eq::entropy_estimator_LZ})) of a corpus of sample texts. Each text is considered in both its original form and in a randomized variant (with random word order), giving two values: the entropy of the original text $H_{\rm orig}$ and the entropy of the randomized text $H_{\rm rand}$. The figure also shows the entropy for the same texts computed with punctuation marks taken into account and included into the analysis on the same terms as words. Data sequences including punctuation marks are in two variants as well -- the original and the randomized one -- giving rise to two values of entropy, $H_{\rm orig}^{\rm punct}$, $H_{\rm rand}^{\rm punct}$, respectively. Additional quantities: $H_{\rm rand}-H_{\rm orig}$, $H_{\rm rand}^{\rm punct} - H_{\rm orig}^{\rm punct}$, $H_{\rm orig}^{\rm punct}-H_{\rm orig}$, $H_{\rm rand}^{\rm punct}-H_{\rm rand}$, are also shown in the figure. The narrow distributions of $H_{\rm rand}-H_{\rm orig}$ and $H_{\rm rand}^{\rm punct} - H_{\rm orig}^{\rm punct}$ confirm the mentioned result reported in the literature -- that the values of relative entropy of word ordering are concentrated in the range between 3 and 4 bits per word. The distribution of $H_{\rm orig}^{\rm punct}-H_{\rm orig}$ (all values slightly below 0) indicates that taking punctuation marks results in a decrease of text's entropy. This could lead to a conclusion that punctuation organizes written language in a manner that lowers the ``randomness'' of a text. However, since practically the same effect is observed for randomized texts ($H_{\rm rand}^{\rm punct}-H_{\rm rand}$ is also slightly below 0), the decrease of entropy can be attributed to changes in frequencies of individual symbols introduced by including punctuation marks into the analysis. This can be understood with the help of results regarding how punctuation influences the shape of word frequency distribution in texts (Fig.~\ref{fig::Zipf_Mandelbrot_with_punctuation}). Treating punctuation marks as words brings the shape of word frequency distribution closer to the one specified by a power law -- the frequencies of the most frequent symbols (words or punctuation marks) become higher than in the original distribution. The distribution becomes thus less ``uniform'' -- as the discrepancy between the highest and the lowest frequencies increases. This results in a decrease of distribution's entropy (as entropy is maximized for maximally uniform distributions) which affects the entropy of a sequence consisting of symbols coming from the considered distribution. Hence, the behaviour of the entropy caused by introducing punctuation marks into analysis can be considered a consequence of the influence that punctuation has on Zipf-Mandelbrot law describing word frequencies in texts.

Entropy allows for calculation of information content of a given text sample. For example, in a simple way information can be parametrized by the word or part-of-speech diversity. For this point of view, the more distinct words or the more parts of speech can be found in a sample, the more informative it is. This can be of much importance in bilingual translation. One may expect that a good translation would preserve the word diversity of a source, while a poor one would have tendency to decrease it. A special case of translation is interpretation, which can be either consecutive if a speaker pauses from time to time in order to allow their statements to be translated into a target language, or simultaneous if no pauses are present and the interpretation is also continuous but delayed by a few seconds. A study of the information content of both types of interpretation was carried out based on Mandarin Chinese and English transcriptions of the official speeches and press conferences of Chinese political leaders~\cite{LinY-2023a}. According to this study which used information entropy as a measure of information content, simultaneous interpretations occur to have decreased entropy than consecutive interpretations. Consistently, the word repeat rate defined as
\begin{equation}
RR = {1 \over N^2} \sum_i n_i^2,
\end{equation}
where $n_i$ is the number of $i$th word occurrences and $N$ is the vocabulary size, for simultaneous interpretation is higher than for consecutive one. This result supports intuition that an extremely fast translation can be done in expense of the word diversity, because interpreters must rely heavily on common words that are the most easily accessible for them. Moreover, the consecutive interpreters can even amplify diversity of the word usage with respect to the source speaker.


\subsection{Long-range correlations in written language}

Time series analysis is a tool well-suited to the study of natural language, as in certain situations language can be treated as a signal, often having the form of time series. There are multiple ways of representing a language sample as a signal in time domain. Different approaches allow to focus on different properties. Spoken language takes the form of auditory signal, which can therefore be considered a basic, ``raw'' representation of language. Extracting individual sounds (phones), words, sentences etc. allows to construct higher-level representations. This applies also to written language (with the distinction that on the most basic level information is carried by appropriate symbols instead of sounds). 

Studying linguistic data having the form of a time series might give an opportunity to reveal patterns of organization which can be universal for language or specific to particular language samples (samples of language typical for particular situations, for instance). Investigating the behaviour of the quantities like grammatical distances between words, word recurrence times, or word lengths (as a function of their positions in text) allows to identify certain statistical regularities which are useful in attempts of characterizing the processes governing language usage~\cite{Montemurro2002,AusloosM-2012a,Altmann2009,DrozdzS-2016a,Montemurro2011,Montemurro2014,Futrell2015,Liu2017}.

\subsubsection{Texts as symbolic sequences}

In~\cite{EbelingW-1992a,AnishchenkoVS-1994a} sample texts written in one of two languages: English or Russian were transformed into symbolic sequences that consisted of three symbols (representing vowels and digits, consonants, and punctuation marks) and analyzed by means of entropy and spectral density. It was shown that spectral density assumes a power-law form $S(f)\sim f^{\beta}$ with $\beta \approx 0.4$ (Russian) and $\beta \approx 0.5$ (English), and $\beta \approx 0.9$ (English). All three texts were thus long-range correlated. Apart from the ternary symbol encoding, the same texts in the original ASCII encoding were studied with roughly the same outcome. Authors of~\cite{AnishchenkoVS-1994a} concluded that the particular choice of encoding did not influence the results, which could allow one to reduce the effective number of symbols subject to analysis. The conclusion about the existence of long-range correlations was supported by calculation of entropy $\hat{H}$ dependence on $n$, which occurred to be also power-law. Authors drew a parallel between the properties of the text-related symbolic sequences and the phenomenon of intermittency in the dynamical systems. A similar analysis based on the autocorrelation function calculated for symbolic sequences representing books in German also supported the previous observations of the long-range memory by giving $R(n) \sim n^{-\gamma}$ with $0.3 \le \gamma \le 0.8$~\cite{EbelingW-1995a}.

\subsubsection{Sentence lengths}

An interesting example of a signal constructed from linguistic data is a time series representing the lengths of sentences in a text, measured by the number of words. It is a sequence of numbers in which the $k$-th number is the number of words in the $k$-th sentence. For practical purposes, a sentence can be understood as a sequence of words between punctuation marks belonging to the following group: period, question mark, exclamation mark, and ellipsis (usually, the text needs to be appropriately pre-processed in order to remove periods denoting abbreviations, for instance). Such a time series allows to investigate organization of language on a level higher than the one corresponding to individual words. Sentences are structures in which the complexity of syntax is manifested and in which words fully acquire their meanings. The content of a sentence is usually linked to the content of neighbouring sentences, which constitute the context. But it turns out that the correlations typically have range larger than a few closest sentences; this effect can be captured by analyzing time series representing sentence lengths.

\begin{figure}[p]
\centering
\begin{minipage}{\figurecustomwidth}
\centering
\subfloat[]{\includegraphics[width=0.48\textwidth]{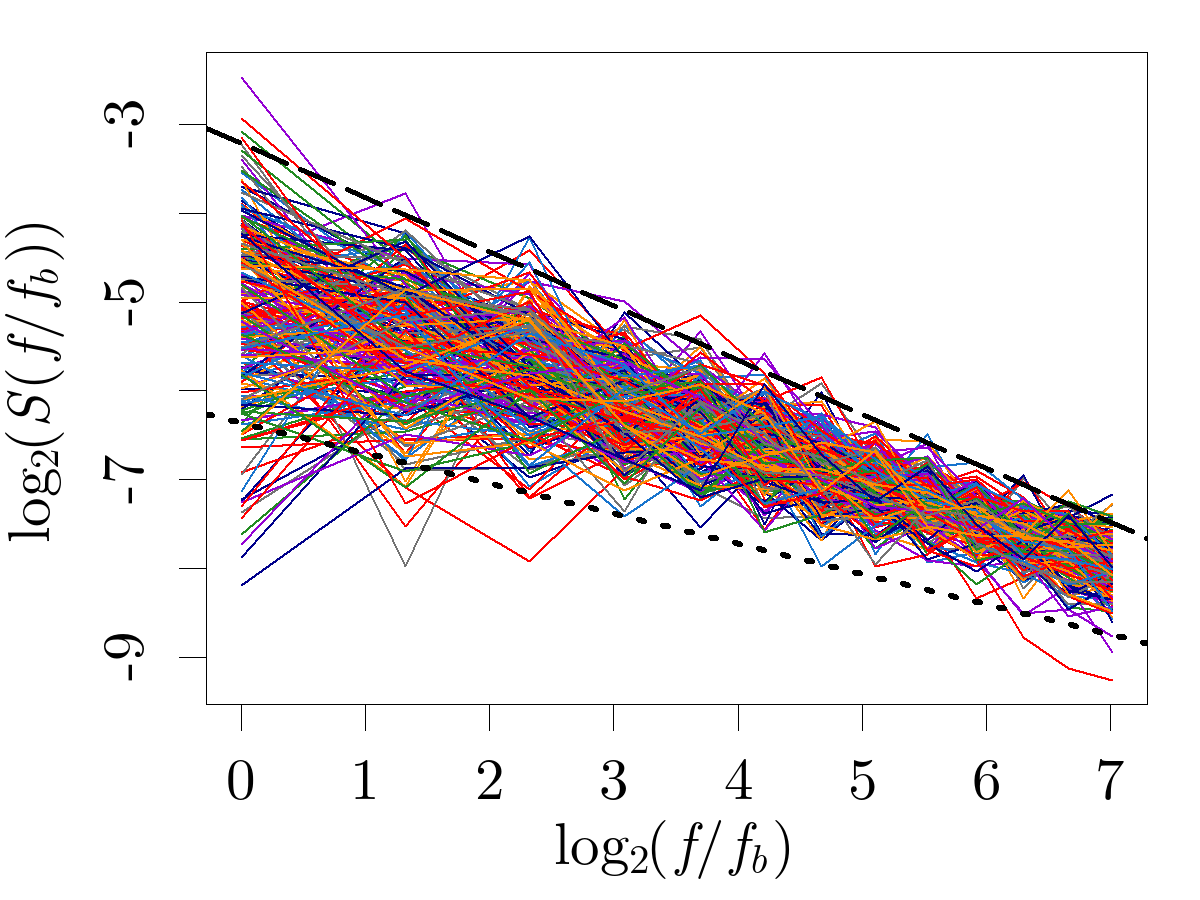} \label{fig_long_range_correlations_sentences_spectra}}
\hfill
\subfloat[]{\includegraphics[width=0.48\textwidth]{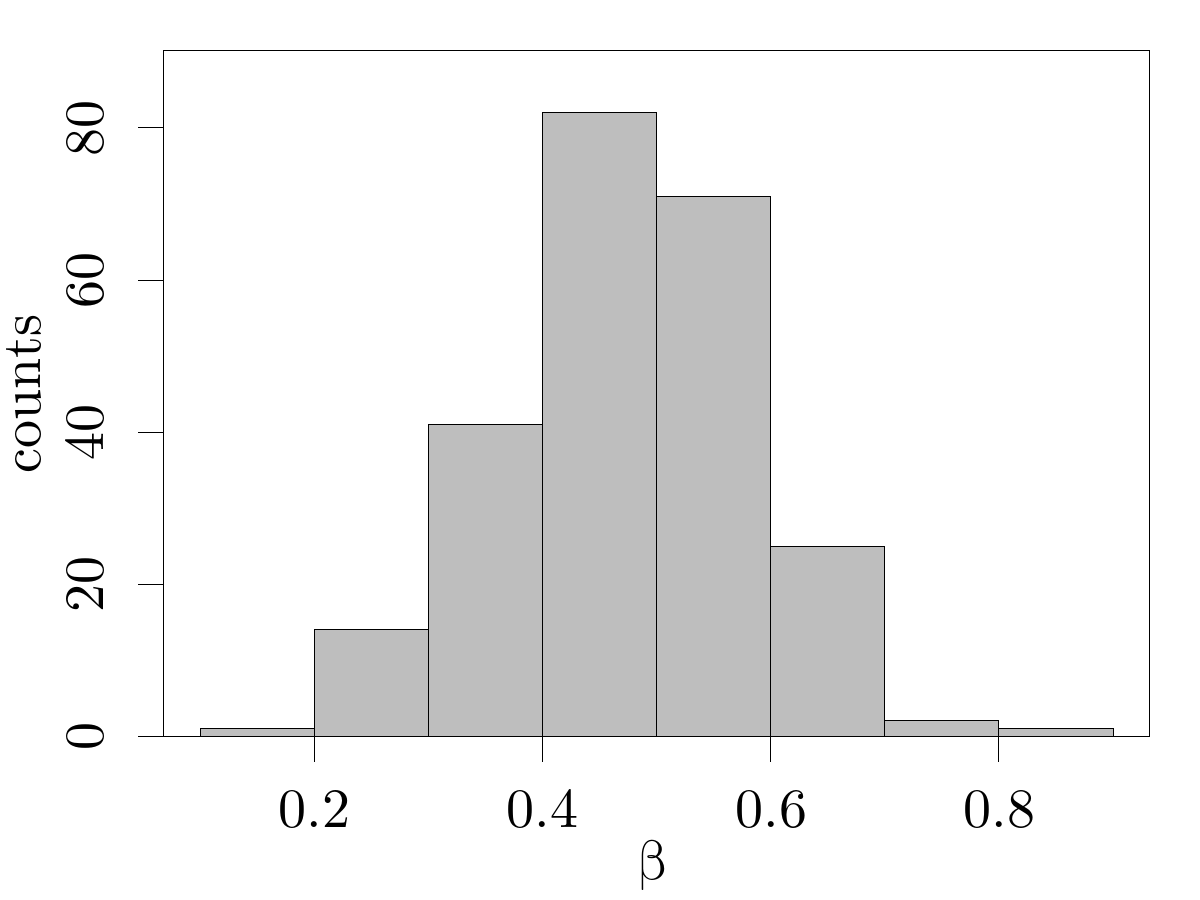}\label{fig_long_range_correlations_sentences_beta_histogram}}
\hfill
\subfloat[]{\includegraphics[width=0.48\textwidth]{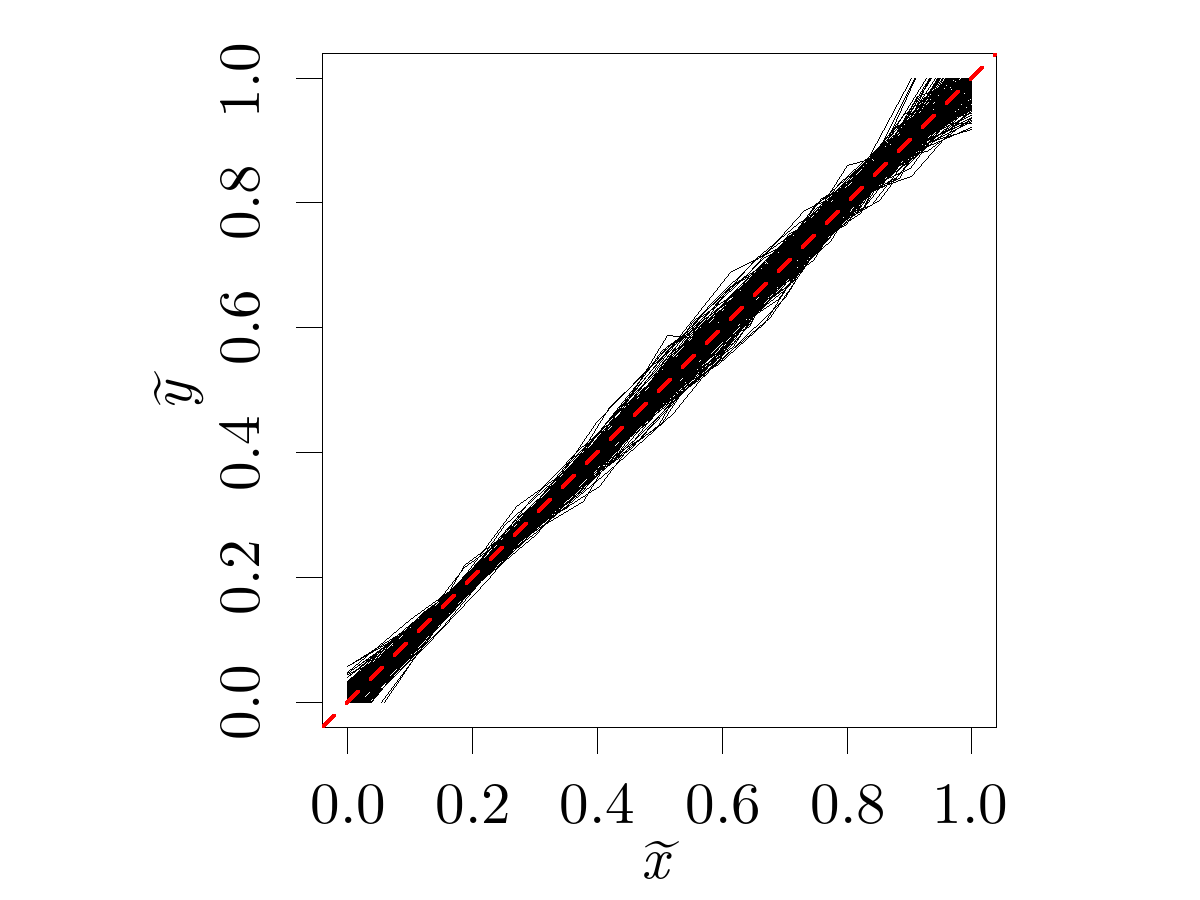}\label{fig_long_range_correlations_sentences_normalized_scaling}}
\hfill
\subfloat[]{\includegraphics[width=0.48\textwidth]{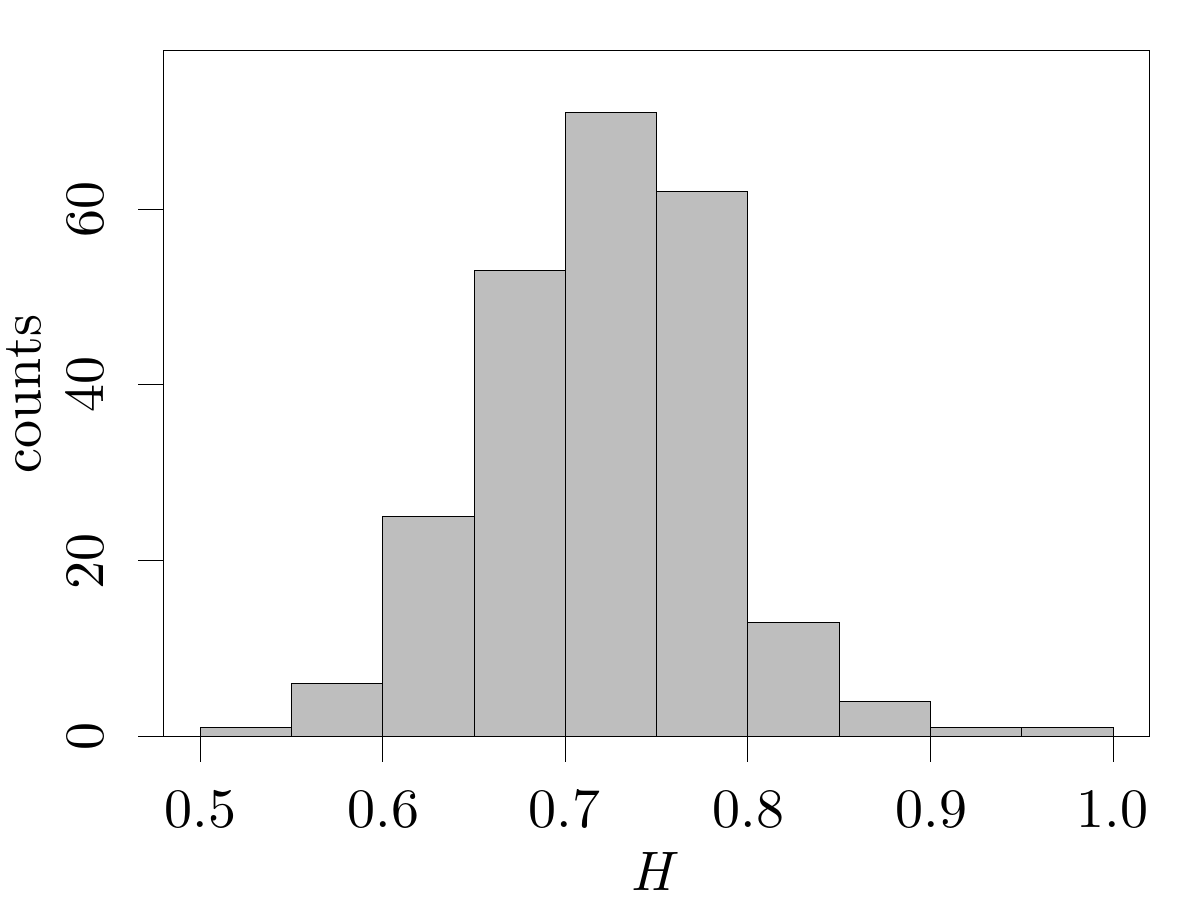}\label{fig_long_range_correlations_sentences_H_histogram}}
\hfill
\subfloat[]{\includegraphics[width=0.48\textwidth]{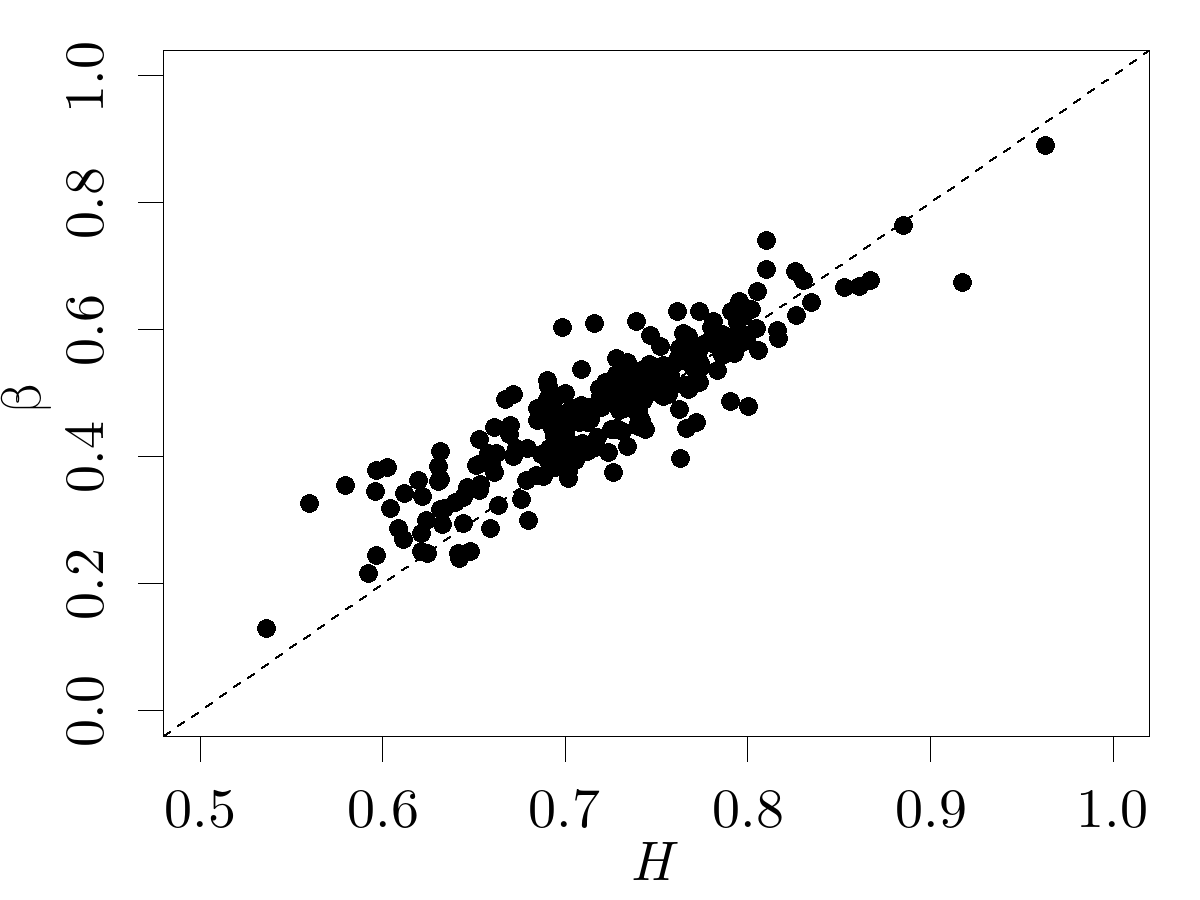}\label{fig_long_range_correlations_sentences_beta_vs_H}}
\hfill
\caption{The properties of time series representing sentence lengths, quantified by spectral densities and Hurst exponents, for sample books. (a) Spectral densities $S(f/f_b)$ for each of the studied texts, plotted in the range of small $f$ in the form of a log-log plot ($f_b$ denotes the fundamental frequency of the DFT). Each solid line corresponds to one text. Determining $S(f/f_b)$ for a given series involves splitting the series into 3 segments of equal length, computing the periodograms within the segments, averaging the results, smoothing, and restricting the range of $f/f_b$ to the one presented in the plot. Series are normalized to have the same power in the considered range of frequencies. It can be seen that for small $f/f_b$ (spanning more than two orders of magnitude), the signals exhibit a $1/f^\beta$ behaviour. The dotted and the dashed line represent, respectively, the slopes determined by the 10th and the 90th percentile of the distribution of estimated $\beta$ (taken with minus sign); the values of the percentiles are 0.34 and 0.61. (b) The histogram of the spectral density exponents $\beta$, obtained by fitting linear relationships to log-log plots of $S(f/f_b)$. (c)~A~plot demonstrating power-law behaviour of fluctuation functions $F_2(s)$, computed with DFA. The plot presents $\widetilde{y}(\widetilde{x})$, where $\widetilde{x}$ and $\widetilde{y}$ are normalized coordinates, obtained by setting $x = \log s$ and $y = \log \left(F_2(s)\right)$ in Eqs.~(\ref{eq_linearity_assessment_norm_coords_1}) and~\ref{eq_linearity_assessment_norm_coords_2}. Each solid line represents one text; its deviation from the $\widetilde{y} = \widetilde{x}$ relationship (dashed line) corresponds to a deviation from a power law. (d) The histogram of the Hurst exponents $H$. (e) The relationship between $H$ and $\beta$. Each point represents one text and dashed line is given by the equation $\beta = 2H-1$.}
\label{fig_long_range_correlations_sentences}
\end{minipage}
\end{figure}

Fig.~\ref{fig_long_range_correlations_sentences_spectra} shows spectral densities of time series representing sentence lengths, for 239 books in 7 languages. The series seem to behave as $1/f^{\beta}$ signals, with $\beta$ depending on the text. The histogram of the values of $\beta$, obtained by fitting lines to log-log plots of spectra $S(f)$, is shown in Fig.~\ref{fig_long_range_correlations_sentences_beta_histogram}. Typical values of $\beta$ lie between 0.2 and 0.8. The presence of long-range correlations is confirmed by observing power-law behaviour of fluctuation functions (Fig.~\ref{fig_long_range_correlations_sentences_normalized_scaling}; see Section \ref{sect::multifractal_analysis_of_time_series} for the definition of fluctuation functions), yielding Hurst exponents $H$ greater than 0.5 (Fig.~\ref{fig_long_range_correlations_sentences_H_histogram}). The correspondence between Hurst exponents $H$ and the exponents of spectral densities $\beta$ is presented in Fig.~\ref{fig_long_range_correlations_sentences_beta_vs_H}; it can be seen that the data conforms to an approximate relationship $\beta = 2H - 1 $. That relationship, mentioned before for fractional Gaussian noises (Eq.~(\ref{eq_powerlaw_fGn_spectrum})), can be considered a more general result, holding approximately for a wider class of signals~\cite{Heneghan2000}.

Assessing the compliance of fluctuation functions $F_2(s)$ with power-law behaviour used to determine Hurst exponents can be done by inspecting the linearity of relevant log-log plots. To present all of them in a single figure, one can use a linear transformation which makes each of the plots fit in the square $[0,1]\!\times\![0,1]$ and in which the linearity of original plot is transformed into the linearity with slope 1 and intercept 0. The transformation having those properties is defined as follows. Let $y(x)$ be the relationship between finite sets of values $x$ and $y$ whose linearity is investigated. Let ($x_{\rm min}$,$x_{\rm max}$,$y_{\rm min}$,$y_{\rm max}$) be the minimum and the maximum values of $x$ and $y$, respectively. Let $y=a+bx$ be the equation describing the assumed linear relationship. The boundaries of the rectangle enclosing the $y(x)$ plot, denoted by $x_{\rm plot.min}$, $x_{\rm plot.max}$, $y_{\rm plot.min}$, $y_{\rm plot.max}$, are defined as follows:
\begin{equation}
\begin{aligned} 
& x_{\rm plot.min} = \min \left\{ x_{\rm min}, \, \frac{y_{\rm min}-a}{b} \right\}, \\
& x_{\rm plot.max} = \max \left\{ x_{\rm max}, \, \frac{y_{\rm max}-a}{b} \right\}, \\
& y_{\rm plot.min} = \min \left\{ y_{\rm min}, \, a + b\,x_{\rm min} \right\}, \\
& y_{\rm plot.max} = \max \left\{ y_{\rm max}, \, a + b\,x_{\rm max} \right\}.
\end{aligned}
\label{eq_linearity_assessment_norm_coords_1}
\end{equation}
The transformation from $(x,y)$ to the normalized coordinates $(\widetilde{x},\widetilde{y})$, which fit in the unit square and in which $y=a+bx$ is transformed into $\widetilde{y} = \widetilde{x}$, is given by:
\begin{equation}
\begin{aligned}
&\widetilde{x}=\frac{x-x_{\rm plot.min}}{x_{\rm plot.max}-x_{\rm plot.min}}, \\
&\widetilde{y}=\frac{y-y_{\rm plot.min}}{y_{\rm plot.max}-y_{\rm plot.min}}.
\end{aligned}
\label{eq_linearity_assessment_norm_coords_2}
\end{equation}
When multiple data sets $(x,y)$, transformed with the given equations, are all presented in one $\widetilde{y}$ vs. $\widetilde{x}$ plot, and all the data points lie close to the line $\widetilde{y} = \widetilde{x}$, then all the original sets $(x,y)$ can be considered to approximately conform to linear relationships (with possibly different slopes and intercepts). So a collective plot of $\widetilde{y}$ vs. $\widetilde{x}$ for multiple linear fits can serve as a tool for a qualitative assessment of the linear relationships' detection validity. The presented idea is applied in Fig.~\ref{fig_long_range_correlations_sentences_normalized_scaling}. The log-log plots of the fluctuation functions $F_2(s)$ computed for the studied texts are transformed to normalized coordinates $(\widetilde{x},\widetilde{y})$ by setting $x = \log s$ and $y = \log \left(F_2(s)\right)$ in the procedure given above.

\subsubsection{Punctuation mark waiting times}

Dividing a text into sentences seems quite natural -- a sentence typically constitutes a complete, closed structure, capable of expressing a concrete thought. Such a partition is also meaningful from a quantitative point of view. A sentence can be treated as a sequence of words between two appropriate punctuation marks. So the length of a sentence can be interpreted as the ``waiting time'' for the next such mark, right after the previous one is encountered. Here, ``time'' is measured by the number of words. If, for example, instead of punctuation marks used to end sentences, one considers some selected words as the delimiters of the sequences of other words, the waiting times are no longer multifractal~\cite{DrozdzS-2016a}. This result in a sense confirms the significance of the multifractal analysis of sentence lengths, as it indicates that their multifractality is not a spurious effect.

Another way of partitioning a text into word sequences and representing it as a signal is based on considering all punctuation marks, instead of only the ones used to end a sentence. A time series can be formed of the punctuation waiting times, that is, the lengths of the word sequences between consecutive punctuation marks. Although it may seem somewhat artificial, from a certain point of view a time series of punctuation waiting times can constitute a representation encoding useful information. The historical origins of the use of punctuation in written language are related to the attempts to split texts into pieces in order to make reading in public more manageable~\cite{Parkes1993}. Punctuation was less specialized and less standardized than today. The classification of punctuation marks and the rules of their usage have been established in modern times. Therefore, an approach in which all punctuation marks are treated as symbols indicating the presence of some kind of a pause seems justified. The ``pauses'' do not have to be related to reading out loud -- they might be necessary to keep the logical consistency of the text or to avoid ambiguity, for instance. So it can be postulated that punctuation marks act as boundaries for word sequences which are separated from others logically, grammatically, or in the way that facilitates comprehension and reading.

\begin{figure}[!htp]
\centering
\begin{minipage}{\figurecustomwidth}
\centering
\subfloat[]{\includegraphics[width=0.48\textwidth]{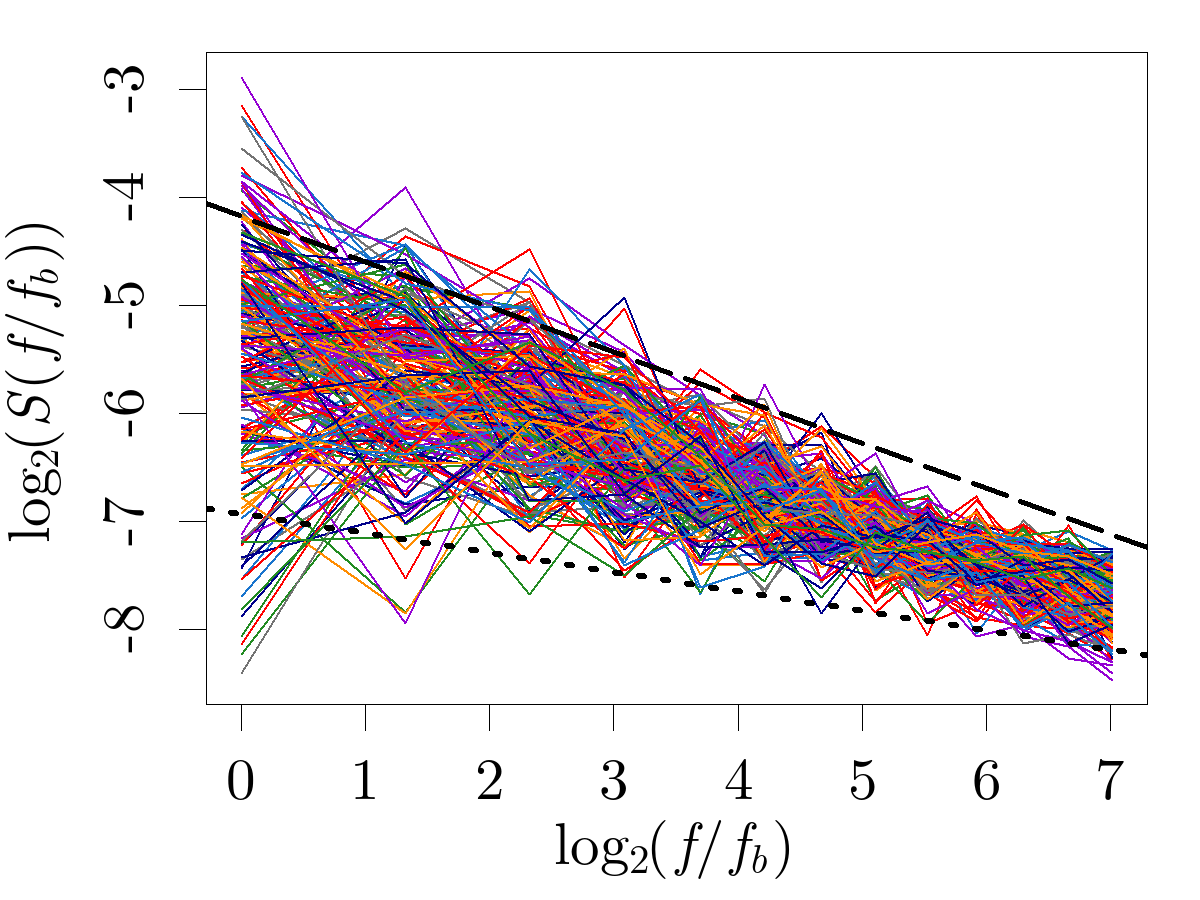} \label{fig_long_range_correlations_all_punctuation_marks_spectra}}
\hfill
\subfloat[]{\includegraphics[width=0.48\textwidth]{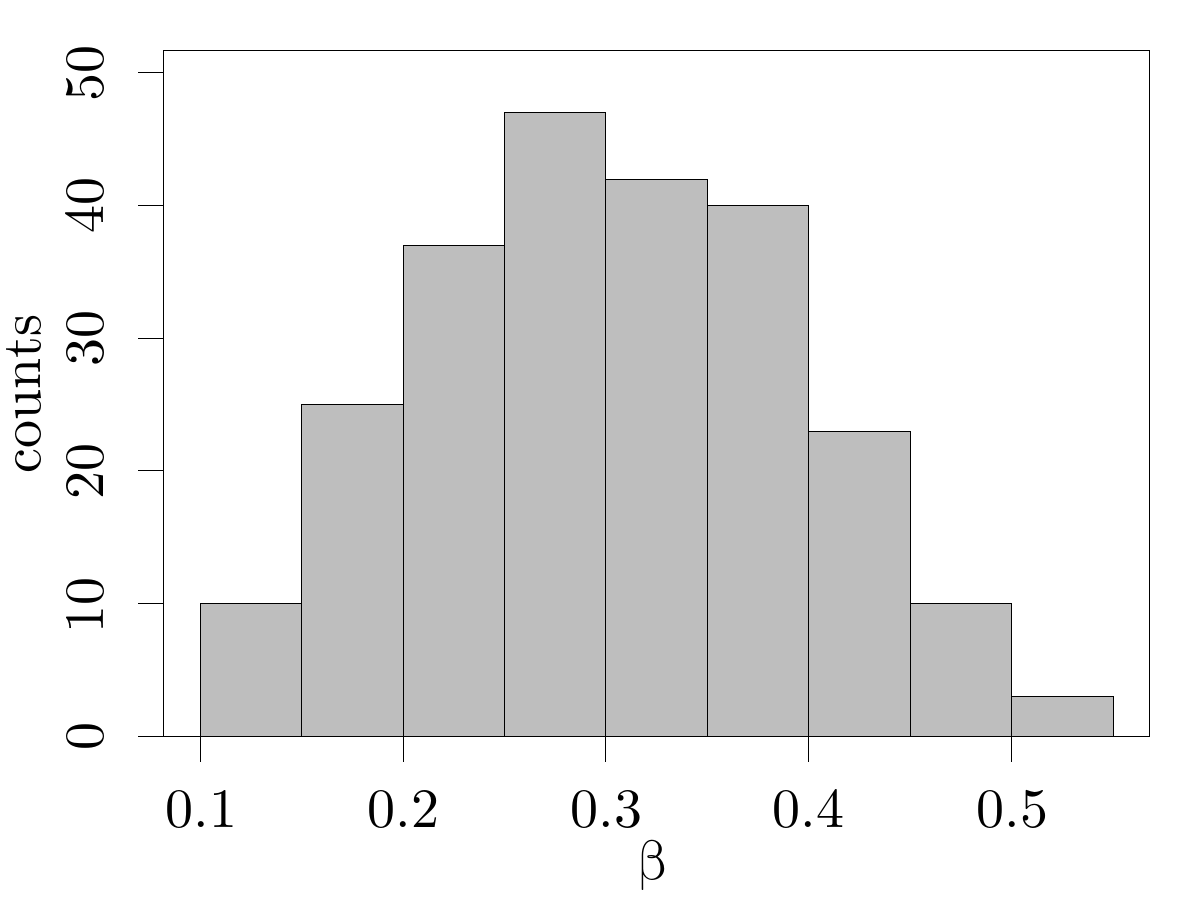}\label{fig_long_range_correlations_all_punctuation_marks_beta_histogram}}
\hfill
\subfloat[]{\includegraphics[width=0.48\textwidth]{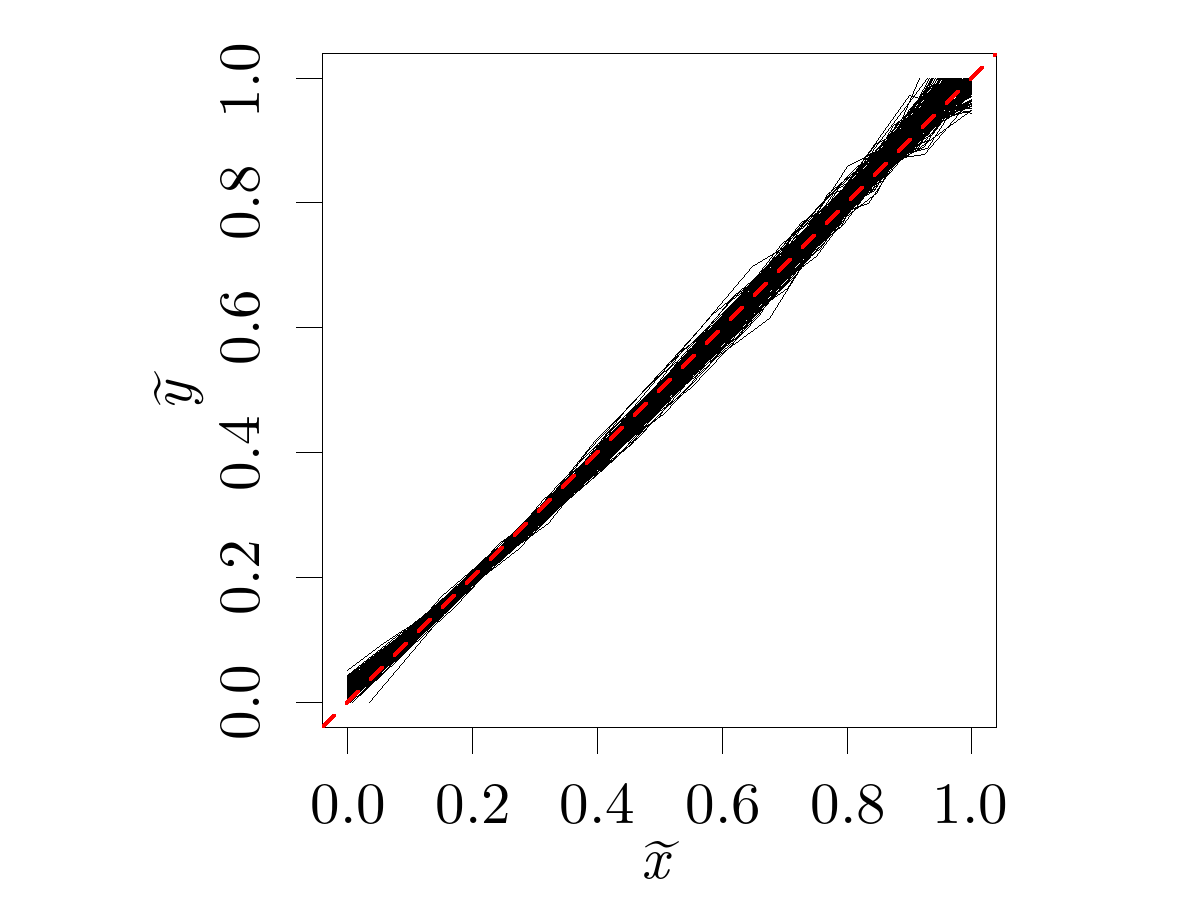}\label{fig_long_range_correlations_all_punctuation_marks_normalized_scaling}}
\hfill
\subfloat[]{\includegraphics[width=0.48\textwidth]{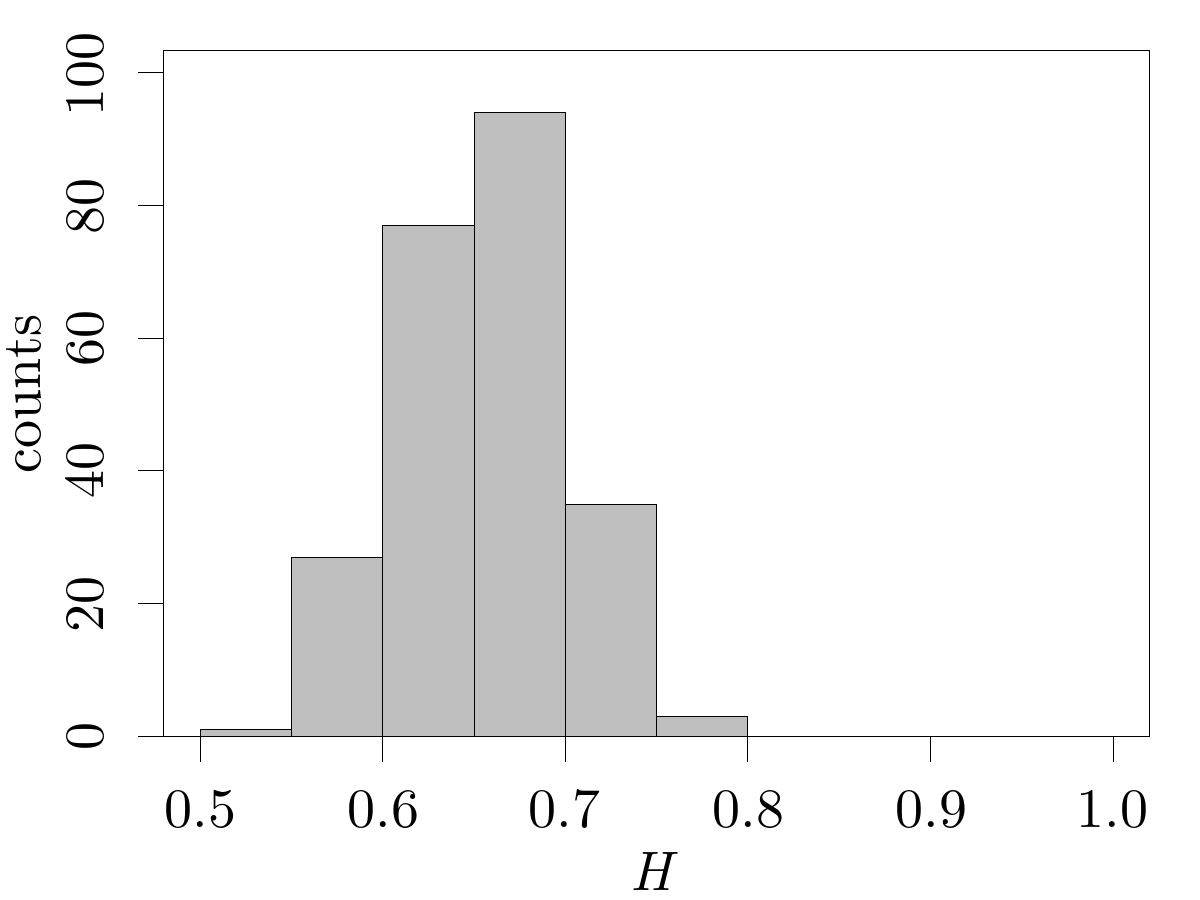}\label{fig_long_range_correlations_all_punctuation_marks_H_histogram}}
\hfill
\subfloat[]{\includegraphics[width=0.48\textwidth]{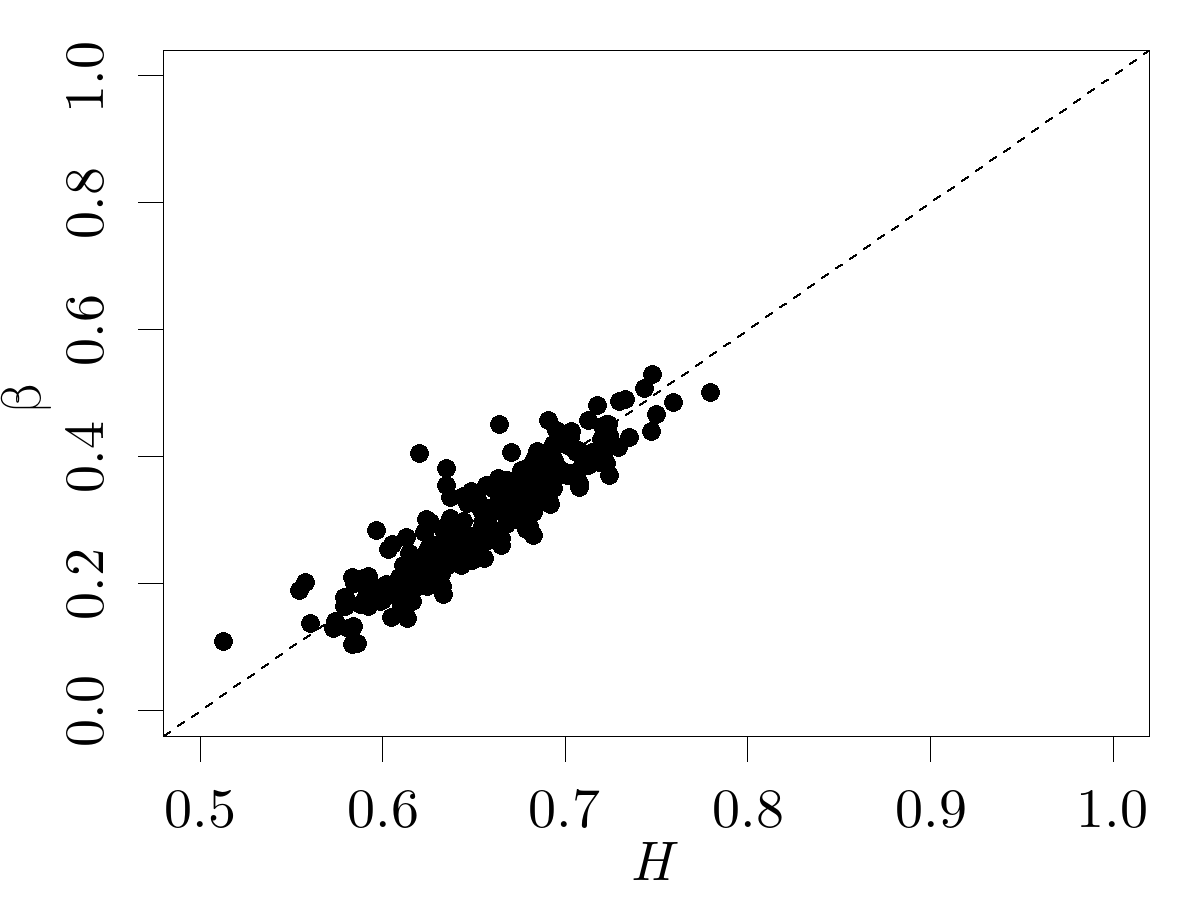}\label{fig_long_range_correlations_all_punctuation_marks_beta_vs_H}}
\hfill
\caption{The properties of time series representing punctuation waiting times (numbers of words between consecutive punctuation marks), quantified by spectral densities and Hurst exponents for sample books. (a) Spectral densities $S(f/f_b)$ for each of the studied texts, plotted in the range of small $f$ in the form of a log-log plot ($f_b$ denotes the fundamental frequency of the DFT). Each solid line corresponds to one text. Determining $S(f/f_b)$ for a given series involves splitting the series into 3 segments of equal length, computing the periodograms within the segments, averaging the results, smoothing, and restricting the range of $f/f_b$ to the one presented in the plot. Series are normalized to have the same power in the considered range of frequencies. It can be seen that for small $f/f_b$ (spanning more than two orders of magnitude), the signals exhibit a $1/f^\beta$ behaviour. The dotted and the dashed line represent, respectively, the slopes determined by the 10th and the 90th percentile of the distribution of estimated $\beta$ (taken with minus sign); the values of the percentiles are 0.18 and 0.42. (b) The histogram of the spectral density exponents $\beta$, obtained by fitting linear relationships to log-log plots of $S(f/f_b)$. (c) A plot demonstrating power-law behaviour of fluctuation functions $F_2(s)$, computed with DFA. The plot presents $\widetilde{y}(\widetilde{x})$, where $\widetilde{x}$ and $\widetilde{y}$ are normalized coordinates, obtained by setting $x = \log s$ and $y = \log \left(F_2(s)\right)$ in Eqs.~(\ref{eq_linearity_assessment_norm_coords_1}) and~(\ref{eq_linearity_assessment_norm_coords_2}). Each solid line represents one text; its deviation from the $\widetilde{y} = \widetilde{x}$ relationship (dashed line) corresponds to a deviation from a power law. (d) The histogram of the Hurst exponents $H$. (e) The relationship between $H$ and $\beta$. Each point represents one text and dashed line is given by the equation $\beta = 2H-1$.}
\label{fig_long_range_correlations_all_punctuation_marks}
\end{minipage}
\end{figure}

\begin{figure}
\centering
\begin{minipage}{\figurecustomwidth}
\centering
\includegraphics[width=0.6\textwidth]{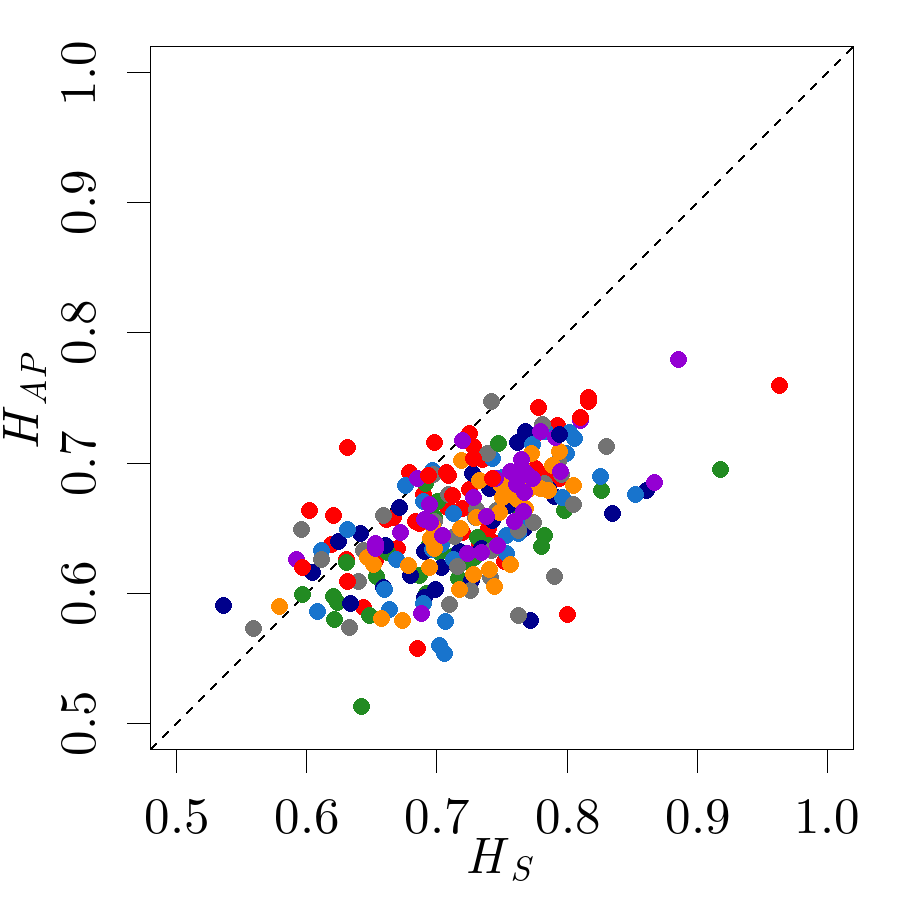}
\caption{The scatterplot of the Hurst exponent of punctuation waiting times $H_{\rm AP}$ versus the Hurst exponent of sentence lengths $H_{\rm S}$ for sample books. Each point represents one text and colors correspond to languages: red - English, green - German, dark blue - French, light blue - Italian, gray - Spanish, orange - Polish, and purple - Russian. Dashed line has the equation $H_{\rm AP} = H_{\rm S}$. The Pearson correlation coefficient between $H_{\rm S}$ and $H_{\rm AP}$ in the whole dataset is equal to 0.59.}
\label{fig_long_range_correlations_H_punctuation_marks_vs_H_sentences}
\end{minipage}
\end{figure}

As is the case with sentence lengths, punctuation waiting times in literary texts exhibit long-range correlations and behave as $1/f^\beta$ signal. Fig.~\ref{fig_long_range_correlations_all_punctuation_marks} shows the spectral densities $S(f/f_b)$, the histogram of the spectra's exponents $\beta$, the scaling of the fluctuation functions $F_2(s)$ (in normalized coordinates defined by Eq.~(\ref{eq_linearity_assessment_norm_coords_1}) and Eq.~(\ref{eq_linearity_assessment_norm_coords_2})), and the relationship between the values of $\beta$ and Hurst exponents $H$ for sample books. The punctuation marks taken into consideration are: period, question mark, exclamation mark, ellipsis, comma, dash, semicolon, colon, left  parenthesis and right parenthesis. Symbols not present on that list and not being words (quotation marks, for instance) are removed from the texts. The time series are formed of non-zero waiting times (a waiting time equal to zero occurs when two punctuation marks are placed next to each other, for example when a question mark is followed by an exclamation mark. Since such cases correspond to a single ``pause'' in a text, they can be disregarded in the construction of the series). It can be observed that punctuation waiting times usually have the Hurst exponent $H$ lower than the Hurst exponent of sentence lengths in the same text (Fig.~\ref{fig_long_range_correlations_H_punctuation_marks_vs_H_sentences}). Nevertheless, punctuation waiting times have the value of $H$ still above 0.5, which indicates their persistence.

\emph{Sentence randomization.} The Hurst exponents of sentence lengths $H_{\rm S}$ and the Hurst exponents of punctuation waiting times $H_{\rm AP}$ have values above 0.5 and are correlated (the Pearson correlation coefficient between $H_{\rm S}$ and $H_{\rm AP}$ is equal to 0.59), as evidenced in Fig.~\ref{fig_long_range_correlations_H_punctuation_marks_vs_H_sentences}. This raises a question about how the properties of these two types of series are related, and whether the relationship between their Hurst exponents is a consequence of the way in which the series are constructed, or whether it can be attributed to other factors. Since sentence-ending punctuation marks constitute a subset of all punctuation marks used in written language, it seems natural that the properties of sentence lengths and of punctuation waiting times are not entirely independent. To way of approaching that issue quantitatively is to investigate the behaviour of both types of series, randomized in the way that keeps the other series unchanged. A randomization of text's sentence lengths which does not alter the series of punctuation waiting times can be done by permuting randomly the positions of all the punctuation marks present in the text. The set of punctuation marks' positions remains unchanged, but which mark occupies which position is decided by chance. Therefore, sentences lose their original structure, as sentence-ending marks are located at random positions allowed by the overall arrangement of punctuation marks. This method of randomization models the situation in which punctuation is placed as in the original text, but sentence lengths, apart from satisfying the condition that they are delimited by symbols belonging to an appropriate subset of punctuation, are completely random.

To perform randomization the other way round, a procedure described below can be used. The sentences in a text can be grouped into ``buckets'', each bucket corresponding to particular range of lengths. For example, in such a partition, one bucket might consist of all the sentences in the text which have lengths equal to 1 or~2, another bucket can contain all the sentences of length 3, yet another bucket can be composed of all the sentences with lengths between 12 and 18, and so on. Each sentence in the text needs to be assigned to a (single) bucket. The range of lengths covered by a bucket should be as narrow as possible, provided that each of the buckets contains at least a few (3-5) sentences. Randomization consists of assigning sentences to buckets, and permuting the positions of sentences inside each bucket. This means that sentences randomly swap positions with other sentences belonging to the same bucket (having the same or similar length). Consequently, the series representing sentence lengths is approximately the same as the original one (the exact level of agreement depends on the length ranges used to define buckets), but the contents of sentences (including punctuation) become randomly scattered across the text. However, it should be noted that the resulting the series of punctuation waiting times can be affected by statistical relationships binding the structure of punctuation inside a sentence with sentence length. An example of such a relationship is the one expressed by the Menzerath-Altmann law. For sentences, the law can summarized by the following statement: the longer a sentence, the smaller the average size of the constituents it is composed of. Under the assumption that sentences can be divided into constituents separated by punctuation marks, the Menzerath-Altmann law results in a tendency of punctuation waiting times to be short in regions where sentences are long, and to be long in the parts of texts in which sentences are short.

\begin{figure}
\centering
\begin{minipage}{\figurecustomwidth}
\centering
\subfloat[]{\includegraphics[width=0.47\textwidth]{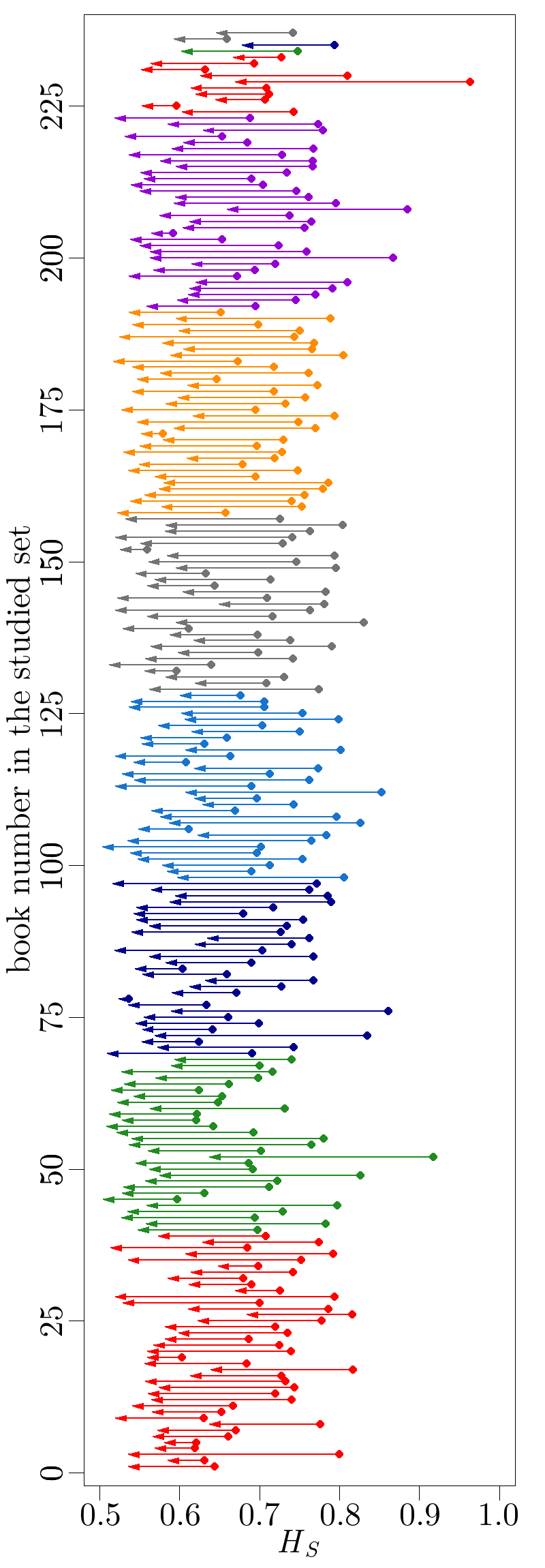}}
\hfill
\subfloat[]{\includegraphics[width=0.47\textwidth]{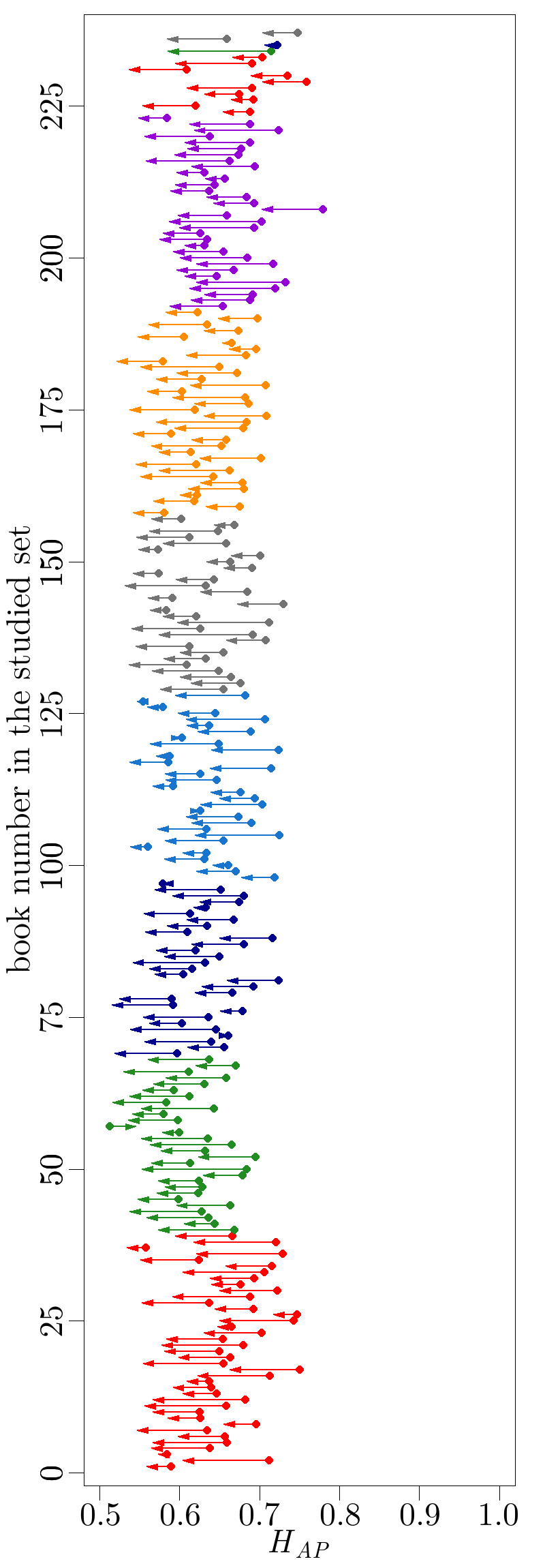}}
\caption{The Hurst exponents of sentence lengths $H_{\rm S}$ and of punctuation waiting times $H_{\rm AP}$, computed for the original and the randomized time series for sample books. (a) Randomized sentence lengths preserving $H_{\rm AP}$. (b) Randomized punctuation waiting times preserving $H_{\rm S}$. Arrows mark the change of the Hurst exponent induced by randomization -- dots denote the Hurst exponents of the original series, arrow heads denote the Hurst exponents of the randomized series (computed as an average over 5 independent randomizations). Consecutive dot-arrow pairs represent consecutive books from the dataset, while colors correspond to languages: red - English, green - German, dark blue - French, light blue - Italian, gray - Spanish, orange - Polish, and purple - Russian.}
\label{fig_long_range_correlations_two_randomizations}
\end{minipage}
\end{figure}

Fig.~\ref{fig_long_range_correlations_two_randomizations} presents how the Hurst exponents of sentence lengths and of punctuation waiting times change when the randomization procedures given above are performed on sample texts. Typically, the Hurst exponents of the appropriately randomized series are substantially lower than the exponents of the corresponding original series, but their value is usually still above 0.5. This means that the persistence of sentence lengths and the persistence of punctuation waiting times can be partially explained by each other -- when for a given text one type of series is randomized and the other is kept unchanged, the randomized one exhibits some degree of persistence due to the persistence of the other one. Also,  performing randomization of any of the presented types does not remove the correlations between the Hurst exponents of sentence lengths and of punctuation waiting times -- that is, texts with high Hurst exponents describing sentence lengths $H_{\rm S}$ also tend to have high Hurst exponents pertaining to punctuation waiting times $H_{\rm AP}$. Conversely, low $H_{\rm S}$ typically co-occurs with low $H_{\rm AP}$. Pearson correlation coefficients between $H_{\rm S}$ and $H_{\rm AP}$, describing that effect, is equal to 0.87 for the randomization of sentence lengths (preserving $H_{\rm AP}$) and equal to 0.62 for the randomization of punctuation waiting times (preserving $H_{\rm S}$). So even when one of the two series is random, it is correlated with the other one, provided that the conditions making sentence lengths and punctuation waiting times consistent with each other are satisfied. Therefore, the correlation between $H_{\rm S}$ and $H_{\rm AP}$ can be seen as an effect caused by the fact that sentence-ending punctuation marks constitute a subset of all punctuation marks.

\subsubsection{Word co-occurrence in a concept space}

A different approach was considered in~\cite{AlvarezLacalleE-2006a} where evolution of vectors in a ``concept space'' spanned by selected vocabulary was studied. After a set of meaningful, frequent words was chosen for each of 12 analyzed English books in order to reduce dimensionality of the problem, these books were divided into a number of 200-word-long chunks and a vector defined by these words was constructed for each chunk. The length of chunks was specified to be as long as a typical ``attention window'' of a reader. This window covered those words that a reader can simultaneously be aware of while reading and an attention vector was constructed of these words. Apart from the attention vectors, a global connectivity matrix ${\bf M}$ was created for each book with entries $m_{ij}$ equal to how many times two words $w_i$ and $w_j$ co-occurred in the same chunk. Singular value decomposition (SVD) of the connectivity matrix provided one with a set of ``concept vectors'' that then were considered as a basis on which the attention vectors were projected. The idea behind such an approach was that, for a text than had been unrelated from the concept-space perspective, the evolution of the attention vectors would have been stochastic. In the opposite case, the vector trajectory could reveal temporal correlations. Autocorrelation function applied to the attention vector projections onto a linear superposition of a number $d$ of SVD vectors exhibited a power-law decay with time for different values of $d$. This outcome indicated that there exist genuine long-term correlations in the occurrence of concepts along texts. The power laws were interpreted as a manifestation of hierarchical structures in the organization of texts. There were only quantitative differences between different books, but qualitative results were similar, which suggested that the observed properties were related to some universality of language rather than to specific characteristics of individual texts, authors, or styles. 

\subsubsection{Sentiment content}

Sentiment value can be attributed to a given text sample through a few different approaches, among which the most popular is to use sentiment dictionaries. This attribution can be done at a word, sentence, paragraph or even higher levels. The sentence level was considered in~\cite{HuQ-2021a}, where the novel \emph{Never Let Me Go} by K.~Ishiguro was studied. In order to suppress possible high-frequency fluctuations of the sentiment time series, the signals were subject to nonlinear adaptive filtering~\cite{GaoJ-2011a} and, then, the Hurst analysis. Based on the Hurst exponent behaviour, which fluctuated between 0.5 and 0.75, which corresponds to a firm persistency, it was concluded that the narrative plot evolves coherently and neither a sudden emotional burst that would elevate $H$ close or above 1 nor emotional bi-polarity that would result in $H$ below 0.5 could be observed in the book. Authors were convinced that the Hurst exponent as a measure of the long-range correlations could be a useful tool for quantitative analysis of literary works.


\section{Hazard function}


\subsection{Discrete Weibull distribution}

The distribution of punctuation waiting times in texts can be characterized by two numbers, being the parameters of the so-called discrete Weibull distribution~\cite{StaniszT-2023a}. The distribution can be introduced with the help of the following reasoning. When a text is considered a sequence of words and punctuation marks occurring between some of them, it can be assumed that distributing punctuation marks across text is governed by some process deciding for each consecutive word whether a punctuation mark is to be placed after that word or not. Assuming that the process is random and it puts a punctuation mark after a word with some constant probability $p$, each such decision is a Bernoulli trial with $p$ being the probability of success. In such a case, the punctuation waiting time (the number of words between consecutive punctuation marks) is the number $k$ of trials required to obtain the first success, after the last one observed ($k=1,2,3,...$). The number of trials until the first success in Bernoulli process follows the geometric distribution. Observing a waiting time longer than $k$ is equivalent to not observing a success in the first $k$ trials. Therefore, one can write:
\begin{equation}
1-F(k)=\left( 1-p \right)^k,
\end{equation}
where $F$ is the cumulative distribution function ($F(k)$ is defined as the probability that a waiting time is less than or equal to $k$). The above relationship pertains to situation when punctuation marks are placed independently of each other, with constant probability. However, it is reasonable to anticipate that the probability of placing a punctuation mark after a particular word depends on the sequence of words and punctuation marks preceding the considered word. Hence, a distribution more general than the geometric distribution is required. One way of generalizing the geometric distribution is introducing an exponent, $\beta>0$, into its survival function:
\begin{equation}
1-F(k)=\left( 1-p \right)^{k^\beta}.
\label{eq_discrete_Weibull_CCDF}
\end{equation}
A distribution specified in such a way is a discrete analogue of the Weibull distribution, therefore called the discrete Weibull distribution~\cite{Nakagawa1975}. Due to its flexibility, Weibull distribution, especially in its continuous version, is widely applied in various fields of science and engineering, for instance in survival analysis, in medicine and health sciences, or in modeling natural phenomena like wind speed or rainfall intensity~\cite{Johnson1994,MillerRG-1998a}. Interestingly, it has also been employed in studies on natural language, namely in investigating the distribution of word recurrence times in textual~data~\cite{Altmann2009}.

\begin{figure}
\centering
\begin{minipage}{\figurecustomwidth}
\centering
\subfloat[]{\includegraphics[width=0.49\textwidth]{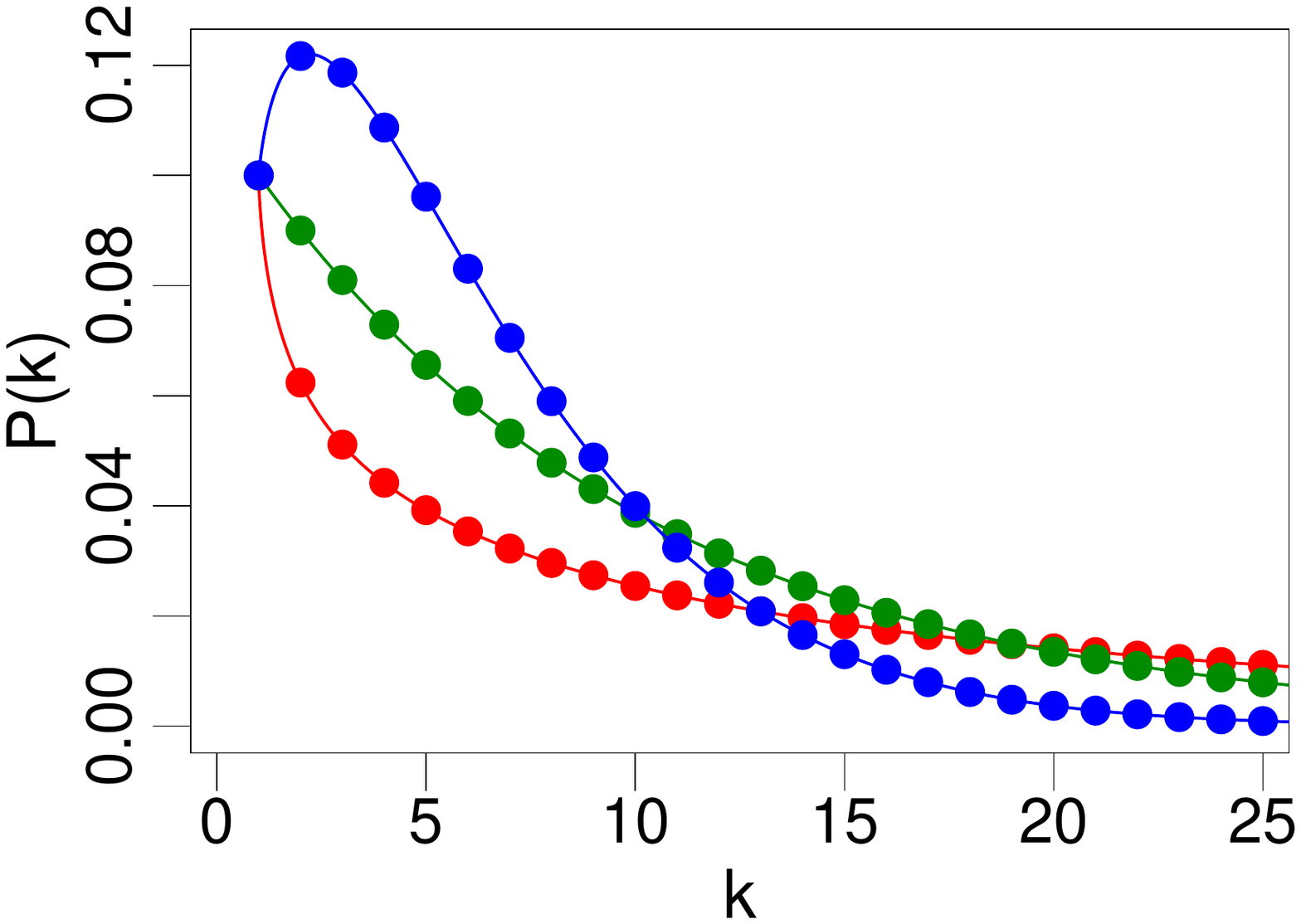}}
\subfloat[]{\includegraphics[width=0.49\textwidth]{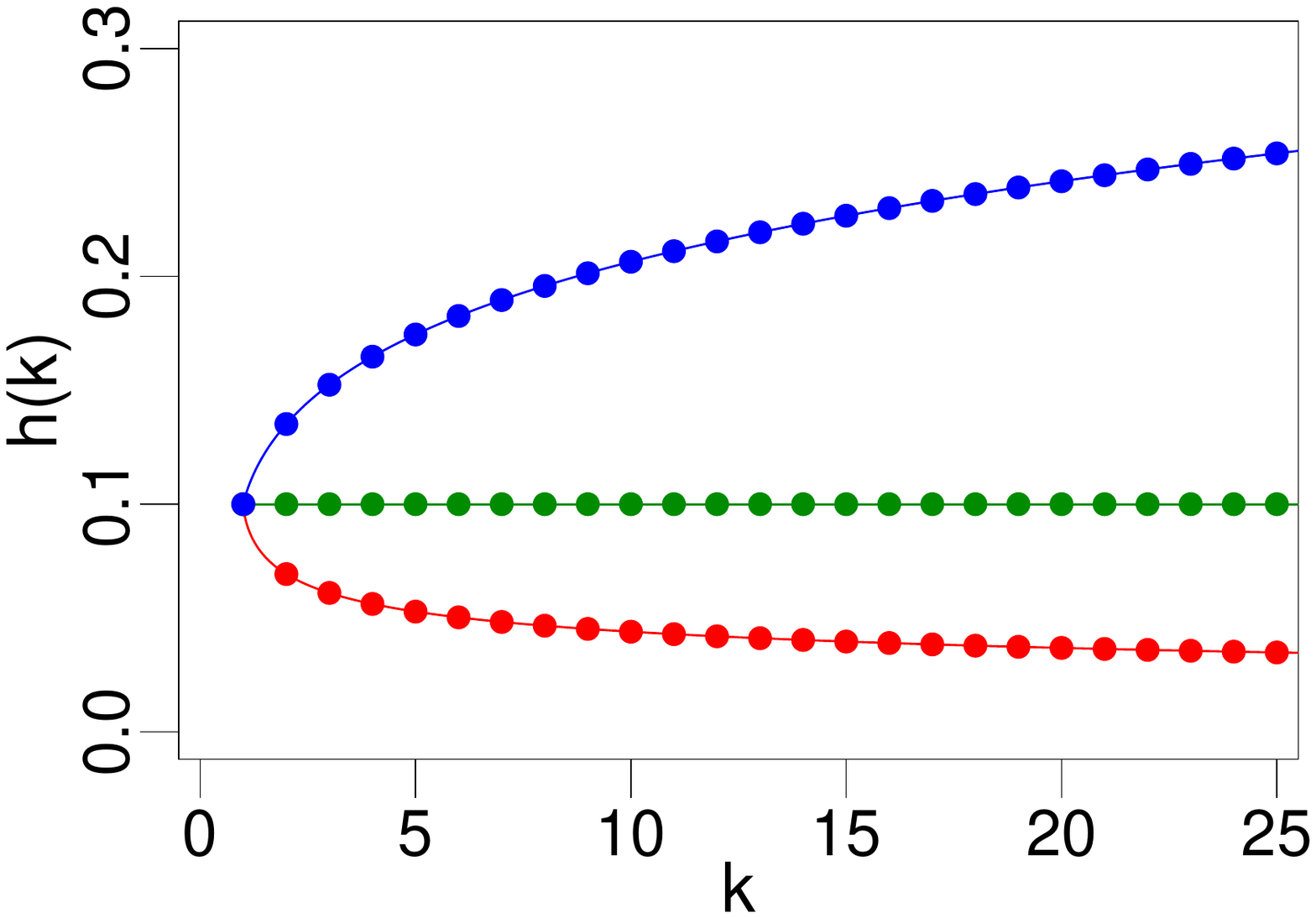}}
\caption{The probability mass functions $P(k)$ (a) and the hazard functions $h(k)$ (b) of the discrete Weibull distribution, for $p=0.1$ and three different values of $\beta$: $\beta=0.75$ (red), $\beta=1$ (green), $\beta=1.25$ (blue). Since the distribution is discrete, $P(k)$ and $h(k)$ are defined only at integer $k$ and their values are represented by dots. The connecting lines are only guide for the eye and do not indicate continuity.}
\label{fig_discrete_Weibull_distribution_examples}
\end{minipage}
\end{figure}

The parameter $\beta$ of the discrete Weibull distribution determines the deviation from the geometric distribution, which is recovered for $\beta=1$. It describes how the probability of obtaining a success depends on how many trials have been performed since the last success observed. This dependence can be characterized by the so-called \textit{hazard function} $h(k)$. Hazard function can be defined as the conditional probability that a success occurs on the $k$-th trial, given that it has not occurred in the preceding $k\!-\!1$ trials. With $P(k)$ denoting the probability mass function, the hazard function of the discrete Weibull distribution is given by:
\begin{equation}
h(k) = \frac{P(k)}{1-F(k-1)} = \; 1-\left(1-p\right)^{k^{\beta} - (k-1)^{\beta}}.
\label{eq_hazard_DWeibull}
\end{equation}
For $\beta<1$, the hazard function is a decreasing function -- the probability of observing a success becomes smaller as the waiting time gets longer. For $\beta>1$, it is increasing with time. For $\beta=1$, the hazard function is a constant -- one obtains the geometric distribution, which is therefore said to be \textit{memoryless}. The parameter $p$ of the discrete Weibull distribution also can be intuitively interpreted -- it is the probability of observing a success in the first trial. The plots presenting the discrete Weibull distribution for selected values of $p$ and $\beta$ are shown in Fig.~\ref{fig_discrete_Weibull_distribution_examples}.

An easy way of assessing how well a given data set fits to a Weibull distribution (both in continuous and discrete case) is constructing the so-called Weibull plot. It is straightforward to show that Eq.~(\ref{eq_discrete_Weibull_CCDF}) can be rewritten as:
\begin{equation}
\log\!\left( -\!\log \!\left( 1\!-\!F(k) \right) \right) = \beta \log k + \log\!\left( -\!\log \left( 1\!-\!p \right) \right).
\end{equation}
Therefore, if the data comes from the discrete Weibull distribution with parameters $(p,\beta)$, then when plotting the empirical cumulative distribution function $F_{\rm emp}(k)$ in coordinates $(x,y)$, where 
\begin{align*} 
&x=\log k \\
&y=\log\left( -\log \left( 1-F_{\rm emp}(k) \right) \right),
\end{align*}
one should observe a straight line with slope $\beta$ and intercept $\log\left( -\log \left( 1-p \right) \right)$. To make comparison between fits to different Weibull distributions easier, one can use the transformation analogous to the one given by Eq.~(\ref{eq_linearity_assessment_norm_coords_1}) and Eq.~(\ref{eq_linearity_assessment_norm_coords_2}) to rescale the coordinates $(x,y)$ to $(\widetilde{x}, \widetilde{y})$, fitting in the square $[0,1]\!\times\![0,1]$. In a plot in rescaled coordinates (here referred to as a \textit{rescaled Weibull plot}), the deviation from the Weibull distribution is observed as the deviation from the line $\widetilde{y} = \widetilde{x}$. 


\subsection{Punctuation mark waiting times}

\begin{figure}[!b]
\centering
\begin{minipage}{\figurecustomwidth}
\centering
\subfloat[\textit{Alice's Adventures in Wonderland}, punctuation waiting times]{\includegraphics[width=0.475\textwidth]{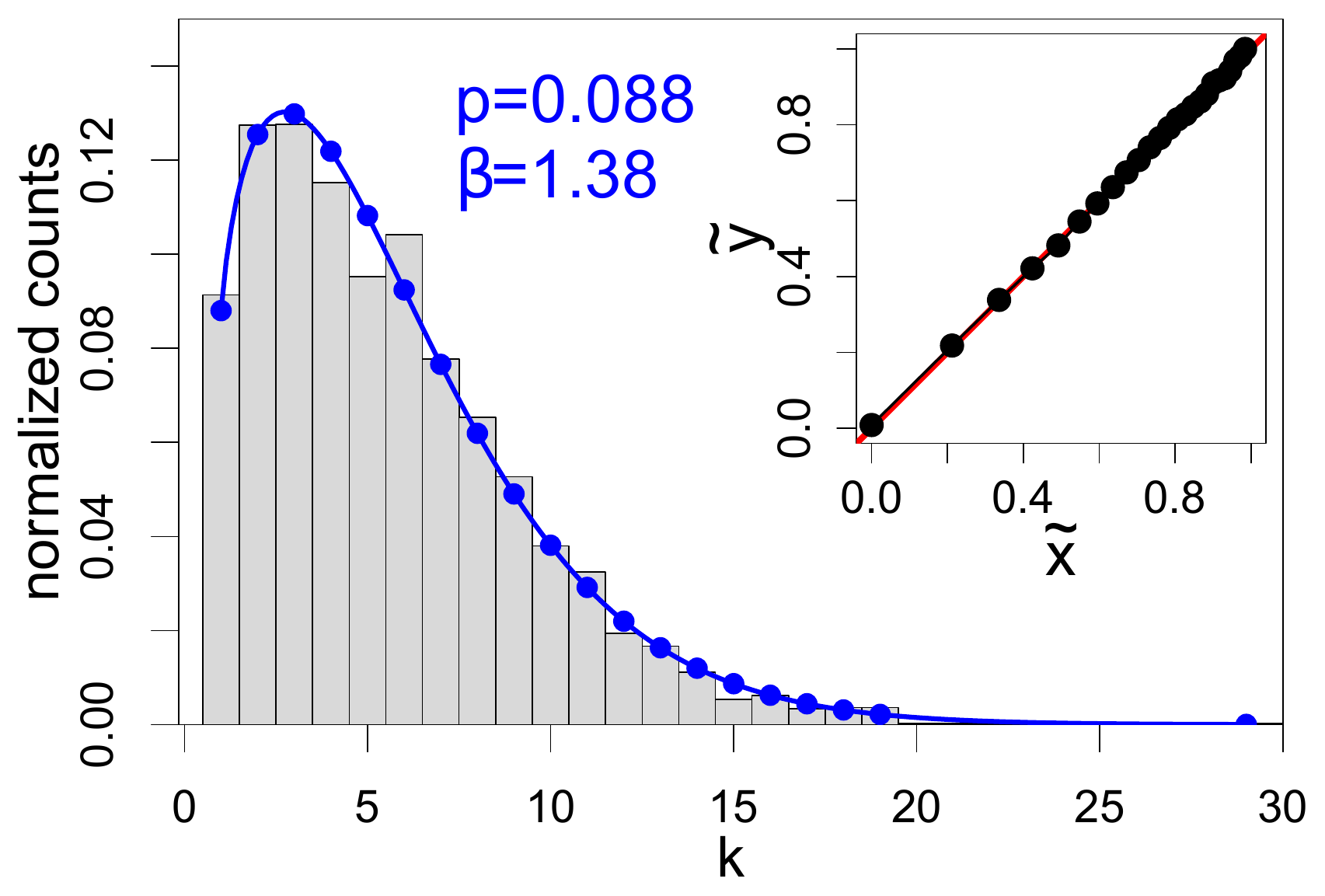}}
\hfill
\subfloat[\textit{David Copperfield}, punctuation waiting times]{\includegraphics[width=0.475\textwidth]{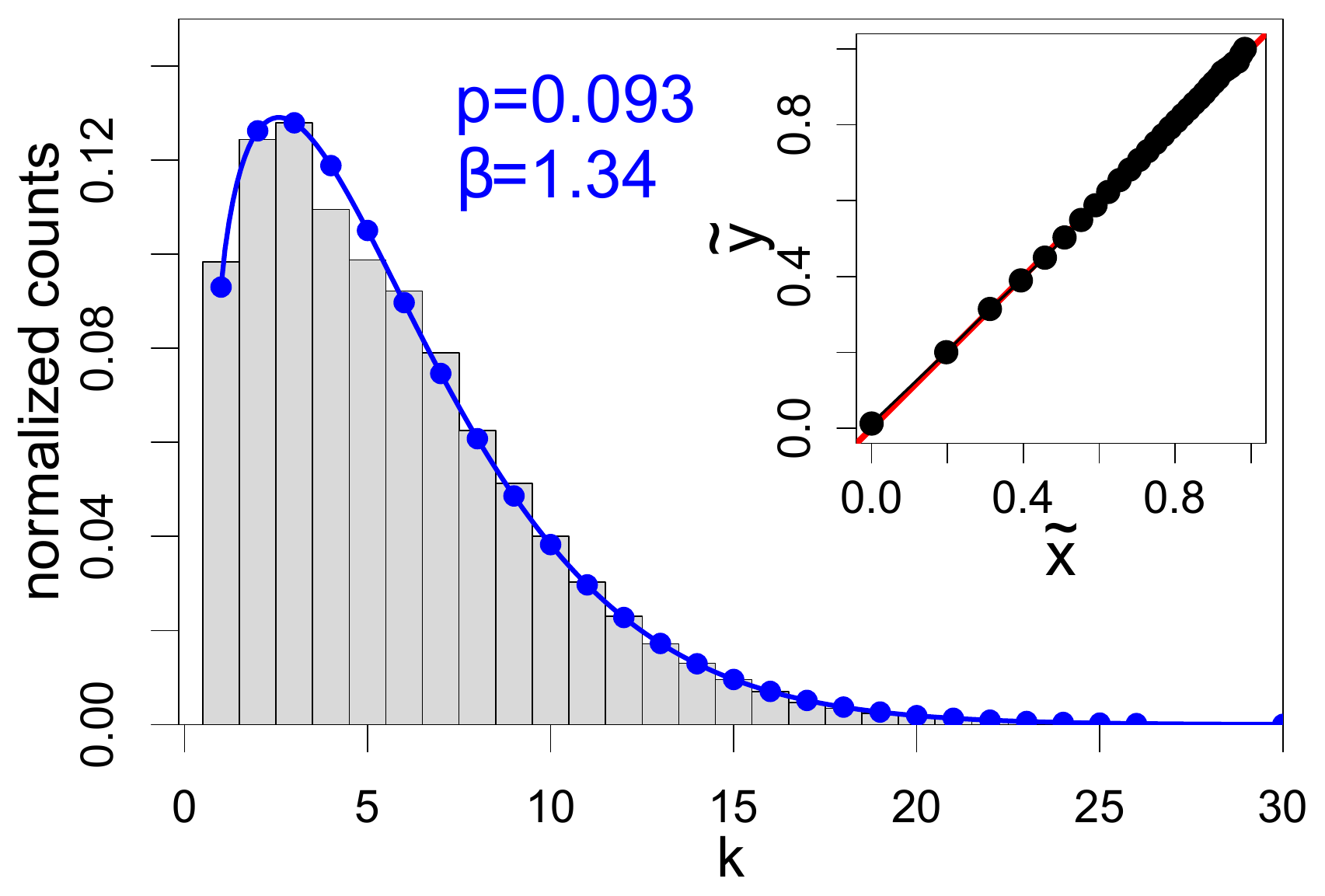}}
\hfill

\vspace{0.5cm}

\subfloat[\textit{Alice's Adventures in Wonderland}, sentence lengths]{\includegraphics[width=0.475\textwidth]{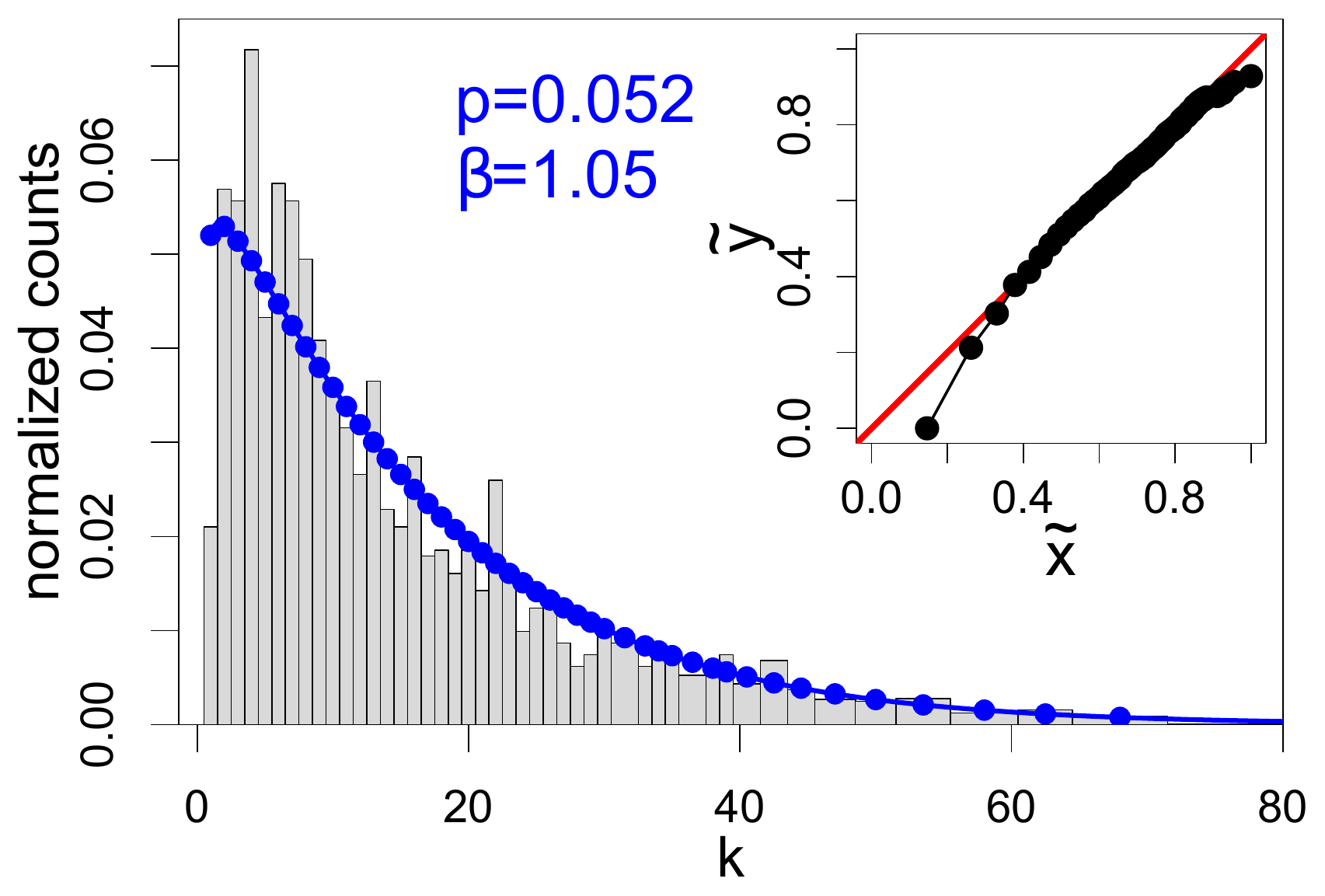}}
\hfill
\subfloat[\textit{David Copperfield}, sentence lengths]{\includegraphics[width=0.475\textwidth]{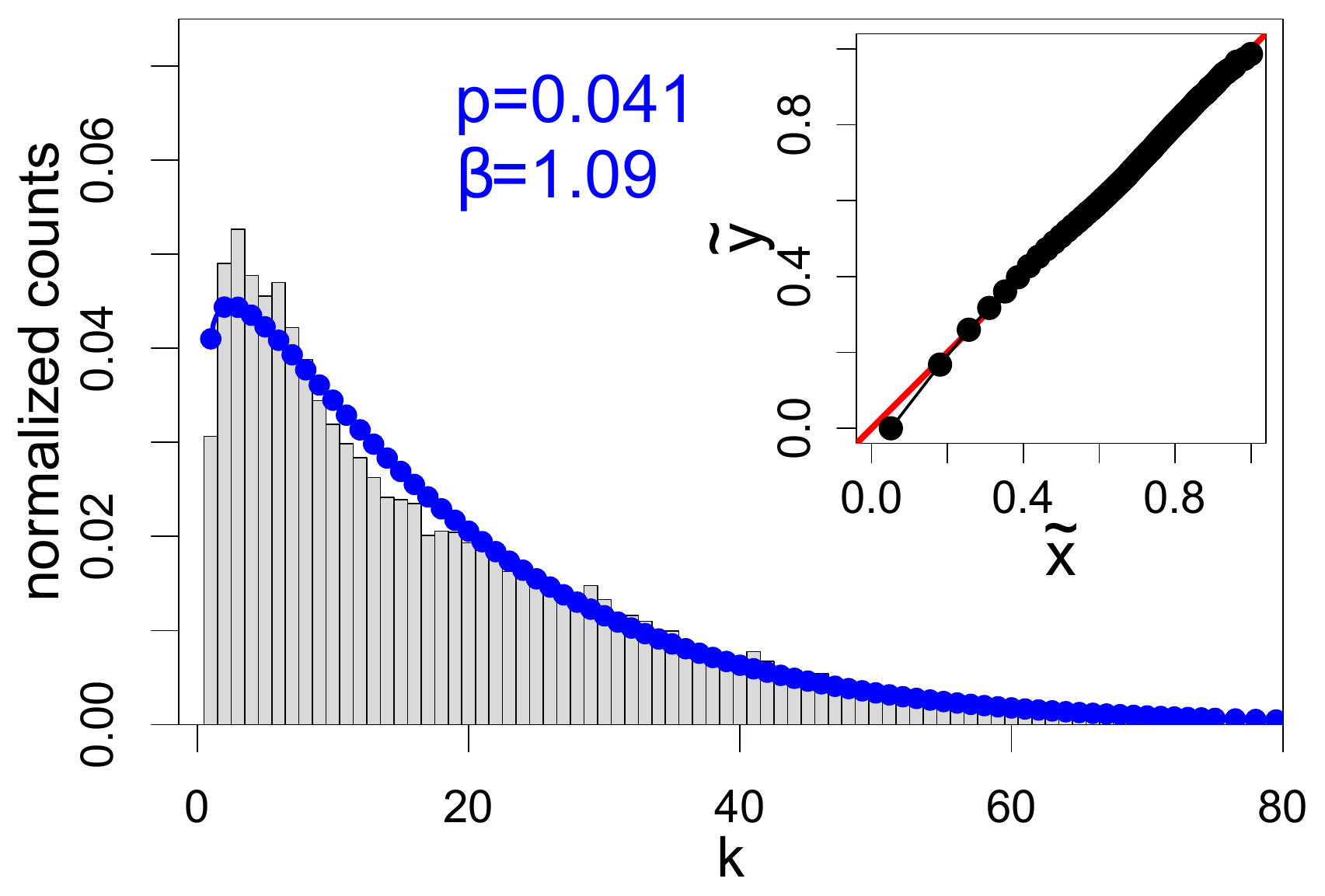}}
\hfill
\caption{Examples of the distributions of punctuation mark waiting times and of sentence lengths, for two books: \textit{Alice's Adventures in Wonderland} by Lewis Carroll and \textit{David Copperfield} by Charles Dickens. In each figure, histogram represents the empirical distribution and blue dots represent the discrete Weibull distribution fitted using maximum likelihood estimation (the obtained parameters $p$ and $\beta$ are given above the plots). Insets show the corresponding rescaled Weibull plots, in which deviations from the line $\widetilde{y} = \widetilde{x}$ correspond to discrepancies between the fitted and the empirical distribution.}
\label{fig_DWeibull_Alice_and_David}
\end{minipage}
\end{figure}

Fig.~\ref{fig_DWeibull_Alice_and_David} shows the empirical distributions of punctuation waiting times and of sentence lengths, for two books: \textit{Alice's Adventures in Wonderland} by Lewis Carroll and \textit{David Copperfield} by Charles Dickens. Discrete Weibull distribution is fitted to the data -- maximum likelihood estimation (MLE) is used to find the parameters of the distribution.
 
\begin{figure}[!b]
\centering
\begin{minipage}{\figurecustomwidth}
\centering
\subfloat[]{\includegraphics[width=0.49\textwidth]{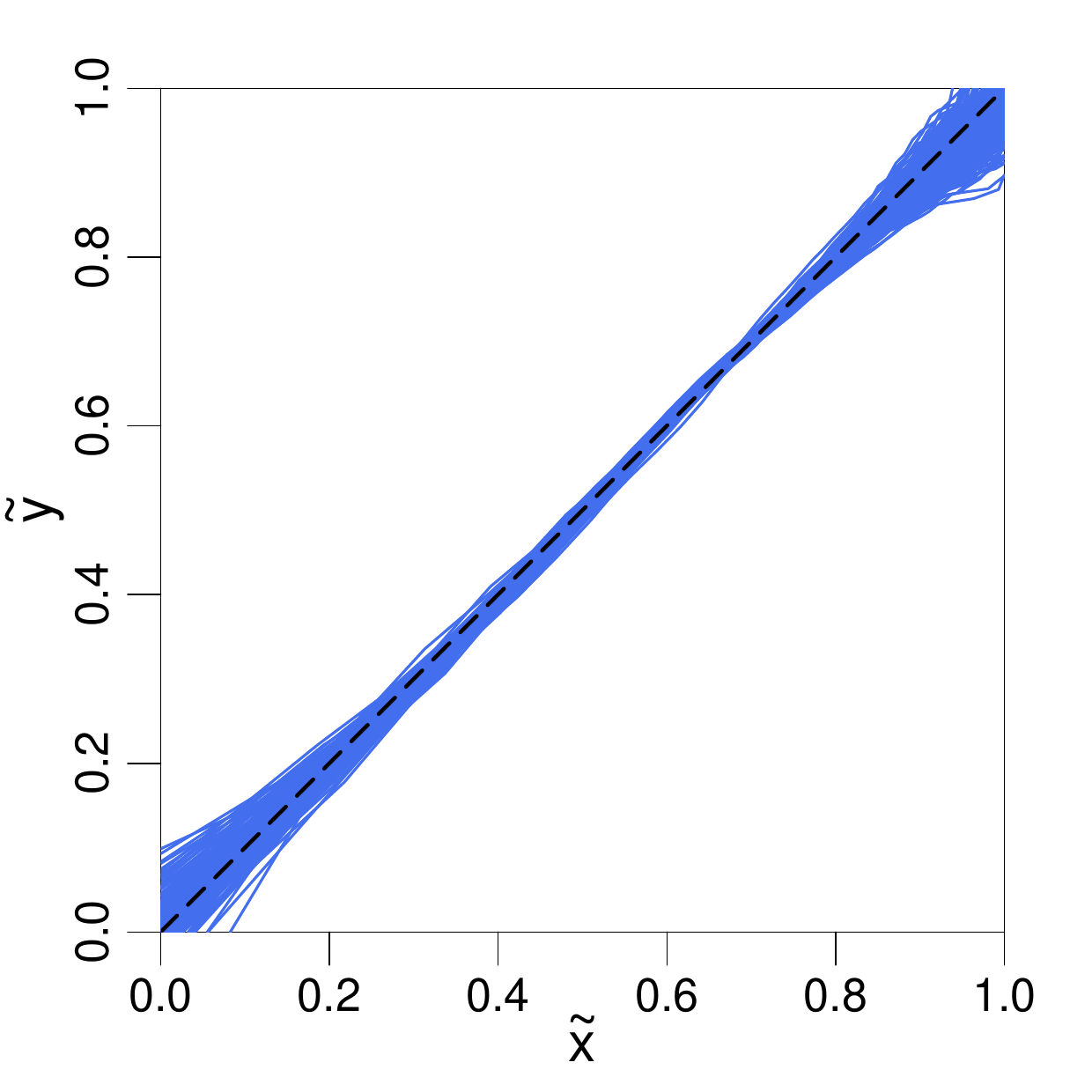}}
\hfill
\subfloat[]{\includegraphics[width=0.49\textwidth]{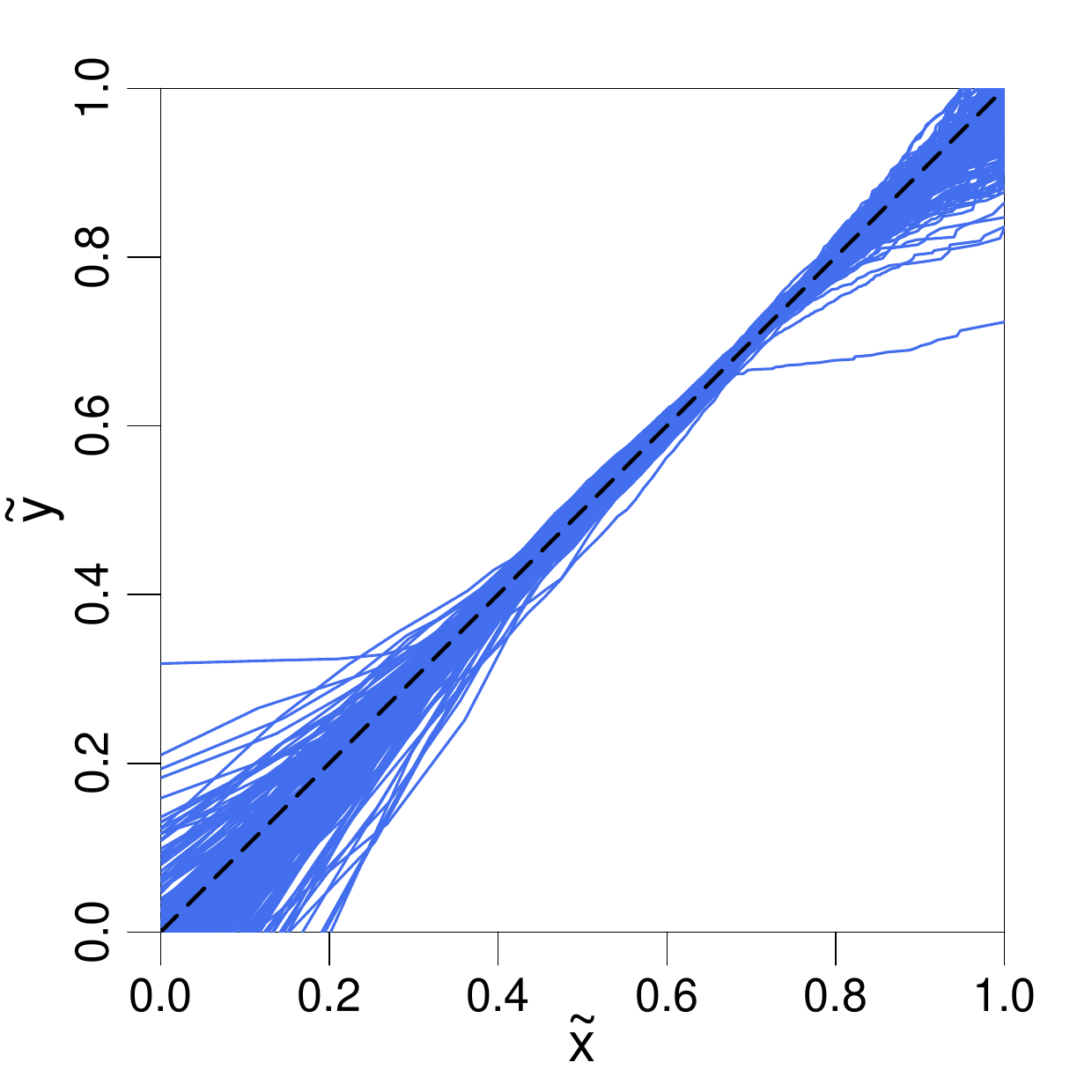}}
\hfill
\caption{The rescaled Weibull plots of punctuation waiting times (a) and sentence lengths (b) for sample books. Each curve on a plot corresponds to one book; dashed line $\widetilde{y} = \widetilde{x}$ represents the ideal fit to the discrete Weibull distribution.}
\label{fig_DWeibull_multiple_books}
\end{minipage}
\end{figure}

It can be seen that punctuation waiting times in both of the books are well described by discrete Weibull distribution, but in case of sentence lengths one of the books (\textit{Alice's Adventures in Wonderland}) exhibits considerably worse agreement between the empirical and the proposed model distribution. It turns out that this  applies also to other texts -- while the distribution of punctuation waiting times can almost universally be modeled by discrete Weibull distribution, the distribution of sentence lengths might either be of the same type or of more ``irregular'' nature (meaning that it is much harder to find a distribution with relatively simple functional form that would accurately represent the data). This fact is demonstrated in Fig.~\ref{fig_DWeibull_multiple_books}, which presents rescaled Weibull plots of punctuation waiting times and of sentence lengths for 223 books in 7 languages. The deviations from the line $\widetilde{y} = \widetilde{x}$ in the rescaled Weibull plots of sentence lengths tend to be significantly larger than the ones observed in the rescaled Weibull plots of punctuation waiting times.

From the viewpoint considering only the probability distribution characterizing punctuation, the process of writing a text can be thought of in terms of a simple mathematical model, based on the properties of the discrete Weibull distribution. The model assumes that a text is generated word by word, and a punctuation mark can be placed after each word, with some probability $h(k)$ which depends only on $k$, the number of words that occurred since the last placed punctuation mark. The relationship between $h$ and $k$ is of the form given in the Eq.~(\ref{eq_hazard_DWeibull}). The resulting distribution of distances between punctuation marks in the text is the discrete Weibull distribution. By adjusting the parameters $p$ and $\beta$ in the function $h(k)$, one can obtain a distribution observed in real texts. The parameters are easily interpreted: $p$ is the probability that a punctuation mark appears right after the first word since the last punctuation mark. $\beta$ describes how fast the probability of the punctuation mark occurrence changes with the growing number of words appearing since the last punctuation mark observed. The assumption that the probability of a punctuation mark occurrence depends only on $k$, which is equivalent to the statement that word sequences between punctuation marks are generated independently, is of course idealized. In real texts, it is obviously violated by the presence of long-range correlations, for instance. However, when only the probability distribution is considered, correlations between punctuation waiting times can be neglected.

\begin{figure}
\centering
\begin{minipage}{\figurecustomwidth}
\centering
\includegraphics[width=0.99\textwidth]{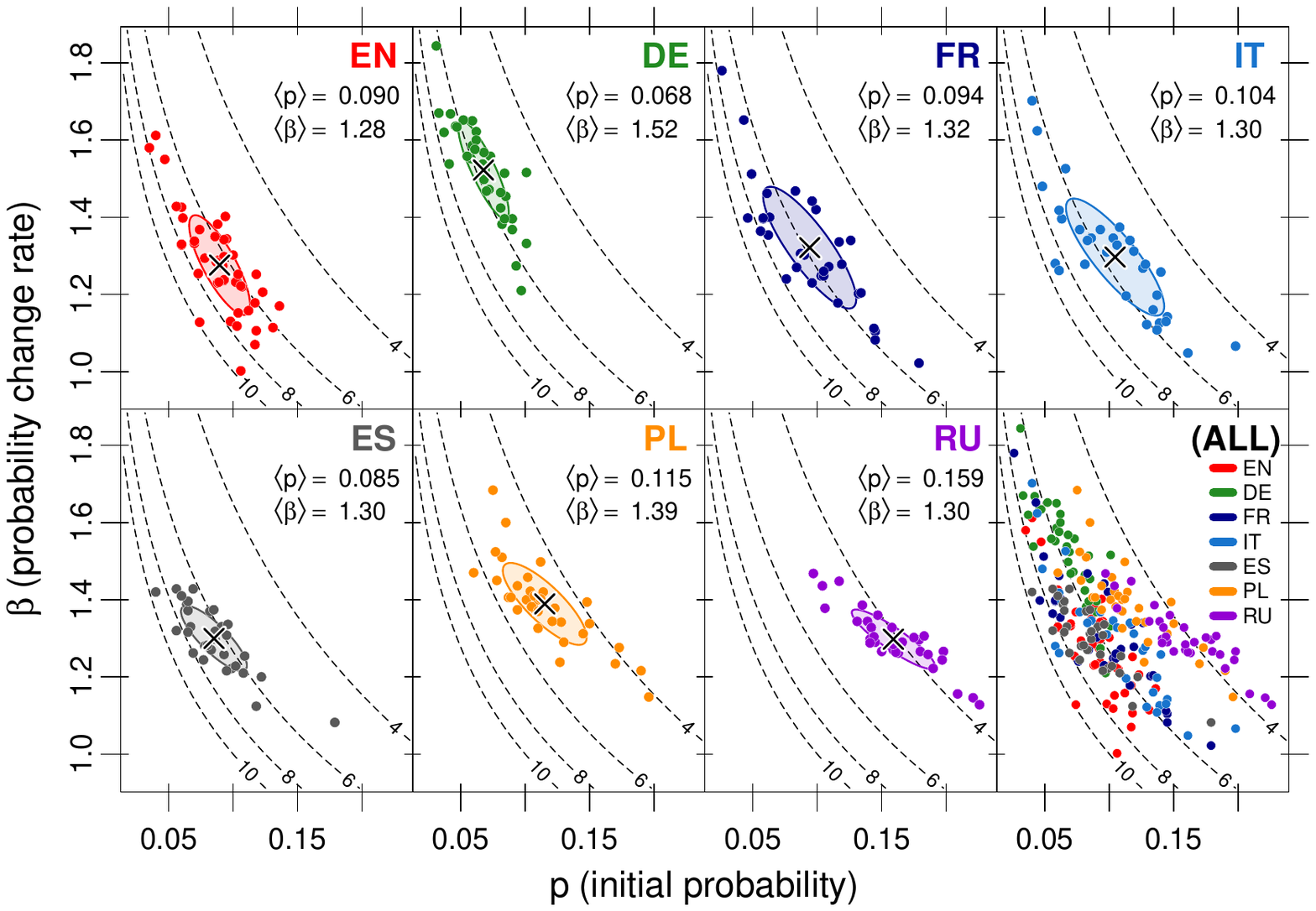}
\caption{Parameters $p$ and $\beta$ of the discrete Weibull distributions fitted to punctuation waiting times for a set of books (for the list of books, see Ref.~\cite{StaniszT-2023a}). The first 7 charts pertain to individual languages, the chart in the lower right corner (ALL) presents the books in all the studied languages collectively. In each plot, a text is represented by a point $(p,\beta)$. All the plots are in the same scale. The dashed lines are isolines of constant expected value of the discrete Weibull distribution -- all distributions with $(p,\beta)$ along one such line have the same expected value. In each plot pertaining to a single language, the quantities $\langle p\rangle$ and $\langle\beta\rangle$, the average values of $p$ and $\beta$, are given, and the centroid of the point cloud, $\left(\langle p\rangle,\langle\beta\rangle\right)$ is marked by ''x''. The ellipses characterize the distributions of points -- the semi-axes of each ellipse are the principal components of the point set in the given language. The major semi-axis of the ellipse gives the direction of the greatest variance and its length is the square root of that variance. The length of the minor semi-axis is the square root of the variance in the perpendicular direction. The ellipses for each language are shown collectively in Fig.~\ref{fig::DWeibull_centroids_and_ellipses}.}
\label{fig::scatterplots.original}
\end{minipage}
\end{figure}

\begin{figure}[!hb]
\centering
\begin{minipage}{\figurecustomwidth}
\centering
\includegraphics[width=0.8\textwidth]{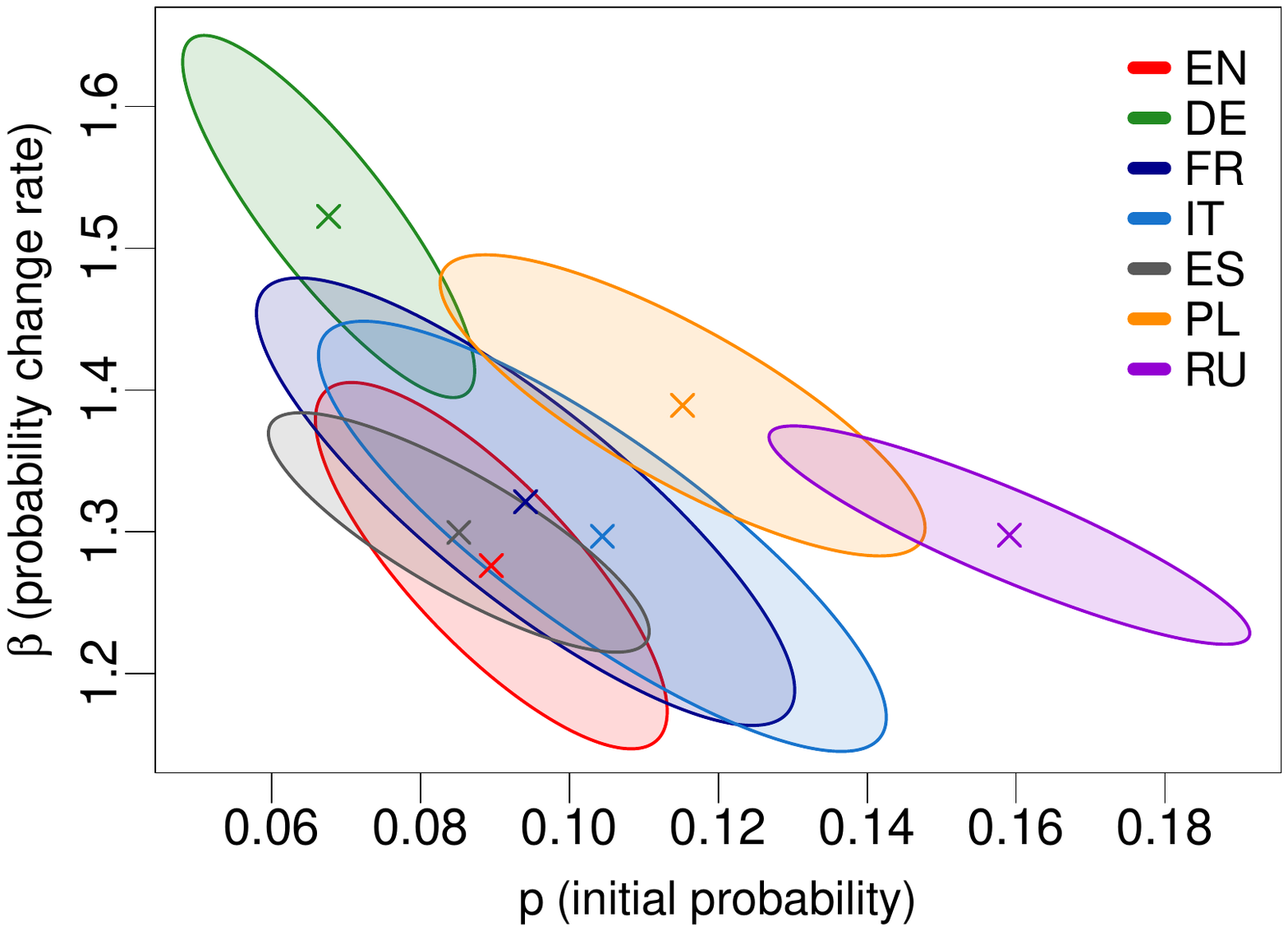}
\caption{Ellipses characterizing the distributions of $(p,\beta)$ and parameters of the discrete Weibull distributions for punctuation waiting times in sample books (these are the ellipses shown in Fig.~\ref{fig::scatterplots.original} collected in a single plot). The centroids of $(p,\beta)$ for each language are marked by ''x''.}
\label{fig::DWeibull_centroids_and_ellipses}
\end{minipage}
\end{figure}

\begin{figure}[!hb]
\centering
\begin{minipage}{\figurecustomwidth}
\centering
\includegraphics[width=0.8\textwidth]{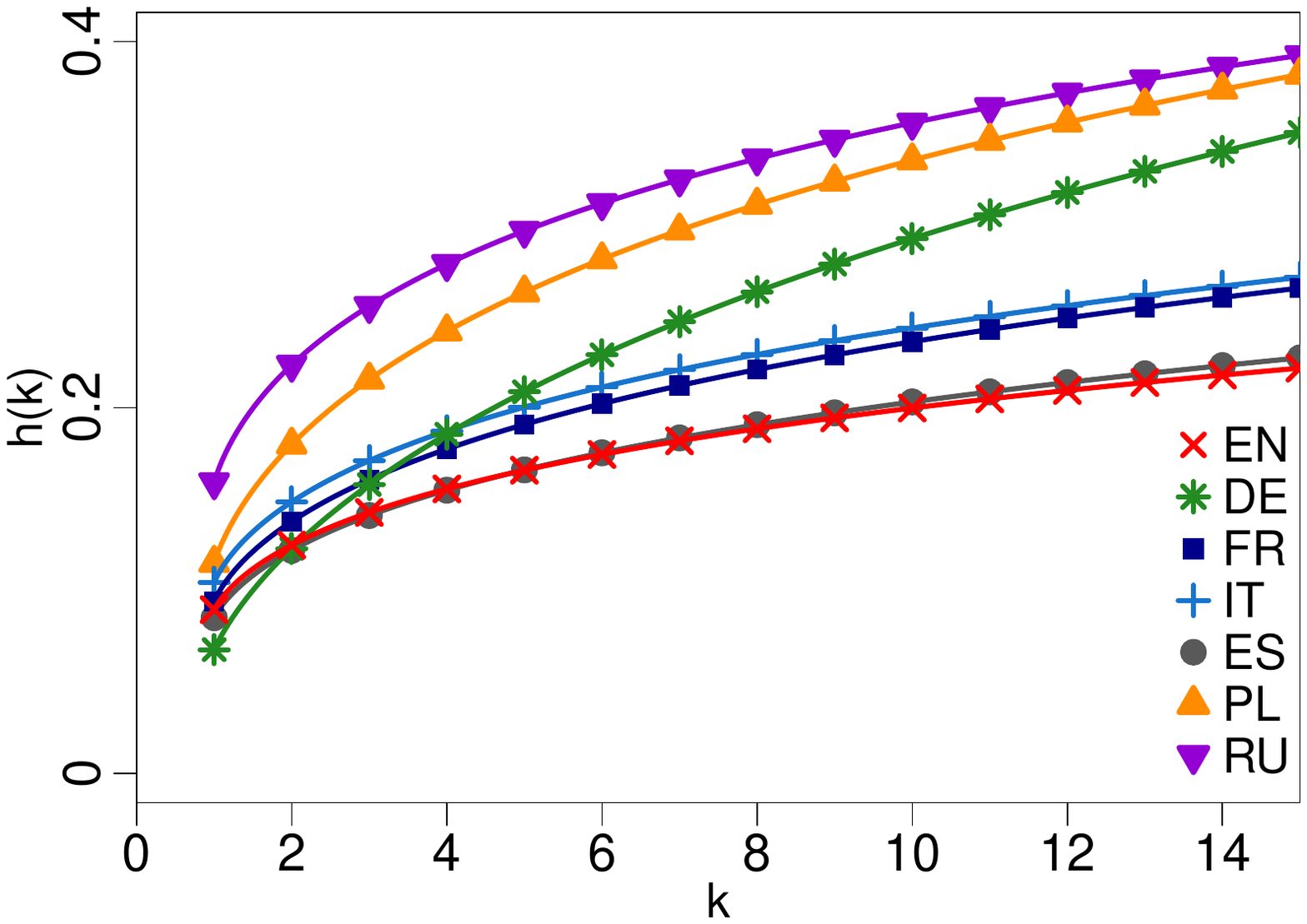}
\caption{Hazard functions $h(k)$ of the discrete Weibull distributions with parameters $(p,\beta)$ corresponding to centroids of the ellipses shown in Fig.~\ref{fig::DWeibull_centroids_and_ellipses} (the average values of $p$ and $\beta$ for individual languages).}
\label{fig::hazard.functions_DWeibull_centroids}
\end{minipage}
\end{figure}

Discrete Weibull distributions characterizing punctuation waiting times in all of the presented texts have the value of $\beta$ between 1 and 2. This means that $h(k)$ is an increasing function. With $k \to \infty$, it converges to 1. It seems to be a reasonable result -- the sequences of words without punctuation should not be infinitely long. The values of $p$ are typically below 0.2. Interestingly, the parameters of the distributions (determining their shape) seem to be to some degree specific to particular languages. When the values of $p$ and $\beta$ related to each book are plotted on a plane (each point represents one book), one can distinguish regions occupied mainly by the texts in the same language (Fig.~\ref{fig::scatterplots.original}, Fig.~\ref{fig::DWeibull_centroids_and_ellipses}). Average values of $p$ and $\beta$ for each language can be calculated to determine the corresponding hazard functions $h(k)$ (Fig.~\ref{fig::hazard.functions_DWeibull_centroids}). Using the concept of random process underlying the arrangement of punctuation in texts, these functions characterize the dynamics of the process. They provide information how "urgent" it is to place a punctuation mark in order to finish an uninterrupted word sequence, depending on the length of that sequence. 

It can be noticed in Fig.~\ref{fig::hazard.functions_DWeibull_centroids} that within the range of waiting times between 1 and 15 (which corresponds to more than 80\% of all observed waiting times in each of the studied texts), the Slavic languages have the highest values of the averaged hazard functions among the studied ones, therefore being the most inclined towards short word sequences between consecutive punctuation marks. Regarding the punctuation distribution properties, two of the Romance languages considered, French and Italian, turn out to be quite similar to each other. They have close average values of $p$ and $\beta$ and their dispersions are overlapping. The averaged hazard function for German, having the lowest $p$ and the highest $\beta$, starts from a low value and increases quickly. The most slowly-varying averaged hazard functions belong to English and Spanish, suggesting that long sequences of words between pauses indicated by punctuation marks are more natural for those languages than for the others. However, a comprehensive description of such properties would require a more detailed investigation. For example, the above-mentioned tendency of Polish and Russian to favour short intervals between punctuation marks may be caused by the lack of articles in these languages; in other languages studied here articles are present. Although they are not stand-alone words, they are treated just as the other ones in the analysis, and therefore they lengthen the sequences of words appearing between punctuation marks.

\subsubsection{Original texts vs. translations}

\begin{figure}[ht]
\centering
\begin{minipage}{\figurecustomwidth}
\centering
\includegraphics[width=1.0\linewidth]{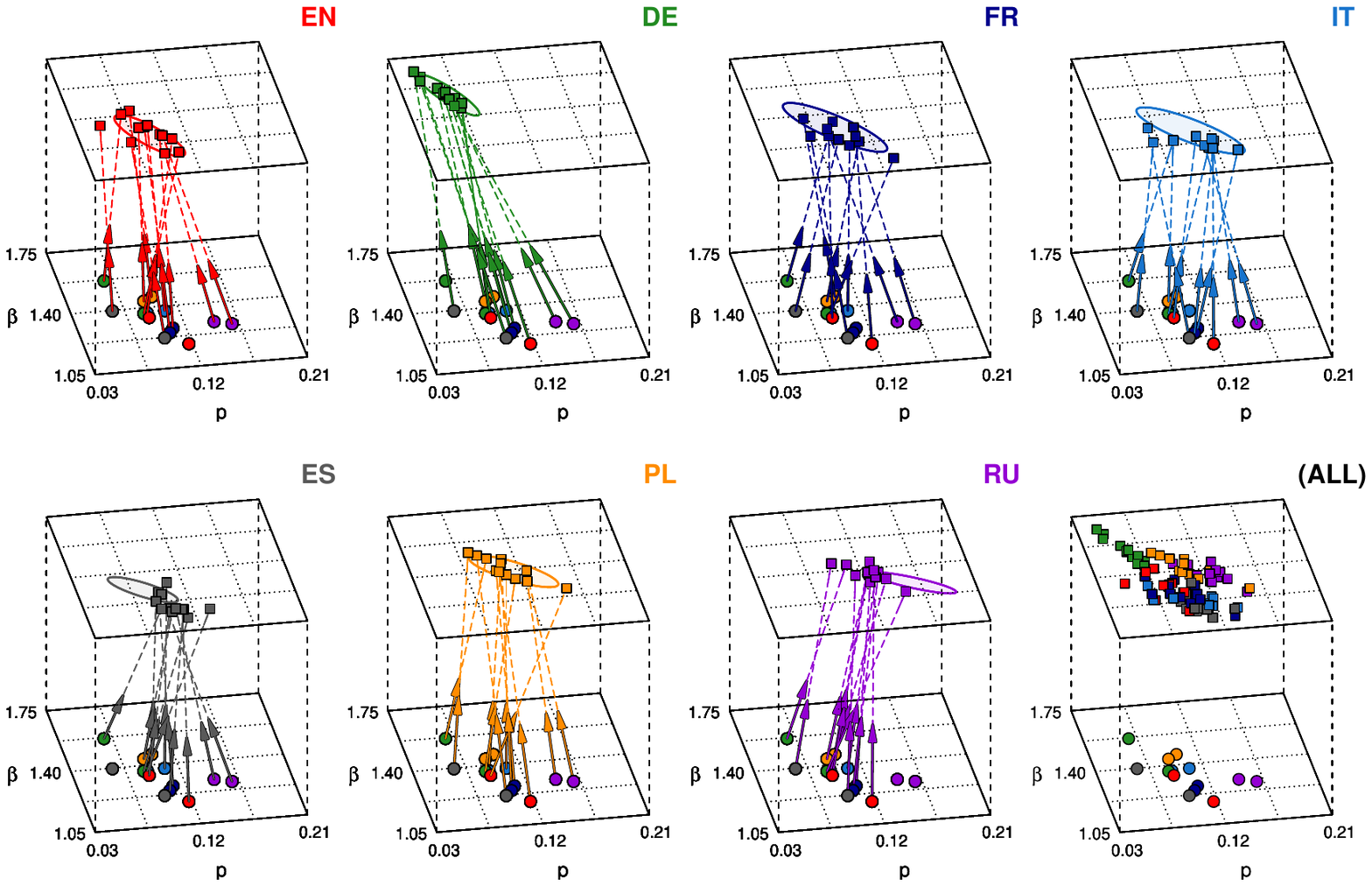}
\caption{Impact of text translation on the parameters $p,\beta$ of the discrete Weibull distribution describing distances between consecutive punctuation marks. The set of texts consists of 14 books and their translations: 2 books written in one of the 7 studied languages were translated into 6 remaining languages. With two translations missing, this gives 96 texts in total. In each plot, the dots on the lower plane mark the values of $p$ and $\beta$ for the texts in their original languages and the squares on the upper plane give $p$ and $\beta$ for the same texts translated into the target language. The correspondence between the original and translated texts is marked by dashed lines and arrows (with the exception of the last plot, presenting translated texts collectively). Colors of dots and squares represent individual languages in the same way as in Figs.~\ref{fig::scatterplots.original},~\ref{fig::DWeibull_centroids_and_ellipses}, and~\ref{fig::hazard.functions_DWeibull_centroids}. The ellipses on the upper planes are the same ellipses that are shown in Fig.~\ref{fig::scatterplots.original}. The list of analyzed books can be found in Ref.~\cite{StaniszT-2023a}.}
\label{fig::3d.planes.scatterplots.translations}
\end{minipage}
\end{figure}

Having observed the presented statistical properties of punctuation in texts in their original languages, one could ask a question about how those properties are affected by text translation. On the one hand, it could be anticipated that some specific patterns of punctuation usage -- related to specific narrative techniques and ways of expressing thoughts in a text -- might be preserved in the translated text, as the translator usually tries to keep the characteristics of the source material. On the other hand, each language has its own structure and imposes specific rules on the elements constituting the text, including punctuation. It turns out that punctuation waiting times in translated texts seem to follow the discrete Weibull distribution with parameters $p$ and $\beta$ corresponding to the target language -- translation tends to shift the position of a text on a $(p,\beta)$ plane towards the region (ellipsoid) occupied by the texts in target language. this effect is shown in Fig.~\ref{fig::3d.planes.scatterplots.translations}.

\section{Fractal structure of language}
\label{sect::fractal_structure_of_language}


\subsection{Basic definitions and relations}

\subsubsection{Fractals and fractal dimension}

Fractals are irregular, rough objects composed of the parts that are copies of the whole shape modified by the so-called contraction mappings -- which can be considered a sub-type of affine transformations. Affine transformations $S: \mathbb{R}^n \rightarrow \mathbb{R}^n$ are defined as
\begin{equation}
S(v) = T(v) + b,
\end{equation}
where $v$ and $b$ are vectors in $\mathbb{R}^n$, while $T:\mathbb{R}^n \rightarrow \mathbb{R}^n$ is a linear transformation usually represented by a $n \times n$ matrix. An affine transformation can consist of scaling, reflection, rotation, and translation. If there exists a number $c \in (0;1)$ such that for all $v_1, v_2 \in \mathbb{R}^n$ an affine transformation $S$ satisfies
\begin{equation}
\left\| S(v_2) - S(v_1) \right\| \leq c \left\| v_2-v_1 \right\|,
\end{equation}
where $\left\| \boldsymbol{\cdot} \right\|$ denotes the norm of a vector, then $S$ is a contraction mapping. The objects that are invariant under a system of contraction mappings are often called self-affine or, less precisely, self-similar. Self-affinity can be exact for mathematical fractals, while only approximate for natural objects. In the latter case, the self-affinity or self-similarity may be understood in a statistical sense: a statistical fractal is an object whose certain statistical properties are the same in all relevant scales or in a wide range of scales. Accordingly, fractals are sometimes categorized into deterministic and random (stochastic) fractals.

Quantitatively, fractals can be characterized by their fractal dimension. It can be thought of as a way of expressing the information about how the characteristics of an object change when inspected at different scales. It is also often interpreted as a quantity expressing the complexity of shape, understood as ``roughness'' or the capacity to fill the embedding space. There are a few ways of defining it, but all the variants express the same idea: the structure of a fractal object is described by a power law. Among the most popular definitions is the Hausdorff dimension $d_{\rm H}$ that for a nonempty set $F$ in a metric space is given by
\begin{equation}
d_{\rm H} = \inf \left\lbrace s : s \geq 0 \wedge \mathcal{H}^{s}(F) = 0 \right\rbrace,
\label{eq::hausdorff.dimension}
\end{equation}
where $\mathcal{H}^{s}(F)$ is the $s$-dimensional Hausdorff measure of $F$ equal to
\begin{equation}
\mathcal{H}^{s}_{\delta}(F) = \inf \left\lbrace \sum_{i} \left( \mathrm{diam}(U_i) \right)^s : \left\lbrace U_i \right\rbrace \, \text{is a $\delta$-cover of } F \right\rbrace,
\label{eq::hausdorff.measure}
\end{equation}
where, for all $i$, diameter of a set $U_i$ is $\mathrm{diam}(U_i) \le \delta$ and $F \subset \bigcup_i U_i$. The Hausdorff dimension $d_{\rm H}$ is the number such that
\begin{equation}
\mathcal{H}^{s}(F) =
  \begin{cases}
  \infty, & \text{ for } s < d_{\rm H}\\
  0, & \text{ for } s > d_{\rm H}.
  \end{cases}
\end{equation}
The value of $\mathcal{H}^{s}(F)$ for $s=d_{\rm H}$ is usually finite and different from zero, but it can also be $\infty$ or 0.

In practical applications, calculating $d_{\rm H}$ is troublesome, so thus another definition of the fractal dimension is commonly used -- the box-counting dimension. Let $N(\delta)$ denote the smallest possible number of balls of diameter $\delta$ needed to cover $F$. The box-counting dimension $d_{\rm C}$ of $F$ is defined as
\begin{equation}
d_{\rm C} = \lim_{\delta \to 0} \frac{\log\left(N(\delta)\right)}{\log(1/\delta)}
\label{eq::box.counting.dimension}
\end{equation}
provided that the limit exists. If it does not, it is sometimes helpful to consider limit inferior and limit superior, but a need to distinguish between them and $d_{\rm C}$ happens rarely in practical applications. The coverings of $F$ do not necessarily have to be done with balls, other types of sets with given diameter or characteristic size directly related to diameter can be used. When the considered space is $\mathbb{R}^n$, one can use (hyper)cubes (boxes) with the side of length $\delta$. Popularity of the box-counting dimension stems from the fact that Eq.~(\ref{eq::box.counting.dimension}) can be transformed into a form that is convenient for numerical computation of $d_{\rm C}$:
\begin{equation}
N(\delta) \sim K(1/\delta)^{d_{\rm C}} \quad \text{ as } \delta \to 0
\label{eq_fractals_box_counting_dimension_power_law}
\end{equation}
where $K$ is constant or varies sufficiently slowly with $\delta$. Similarly to other definitions of the fractal dimension, $d_{\rm C}$ can be fractional or integer, depending on the object structure.

In most cases relevant in practical applications, $d_{\rm C}$ and $d_{\rm H}$ are equal, but in principle they satisfy the following inequality~\cite{Muzy1994}:
\begin{equation}
d_{\rm T} \leq d_{\rm H} \leq d_{\rm C},
\label{eq::fractal.dimensions.inequality}
\end{equation}
where $d_{\rm T}$ stands for the topological dimension.

\subsubsection{Multifractals}

The fractal dimensions like $d_{\rm H}$ and $d_{\rm C}$ characterize an object's shape as a whole. However, there exist objects (multifractals) whose different parts have different local scaling properties. More general tools are required in order to describe such objects. While the mathematical description of fractals utilizes the notion of a set, the description of multifractals is formalized in terms of measures~\cite{Halsey1986, Feder1988, Falconer2014, Mandelbrot1988, Muzy1994}. Let $\mu$ be a measure in $\mathbb{R}^n$ and let $\mathrm{supp}(\mu)$ denote the support of this measure. Let $\mu$ be power-law distributed around an arbitrary point $x_0 \in \mathrm{supp}(\mu)$:
\begin{equation}
\mu\left(K(x_0,\varepsilon)\right) = C_{\mu} \, \varepsilon^{\alpha(x_0)} \quad \text{ for } \varepsilon \to 0,
\label{eq::multifractal.measure}
\end{equation}
where $K(x_0, \varepsilon)$ is the (hyper)cube of side length $\varepsilon$ centered at $x_0$, the variable $\alpha(x_0) \ge 0$ is called the H\"older exponent, and $C_{\mu}$ is a constant independent of $x_0$ and $\varepsilon$. The Hölder exponent describes the ``strength'' of singularity of $\mu$ around $x_0$: the lower $\alpha(x_0)$ is, the more singular is the measure $\mu$ at $x_0$. The limiting value of $\alpha(x_0) = 0$ corresponds to a behaviour that resembles a Dirac delta at $x_0$. Conversely, the greater the $\alpha(x_0)$ is, the more uniform is the measure around $x_0$. From Eq.~(\ref{eq::multifractal.measure}) it can be seen that the H\"older exponent can be expressed as
\begin{equation}
\alpha(x_0) = \lim_{\varepsilon \to 0} \frac{\log \:\! \mu \! \left(K(x_0,\varepsilon)\right)}{\log \varepsilon}.
\label{eq::holder.exponent}
\end{equation}
For any value of $\alpha$, one can define a set $E_\alpha \subset \mathrm{supp}(\mu)$ that contains all the points $x$ for which $\alpha(x)=\alpha$. This set $E_\alpha$ can be characterized by its Hausdorff dimension:
\begin{equation}
f(\alpha) = d_{\rm H}(E_\alpha).
\label{eq::singularity.spectrum}
\end{equation}
It means that $f(\alpha)$ is a function that assigns to each $\alpha$ the fractal dimension of the set of points having the H\"older exponent equal to $\alpha$. The set of pairs $\left(\alpha,f(\alpha)\right)$ for all $\alpha$s occurring in the system (i.e., $\alpha \in \left[\alpha_{\min},\alpha_{\max} \right]$) is called the \textit{singularity spectrum}. When the whole system is characterized by a unique $\alpha=\alpha_0$, the singularity spectrum reduces to a single point $\left(\alpha_0,f(\alpha_0)\right)$. The singularity spectrum reduced to a single point corresponds to a measure that is called \textit{monofractal}. Conversely, a measure whose singularity spectrum comprises some range of the H\"older exponents is called \textit{multifractal}. A term often appearing in relation to multifractality is \textit{multiscaling}. It refers to the fact that different parts of a multifractal object exhibit different types of scaling behaviour.

Typically, $f(\alpha)$ is a concave function resembling an inverted parabola, which is spanned between $\alpha_{\min}$ and $\alpha_{\max}$ and its maximum value is $\max \lbrace f(\alpha) \rbrace=d_{\rm H}(\mathrm{supp}(\mu))$~\cite{Feder1988,Falconer2014,Halsey1986,Muzy1994,Stanley1988}. However, it is worth mentioning that the singularity spectra of different shapes are also possible~\cite{Mandelbrot1990,Mandelbrot1990a,Riedi1995,Riedi1997}. The width of the singularity spectrum
\begin{equation}
\Delta \alpha := \alpha_{\max} - \alpha_{\min}
\label{eq::singularity.spectrum.width}
\end{equation}
expresses the variety of the H\"older exponents and, therefore, a wide spectrum $f(\alpha)$ is often considered as a sign of a certain type of complexity.

Detecting and quantifying multifractality in empirical data based directly on the definitions given above suffers typically from large errors. As a countermeasure, an approach based on the so-called \textit{partition function} is applied. The space $\mathbb{R}^n$ can be divided into (hyper)cubic cells of side length $\varepsilon$ numbered by $i = 1,2,3,..., N(\varepsilon)$. Let the measure $\mu$ contained in the $i$th cell be denoted by $\mu_i(\varepsilon)$. The partition function $Z(q,\varepsilon)$ for $q \in \mathbb{R}$ is defined as
\begin{equation}
Z(q,\varepsilon) = \sum_{i=1}^{N(\varepsilon)} \mu_i(\varepsilon)^q
\label{eq::partition.function}
\end{equation}
and its behaviour in the limit $\varepsilon \to 0$ for fixed $q$ is given by a power law:
\begin{equation}
Z(q,\varepsilon) = C_Z \, \varepsilon^{\tau(q)},
\label{eq::partition.function.tau}
\end{equation}
where $C_Z$ is a constant independent of $\varepsilon$. The exponent $\tau(q)$ is called the \textit{generalized scaling exponent} or the \textit{mass exponent}. The spectra $(q,\tau(q))$ and $(\alpha,f(\alpha))$ are equivalent and related with each other by the following formula:
\begin{equation}
\tau(q) = q \alpha(q) - f(\alpha(q)).
\label{eq::multifractal.spectrum.vs.singularity.spectrum}
\end{equation}
This interchangeability of descriptions of the multifractal objects in terms of $(\alpha, f(\alpha))$ and $(q, \tau(q))$ facilitates practical calculations since $\tau(q)$ is often easier to compute than $f(\alpha)$.

Based on the multifractal exponent $\tau(q)$, one can calculate the so-called \textit{generalized fractal dimensions} $D_q$ defined as
\begin{equation}
D_q =
  \begin{cases}
  \dfrac{\tau(q)}{q-1}, & \text{ for } q \neq 1 \\
  \lim \limits_{q \to 1} \dfrac{\tau(q)}{q-1} = \left. \dfrac{d \tau}{d q}\right|_{q=1}, & \text{ for } q = 1.
  \end{cases}
\label{eq::generalized.fractal.dimension}
\end{equation}
There are some properties of a fractal object that can be demonstrated by Eqs.~(\ref{eq::multifractal.spectrum.vs.singularity.spectrum}) and (\ref{eq::generalized.fractal.dimension}). If the measure $\mu$ is monofractal, then  $d \tau / dq$ is constant and $\tau(q)$ is linear, while nonlinearity of $\tau(q)$ implies the multifractal measure. Moreover, the following relations are valid:
\begin{equation}
\begin{aligned}
& D_0 = d_{\rm H}({\rm supp}(\mu)) = \max\left\lbrace f(\alpha)\right\rbrace,\\
& D_{+\infty} = \alpha(q\!=\!+\infty) = \alpha_{\min}, \\
& D_{-\infty} = \alpha(q\!=\!-\infty) = \alpha_{\max}. \\
\end{aligned}
\end{equation}

\subsubsection{Multifractal analysis of time series}

\label{sect::multifractal_analysis_of_time_series}

There are a few methods of detecting and quantifying fractality and multifractality in time series, but the most common one among them is the multifractal detrended fluctuation analysis (MFDFA)~\cite{Kantelhardt2002}. MFDFA is a generalization of the detrended fluctuation analysis (DFA) designed to estimate the Hurst exponent of a time series~\cite{Peng1994,Kantelhardt2001}. It allows to estimate both the singularity spectrum and the Hurst exponent of a time series. An important feature of DFA/MFDFA is that it allows to analyze non-stationary series by removing trends from the data and focusing on the fluctuations around the trends.

Let $\{x(i)\}$ be a real-value time series of length $T$, which is divided into $M_s$ segments of size $s$. In each segment $\nu$, it is integrated and then locally detrended by subtracting a fitted polynomial trend (typically, the polynomial is of the second order). Next the so called $q$th-order fluctuation function $F_q(s)$ is calculated as a power mean of the variance $F^2(s,\nu)$ of the segment-wise detrended signal, where the averaging is taken over the segments:
\begin{equation}
F_q(s) =
  \begin{dcases}
  \left( \dfrac{1}{2M_s} \sum\limits_{\nu=1}^{2M_s} \left(F^2(\nu,s)\right)^{q/2}\right)^{1/q}, & \text{ for } q \neq 0\\
  \exp \left( \dfrac{1}{4M_s} \sum\limits_{\nu=1}^{2M_s} \log \left(F^2(\nu,s)\right)\right), & \text{ for } q=0.
  \end{dcases}
  \label{eq::fluctuation.function}
\end{equation}
(The second formula in Eq.~(\ref{eq::fluctuation.function}) is obtained as a limit $q \to 0$ of the first formula.) Typically, a family of the fluctuation functions $F_q(s)$ for various values of $s$ and $q$ is calculated in some interval (for instance, $q \in [-4,4]$ and $s \in [10,T/5]$~\cite{Oswiecimka2006}). $F_q(s)$ characterizes the fluctuations of given magnitude at a given scale. The main contribution to $F_q(s)$ for strongly negative $q$ comes from small fluctuations, while the largest fluctuations are ``amplified'' for large positive $q$s. As a final step of the MFDFA procedure, scaling behaviour of $F_q(s)$ is investigated:
\begin{equation}
F_q(s) \sim s^{h(q)},
\label{eq::fluctuation.function.scaling}
\end{equation}
where $h(q)$ is the \textit{generalized Hurst exponent} (its special case for $q=2$ is the Hurst exponent $H$ given by Eq.~(\ref{eq::Hurst.exponent})). The sequence of generalized Hurst exponents $h(q)$ characterizes the scaling of the fluctuations of different magnitudes. If $h(q)=H$ for all values of $q$, a time series $\{x(i)\}$ can be characterized by a single scaling relationship but if there is a significant dependence of $h$ on $q$, then the time series exhibits multiscaling. There is a relation coupling $\tau(q)$ and $h(q)$ that is given by $\tau(q)=qh(q)-1$, from which the formulas for computing the singularity spectrum $(\alpha,f(\alpha))$ can be derived:
\begin{equation}
  \begin{cases}
  \alpha = h(q) + q \dfrac{d h}{dq} \\
  f(\alpha) = q (\alpha - h(q)) + 1.
  \end{cases}
\end{equation}


\subsection{Multiscaling in written language}

\subsubsection{Texts as symbolic sequences}

The first approach to multiscale analysis of texts was carried out in~\cite{PavlovA-2001a}, where a complete text of \emph{Moby Dick} by H.~Melville (and some other English texts, analysis of which was not presented explicitly) was transformed into point processes representing occurrence of each character or a short sequence of characters (like \emph{th}). It was reported that the results for the waiting times between the consecutive occurrences were multifractal. However, the singularity spectra $f(\alpha)$ presented there, which were calculated by means of the WTMM approach, might not be interpreted in this way with the today's knowledge and methodological recommendations (see~\cite{Oswiecimka2006,Drozdz2009,KwapienJ-2023a}).

\subsubsection{Word lengths and word frequencies}

A word is the basic carrier of information in any language sample. As it enters into complex interactions with other words that occur in its neighbourhood and as these interactions are constrained by the rules of grammar, style, and context, one may expect that word sequences (e.g., literary texts) show a high degree of intrinsic complexity. Indeed, a series of studies of word-length time series and word-frequency time series representing selected literary texts (\emph{Alice in Wonderland} and \emph{Through a Looking Glass} by L. Caroll) reported that both types of data reveal broad singularity spectra, suggesting multifractality~\cite{AusloosM-2012b,AusloosM-2012a}. The word-length time series were created by replacing each consecutive word by the number representing its length in letters, while in the word-frequency time series each word was replaced by its rank in the Zipf plot. Apart from the original English texts, an Esperanto translation of \emph{Alice in Wonderland} was studied for comparison. It showed a broader $f(\alpha)$ that the original, mainly for the small rank words, which may illustrate differences in the vocabulary use and grammar rules between these two languages. Despite this difference, all three texts were characterized by significant left-right asymmetry with a broader high-$\alpha$ part, which did not disappear after randomizing the time series. It has to be noted that widths of the singularity spectra for the randomized time series decreased respective to the original time series, which is a natural effect of destroying the temporal correlations. Although a multiplicative cascade-like dynamics can govern words of all lengths and frequencies in this case, such an asymmetric structure of $f(\alpha)$ indicates~\cite{DrozdzS-2015a} that the dynamics of both the short words and the low-rank (most frequent) words is richer and more complex than the dynamics of long words and low-frequency words. Such a result for the word-frequency time series can be explained straightforward by a random character of rank attribution for rare words, while its counterpart for the word-length time series can be explained in the same way if one realizes that, statistically, the longer is a word, the less frequent it is used~\cite{ZipfGK-1949a}.

A rare dynamical perspective for the same literary works was offered in~\cite{AusloosM-2008a} where the evolution of time series of word lengths was studied by means of the correlation integrals~\cite{GrassbergerP-1983b}. Correlation integrals can be used to calculate correlation dimension that is equivalent to $D_2$. By reconstructing the phase space in which the related dynamical system is embedded, it was shown that although the subspace occupied by the trajectory is characterized by a smaller correlation dimension than the whole phase space, this dimension could not be measured as no saturation effect was observed while increasing the embedding dimension. Such a result was interpreted that this particular linguistic process is high-dimensional yet not completely random.

A much broader study was reported in~\cite{ChatzigeorgiouM-2017a}, in which a large corpus of parallel multi-language proceedings of the European Parliament was transformed to word-length time series. A set of time series representing 10 western-European languages was subject to the multiscale box-counting algorithm that allowed for calculation of the generalized dimensions $D(q)$ and the singularity spectra $f(\alpha)$. The Germanic languages were shown to be more sensitive to varying scale $s$ than the Romance languages, which translates to their richer multifractality. Somehow surprisingly English was more aligned with the Romance group than with the Germanic one. If $D(q)$ was compared with its counterpart for the randomized time series, it was found that the Germanic languages showed poorer genuine multifractality (i.e., less variable $D(q)$ and narrower $f(\alpha)$) than the Romance languages. The latter were characterized by weaker spurious effects related to the heavy tail probability distribution functions, though (see~\cite{Drozdz2009,KwapienJ-2023a} for a detailed discussion of such effects).
Authors of that study reported some peculiar multifractal properties of Finnish that do not resemble any other language from the considered set, but a closer inspection of their methodology suggests that it could be a methodological issue in a power-law fitting procedure rather than a genuine observation, so here this case shall not be discussed.

\subsubsection{Sentence lengths}

Fluctuation scaling having the form of a power law with Hurst exponent $H>1/2$ indicates that sentence lengths are arranged into a specific scale-free structure. However, Hurst exponent provides information which is in some sense averaged over the whole text. Complex patterns of organization in some texts can be investigated in more detail with the use of multifractal formalism. Multifractality of sentence lengths in literary texts has been studied in~\cite{DrozdzS-2016a}. The analysis has revealed that while fractality is a rather general property, the degree of multifractality is more rare and specific to individual texts. Among the studied books, the ones with the richest multifractal structure are quite often the ones that use the narrative technique known as the stream of consciousness. On one hand, this technique can be considered as ``natural'' in certain sense, as it attempts to mimic the natural flow of thoughts and feelings passing through a character's mind, which often results in the presence of incomplete thoughts, sensory impressions, unusual grammar, and in general, certain degree of disorganization. On the other hand, it can be considered ``unnatural'', as it is clearly different from how the majority of written language looks like. The distribution of sentence lengths can be highly inhomogeneous, with intermittent bursts of long sentences clustered together. This effect can be captured by multifractal analysis.

Multifractality of a time series, which is usually quantified in terms of the width~$\Delta\alpha$ of series' singularity spectrum $f(\alpha)$, is typically associated with two factors: heavy-tailed distributions and nonlinear correlations. While the presence of correlations of specific type is essential for the emergence of a multifractal structure, a heavy-tailed distribution of the series' values might increase the width of the spectrum both in case when the series is truly multifractal and in case when the nonzero width of the spectrum is an artifact being a result of the finite length of the series. The latter case is related to the fact that a spectrum of a finite-length, uncorrelated random series can spuriously indicate multifractality, although such a series does not have any specific organization and in the limit of infinite length is either monofractal or bifractal (having two distinct values of singularity exponents) -- depending on the distribution of series' values~\cite{Drozdz2009}. To clarify whether a spectrum of nonzero width is indeed related to series' multiscaling, it is profitable to confirm that the values of the series are correlated with each other in some way, using tools like autocorrelation function~\cite{Drozdz2009}. To demonstrate that the correlations responsible for multifractality are of nonlinear character, one can investigate the singularity spectrum of a specifically constructed surrogate series -- a series which is randomized in a way that removes all correlations except the linear ones. Constructing such a surrogate relies on phase randomization of the Fourier transform: for a series $x_n = x_0, x_1, x_2,... x_{N-1}$ the (discrete) Fourier transform $ \widehat{x}_k = \widehat{x}_0, \widehat{x}_1, \widehat{x}_2,... \widehat{x}_{N-1}$ is computed. Then the phases $\arg(\widehat{x}_k)$ of the transform's coefficients are randomized, by multiplying each of them by an uniformly distributed random number from the interval $[0;2 \pi]$. Then the inverse Fourier transform gives the desired surrogate. Such a surrogate has the same spectral density as the original series (as spectral density depends only on the modulus of the Fourier transform), and, consequently, the same linear correlations. Correlations of other types are destroyed. It is worth noting that this procedure in general alters the distribution of the series' values. The expected result of the multifractal analysis of such a surrogate series is a singularity spectrum practically reduced to one point, corresponding to singularity exponent determined by linear correlations.

\begin{figure}[htp]
\centering
\begin{minipage}{\figurecustomwidth}
\centering
\subfloat{\fbox{\includegraphics[width=0.925\textwidth]{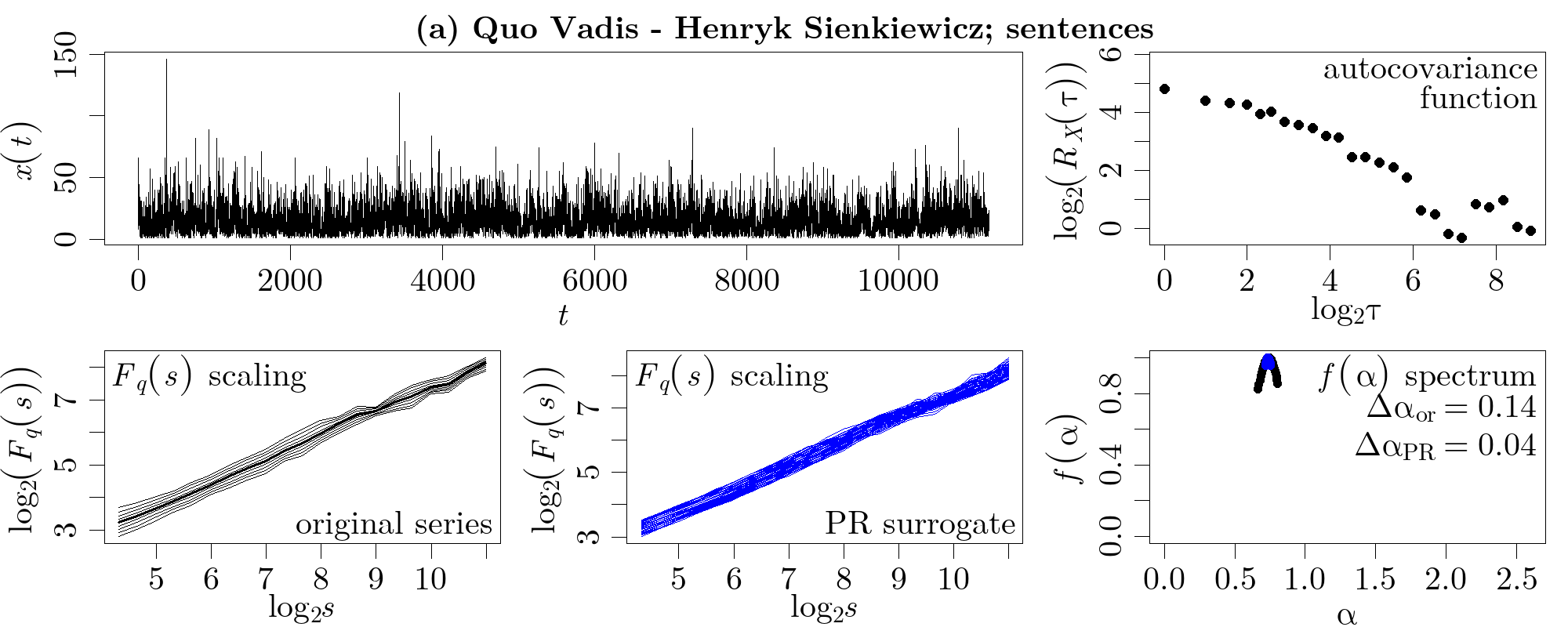}}}
\vspace{0.1cm}
\subfloat{\fbox{\includegraphics[width=0.925\textwidth]{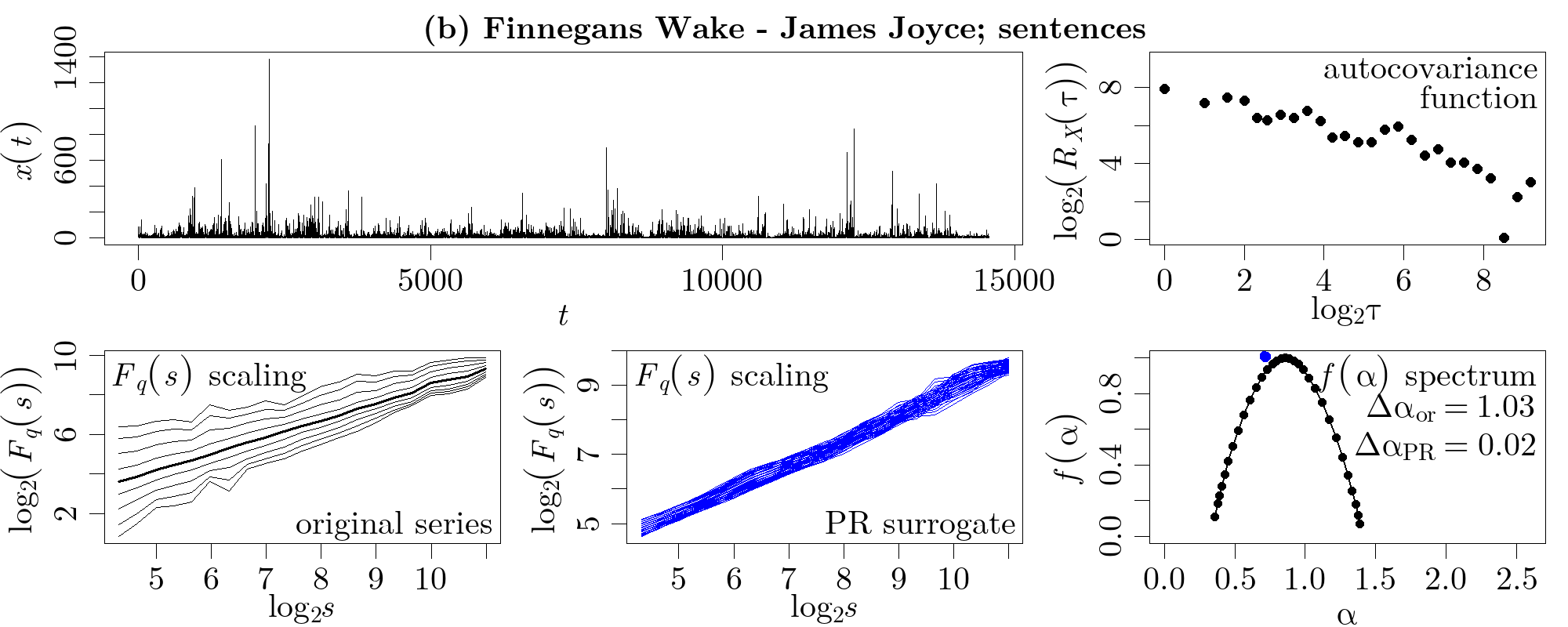}}}
\vspace{0.1cm}
\subfloat{\fbox{\includegraphics[width=0.925\textwidth]{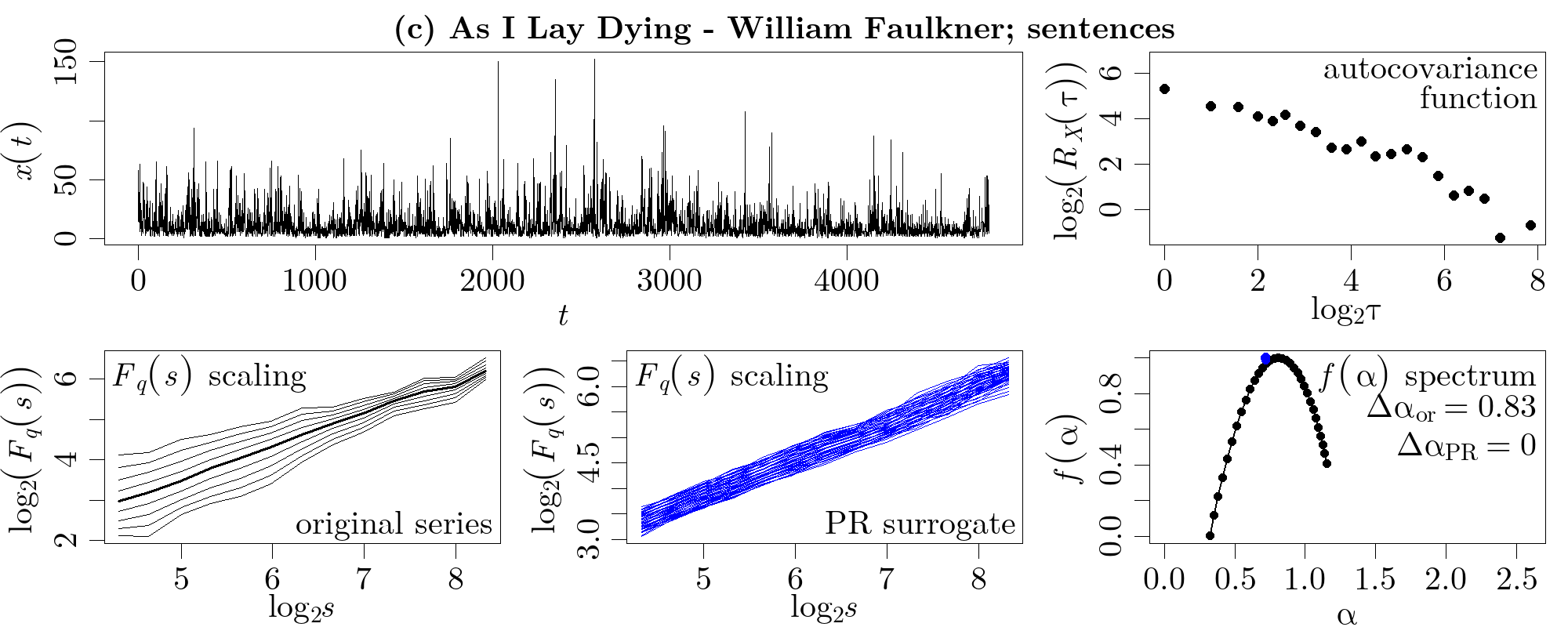}}}
\caption{Multifractal analysis (MFDFA) of time series representing sentence lengths for three books: \textit{Quo Vadis} (a), \textit{Finnegans Wake} (b), and \textit{As I Lay Dying} (c). For each book, the upper row of plots contains a run chart $x(t)$ of the time series and a log-log plot of the autocorrelation function $R_X(\tau)$, while the lower row contains two log-log plots of the fluctuation functions $F_q(s)$ with integer $q$ in the range $[-4;4]$, for the original series (black) and for 5 phase-randomized (''PR'') surrogates (blue), and the respective singularity spectra $f(\alpha)$. For a time series of length $N$, the range of scales $s$ was chosen such that the log-log plots of $F_q(s)$ are approximately linear and $20 \le s \le N/5$, simultaneously. Width of the singularity spectra $\Delta \alpha_{\rm or}$ and $\Delta \alpha_{\rm PR}$ are given in the upper right corner of the $f(\alpha)$ plot, for the original and the PR surrogate series, respectively. Calculation of $f(\alpha)$ for the surrogate series involved calculation of the generalized Hurst exponents $h(q)$ and averaging them over 5 independent realizations. As expected, $f(\alpha)$ for the PR surrogate series nearly collapsed to single points as nonlinear correlations responsible for multifractality were destroyed.}
\label{fig_MFDFA_sentence_lengths}
\end{minipage}
\end{figure}

Fig.~\ref{fig_MFDFA_sentence_lengths} shows the run charts $x(t)$, the autocorrelation functions $R_X(\tau)$, the fluctuation functions $F_q(s)$ and the singularity spectra $f(\alpha)$, for sentence lengths in three books: \textit{Quo Vadis}, \textit{Finnegans Wake}, and \textit{As I Lay Dying} (each in its original language -- Polish or English). Power-law decay of the books' autocorrelation functions confirms the presence of long-range correlations, also detected by the analysis of spectral density and of the Hurst exponents (Fig.~\ref{fig_long_range_correlations_sentences}). The latter two of the books are examples of works utilizing stream of consciousness writing style, and \textit{Finnegans Wake} is additionally known for its highly experimental, unusual language with uncommon grammar and vocabulary. The first of the books is more typical in terms of narrative techniques and does not rely on experimental linguistic constructions. The books exhibit different kinds of fluctuation scaling. While \textit{Quo Vadis} is clearly a monofractal (as evidenced by singularity spectrum $f(\alpha)$ collapsed to a narrow range of $\alpha$), \textit{As I Lay Dying} and \textit{Finnegans Wake} have a multifractal structure. \textit{As I Lay Dying} and \textit{Finnegans Wake} are in a sense extreme -- their singularity spectra are significantly wider than the spectra of typical examples of literary texts in prose (as mentioned before, this is to some degree characteristic of some forms of experimental writing). Sentence lengths in a literary text in prose are often either monofractal or have only some trace of multifractality, manifested by spectra of moderate width.

The situation becomes much more complicated in the case of languages whose punctuation differs from the Western languages. For example, in Chinese texts punctuation in the present form did not appear organically like in the Western languages, but relatively lately it was transplanted from those languages in order to avoid the necessity of guessing the structure of sentences predominantly from context. As a non-native system the Chinese punctuation marks have to comply with the distinct character of the written Chinese, so it is impossible to map sentences constructed in this language to sentences written in any Western language in a 1:1 manner. This leads to some ambiguity in deriving sentence lengths in Chinese texts as different punctuation marks can play a similar role as full stops in the Western languages. Therefore in a study of a broad set of Chinese texts that was reported in~\cite{LiuJ-2023a} two cases were considered in parallel. Without diving into details they differed in that the first one was more restrictive (fewer marks were allowed to end a sentence) and the second one was more inclusive (all marks down to the Chinese comma were allowed to end it).

A Hurst analysis of almost 100 novels from different literary epochs indicated that all texts are long-range correlated ($H>0.5$) but exact values of $H$ varied and occurred to be idiosyncratic without any systematic dependence on author, genre or geographical region. It was thus impossible to use the Hurst exponents as a stylometric tool. The only dependence was seen on literary epoch as the smallest values of $H$ were attributed to texts written not later than in the early Revolution period. In general, the sentence-length persistence was smaller in Chinese than it was observed in the Western languages, including English. However, the difference was small enough that one could not consider it qualitative. More heterogeneity among the texts were observed in the singularity spectrum width $\Delta \alpha$ (Eq.~(\ref{eq::singularity.spectrum.width})). Like in~\cite{DrozdzS-2016a} a vast majority of texts were classified as monofractal and only 12\% as multifractal. However, it is difficult to assess this result as no criterion for identifying scaling regions in fluctuation function plots were shown in~\cite{LiuJ-2023a}. Unlike the Western language books, in Chinese there was only a weak correlation between the stream-of-consciousness (SoC) narrative style and multifractality of the sentence length time series. One interesting example was Gao Xingjian’s novel “Soul Mountain”, being written as an internal monologue, that exhibited a richer multifractality than Finnegans Wake by James Joyce, the Western book that is the record holder in this category~\cite{DrozdzS-2016a}. Another interesting example are Maoist-epoch realist novels like “Sanliwan Village” by Shu-li Chao that showed broad $\Delta \alpha$ without any SoC signature. It is worth mentioning that there were only minor quantitative differences between the Hurst exponents for time series that represented sentence lengths defined by full stops vs. those defined by all punctuation marks (with the former giving on average a little higher $H$ than the latter) and time series that represented sentence lengths measured in characters vs. those measured in words (with a similar average relation between the former and the latter)~\cite{LiuJ-2023a}.

Multifractal analysis (MFDFA) of sentence lengths and its variant, in which only the tokens representing a specific part of speech (verbs, nouns, adjectives, etc.) were considered, was subject of a systematic study communicated in~\cite{MohseniM-2021a}. The analyzed texts were divided into two categories based on genre: fiction and non-fiction and two categories based on literary quality: canonical fiction and non-canonical fiction. Authors concluded that all the texts appeared fractal in both the sentence-length time series and the part-of-speech-frequency time series with over 90\% of the texts revealing genuine multifractality related to the long-range correlations. Based on the width of the $f(\alpha)$ spectra and their asymmetry, it was possible to categorize the texts representing fiction and non-fiction with a slightly larger success ratio than the texts representing canonical and non-canonical literature, but nevertheless in both cases the categorization based on fractal properties was feasible. A suggested origin of this feasibility was attributed to a richer variability of narrative modes, which could be associated with a stronger heterogeneity of sentence constructions and part-of-speech frequencies, observed in both fiction (vs. non-fiction) and canonical fiction (vs. non-canonical fiction). Interestingly, some higher-level measures of the structural diversity of texts and the probability of topics were found less successful~\cite{MohseniM-2021a}.

Sentence lengths were also used as a time series representation of the poems of R.~Tagore written in Bengali and then translated (partially by Tagore himself) to English~\cite{GhoshD-2019a}. By using the MFDCCA method, it was possible to analyze the multiscale detrended cross-correlations present in both time series. Although the results of that work could be viewed as controversial, because its authors applied a multifractal algorithm that had already been known to be flawed (i.e., the unsigned version of MFDCCA), the study deserved to be mentioned as it was the first attempt to look into the structure of the multiscale detrended cross-correlations in natural language samples.

\subsubsection{Punctuation mark waiting times}

\begin{figure}[htp]
\centering
\begin{minipage}{\figurecustomwidth}
\centering
\subfloat{\fbox{\includegraphics[width=0.925\textwidth]{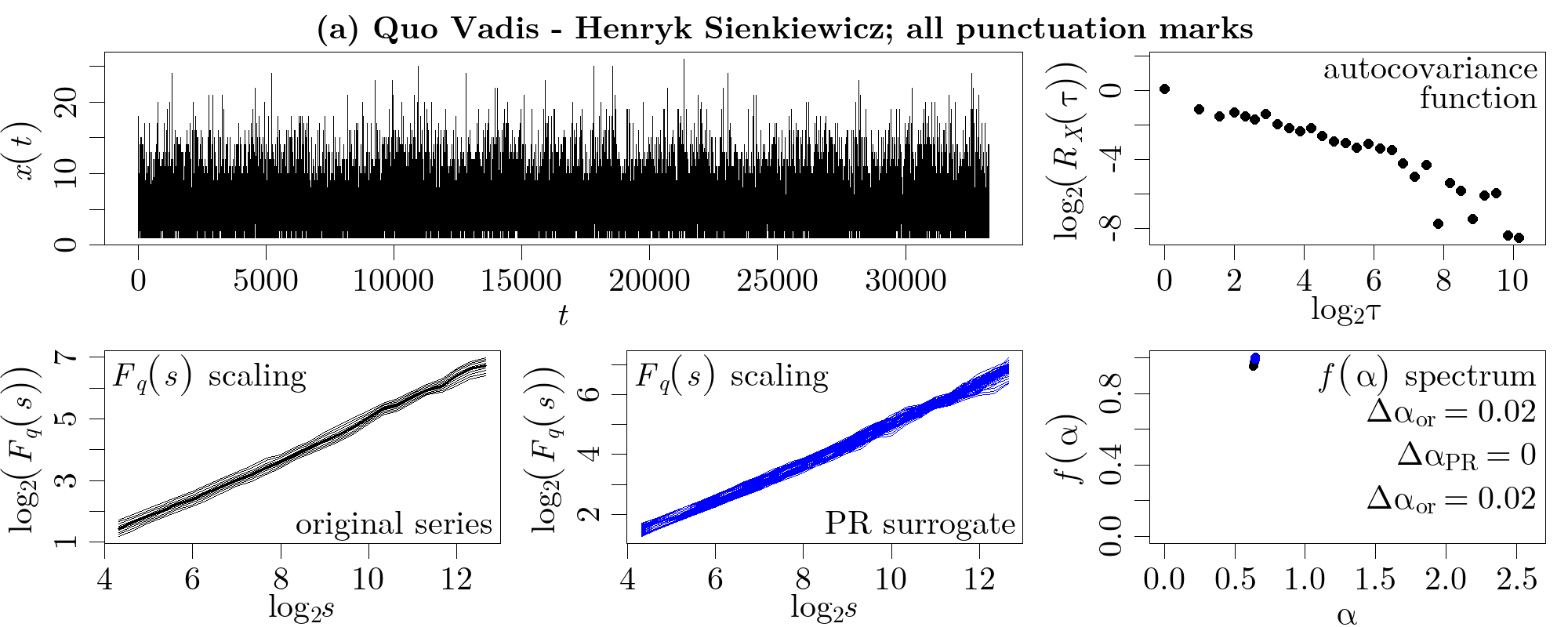}}}
\vspace{0.1cm}
\subfloat{\fbox{\includegraphics[width=0.925\textwidth]{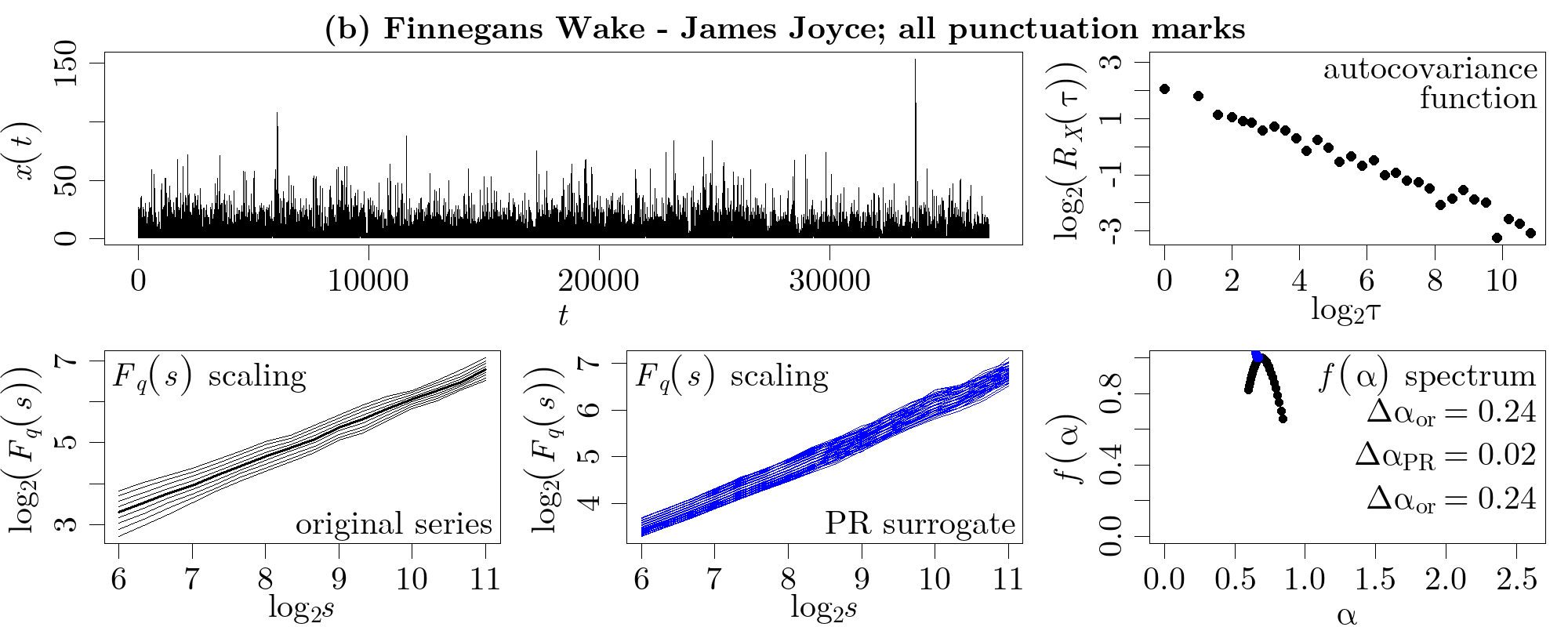}}}
\vspace{0.1cm}
\subfloat{\fbox{\includegraphics[width=0.925\textwidth]{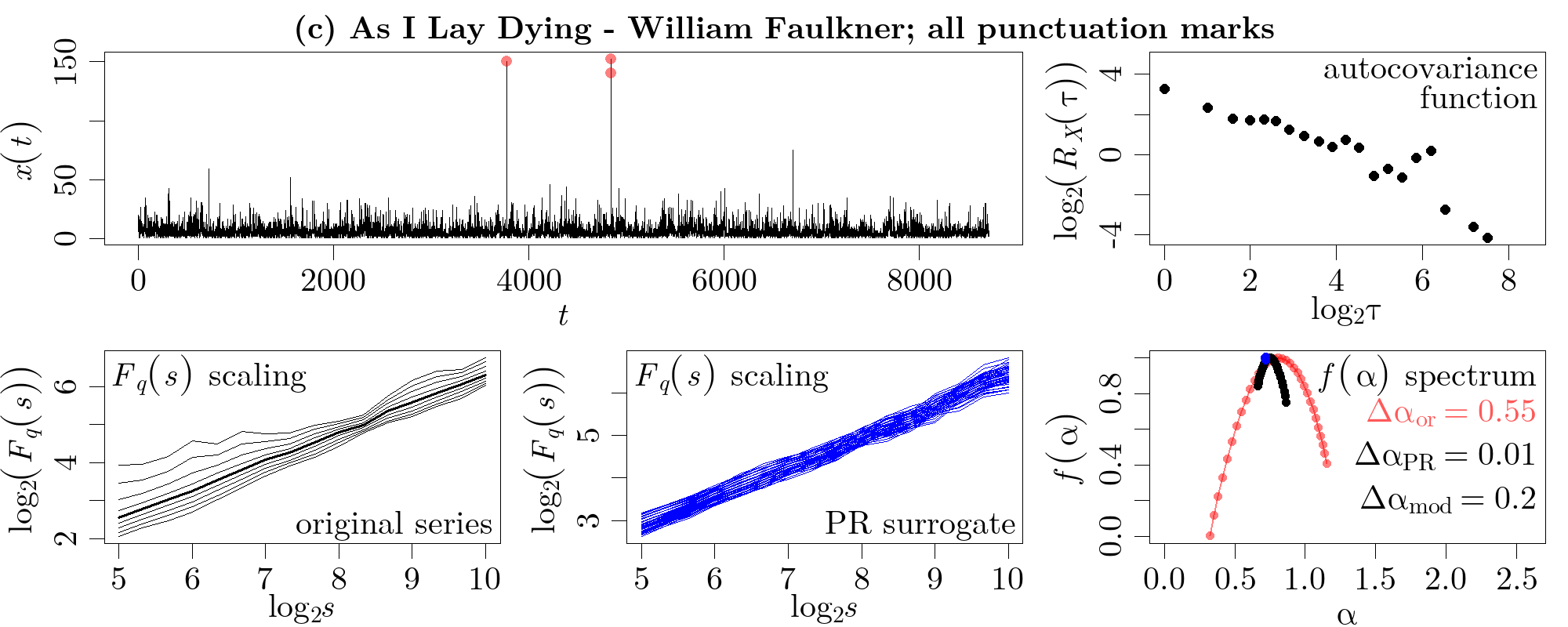}}}
\caption{Multifractal analysis (MFDFA) of time series representing punctuation waiting times for the same books as in Fig.~\ref{fig_MFDFA_sentence_lengths}. For each book, the upper row of plots contains a run chart $x(t)$ of the time series and a log-log plot of the autocorrelation function $R_X(\tau)$, while the lower row contains two log-log plots of the fluctuation functions $F_q(s)$ with integer $q$ in the range $[-4;4]$, for the original series (black) and for 5 phase-randomized (''PR'') surrogates (blue), and the respective singularity spectra $f(\alpha)$. For a time series of length $N$, the range of scales $s$ was chosen such that the log-log plots of $F_q(s)$ are approximately linear and $20 \le s \le N/5$, simultaneously. Width of the singularity spectra $\Delta \alpha_{\rm or}$ and $\Delta \alpha_{\rm PR}$ are given in the upper right corner of the $f(\alpha)$ plot, for the original and the PR surrogate series, respectively. Calculation of $f(\alpha)$ for the surrogate series involved calculation of the generalized Hurst exponents $h(q)$ and averaging them over 5 independent realizations. As expected, $f(\alpha)$ for the PR surrogate series nearly collapsed to single points as nonlinear correlations responsible for multifractality were destroyed. In case of \textit{As I Lay Dying}, the original time series $x(t)$ contains 3 values significantly larger than the remaining ones -- they are marked by red dots, and the singularity spectrum $f(\alpha)$ corresponding to the series containing these values is also plotted in red (its width is labeled $\Delta \alpha_{\text{or}}$). However, since there are only 3 such observations, they are treated as outliers, and therefore the plots of $R_X(\tau)$, $F_q(s)$ and $f(\alpha)$ pertain to the series with these extreme values removed (the width of the relevant $f(\alpha)$ spectrum is labeled $\Delta \alpha_{\text{mod}}$).}
\label{fig_MFDFA_punctuation_waiting_times}
\end{minipage}
\end{figure}

\begin{figure}
\centering
\begin{minipage}{\figurecustomwidth}
\centering
\includegraphics[width=0.75\textwidth]{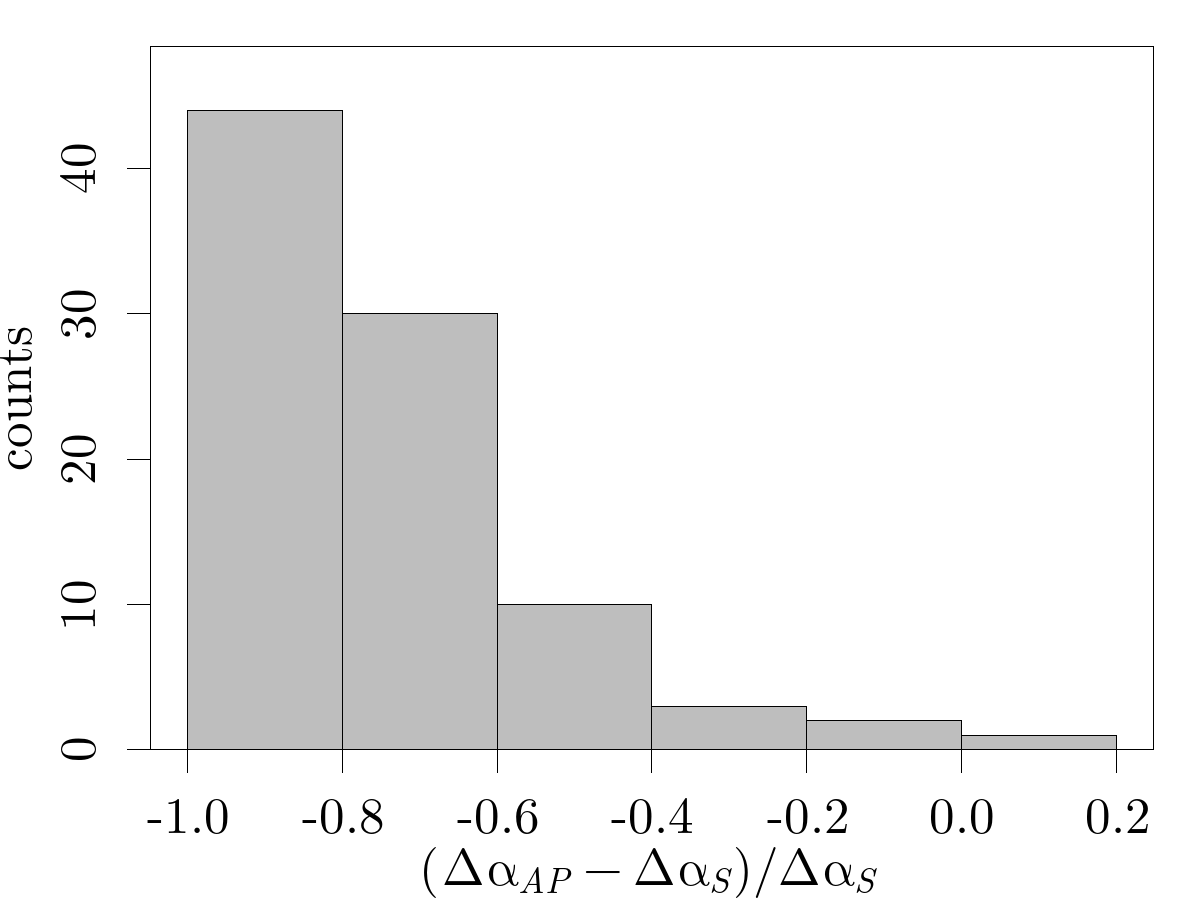}
\caption{Histogram of $(\Delta \alpha_{\rm AP} - \Delta \alpha_{\rm S)}/\Delta \alpha_{\rm S}$, where $\Delta \alpha_{\rm S}$ and $\Delta \alpha_{\rm AP}$ are widths of the singularity spectra for sentence lengths and punctuation waiting times, respectively. Results for sample books of at least 3000 sentences, having $\Delta \alpha_{\rm S}$ greater or equal to 0.2 are shown. The quantity $(\Delta \alpha_{\rm AP} - \Delta \alpha_{\rm S})/\Delta \alpha_{\rm S}$ is a relative change of the singularity spectrum width observed while changing the text representation from sentence lengths to punctuation waiting times.}
\label{fig_MFDFA_from_sentences_to_all_punctuation_spectrum_width_change}
\end{minipage}
\end{figure}

In terms of fluctuation scaling, punctuation waiting times typically display more uniform behaviour than sentence lengths (Fig.~\ref{fig_MFDFA_punctuation_waiting_times}) -- their singularity spectra are usually significantly narrower than the spectra of sentence lengths, although non-negligible spectrum width can sometimes be observed. However, it should be noted that in some cases, non-zero width of $f(\alpha)$ spectrum might be a largely spurious effect, resulting from the impact of a few outliers in the studied time series on the numerical procedures of determining $f(\alpha)$ -- this can be observed for \textit{As I Lay Dying} in Fig.~\ref{fig_MFDFA_punctuation_waiting_times}. The histogram in Fig.~\ref{fig_MFDFA_from_sentences_to_all_punctuation_spectrum_width_change} illustrates how much the singularity spectrum width decreases when the representation of a text changes from sentence lengths to punctuation waiting times. The quantity presented in the histogram is the relative change of singularity spectrum width, that is $(\Delta \alpha_{\rm AP} - \Delta \alpha_{\rm S})/ \Delta \alpha_{\rm S}$, where $\Delta \alpha_{\rm S}$ is the spectrum width for sentence lengths and $\Delta \alpha_{\rm AP}$ is the spectrum width for punctuation waiting times (the subscript ''AP'' is derived from ``all punctuation marks''). The books used here satisfy three additional conditions. First, they have no less than 3000 sentences each. Second, the log-log plots of their sentence lengths' fluctuation functions $F_q(s)$ for all $q \in [-4;4]$ are approximately linear for $s$ in the range $[20; N/5]$, where $N$ is the overall number of sentences in a text. Third, the width of their singularity spectra of sentence lengths is not less than 0.2. The presented conditions aim to ensure that the books are sufficiently long and have a range of $F_q(s)$ scaling wide enough to provide reasonable amount of data for the automated estimation of multifractal properties and that their sentence lengths exhibit at least weak multifractality.

Comparing sentence lengths and punctuation waiting times in terms of long-range correlations, fractality, and multifractality provides an insight into the significance of punctuation's role in language. In a sense, the properties of punctuation waiting times seem more universal than the corresponding properties of sentence lengths  -- for example, the variability of Hurst exponents among different texts is lower when all punctuation marks are considered instead of only the marks which divide texts into sentences. Also, in terms of fluctuation scaling, punctuation marks treated collectively determine a structure more homogeneous than the one constituted by sentences -- this fact is reflected by a stronger inclination of punctuation waiting times towards monofractality. The presented results can be viewed as being in agreement with a common intuition that the division of a text into sentences involves some degree of arbitrariness. A message or thought which is expressed by a long sentence, composed of several components, usually can also be expressed by a few short sentences, each of which corresponds to some component of the long sentence. Therefore, the number and the lengths of the used sentences depend on author's choice. However, when the first of the two options is chosen (one long sentence), the components of the sentence usually have to be separated by punctuation marks (comma, for instance). Hence, a certain number of such marks has to appear inside the sentence and punctuation waiting times are not arranged entirely freely.

\subsubsection{Parts-of-speech ranks}

A text representation that differs from all the above-discussed ones is the part-of-speech rank time series, in which each word is mapped to a rank that the associated part of speech describing this word has got in the part-of-speech rank plot. It is related to the grammar properties of a given language more than to its lexical ones. Exactly this representation was analyzed in~\cite{DeSantisE-2023a} based on selected ancient and contemporary texts written in Ancient Greek, Coptic, Arabic, Neo-Latin, Italian, French, Catalan, Spanish, German, and English. These languages belong to a few distinct language families, which made the analysis exceptionally broad. The results showed that the considered time series display various degrees of multifractal strength, among which the Ancient Greek texts (mainly the Biblical texts) are the richest in this context. This seems to be a property of the Greek language rather than a property of writing style, because the same Biblical texts show relatively suppressed multifractality if written in Coptic. On the opposite pole one can find English, whose multifractal strength expressed by the $f(\alpha)$ widths is the smallest on average. A noticeably higher results are shown by the analyzed Romance languages. Interestingly, this representation of \emph{Ulysses} by J.~Joyce is more multifractal than the other typical texts written in this language, which goes in parallel with an earlier work~\cite{DrozdzS-2016a}, where time series of sentence lengths were studied. As an additional result of~\cite{DeSantisE-2023a}, it was shown that one can successfully classify languages into their higher-level families.


\section{Linguistic networks}
\label{sect::linguistic_networks}

A system that consists of a number of constituents interacting or having some relationship with each other can be represented by a graph or a network (these notions are used interchangeably). Natural language with its structure consisting of basic units (phonemes, characters, words, etc.), the general rules governing their use and arrangement (syntax and grammar), and meaning can serve as an example of such a system. This is why network analysis has become popular in quantitative linguistics, mostly owing the accelerated development the network science has been undergoing in recent decades.


\subsection{Basic concepts in network theory}

\begin{figure}[!b]
\centering
\begin{minipage}{\figurecustomwidth}
\centering
\subfloat[]{\includegraphics[width=0.4\textwidth]{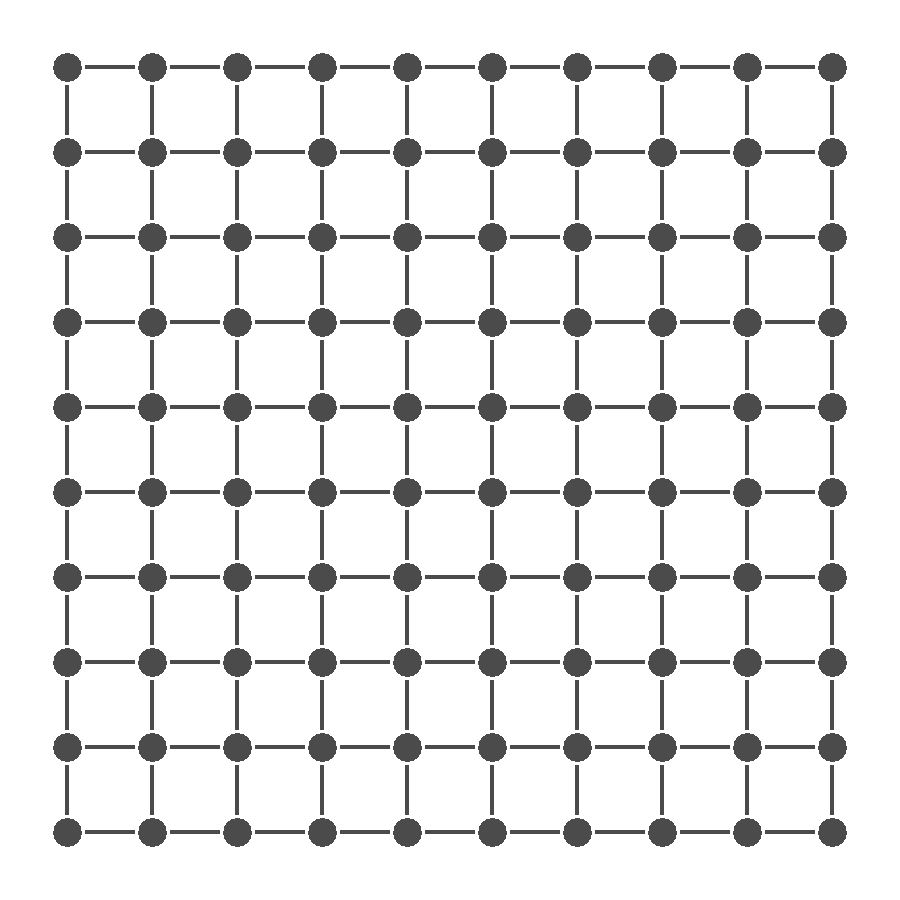}}
\hfill
\subfloat[]{\includegraphics[width=0.5\textwidth]{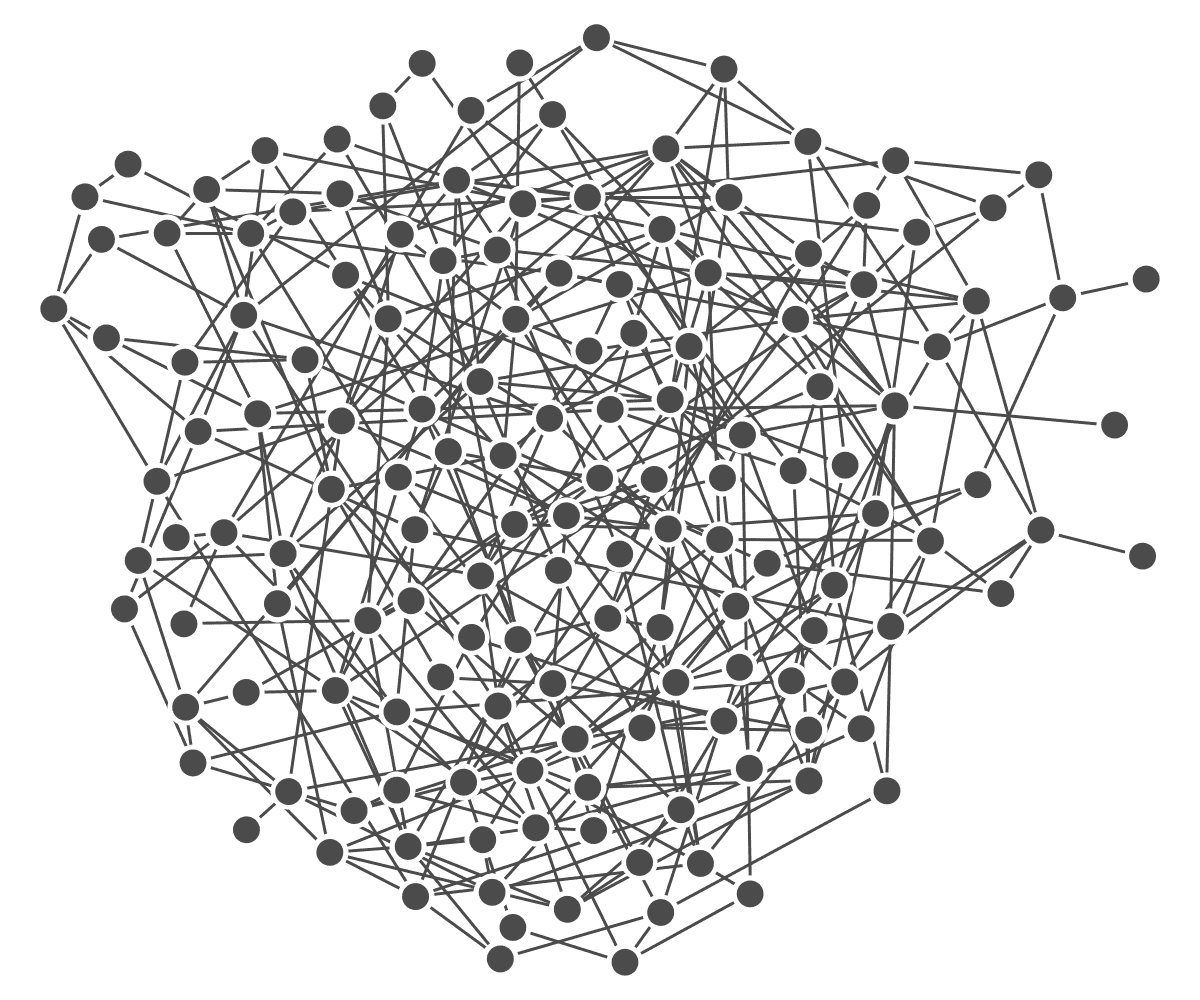}}
\hfill
\subfloat[]{\includegraphics[width=0.7\textwidth]{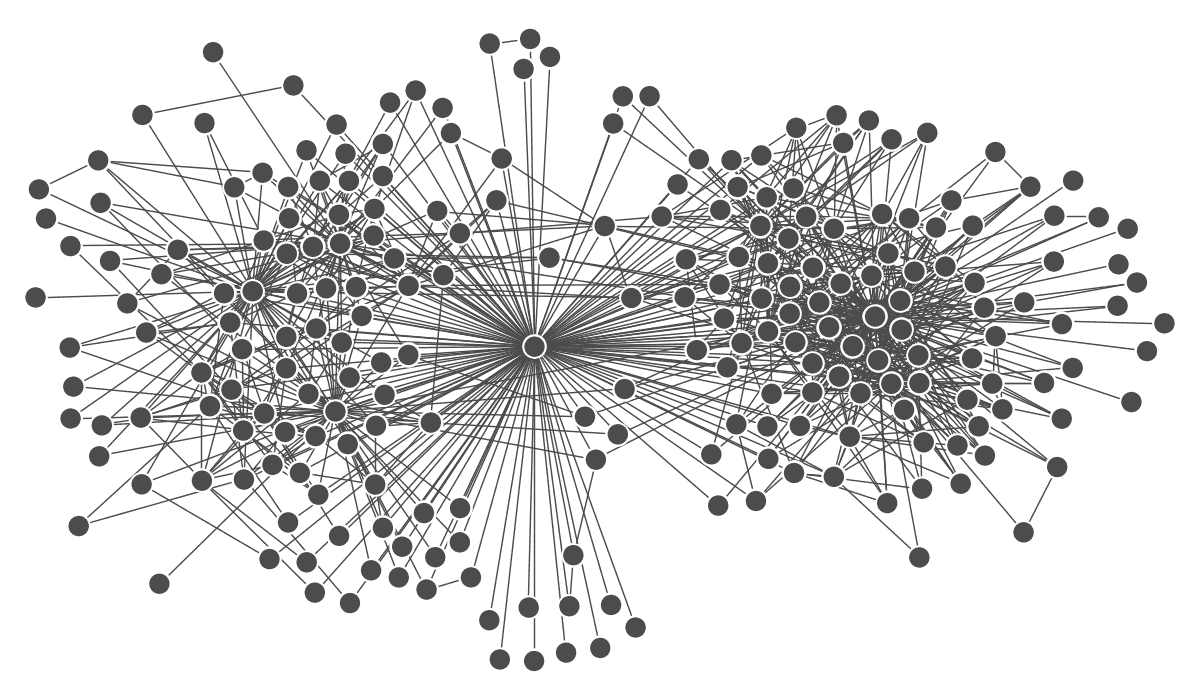}}
\hfill
\caption{Examples of networks: (a) a network with simple, regular structure, (b) a random network, (c) a network with nontrivial organization.}
\label{fig::network_examples}
\end{minipage}
\end{figure}

From a formal point of view, a graph is a pair of sets $G=(V,E)$, where $V$ is a set of \textit{vertices} (or \textit{nodes}) and $E$ is a set of two-element subsets of $V$ called \textit{edges} or \textit{links}. According to this definition, a graph consists of objects and connections between these objects. Connections are binary in their nature, that is, a pair of distinct vertices is either connected by an edge or not (Fig.~\ref{fig::network_examples}). However, a number of modifications can be introduced to generalize this concept. If pairs of vertices are allowed to be connected by more than one edge or edges are allowed to connect a vertex with itself (forming a \textit{loop}), then a \textit{multigraph} is obtained. If edges are assigned numbers (called \textit{weights}), then the graph becomes a \textit{weighted graph}. If, instead of being two-element unordered subsets of $V$, edges are ordered pairs, then an edge direction can be assigned and the graph is a \textit{directed graph}. Graphs that do not possess these attributes are called \textit{unweighted} and \textit{undirected}, respectively.

Let $N$ be the number of nodes of a graph $G$ and let the nodes be numbered by consecutive positive natural numbers. Then this graph can be completely described by an $N \times N$ \textit{adjacency matrix} ${\bf A}$ whose elements $a_{ij}$ ($i,j = 1,...,N$) are defined as
\begin{equation}
a_{ij} =
\begin{cases}
  1, & \text{if } \lbrace i,j \rbrace \in E(G),\\
  0, & \text{otherwise.}
\end{cases}
\label{eq_networks_adjmatrix_element_def}
\end{equation}
In a weighted graph, the elements of the adjacency matrix represent edge weights and they can be different from 0 and 1. In a directed graph, the adjacency matrix does not have to be symmetric since in such a graph the presence of an edge from $i$ to $j$ does not imply the presence of an edge from $j$ to $i$. If loops are allowed, then the diagonal of the adjacency matrix might contain elements not equal to zero.

Important concepts are \textit{paths} and \textit{connected graphs}. A path from a vertex $u \in V$ to another vertex $v \in V$ is a sequence $(e_1, e_2, e_3, ..., e_n)$ of edges of $G$ such that:
\begin{itemize}
\item $u$ is the first endpoint of $e_1$;
\item $v$ is the second endpoint of $e_n$;
\item for each $k = 1,...,n$, the second endpoint of $e_k$ is the first endpoint of $e_{k+1}$.
\end{itemize}
The distinction between the first and the second endpoint of an edge is important only in a directed graph; in an undirected graph, they are interchangeable. A path which starts and ends in the same vertex is called a \textit{cycle}. A graph is a connected graph when any vertex can be reached from any other vertex, that is, when for any pair of vertices $(u,v)$ there exists a path from $u$ to $v$.

Let $G' \subset G$, then its dilation $\delta(G') \subset G$ contains $G'$ and all the nodes that are direct neighbours of at least one node of $G'$. A $d$-dilation is obtained by repeating a dilation operation $d$ times: $\delta_d(G')=\delta(\delta(...\delta(G')...))$ with $\delta_0 \equiv G'$. A $d$-ring $R_d(G')$ is then a subset of vertices of $G$ such that $R_d(G')=V(\delta_d(G')) \setminus V(\delta_{d-1}(G'))$. It is a generalization of the concept of neighbourhood: a set of all $k_i$ direct neighbours of a node $v_i$ can be denoted by $R_1(v_i)$. Yet another concept is $k$-core of $G$: it is a subgraph $G'={\rm core}_k(G)$ such that $k_i \ge k$ for all vertices $v_i \in G'$ and it is the largest such a subset of $G$.

\subsubsection{Network characteristics}

\textit{Vertex degree and strength.} In an unweighted network, the degree of a node $v$ is the number of edges incident to $v$, that is, the number of edges that $v$ belongs to. The degree of $v$ is denoted by $\deg(v)$. In weighted networks, the degree can be generalized to the weighted degree (also called \textit{strength} and denoted by $\text{str}(v)$), which is the sum of weights of the edges incident to $v$. An important relationship regarding the (unweighted) node degrees is the degree sum formula: $\sum_{v \in V} \deg(v) = 2M$, where $M$ denotes the number of edges in the network. In a directed network, incoming and outcoming edges are distinguished, so a node may have different in-degree and out-degree.

\textit{Sparsity.} It can be defined for unweighted networks as a global measure of edge density. It is a ratio between the number of edges $M$ that exist in a network and a total number of possible edges $M_{\rm max}$, which is equal to $N(N-1)/2$ for an undirected network.

\textit{Clustering coefficient.} In unweighted networks, the clustering coefficient of a given vertex represents the probability that two randomly chosen direct neighbours of that vertex are also direct neighbours of each other. A \emph{direct neighbour} of a vertex $v$ is here understood as a vertex connected with $v$ by an edge. Let $m_{v}$ be the number of edges in the network that link the direct neighbours of $v$ with other direct neighbours of $v$. Then the clustering coefficient $C_{\rm u}$ (the subscript ``${\rm u}$'' comes from the word ``unweighted'') of the node $v$ is given by:
\begin{equation}
C_{\rm u}(v) = \frac{2m_{v}}{\deg(v)\cdot(\deg(v)-1)}.
\end{equation}
Generalization of the clustering coefficient onto the weighted networks can be done in multiple ways. For example, a definition proposed by Barrat et al.~\cite{Barrat2004} pertains to individual vertices of a network and is formulated as follows. Let $S(v)$ denote the set of direct neighbours of a vertex $v$, and let $w_{uv}$ denote the weight of the edge connecting vertices $u$ and $v$ (if there is no such edge, then $w_{uv}=0$). Let $a_{uv}$ denote an unweighted adjacency matrix element, i.e.\ a number defined in Eq.~(\ref{eq_networks_adjmatrix_element_def}). The weighted clustering coefficient of $v$ is written as:
\begin{equation}
C_{\rm w}(v) = \frac{1}{\text{str}(v) \cdot \left(\deg(v)-1\right)} \sum_{u,t \in S(v)} \! \frac{w_{vu} + w_{vt} }{2} a_{vu}a_{ut}a_{tv},
\end{equation}
where summation is over all pairs $(u,t)$ of neighbours of $v$. It is worth noting that if $\deg{v}=0$ or $\deg{v}=1$, the clustering coefficient cannot be determined from the above-given formulas; in such cases, it is often assumed to be equal to 0. Global clustering coefficient can be defined also in more than one way. However, it is often based on averaging of the local clustering coefficients:
\begin{equation}
C = \frac{1}{N} \sum_{v \in V} C(v).
\end{equation}
This formula is identical for both the unweighted and weighted networks.

\textit{Average shortest path length.} In unweighted networks, the length of a path between two vertices is the number of edges constituting that path. In weighted networks, the length of a path can be defined as the sum of the reciprocals of edge weights on that path. The length of the shortest path between vertices $u$ and $v$ is also called the \emph{distance} between $u$ and $v$ and is denoted by $d(u,v)$. The average shortest path length $\ell(v)$ of a vertex $v$ is the average distance from $v$ to every other vertex in the network. It is one of the measures of the centrality of a vertex in the network, and is given by the formula:
\begin{equation}
\ell(v) = \frac{1}{N-1} \sum_{ u \in V\backslash \{v\}}{d(v,u)}.
\label{eq_networks_mean_shortest_path_1}
\end{equation}
This quantity has finite values only in connected networks. If there are at least two vertices that are not connected by any path, the distance between them is not defined; usually it is treated as infinite and $\ell(v)$ cannot be~calculated. If local average distances $\ell(v)$ for all $v \in V$ are given, then the global average distance in the whole network can be expressed by
\begin{equation}
\ell = \frac{1}{N} \sum_{v \in V} \ell(v).
\label{eq_networks_mean_shortest_path_2}
\end{equation}
Eqs.~(\ref{eq_networks_mean_shortest_path_1}) and~(\ref{eq_networks_mean_shortest_path_2}) apply to both unweighted and weighted networks. A difference between the unweighted and the weighted average shortest path length arises as a consequence of different definitions of a distance in these two types of~networks.

\begin{figure}[!b]
\centering
\begin{minipage}{\figurecustomwidth}
\centering
\subfloat[]{\includegraphics[width=0.3\textwidth]{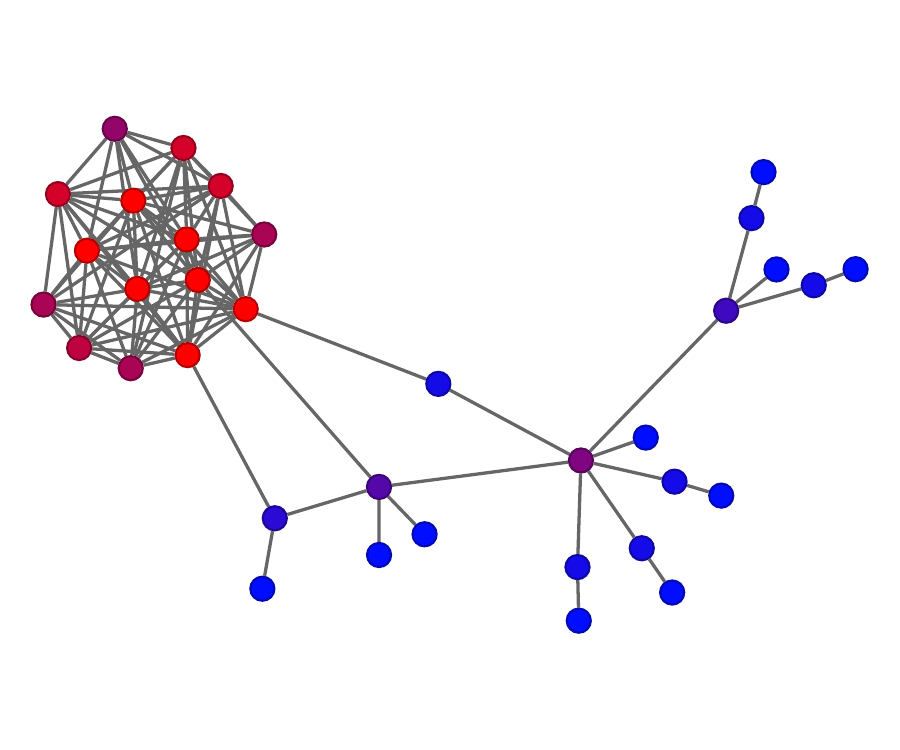}}
\hfill
\subfloat[]{\includegraphics[width=0.3\textwidth]{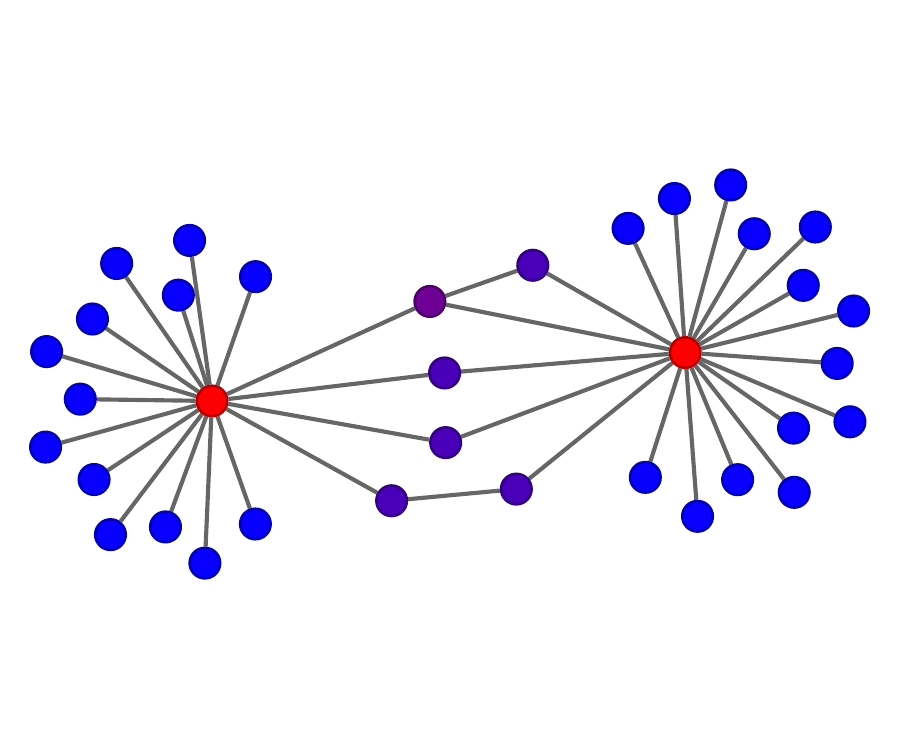}}
\hfill
\subfloat[]{\includegraphics[width=0.3\textwidth]{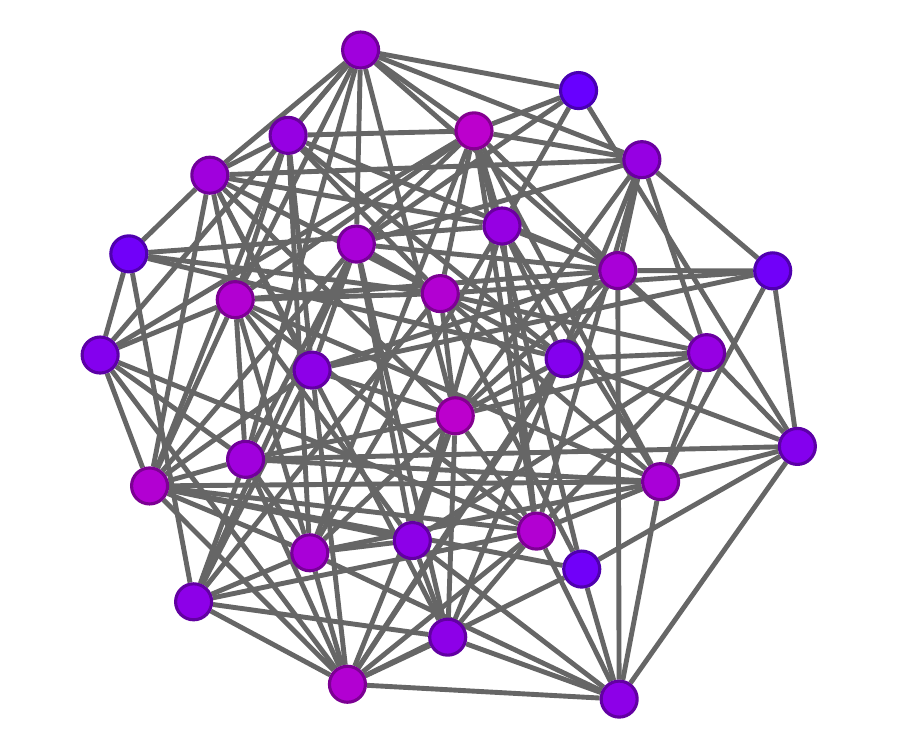}}
\hfill
\caption{Examples of unweighted networks with different values of assortativity coefficient. The network in (a) is assortative ($r=0.63$), the network in (b) is disassortative ($r=-0.91$), and in the network in (c), the degrees of directly linked vertices are not correlated ($|r|<0.01$). In each network, the vertices are colored according to their degree: blue, purple, and red corresponds to low, medium, and high degree, respectively.}
\label{fig_network_characteristics_assortativity_example}
\end{minipage}
\end{figure}

\textit{Assortativity.} It is a global characteristic of a network, describing the preference of vertices to attach to their pars in terms of vertex degree. A network is called assortative if vertices with high degree tend to be directly connected with other vertices with high degree and low-degree vertices are typically directly connected to vertices, which also have low degree. In disassortative networks, high-degree nodes are typically directly connected to nodes with small degree. Network assortativity is often quantified by assortatvity coefficient. In unweighted networks, it can be defined as the Pearson correlation coefficient between degrees of the nodes that are connected by an edge. Let $(u,v)$ denote an ordered pair of vertices that are connected by an edge. Since edges are undirected and the pair $(u,v)$ is ordered, two such pairs can be assigned to each edge in the network. For each pair one can determine the degrees of vertices $u$ and $v$, and form a pair $\left(\deg(u), \deg(v)\right)$. The set of all pairs $\left(\deg(u), \deg(v)\right)$ for all edges can be treated as the set of values of a two-dimensional random variable $(X,Y)$. With such a notation, the assortativity coefficient $r_{\rm u}$ is expressed by the Pearson correlation coefficient of variables $X$~and~$Y$:
\begin{equation}
r_{\rm u} = \text{corr}(X,Y), \quad -1 \le r_{\rm u} \le 1.
\label{eq_networks_unweighted_assortativity_def}
\end{equation}
A generalization of this formula to weighted networks can be done by replacing vertex degrees by their strengths and calculating the weighted correlation coefficient instead of the usual one. Let $(X,Y)$ be a pair of random variables with the values $(x,y)=\left(\text{str}(u), \text{str}(v)\right)$ for all pairs of vertices $(u, v)$ connected by an edge. Let $w$ be a function that assigns to every such pair the weight of an edge connecting $u$ and $v$. Then the weighted assortativity coefficient $r_{\rm w}$ can be written as
\begin{equation}
r_{\rm w} = \text{wcorr}(X,Y;w),  \quad -1 \le r_{\rm w} \le 1.
\label{eq_networks_weighted_assortativity_def}
\end{equation}
where $\text{wcorr}(X,Y;w)$ denotes the weighted Pearson correlation coefficient of variables $X$ and $Y$ with the weighing function $w$. Some of the definitions encountered in literature, for example in~\cite{Leung2007}, are equivalent to the one given above. Networks with positive $r$ are assortative, while networks with negative $r$ are disassortative. Examples of networks with different values of assortativity coefficient are presented in Fig.~\ref{fig_network_characteristics_assortativity_example}.

The assortativity coefficient may also be defined by using ranks instead of values and the Spearman coefficient instead of the Pearson one. In this case it is referred to as \textit{rank assortativity coefficient} and denoted by $\rho$ to avoid ambiguity. The reason behind this idea is the fact that, unlike the Pearson coefficient, the Spearman correlation coefficient allows one to detect monotonic relationships whether they are linear or nonlinear. To obtain unweighted and weighted rank assortativity coefficients $\rho_{\rm u}$ and $\rho_{\rm w}$, it is sufficient to replace $X$ and $Y$ with their ranks, $R(X)$, $R(Y)$, in Eqs.~(\ref{eq_networks_unweighted_assortativity_def}) and~(\ref{eq_networks_weighted_assortativity_def}), respectively.

\textit{Modularity.} Modularity is a global characteristic measuring the extent to which a set of vertices can be divided into disjunctive subsets that maximize edge density within them and minimize the number of inter-subset edges. For an unweighted network $G(V,E)$, \textit{partition of $G$} is a division of $V$ into disjoint subsets called \textit{modules}, \textit{clusters}, or \textit{communities}. Let $a_{uv}$ denote an adjacency matrix element defined by Eq.~(\ref{eq_networks_adjmatrix_element_def}) and let $c_{v}$ denote a module to which a vertex $v$ is assigned by a given partition. \textit{Modularity of the partition} is defined as:
\begin{equation}
q_{\rm u} = \frac{1}{2M} \sum_{u,v \in V} \left( \left[ a_{uv}-\frac{\deg(u) \deg(v)}{2M} \right] \delta(c_u,c_v) \right), \label{eq_networks_unweighted_modularity_def}
\end{equation}
where $M$ is the number of network edges and $\delta(c_u,c_v)=1$ if $c_u=c_v$ and 0 otherwise. Modularity is limited to the range $-1 \le q_{\rm u} \le 1$ and indicates whether the edge density within the modules is larger or smaller than it would be if the edges were distributed at random. A random network that serves as a reference is constructed by using the \emph{configuration model} (Sect.~\ref{subsec_random_network_models}). 

\begin{figure}[!b]
\centering
\begin{minipage}{\figurecustomwidth}
\centering
\subfloat[]{\includegraphics[width=0.4\textwidth]{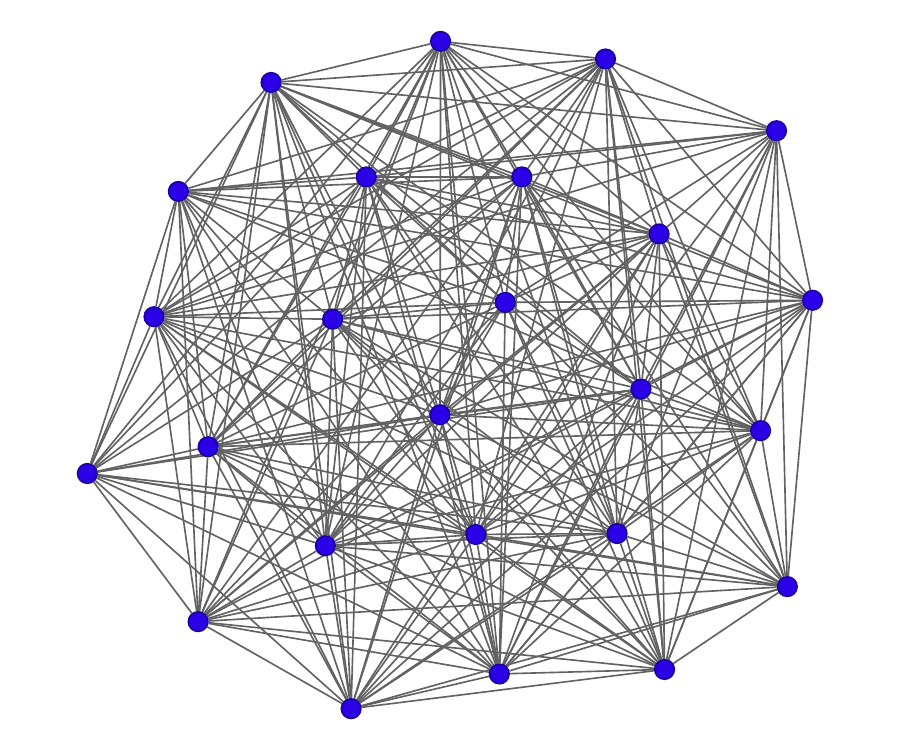}}
\qquad
\subfloat[]{\includegraphics[width=0.4\textwidth]{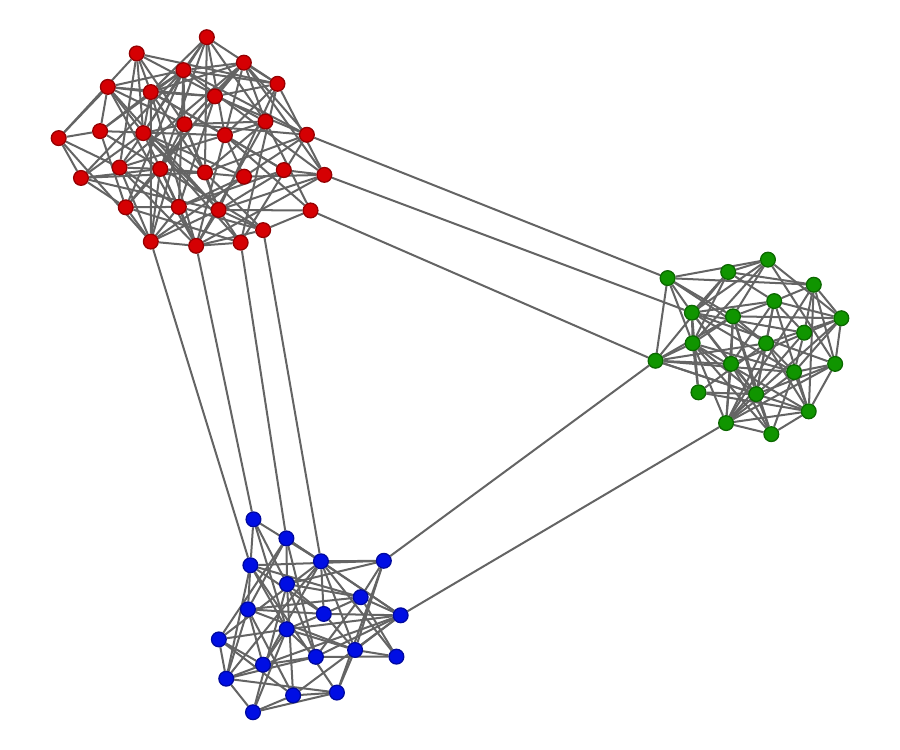}}
\hfill
\caption{Examples of unweighted networks with (a) small modularity ($Q_{\rm u}=0.04$) and (b) large modularity ($Q_{\rm u}=0.64$). Colors in (b) represent the partition which leads to the given modularity value.}
\label{fig_network_characteristics_modularity_example}
\end{minipage}
\end{figure}

The \textit{modularity of a network}, denoted by $Q_{\rm u}$, is the maximum value among the modularities $Q_{\rm u}$ of all possible partitions. Determining the network modularity precisely is computationally intractable, hence a number of heuristic algorithms have been proposed, e.g., the Louvain algorithm~\cite{Blondel2008}. A generalization of modularity onto weighted networks can be done by replacing the quantities appearing in Eq.~(\ref{eq_networks_unweighted_modularity_def}) by their weighted counterparts. If $W$ is the sum of all edge weights, the modularity of a given partition is expressed by
\begin{equation}
q_{\rm w} = \frac{1}{2W} \sum_{u,v \in V} \left( \left[ w_{uv}-\frac{\text{str}(u) \, \text{str}(v)}{2W} \right] \delta(c_u,c_v) \right).
\end{equation}
Again, the weighted network modularity $Q_{\rm w}$ is the maximum modularity obtained from all possible partitions of the network. Examples of the networks with different values of modularity are presented in Fig.~\ref{fig_network_characteristics_modularity_example}.

\subsubsection{Random network models \label{subsec_random_network_models}}

A number of network properties can be considered universal to some extent since they are shared among networks representing many different systems. This led to the development of random network models, i.e., the numerical procedures generating networks with specific properties predefined but being random in terms of the remaining characteristics~\cite{Hofstad2016}. These models allow one to investigate the origin of certain phenomena and the organization patterns observed in networks representing natural systems.

\textit{Erdős–Rényi networks.} A model that is considered the most referential one is the Erdős–Rényi model~\cite{Erdos1959,Erdos1960,Gilbert1959}. The model generates unweighted and undirected graphs (however, a generalization into directed graphs is straightforward) in two slightly different ways. The first one, denoted by $G(N,M)$, is equivalent to choosing a network at random from the set of all graphs with $N$ vertices and $M$ edges. It is constructed by defining an $N$-element vertex set and connecting $M$ vertex pairs chosen at random. The second one, denoted by $G(N,p)$, is constructed by connecting each possible pair of vertices by an edge with a fixed probability $p$. Both ways of generating the random networks provide one with networks that behave similar to each other in certain aspects for large enough $N$~\cite{Hofstad2016}. The Erdős–Rényi networks have no specific patterns of organization beyond the ones arising from distributing edges randomly over the graph. Their capability of modeling real-world networks is severely limited, however. One of the fundamental reasons for this is their binomial distribution of node degrees $k$: if $P(k)$ denotes the probability mass function, then for a large enough $G(N,p)$ network, one can write:
\begin{equation}
P(k) = \binom{N-1}{k} p^k(1-p)^{N-1-k}.
\end{equation}
As the relative dispersion of the the binomial distribution (defined as standard deviation divided by mean) decreases with increasing $N$, node degrees are concentrated around their mean in an Erdős–Rényi network. In the real-world networks, on the other hand, $k$ usually spans many orders of magnitude, which is often a consequence of a power-law distribution of $k$. The fact that certain properties of the Erdős–Rényi networks (e.g., the shape of a node degree distribution) seem to be unrealistic in many situations led to the development of other random network models that attempt to mimic at least some of the characteristics of the empirical networks.

\textit{Configuration model.} Configuration model~\cite{Newman2010} generates networks with an explicitly prescribed node degree distribution. The distribution is specified by a sequence of numbers, in which each number is degree of a node. If a given sequence of numbers $k_1,...,k_N$ satisfies the conditions required to constitute a valid degree sequence (the sum of all the numbers in the sequence has to be even, for instance), then, in the simplest variant of the model, an undirected and unweighted network is generated as follows. A set of $N$ nodes is created with each node $i$ being given the number of edge stubs equal to its target degree $k_i$. In each step two stubs from different nodes are chosen at random and connected by an edge. This is repeated until there is no unconnected stub left. The resulting network has nodes with degrees specified by the sequence $k_1,...,k_N$.

The configuration model is often used in situations when there is need to determine whether certain properties of a network are directly related to a respective node degree distribution. The model allows one to construct a randomized version of the studied network (using a degree sequence taken from that network), which can be expected to preserve the properties resulting from the node degree distribution and to be random in other aspects. Just as the Erdős–Rényi model represents networks whose properties are the result of a random arrangement of edges without any specific restrictions, the configuration model can represent networks whose structure is random, but preserves the prescribed node degrees. It is important to note that the generating procedure presented above is the simplest algorithmic approach to the configuration model but it has certain undesirable properties like a possibility of multiple edges and loops, which makes it a multigraph, or a network might not be connected. However, there exist methods that overcome these problems (e.g.,~\cite{Viger2005a}).

\begin{figure}
\centering
\begin{minipage}{\figurecustomwidth}
\centering
\subfloat[]{\includegraphics[width=0.32\textwidth]{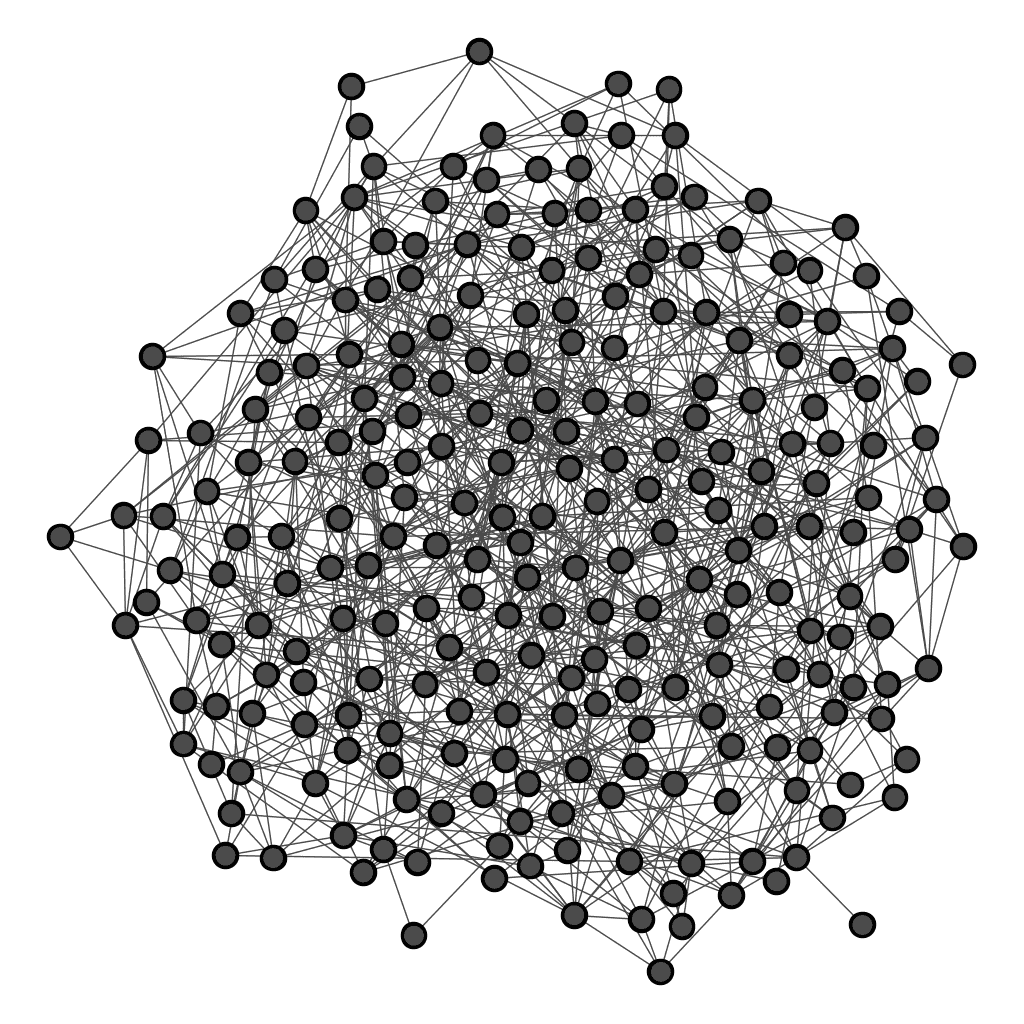}}
\hfill
\subfloat[]{\includegraphics[width=0.32\textwidth]{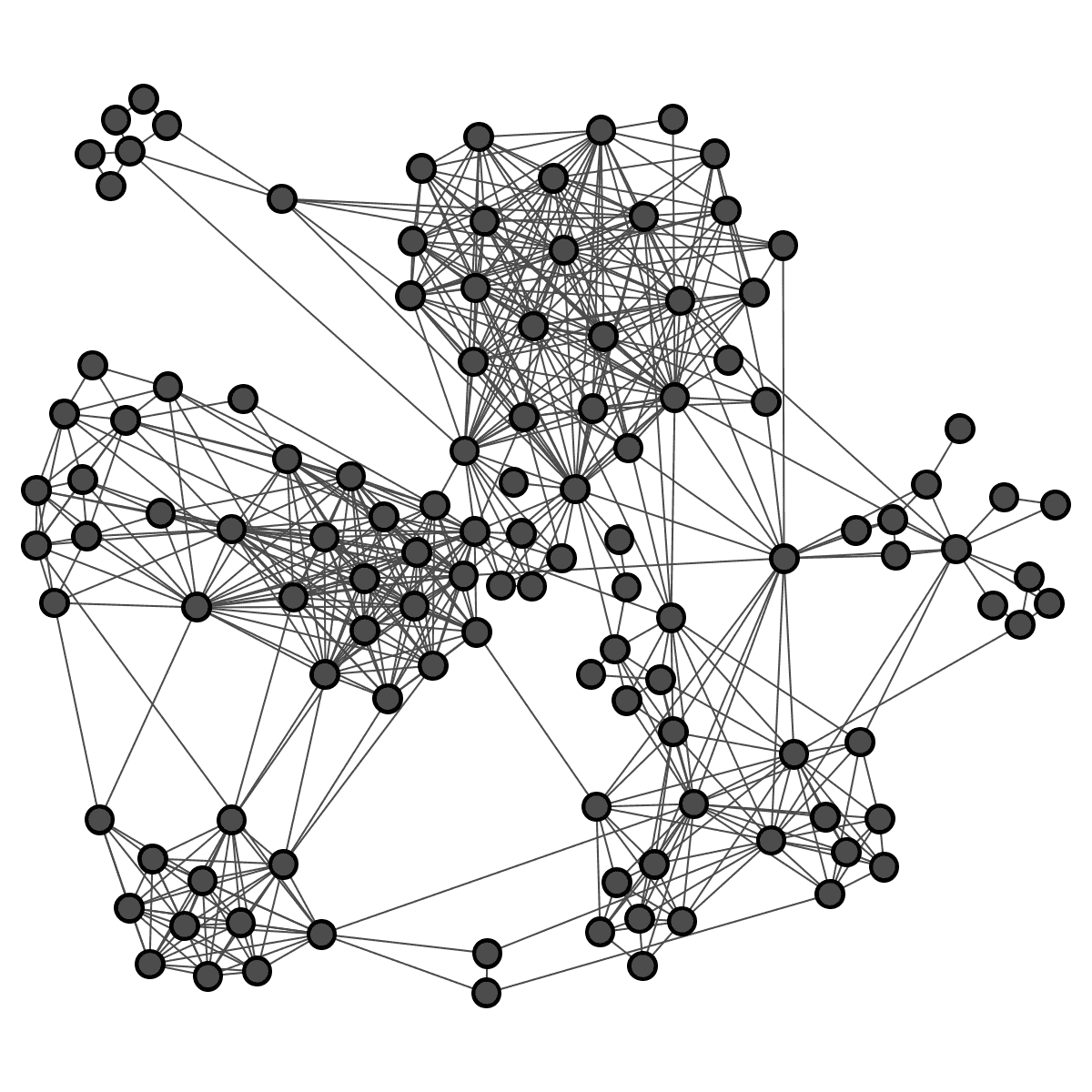}}
\hfill
\subfloat[]{\includegraphics[width=0.32\textwidth]{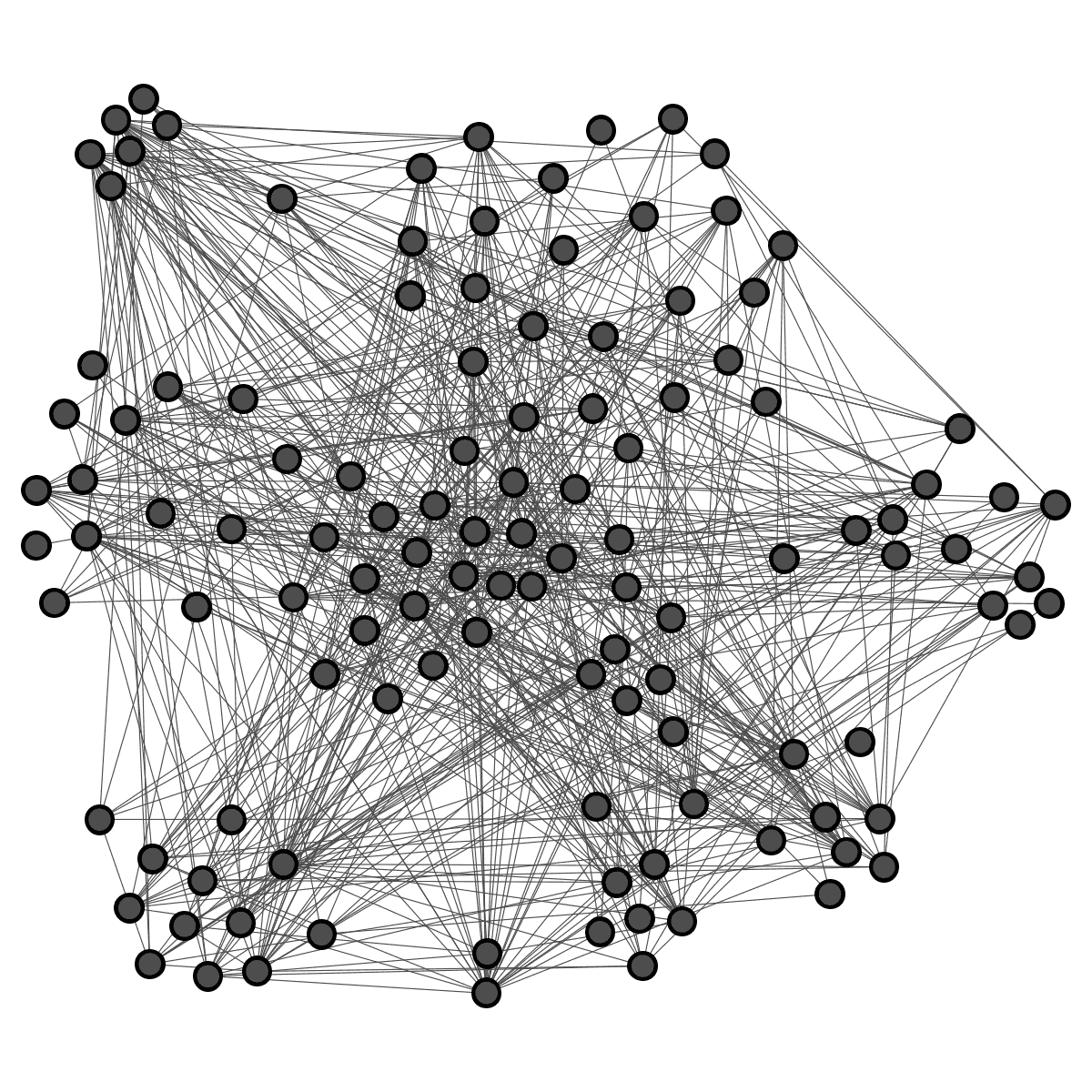}}
\hfill
\caption{Examples of random network models. An Erdős–Rényi network $G(N,p)$ with parameters $N=250$ and $p=0.03$ is shown in (a), a network with moderately modular organization is presented in (b), a randomization of the network (b) based on the configuration model -- that is, a network with the same number of nodes and edges and having the same degree distribution as in the original network -- is presented in (c).}
\label{fig::random_network_models_examples}
\end{minipage}
\end{figure}

The configuration model can serve as a starting point for more complicated network randomization procedures. For example, when a weighted network needs to be randomized in such a way that the unweighted node degrees are kept unchanged, one can generate a random network according to the configuration model by using the unweighted degrees as an input degree sequence and then randomly assign the edge weights from the original network to the edges of the generated network. The distributions of node degrees and of edge weights in the obtained network are identical to the distributions describing the original network (see Fig.~\ref{fig::random_network_models_examples}).

The two presented models certainly do not constitute an exhaustive list. There exist many other random network models designed for various applications, like testing computational methods of network analysis or explaining phenomena observed in systems having a network representation. An example worth mentioning in that context is the \textit{Barabási-Albert (BA) model}~\cite{Barabasi1999}, which uses a mechanism based on a Yule process to generate networks with power-law distributions of node degrees.

\subsubsection{Minimum spanning trees}

A problem that sometimes is encountered in network research is that an analysis of highly complicated structures might suffer from the presence of a large number of details that can make it difficult to get the general understanding of the essential properties. An investigation of large, densely connected networks can benefit from removing certain elements, which can be considered redundant from some perspective. A particularly useful concept in this context is a so-called minimum spanning tree (MST) graph. In order to define it, it is convenient to define the notions of a tree and a spanning tree first. A \textit{tree} is a graph that is connected, undirected, and acyclic (without cycles or loops). In a tree, exactly one path connects each pair of vertices. A \textit{spanning tree} of a connected graph $G=(V,E)$ is a tree with the same vertex set $V$ and a subset of the edge set $E$.

\begin{figure}[ht]
\centering
\begin{minipage}{\figurecustomwidth}
\centering
\subfloat[]{\includegraphics[width=0.47\textwidth]{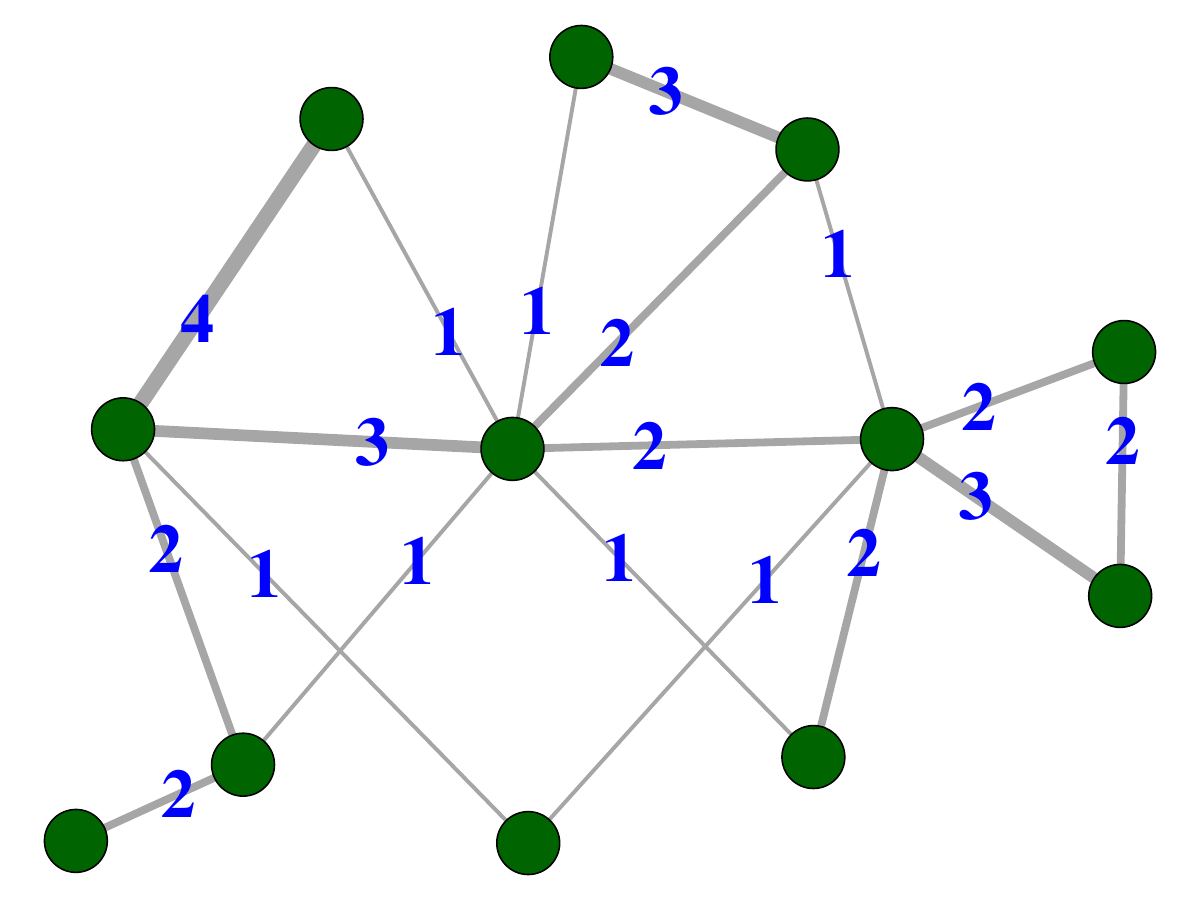}}
\hfill
\subfloat[]{\includegraphics[width=0.47\textwidth]{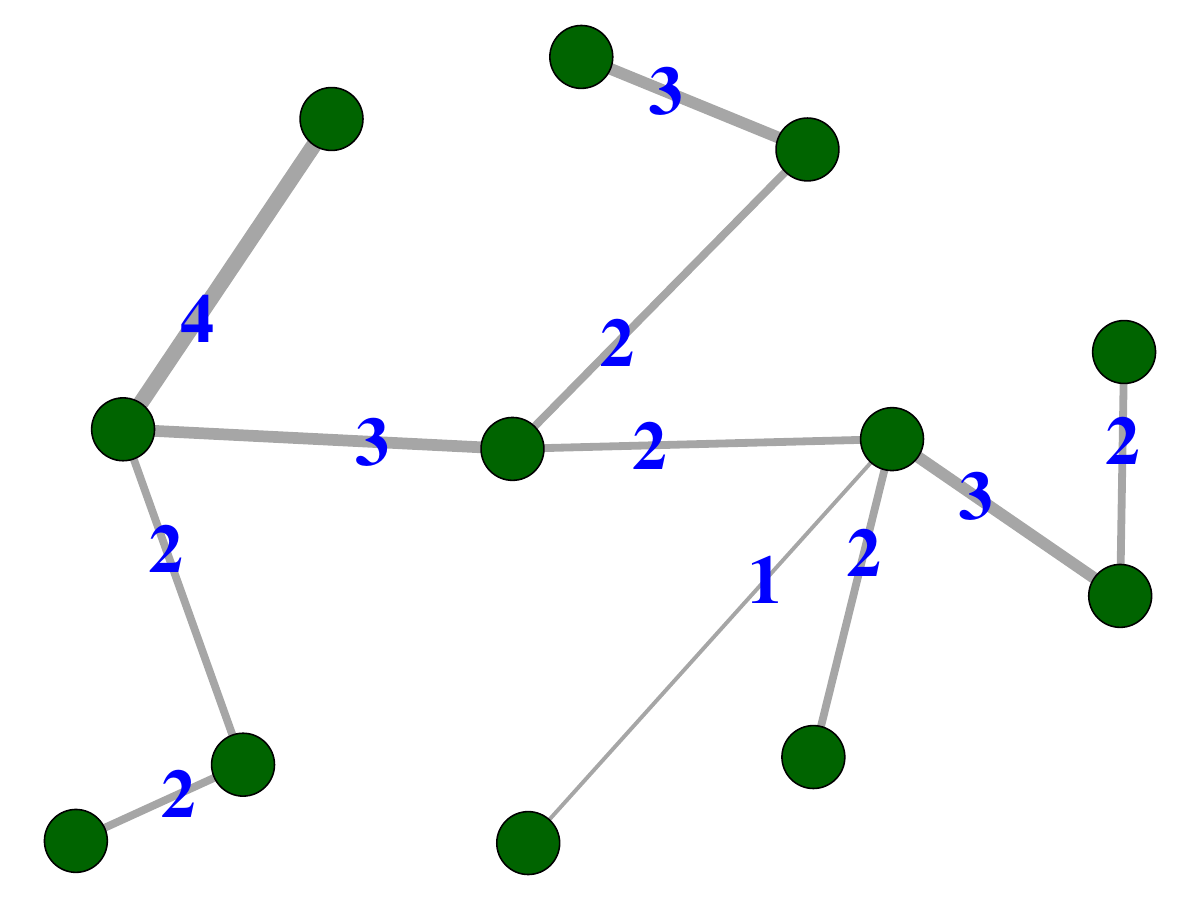}}
\hfill
\caption{(a) A weighted network and (b) a minimum spanning tree for that network. The numbers labeling edges are edge weights. Edge costs that are minimized by the MST are equal to reciprocal weights.}
\label{fig::MST_example}
\end{minipage}
\end{figure}

Let $G$ be an undirected, weighted connected graph, in which every edge has a real number called \textit{edge cost} assigned to it. A \textit{minimum spanning tree} of $G$ is defined as the spanning tree of $G$ having the minimum possible sum of edge costs. Edge costs can be identified with weights, but sometimes it is useful to distinguish these concepts. For example, if the edge weights in a network can be interpreted in terms of connection intensity -- the larger the weight, the stronger the relationship between the connected vertices -- introducing edge costs equal to inverse edge weights allows one to construct an MST that keeps only the most important edges of the original network. There are several algorithms finding the minimum spanning tree of a graph: Borůvka's algorithm, Kruskal's algorithm, and Prim's algorithm are the most popular ones~\cite{Tarjan1987}. A minimum spanning tree constructed from a sample network is presented in Fig.~\ref{fig::MST_example}. Minimum spanning trees have a number of applications in various fields, like financial market analysis~\cite{Drozdz2020,Watorek2021}, image processing~\cite{Peng2013}, analysis of brain networks~\cite{Tewarie2015}, and quantitative linguistics (see below). It is worth mentioning that although the concept of MST is inherently related to weighted networks, there exist methods of applying MST to unweighted networks by exploiting edge centrality measures~\cite{Kim2004}.

\subsubsection{Fractal properties of networks}

Scale-free networks (like the BA networks) owe their name to the power-law probability mass distribution of node degrees and the resulting lack of a characteristic scale. However, this is not the only aspect, in which organization of a network can be scale-free. It can also be organized into a hierarchical, statistically self-similar structure that can be quantified by using fractal analysis. For example, the box-counting method applied to a network relies on partitioning of a node set $V$ into possibly a small number of disjoint subsets $V_i$ (boxes) such that $\bigcup_{i} V_i = V$ and, for every pair of nodes $v_1, v_2$ belonging to the same $V_i$, the inter-node distance meets the condition $d(v_1,v_2) \le s$, where $s$ is a predefined box size. If the number of boxes obtained in such a partition is denoted by $N_s$, making partitions with many different values of $s$ gives a function $N(s)$, which can be power-law $N(s) \propto s^{-d_{\rm C}}$ for a fractal-structure network. In this case $d_{\rm C}$ can be interpreted as the box-counting dimension of the network~\cite{Song2005}. In general, the minimum number of boxes of given size required to cover a network cannot be computed exactly except for very small networks (it is an NP-hard problem). For that reason, a number of algorithms finding an approximate solution have been proposed, for example, the algorithm based on \textit{greedy coloring}~\cite{Song2007}.

Among the structural properties of a network that its fractal organization is related to is a tendency that large-degree nodes are separated from each other rather than being directly connected (this effect, occasionally called \textit{hub repulsion}, is expressed in disassortativity of such a network). Fractality also decreases network's vulnerability to attacks~\cite{Song2006}. Fractal analysis is sometimes applied to transformed or filtered versions of a network -- for example, the so-called \textit{network skeleton}~\cite{Goh2006} -- instead of the original one. In some networks, fractal properties of the original structure and of the network skeleton are approximately the same. However, there exist networks whose self-similarity can be detected only after applying a transformation which removes edges of little importance~\cite{Zhang2013}.


\subsection{Word-adjacency networks}

A number of problems related to natural language can be studied with the use of network theory. Networks allow to represent language on various levels of its structure -- they can represent word co-occurrences, semantic similarities, or grammatical relationships, for instance. Such networks, collectively called linguistic networks, often consist of a large number of nodes and edges and exhibit complex patterns of organization, but graphs with relatively simple structure (for example consisting of about a dozen of vertices), also have their applications in language-related areas of research (examples of such graphs are parse trees, presented in Appendix~\ref{appendix::formal.grammars}). Graphs and networks have been used to approach various practical problems related to natural language processing, like keyword selection, document summation, word-sense disambiguation, or machine translation~\cite{AntiqueiraL-2009a,Navigli2010,Amancio2011,MIHALCEA_2011}. Also, network formalism is used in research areas at the interface between linguistics and other scientific fields, for example in sociolinguistics, which investigates human language usage and evolution by studying social networks~\cite{DallAsta2006,Kalampokis2007}.

\begin{figure}
\centering
\begin{minipage}{\figurecustomwidth}
\centering
\subfloat[]{\includegraphics[width=0.43\textwidth,valign=c]{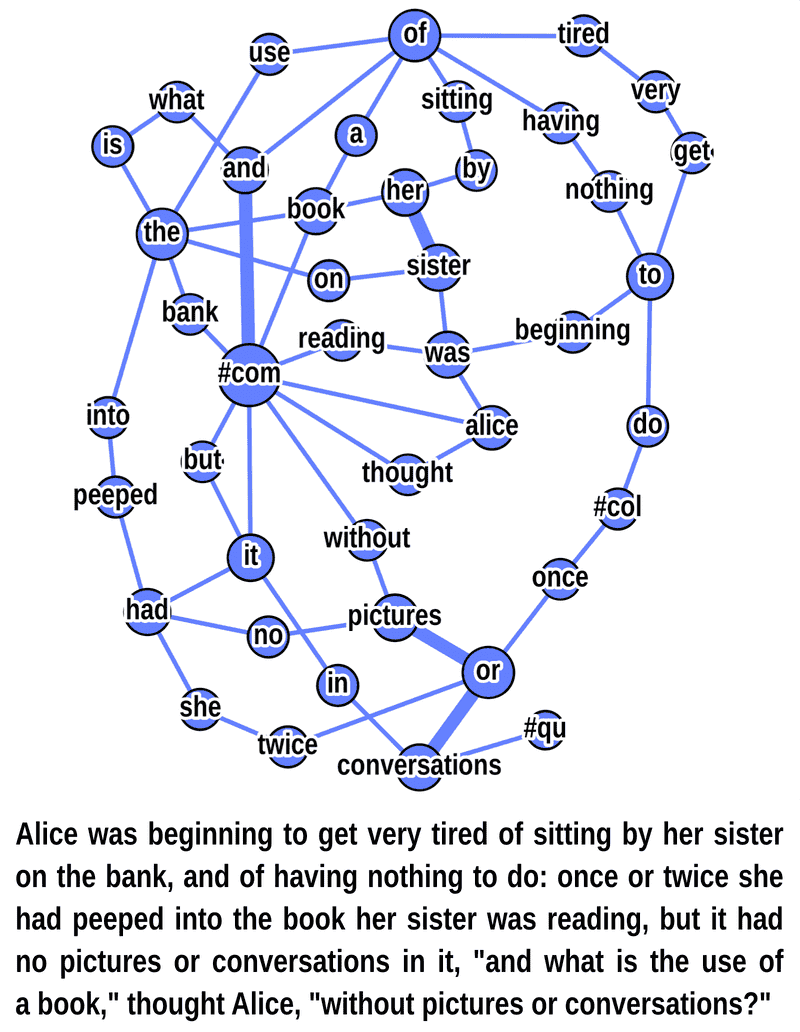}%
\vphantom{\includegraphics[width=0.56\textwidth,valign=c]{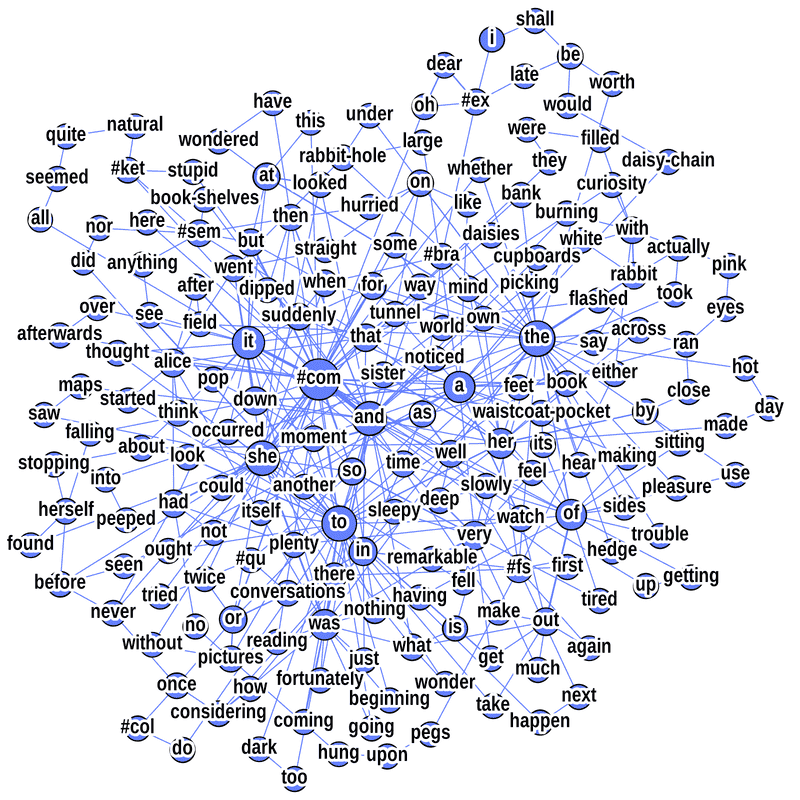}}}
\hfill
\subfloat[]{\includegraphics[width=0.56\textwidth,valign=c]{insertions_dir/Alice_network_2}}

\vspace{0.5cm}

\subfloat[]{\includegraphics[width=0.495\textwidth]{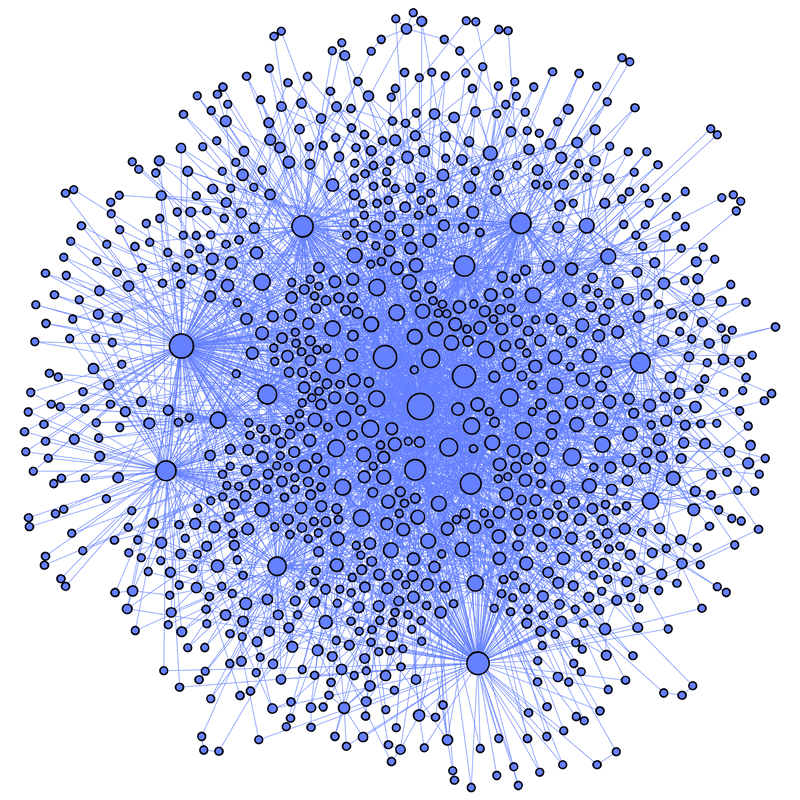}}
\hfill
\subfloat[]{\includegraphics[width=0.495\textwidth]{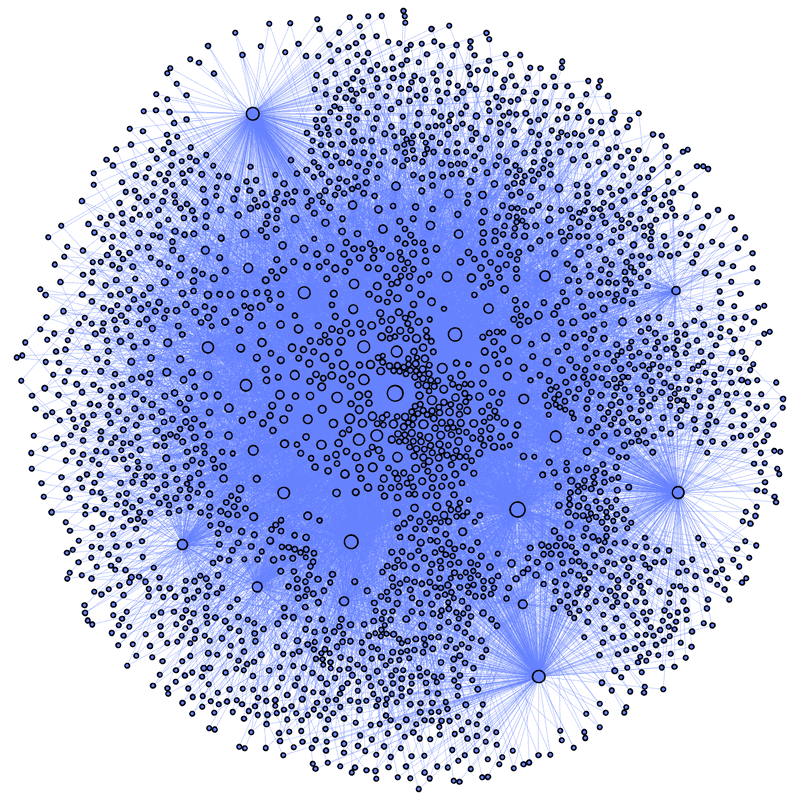}}
\caption{Word-adjacency networks constructed from text samples of different lengths excerpted from \textit{Alice's Adventures in Wonderland} by Lewis Carroll. The samples used to construct the networks are: the first sentence of the book (a), the first 10 sentences (b), the first 5000 words (c), the whole book (d). In (c) and (d) node labels are not shown because of the network size. In all the cases, punctuation marks are treated as words; their labels start from ``\#'' symbol. The first sentence is printed below the network in (a).}
\label{word_adjacency_networks_examples_Alice}
\end{minipage}
\end{figure}

An example of a linguistic network with a very simple construction procedure is a word-adjacency (or word co-occurrence) network. It is created from a text or a corpus of texts. Each unique word (or lemma) becomes a vertex of the network. If two words appear next to each other in the text at least once, the nodes corresponding to those words are connected by an edge. Edges can be binary or weighted if the number of word co-occurrences is considered as a weight. A word-adjacency network can be directed or undirected depending on whether the ordering of word pairs is taken into consideration. Examples of the word-adjacency networks are shown in Fig.~\ref{word_adjacency_networks_examples_Alice}.

Despite their simplicity, word-adjacency networks are able to capture a number characteristics of the underlying text. Since each occurrence of a particular word (except for the first and the last one) adds unity to the weights of the edge between that word and the previous word as well as the edge between that word and the following word, $\mathrm{str}(v) \approx 2\omega(v)$, where $\omega(v)$ is the frequency of the word represented by $v$. If the same word appears twice or more in a row, such a case is ignored in the process of network construction (self-loops are prohibited). Vertex degree (which is strongly correlated but in general not equal to vertex strength~\cite{Kulig2017}) gives information about how many different co-occurrence pairs a word forms with other words in the corpus. Clustering coefficient describes the structure of a node's neighbourhood -- it reveals how often the words being direct neighbours of a word $v$ are also direct neighbours of each other. Measuring network assortativity provides one with information about the correlations between the quantities describing words occurring next to each other (degrees and strengths), and modularity gives an insight into the extent, to which the vocabulary of a text can be divided into clusters of words frequently appearing together.

\subsubsection{Comparing networks of different size}

\begin{figure}
\centering
\begin{minipage}{\figurecustomwidth}
\centering
\subfloat[]{\includegraphics[width=0.4995\textwidth]{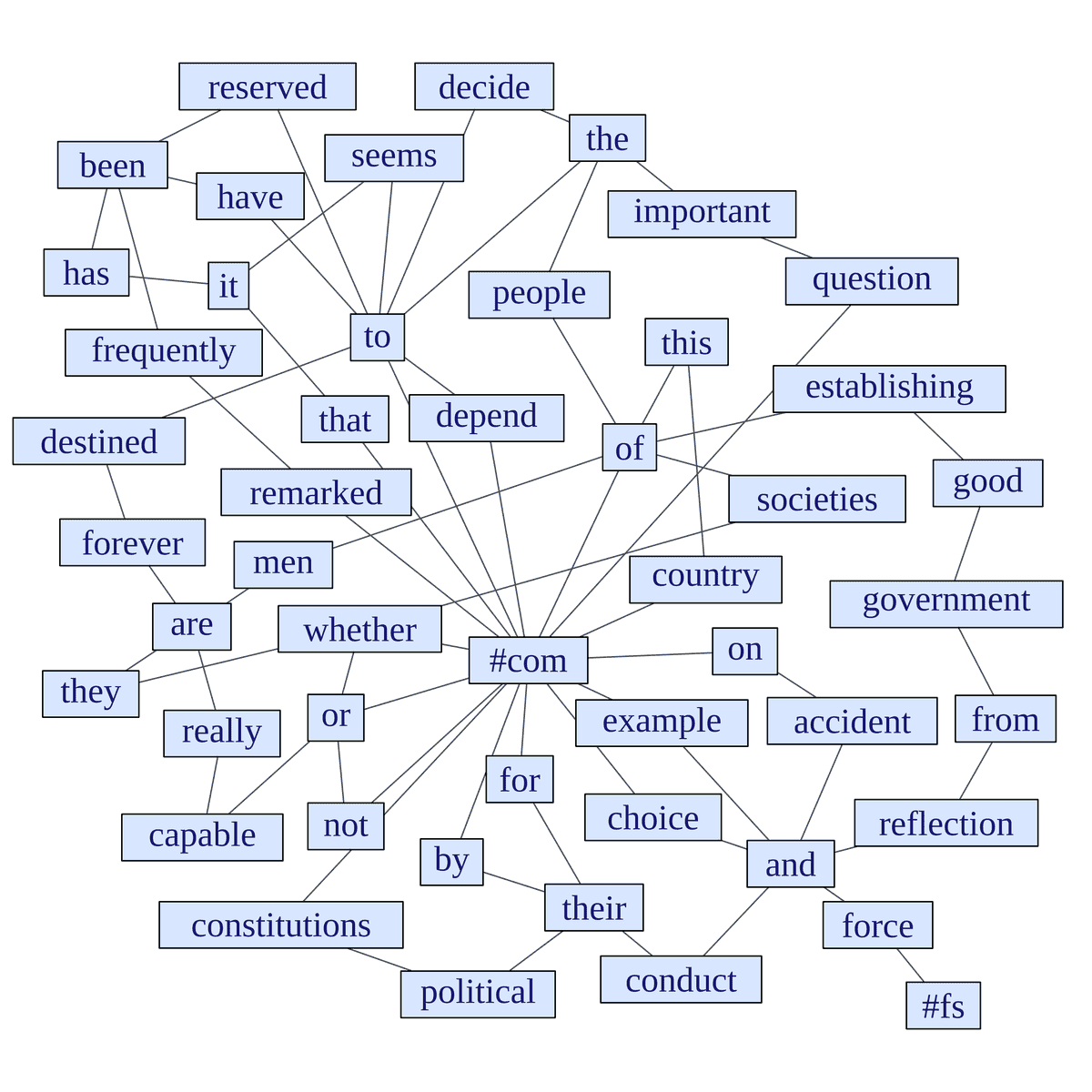}}
\hfill
\subfloat[]{\includegraphics[width=0.4995\textwidth]{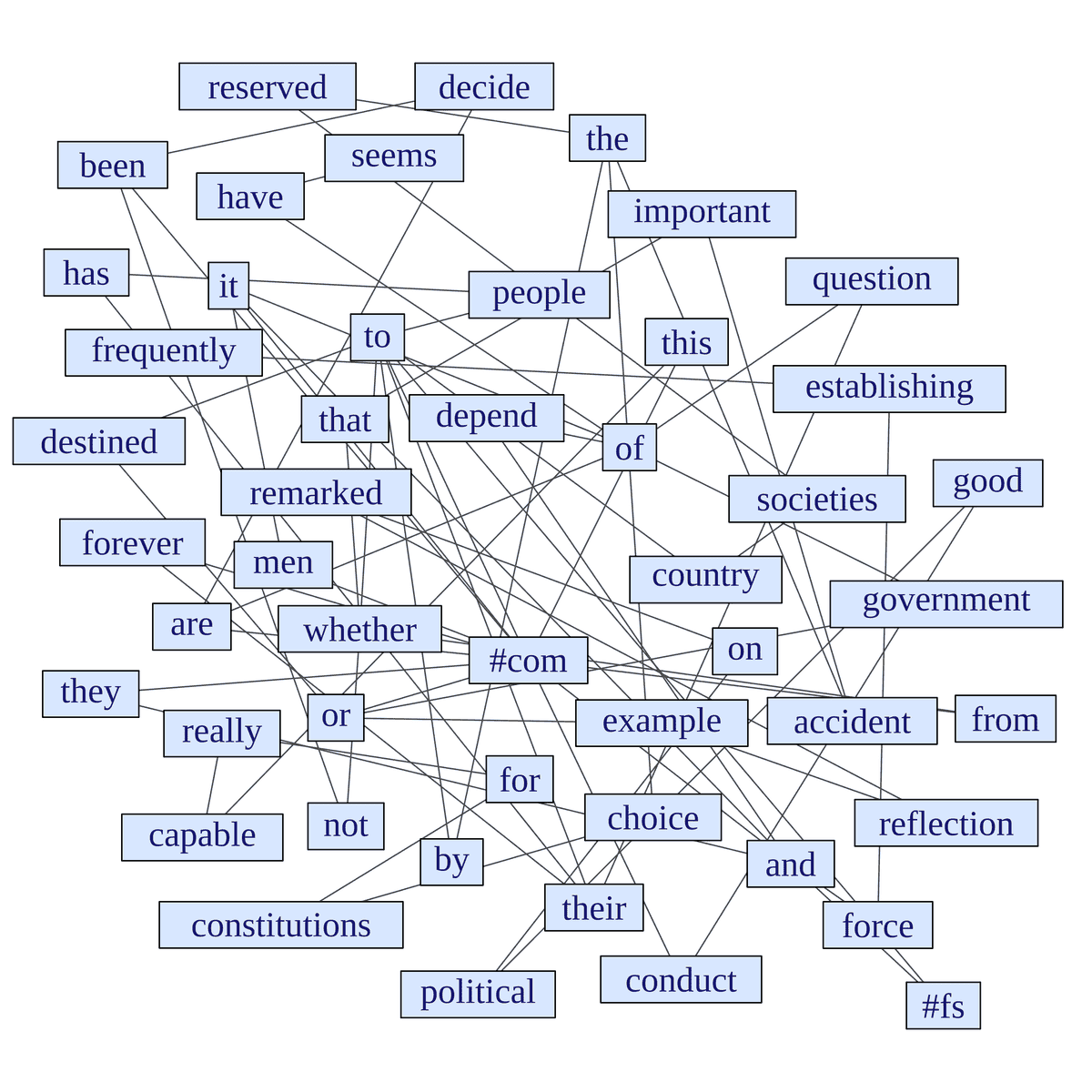}}
\caption{An unweighted word-adjacency network and its randomization. Figure (a) presents a network created from one sentence from \textit{The Federalist Paper No.~1} by Alexander Hamilton. Figure (b) shows a network created from the same text sample, but with shuffled words. Punctuation is taken into consideration in the construction of the network: comma and full stop are denoted by ''\#com''
and ''\#fs'', respectively.}
\label{fig_word_adjacency_network_example_and_randomization}
\end{minipage}
\end{figure}

Word-adjacency networks constructed from texts of different lengths in general have different sizes -- they differ in the numbers of nodes and edges and in magnitudes of edge weights. To compare the properties of word-adjacency networks representing different texts, it is useful to perform some type of normalization. It can be done by creating ``network surrogates'' through randomization. For instance, the words in a text can be shuffled at random a number of times independently and then the respective surrogate word-adjacency networks can be created (Fig.~\ref{fig_word_adjacency_network_example_and_randomization}). This approach preserves node strengths, because word frequencies remain unchanged. The characteristics of interest, either the global ones (pertaining to the whole network) or the local ones (describing specific words), are determined for all the randomized networks and the results are averaged across the network realizations. Then they are compared with their counterparts calculated for the original network. Consequently, an investigated network characteristic $g$ can be transformed to a ``normalized'' form $g^{\rm norm} = g - g^{\rm rand}$ or $g^{\rm norm} = g/g^{\rm rand}$, where $g^{\rm rand}$ is the corresponding average value obtained from randomization.

\begin{figure}
\centering
\begin{minipage}{\figurecustomwidth}
\centering
\subfloat[]{\includegraphics[width=0.495\textwidth]{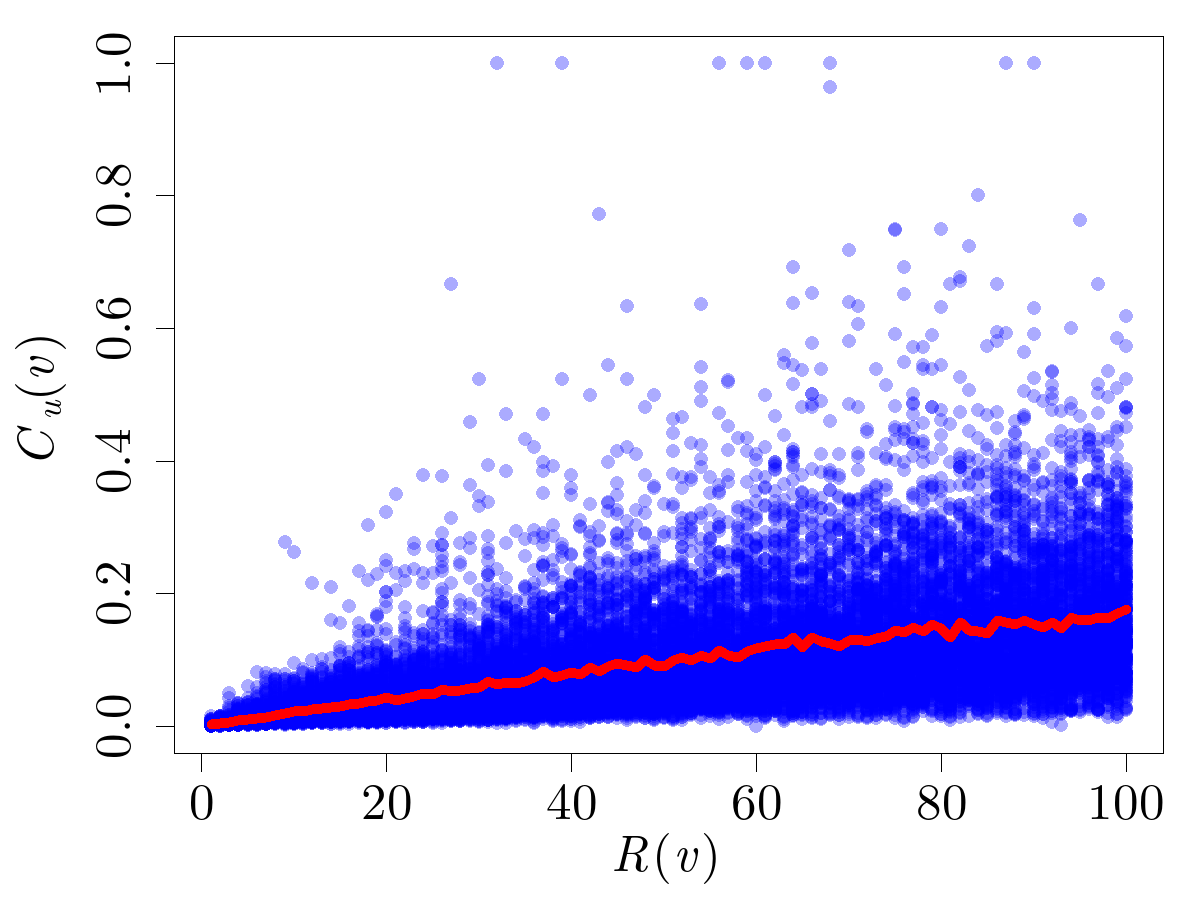}}
\hfill
\subfloat[]{\includegraphics[width=0.495\textwidth]{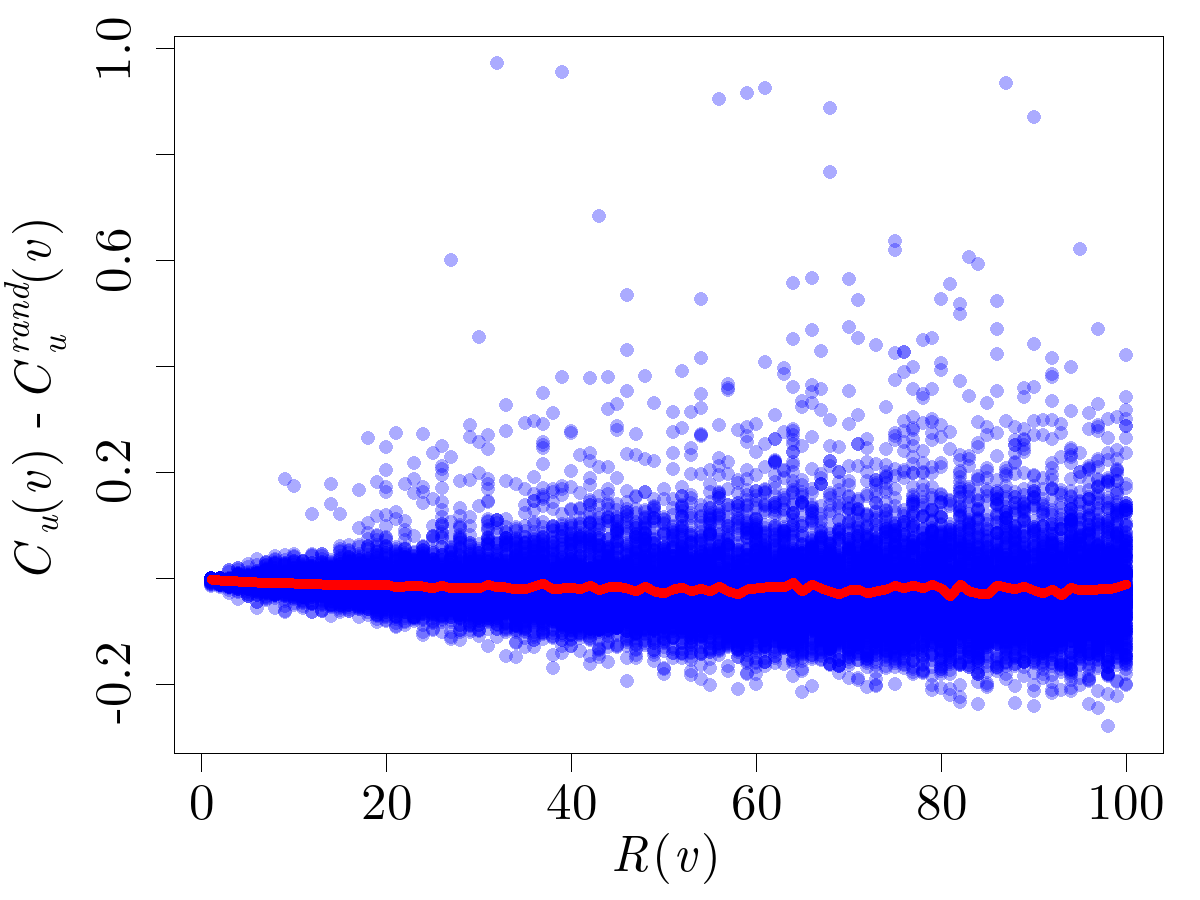}}
\caption{Influence of word frequencies on local unweighted clustering coefficient $C_{\rm u}(v)$ for word-adjacency networks. For each network representing an individual text sample, the unnormalized $C_{\rm u}(v)$ (a) and normalized $C_{\rm u}^{\rm norm}=C_{\rm u}(v) - C_{\rm u}^{\rm rand}(v)$ (b) coefficient is plotted against the word rank for the 100 most frequent words. Red line represents the average over the networks for each rank. While the average unnormalized coefficient depends on word frequency, its average normalized version does not exhibit such a dependence. Therefore, it can be concluded that the observed variability of $C_{\rm u}^{\rm norm}(v)$ may be attributed to the genuine traits of the individual words and text samples.}
\label{fig_word_rank_and_CC_correlation}
\end{minipage}
\end{figure}

Such normalization of the network parameters is carried out in order to allow one for a meaningful comparison between networks of different size and compensation for direct effects of different word frequencies. Since a randomized text has the same length and word frequency distribution as its original source, it can be anticipated that the normalized characteristics neglect the purely frequency-based effects and capture the network's specific organization. An example of how normalization affects the local clustering coefficient $C_{\rm u}(v)$ for individual words can be seen in Fig.~\ref{fig_word_rank_and_CC_correlation}, where the unnormalized and normalized coefficient values for the 100 most frequent words taken from a number of different text samples are displayed. A tendency of $C_{\rm u}(v)$ to increase with increasing word rank $R(v)$ is not observed for the normalized coefficient $C_{\rm u}^{\rm norm} = C_{\rm u}(v) - C_{\rm u}^{\rm rand}(v)$, expressing the genuine differences between the words and texts.

\subsubsection{Punctuation in word-adjacency networks}

\begin{figure}
\centering
\begin{minipage}{\figurecustomwidth}
\centering
\includegraphics[width=0.495\textwidth]{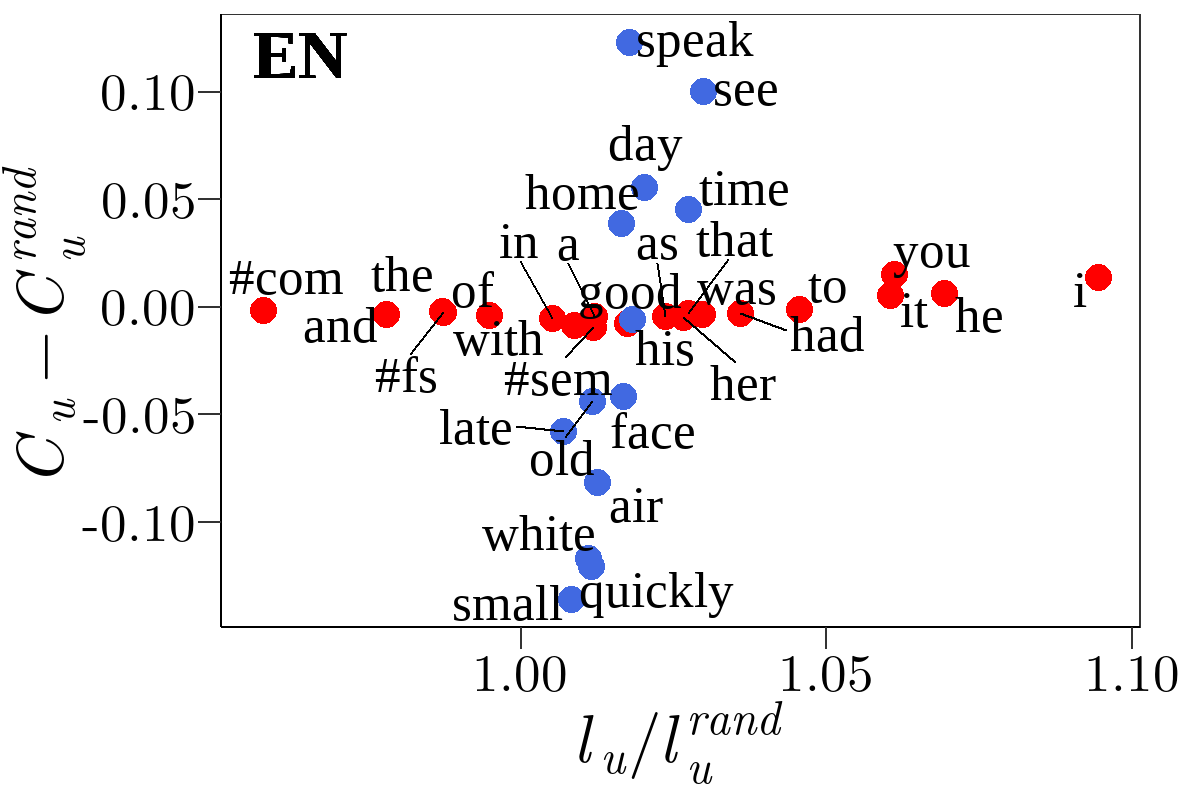}
\hfill
\includegraphics[width=0.495\textwidth]{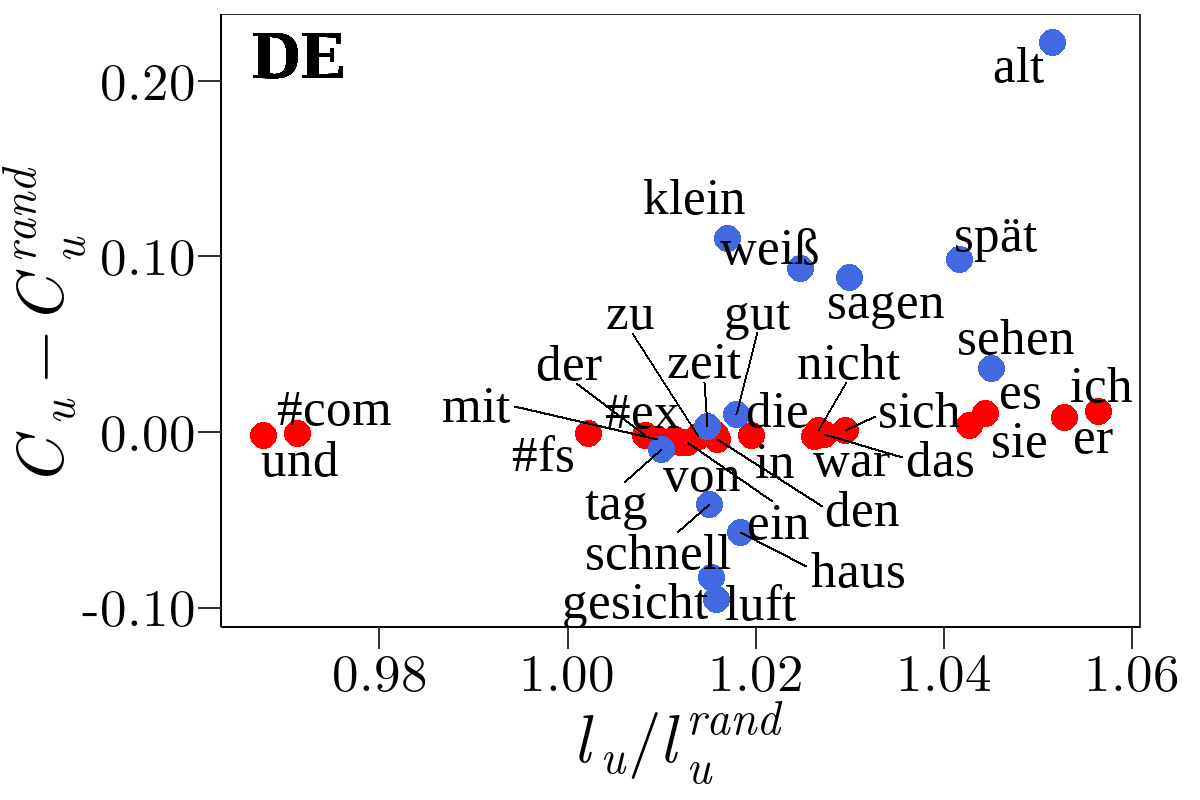}
\hfill
\includegraphics[width=0.495\textwidth]{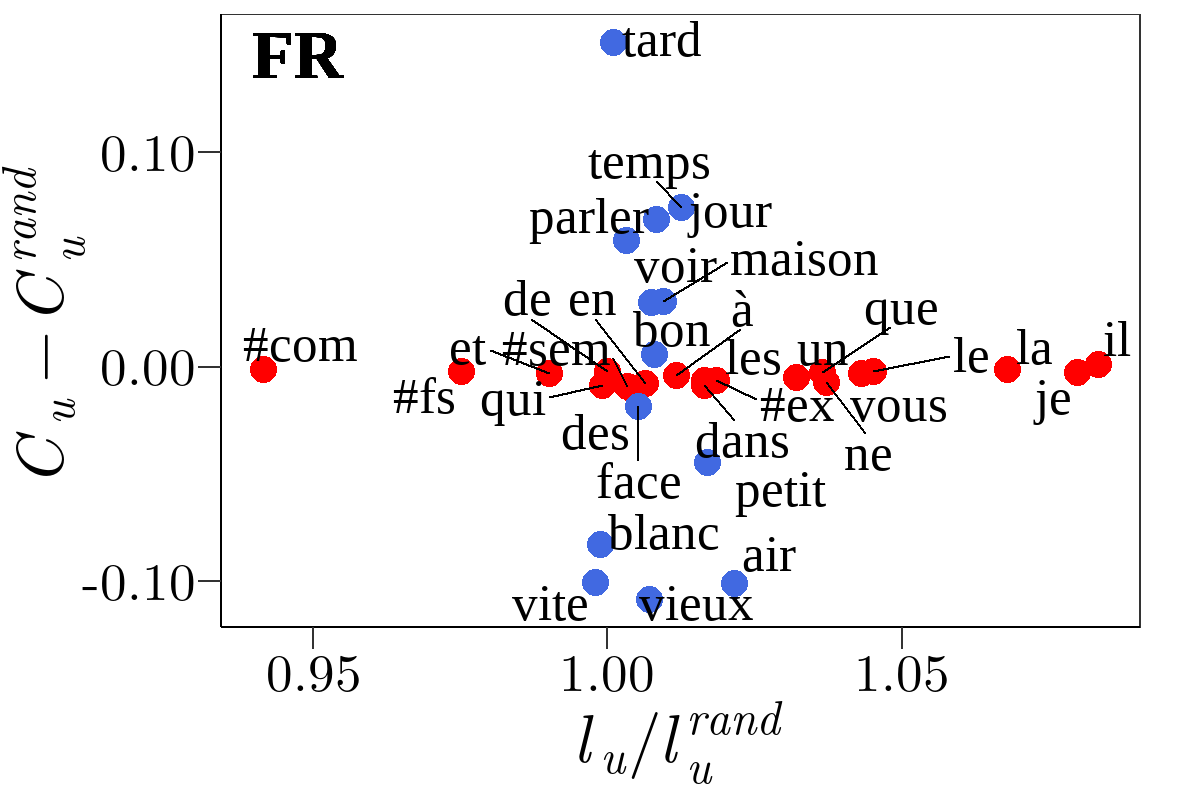}
\hfill
\includegraphics[width=0.495\textwidth]{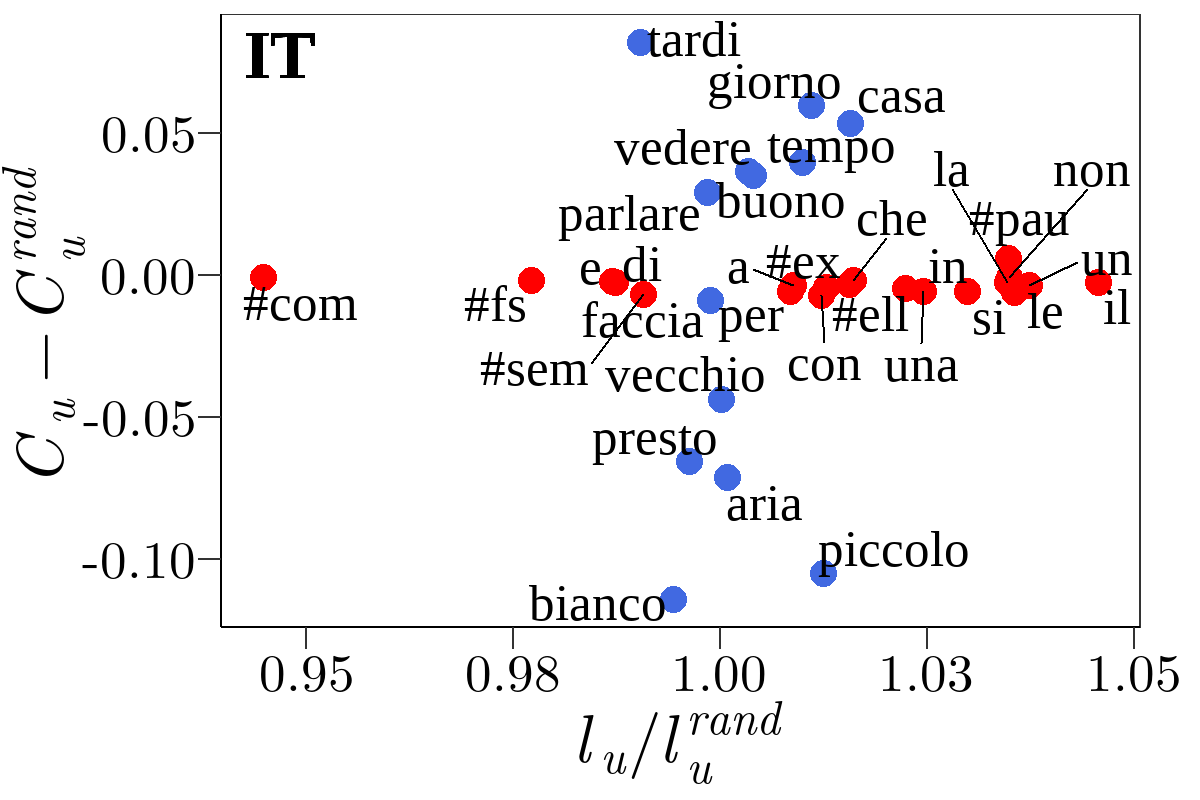}
\hfill
\includegraphics[width=0.495\textwidth]{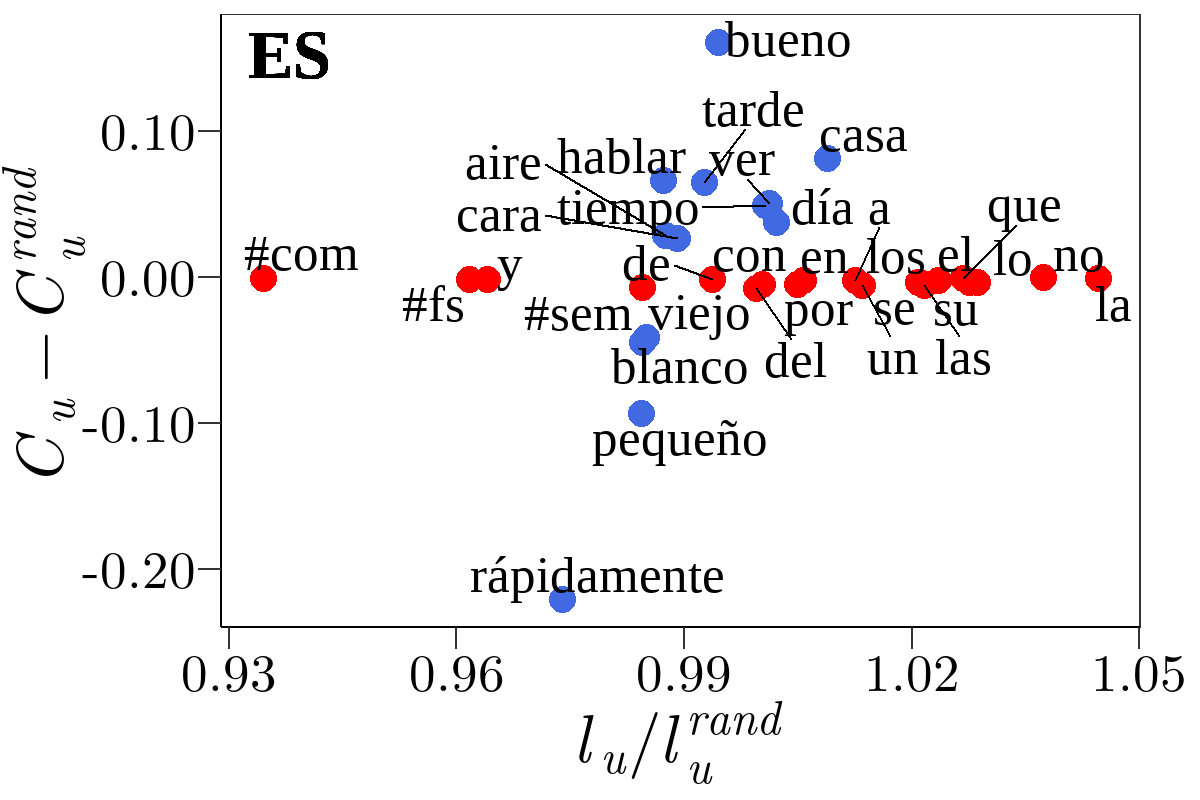}
\hfill
\includegraphics[width=0.495\textwidth]{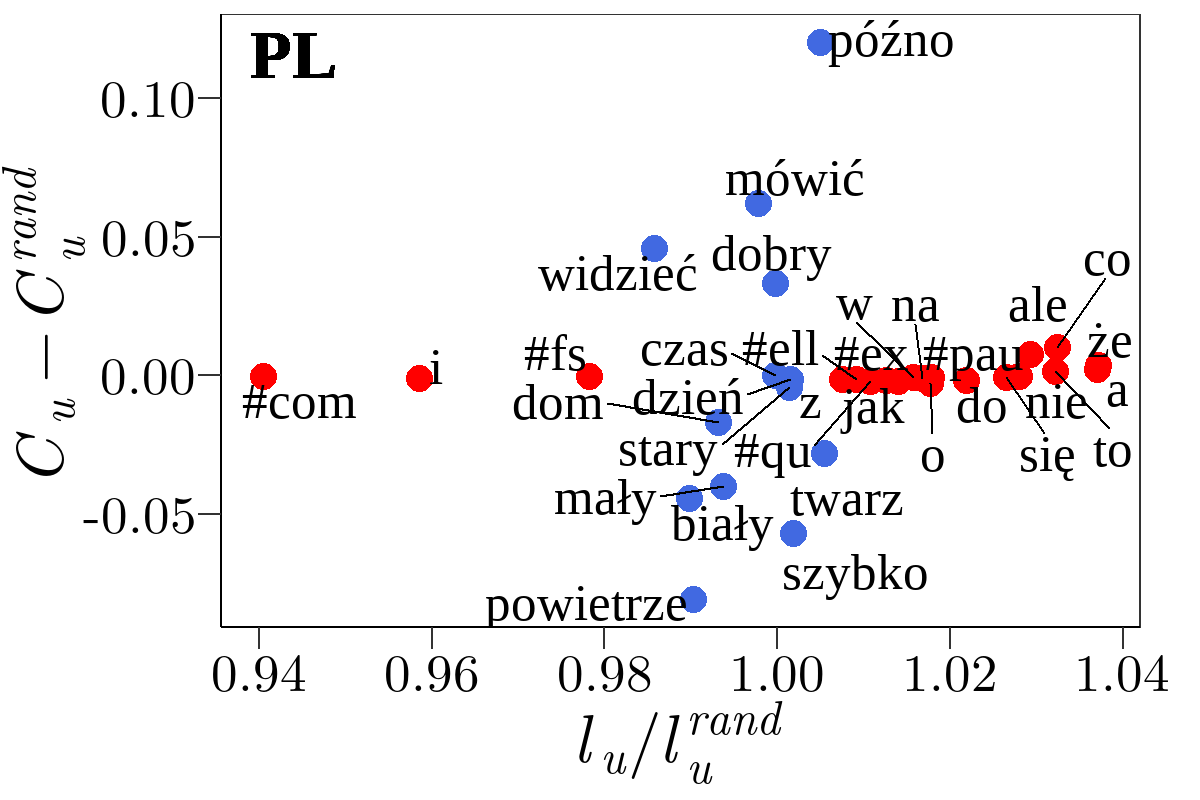}
\hfill
\includegraphics[width=0.495\textwidth]{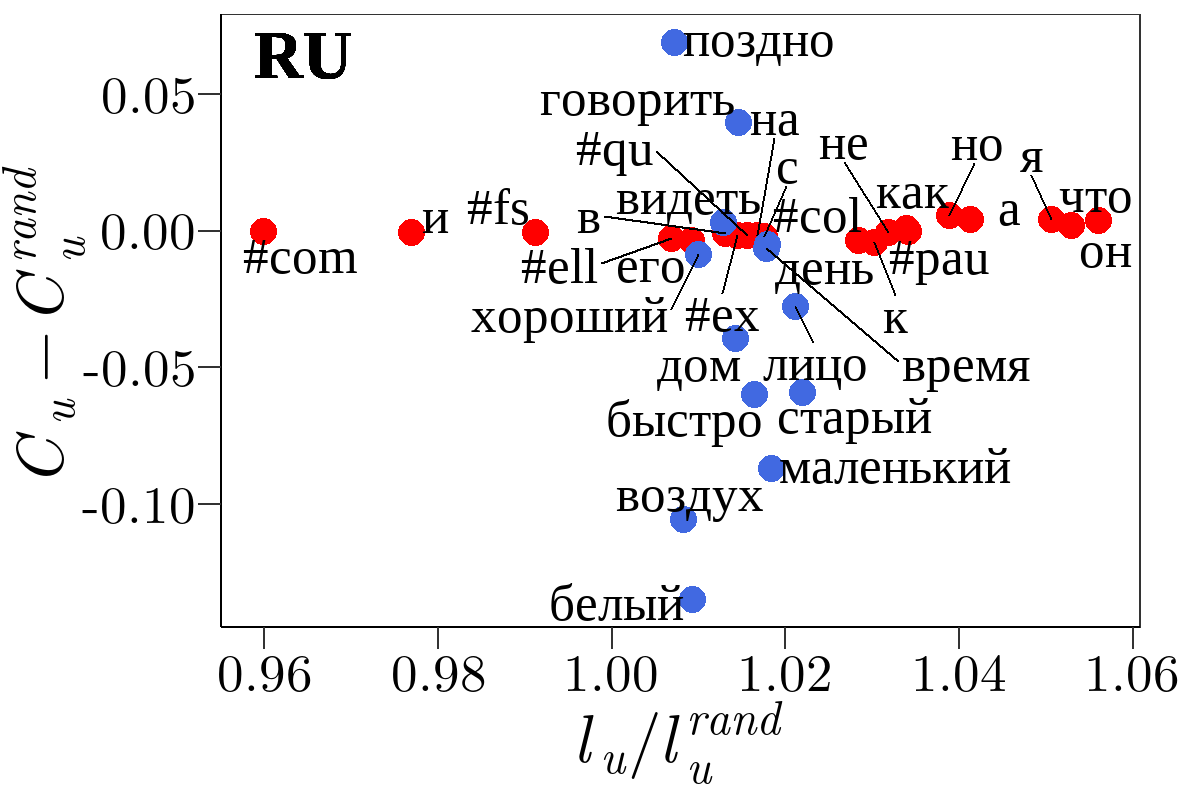}
\caption{The normalized average shortest path lengths $\ell_{\rm u}^{\rm norm}=\ell_{\rm u} / \ell_{\rm u}^{\rm rand}$ and the normalized local clustering coefficients $C_{\rm u}^{\rm norm} =C_{\rm u} - C_{\rm u}^{\rm rand}$ of selected words and punctuation marks in the word-adjacency networks constructed from text corpora in 7 languages: English (EN), German (DE), French (FR), Italian (IT), Spanish (ES), Polish (PL), and Russian (RU). Two groups of words are considered in each language: the 20 most frequent words including the punctuation marks -- period \#fs, question mark \#qu, exclamation mark \#ex, ellipsis \#ell, comma \#com, dash \#pau, semicolon \#sem, colon \#col, left parenthesis \#bra and right parenthesis \#ket (marked in red) and the selected words of medium frequency (marked in blue). In terms of the presented characteristics, the punctuation marks behave like ordinary words with large frequency.}
\label{fig_two_types_of_words_in_word_adjacency_networks}
\end{minipage}
\end{figure}

The word-adjacency network construction procedure can be extended to include objects other than words. A natural extension is to take punctuation marks into consideration and represent them by additional vertices in a network, for instance: period, question mark, exclamation mark, ellipsis, comma, dash, semicolon, colon, left and right parentheses. It turns out that these nodes behave in the same way as the nodes corresponding to words of comparable frequency~\cite{Kulig2017} -- see Fig.~\ref{fig_two_types_of_words_in_word_adjacency_networks} where the normalized local clustering coefficients $C_{\rm u}^{\rm norm} =C_{\rm u} - C_{\rm u}^{\rm rand}$ and the normalized average shortest path lengths $\ell_{\rm u}^{\rm norm}=\ell_{\rm u} / \ell_{\rm u}^{\rm rand}$ are shown for selected words from the language-specific corpora. For each language, a set of words is composed of two parts: (1) a few the most frequent words and punctuation marks and (2) sample less frequent words (with equivalent meaning in each of the considered languages). These two groups display different patterns of variability as it can be seen in Fig.~\ref{fig_two_types_of_words_in_word_adjacency_networks}. While the words from part (1) are placed along the horizontal axis on the $(\ell_{\rm u}^{\rm norm}, C_{\rm u}^{\rm norm})$ plane and their do not show clustering, the words from part (2) are scattered across different values of $C_{\rm u}^{\rm norm}$. In all the studied languages, punctuation marks belong to the regime determined by the most frequent words. This is in accordance with the frequency analysis, which supports an idea that punctuation marks can be treated as words from at least certain points of view. Not only their frequencies fit into the power-law regime of the rank-frequency distribution, but also their properties in the word-adjacency networks resemble the properties of the high-ranked words.

\subsubsection{Word-adjacency networks in different languages}

\begin{figure}
\centering
\begin{minipage}{\figurecustomwidth}
\centering
\subfloat[]{\includegraphics[width=0.49\textwidth]{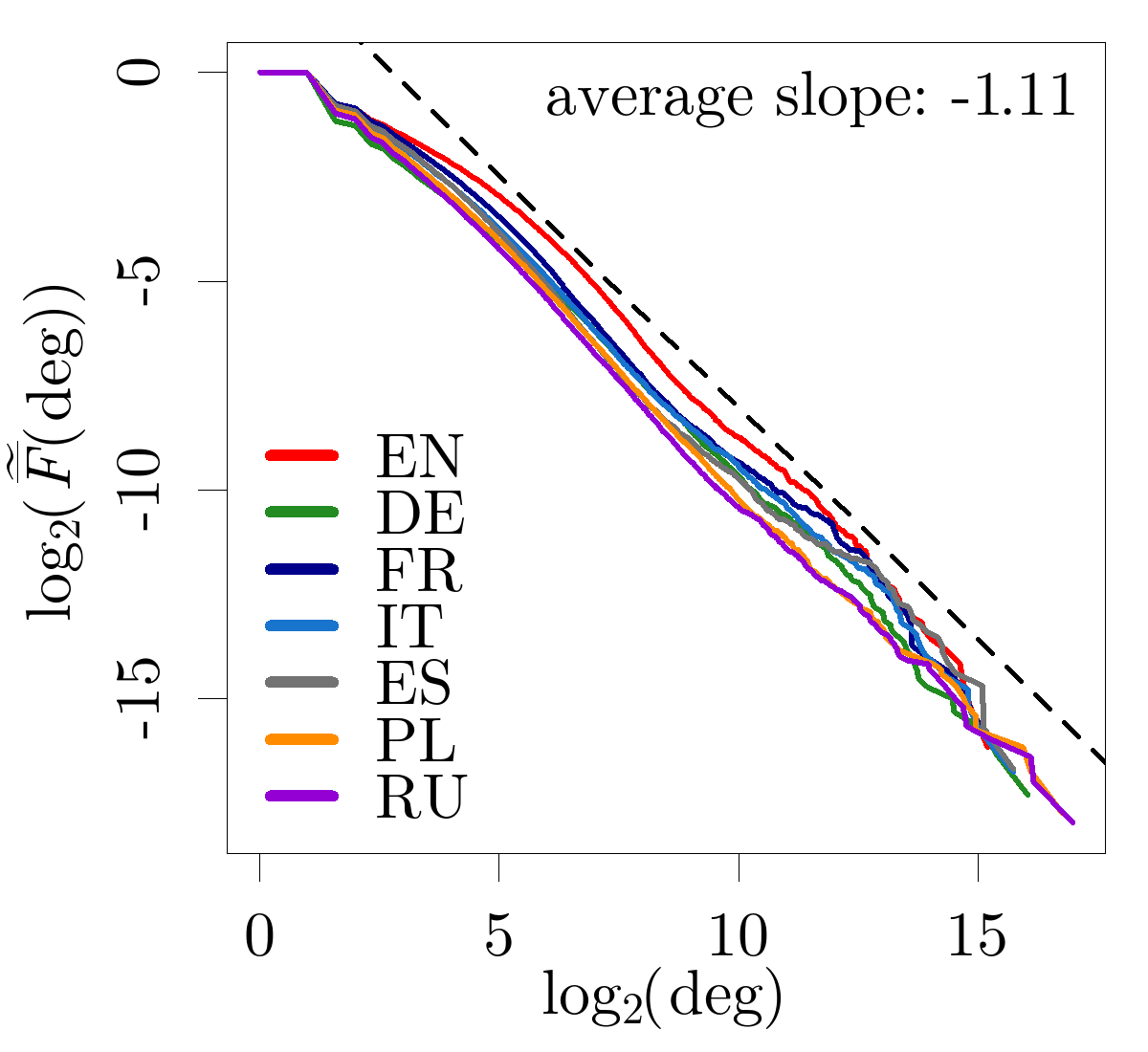}}
\hfill
\subfloat[]{\includegraphics[width=0.49\textwidth]{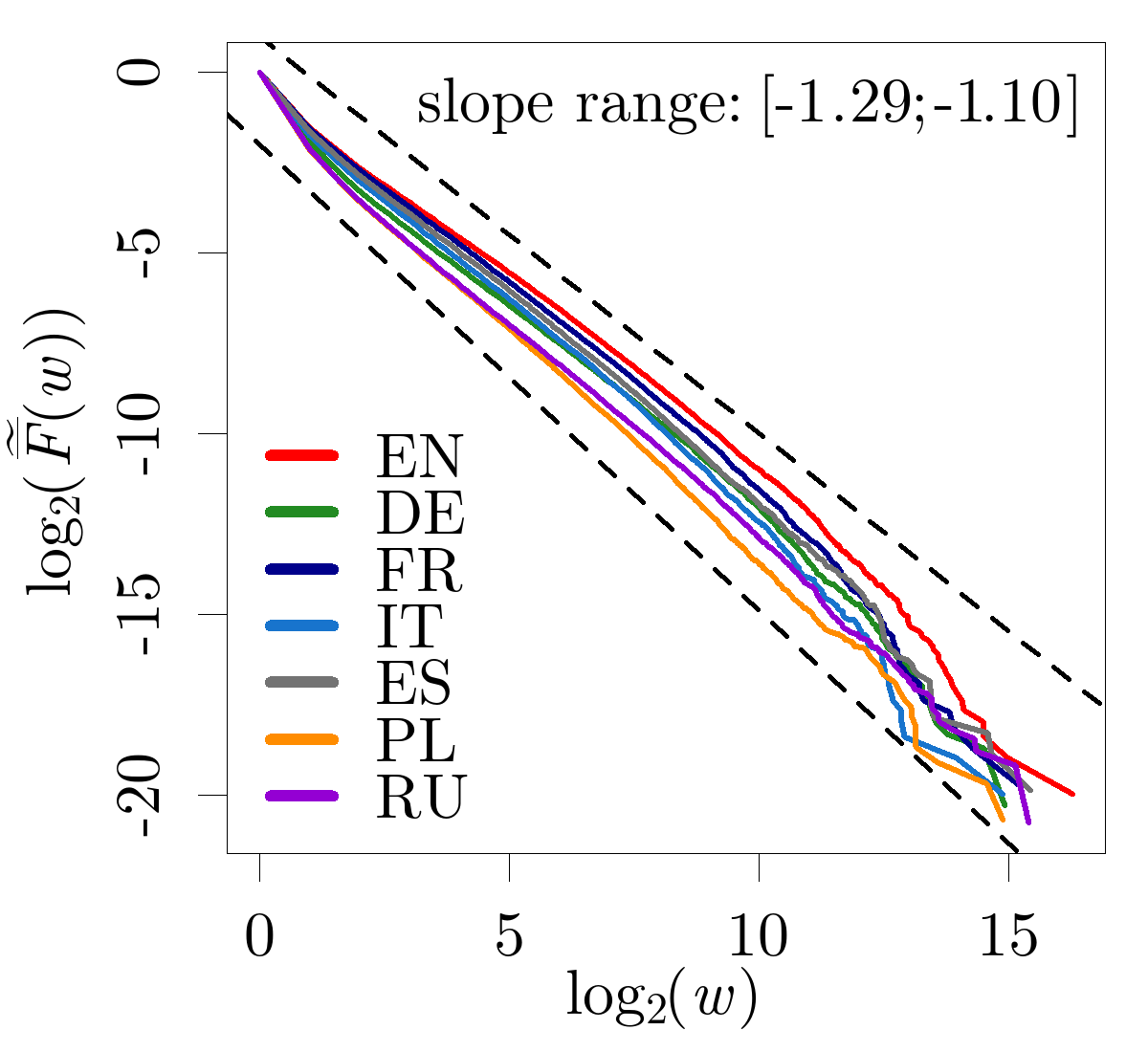}}
\caption{Empirical survival functions $\tilde{\bar{F}}$ representing the node degree distributions (a) and the edge weight distributions (b) for word-adjacency networks constructed from corpora consisting of a monolingual set of books. Slope of the black dashed line in (a) is equal to the average slope of straight lines fitted to each distribution ($-1.11$). Slope of the black dashed lines in (b) is equal to the minimum and maximum slope of straight lines fitted to the edge weight distributions ($-1.29$ and $-1.10$, respectively).}
\label{fig_dataset_1_degree_and_edge_weight_distribution}
\end{minipage}
\end{figure}

Fig.~\ref{fig_dataset_1_degree_and_edge_weight_distribution} shows the log-log plots of degree distributions and edge weight distributions of word-adjacency networks. The form of degree distributions indicates that the networks can be considered approximately scale-free, with the power exponents of the survival function being slightly above 1. This result is in a sense expected: as Zipf's law ensures that the word frequency distributions are described by a power law and as word frequencies are approximately equal to node strengths, the node degree distributions are significantly influenced by word frequency distributions. The edge weight distributions can also be approximated by power laws. It can be associated with the fact that edge weights correspond to the frequencies of 2-grams (pairs of words) and the frequencies of certain linguistic constructs larger than words also seem to be conforming to power-law distributions~\cite{Ha2009,Williams2015}. It should be noted, however, that the effect is not as evident as in case of individual words, described by Zipf's law~\cite{PIANTADOSI_2014,Egghe1999,Egghe2000}.

\begin{figure}
\centering
\begin{minipage}{\figurecustomwidth}
\centering
\subfloat[Clustering coefficient, unweighted]{\includegraphics[width=0.485\textwidth]{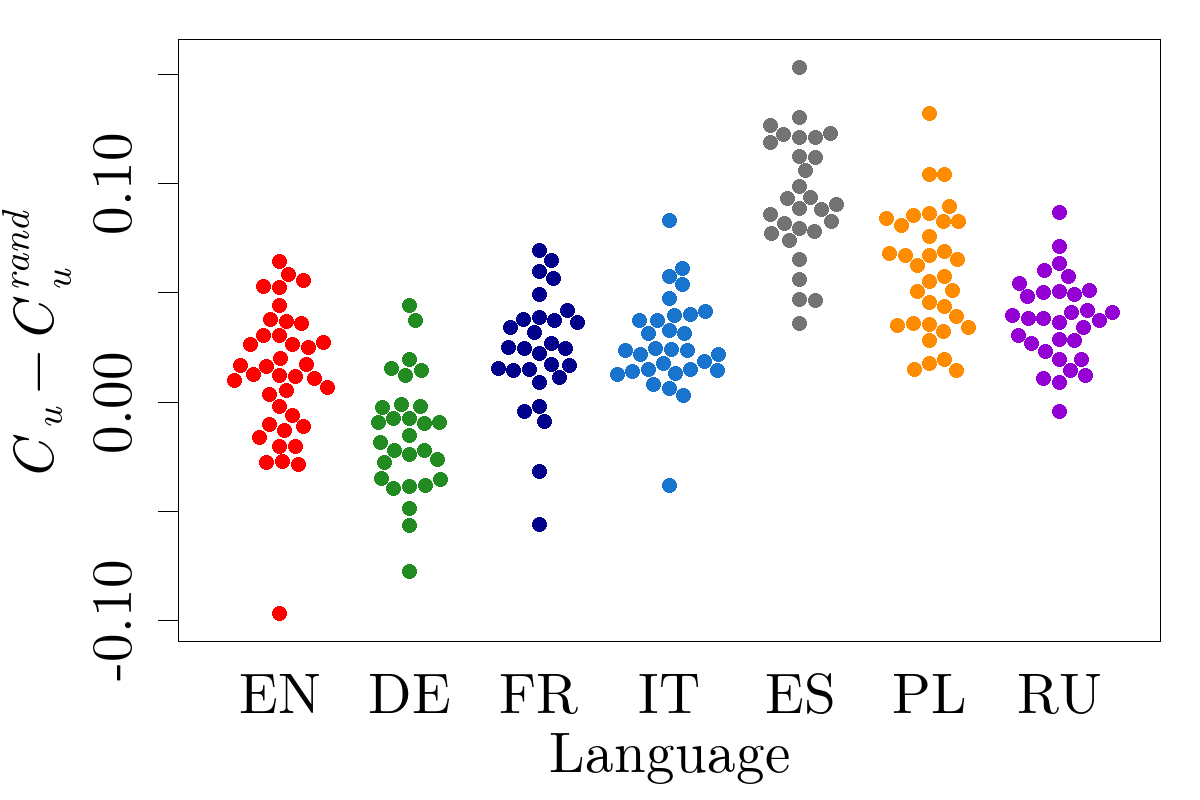}}
\hfill
\subfloat[Clustering coefficient, weighted]{\includegraphics[width=0.485\textwidth]{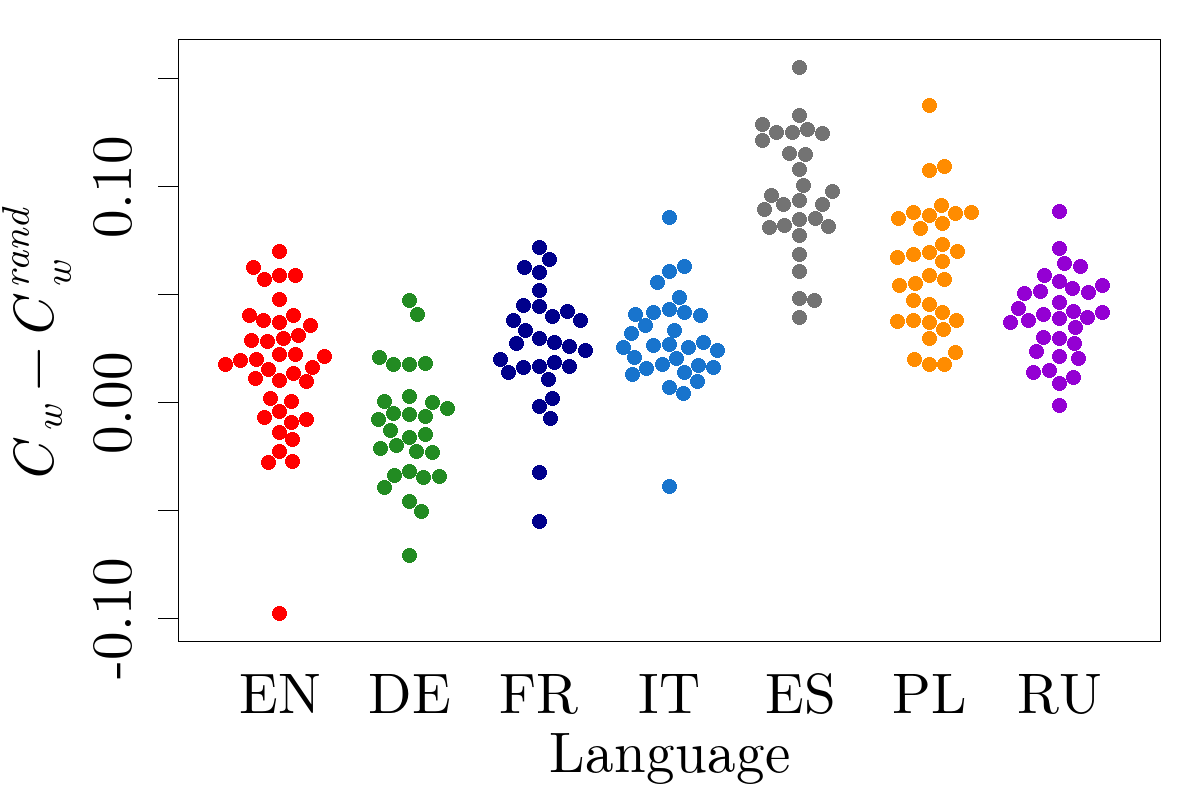}}
\hfill
\vspace{0.5cm}
\subfloat[Assortativity coefficient, unweighted]{\includegraphics[width=0.485\textwidth]{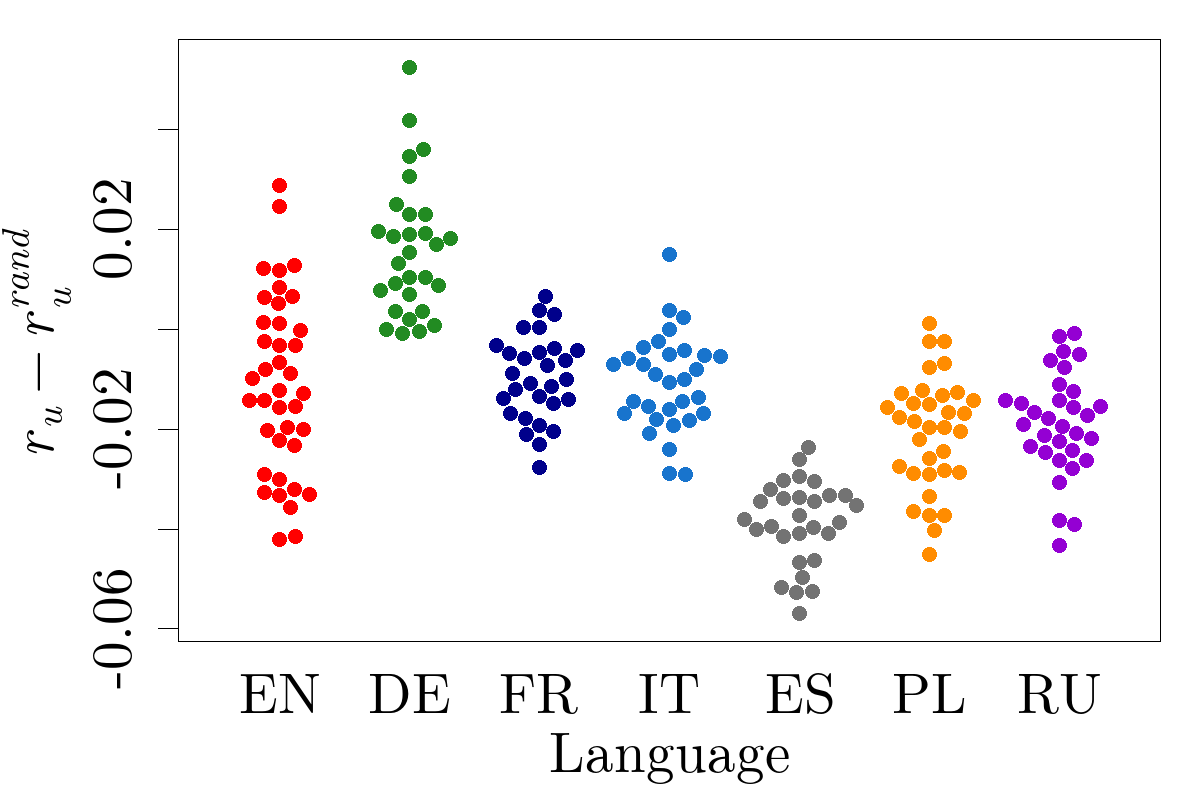}}
\hfill
\subfloat[Assortativity coefficient, weighted]{\includegraphics[width=0.485\textwidth]{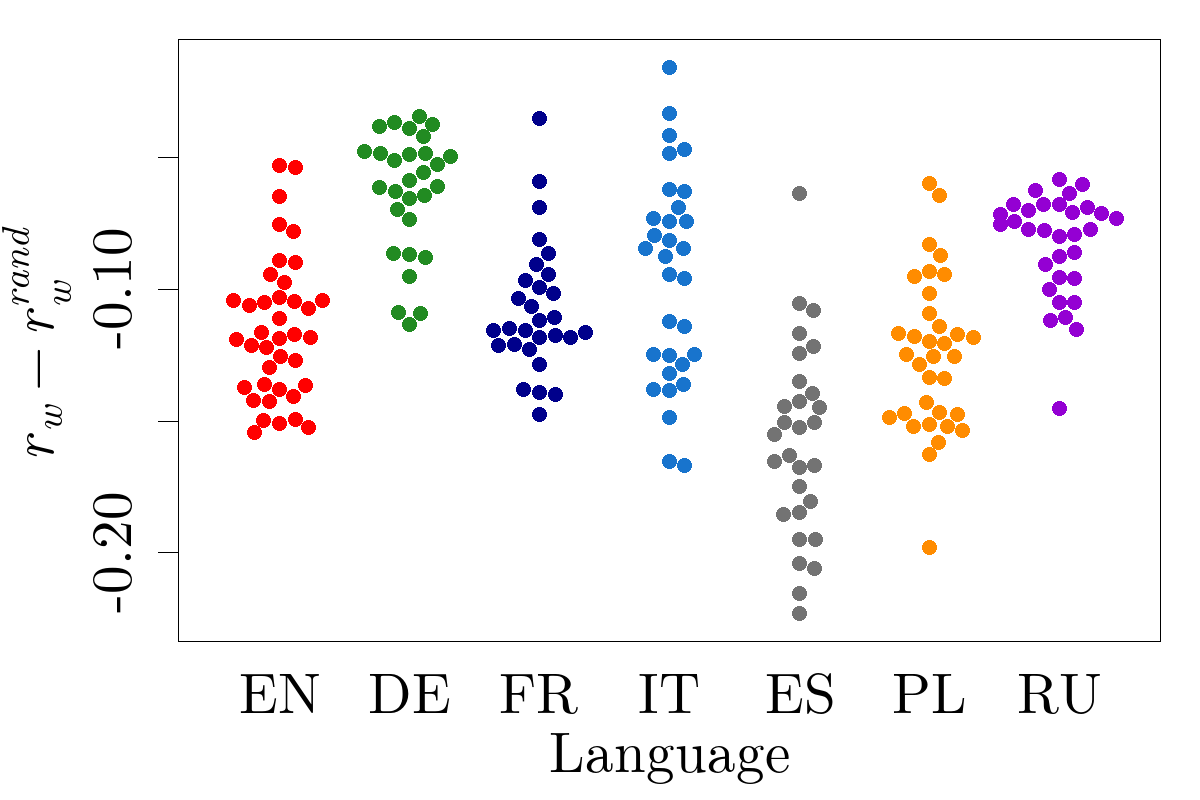}}
\hfill
\vspace{0.5cm}
\subfloat[Rank assortativity coefficient, unweighted]{\includegraphics[width=0.485\textwidth]{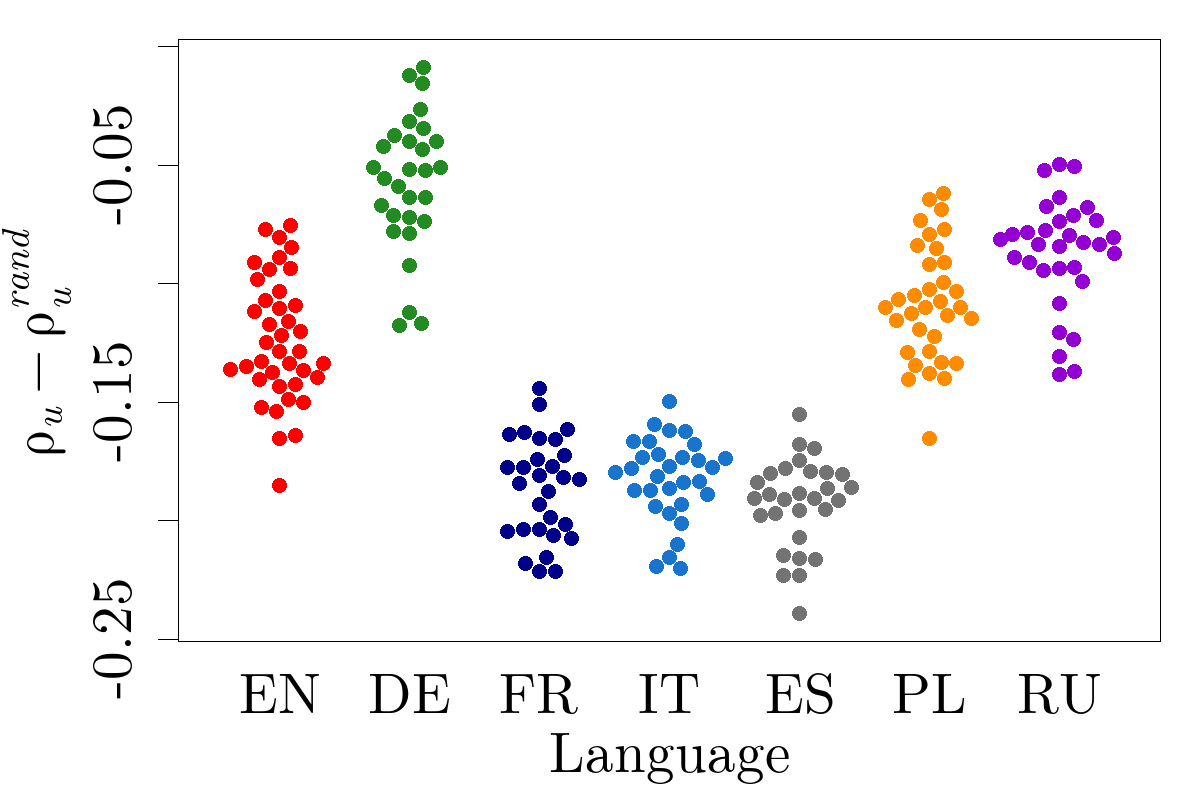}}
\hfill
\subfloat[Rank assortativity coefficient, weighted]{\includegraphics[width=0.485\textwidth]{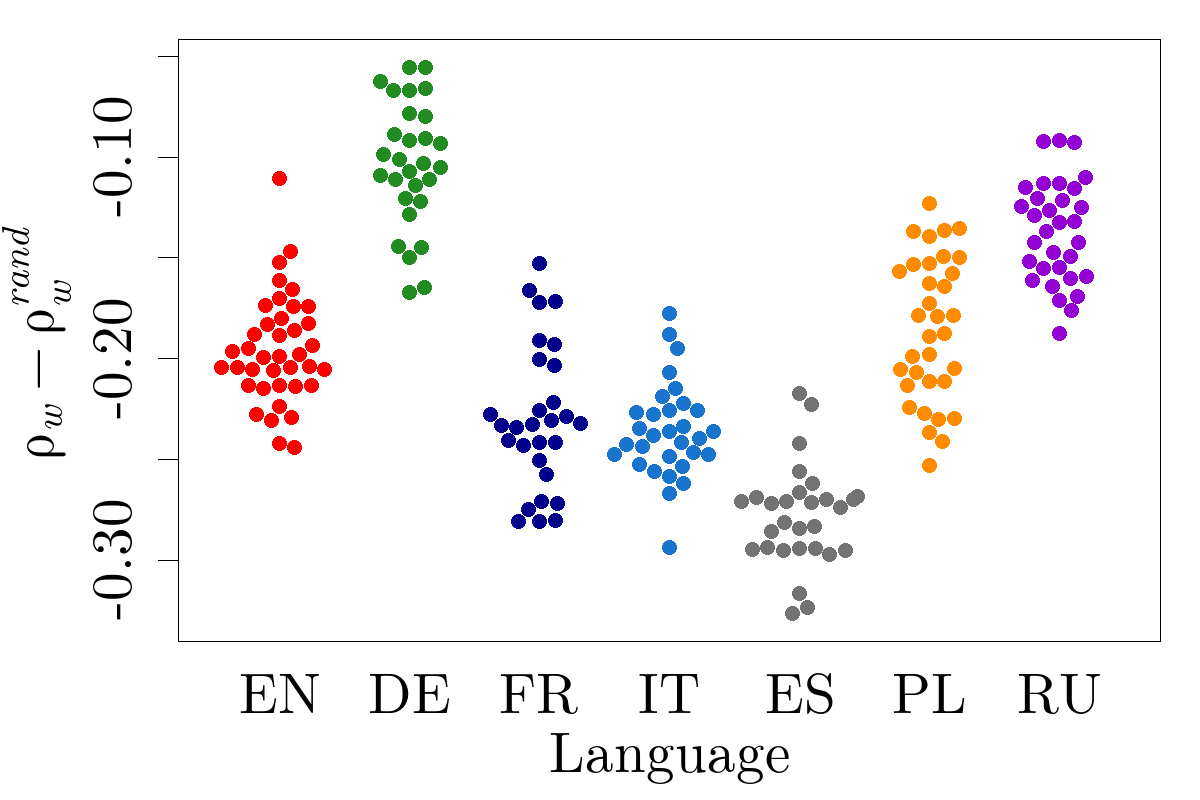}}
\hfill
\vspace{0.5cm}
\subfloat[Modularity, unweighted]{\includegraphics[width=0.485\textwidth]{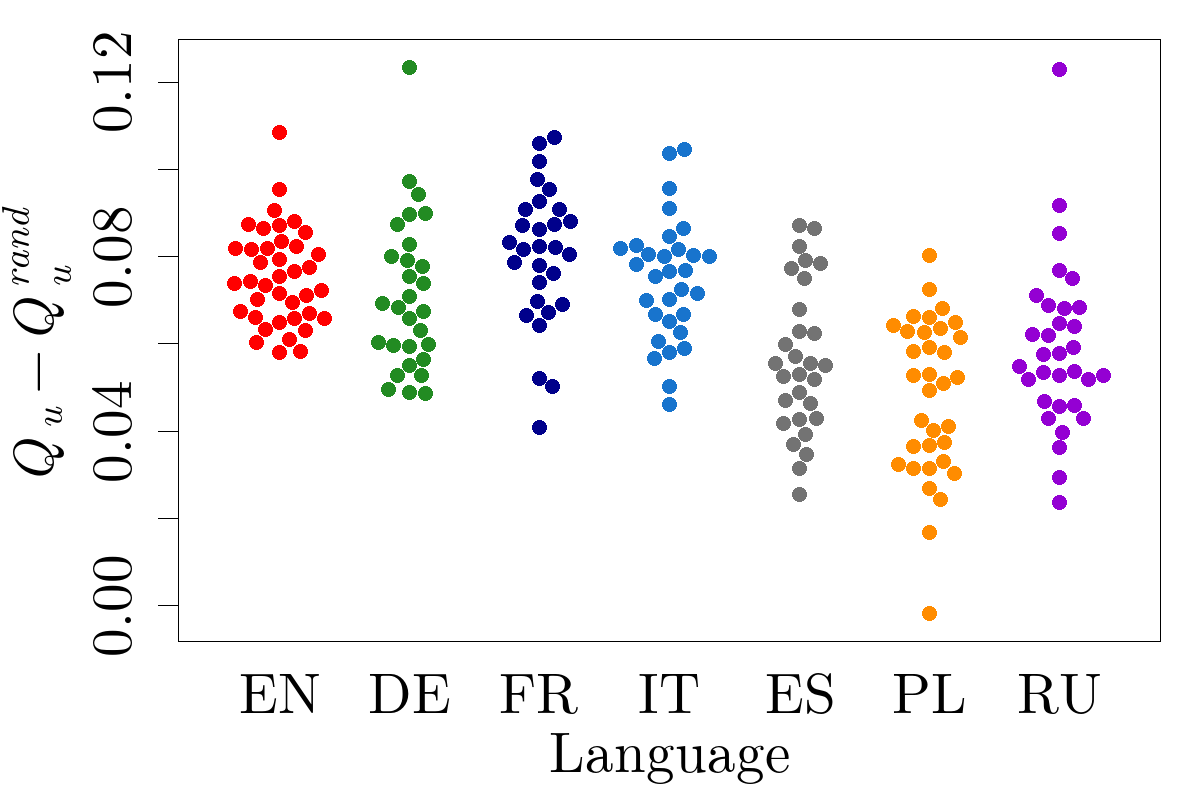}}
\hfill
\subfloat[Modularity, weighted]{\includegraphics[width=0.485\textwidth]{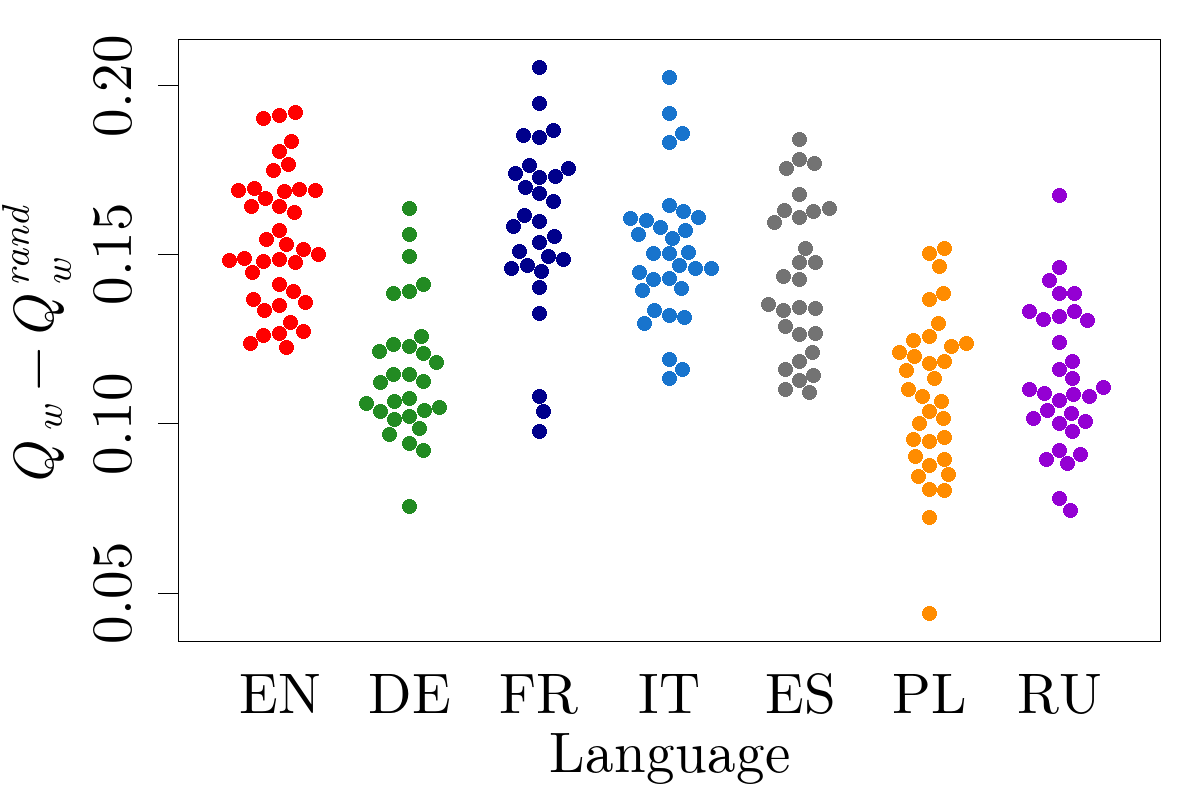}}
\caption{Global characteristics of the word-adjacency networks constructed from texts in different languages. Each plot pertains to a normalized characteristic and each dot represents a sample text. It can be seen that distributions of individual characteristics can be substantially different for different languages with the presence and the strength of that effect varying among the languages.}
\label{fig_networks_global_characteristics_beeswarm}
\end{minipage}
\end{figure}

\begin{figure}[!t]
\centering
\begin{minipage}{\figurecustomwidth}
\centering
\includegraphics[width=0.65\textwidth]{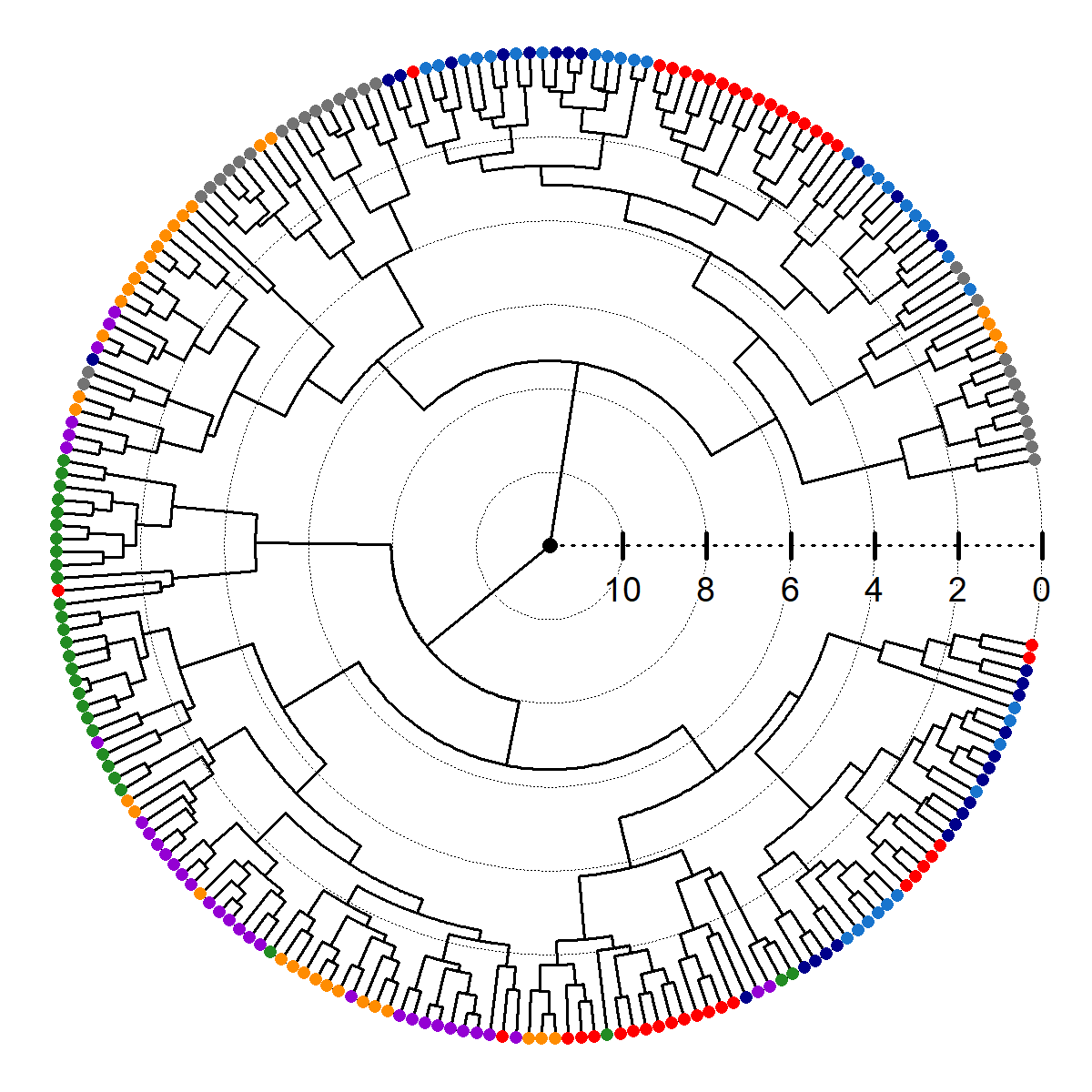}
\vspace{-1em}
\caption{Dendrogram of the hierarchical clustering of sample texts in the parameter space spanned by the normalized global characteristics of the word-adjacency networks presented in Fig.~\ref{fig_networks_global_characteristics_beeswarm}. The scale on the dendrogram radius gives a distance between merged clusters. Each colored dot on the edge of the plot represents a single book. Colors correspond to languages: red - English, green - German, dark blue - French, light blue - Italian, gray - Spanish, orange - Polish, and purple - Russian.}
\label{fig_dendrogram_global_characteristics_dataset_1}
\end{minipage}
\end{figure}

While a number of properties of word-adjacency networks can be considered general and possibly universal, some properties seem to be specific to particular languages, at least to some degree. It is illustrated in Fig.~\ref{fig_networks_global_characteristics_beeswarm} showing that the texts sharing the same language group together in the space spanned by different parameters describing network structure. Based on the hierarchical clustering algorithm linking points that are close to each other according to some metric (e.g., the Euclidean one)~\cite{Tan2005}, Fig.~\ref{fig_dendrogram_global_characteristics_dataset_1} shows a dendrogram of the text proximity in a space of the parameters presented in Fig.~\ref{fig_networks_global_characteristics_beeswarm}. Even though a separation between texts in different languages is far from perfect, these texts do reveal tendency to be adjacent to other texts written in the same language.

\begin{figure}[!b]
\centering
\begin{minipage}{\figurecustomwidth}
\centering
\subfloat[]{\includegraphics[width=0.475\textwidth]{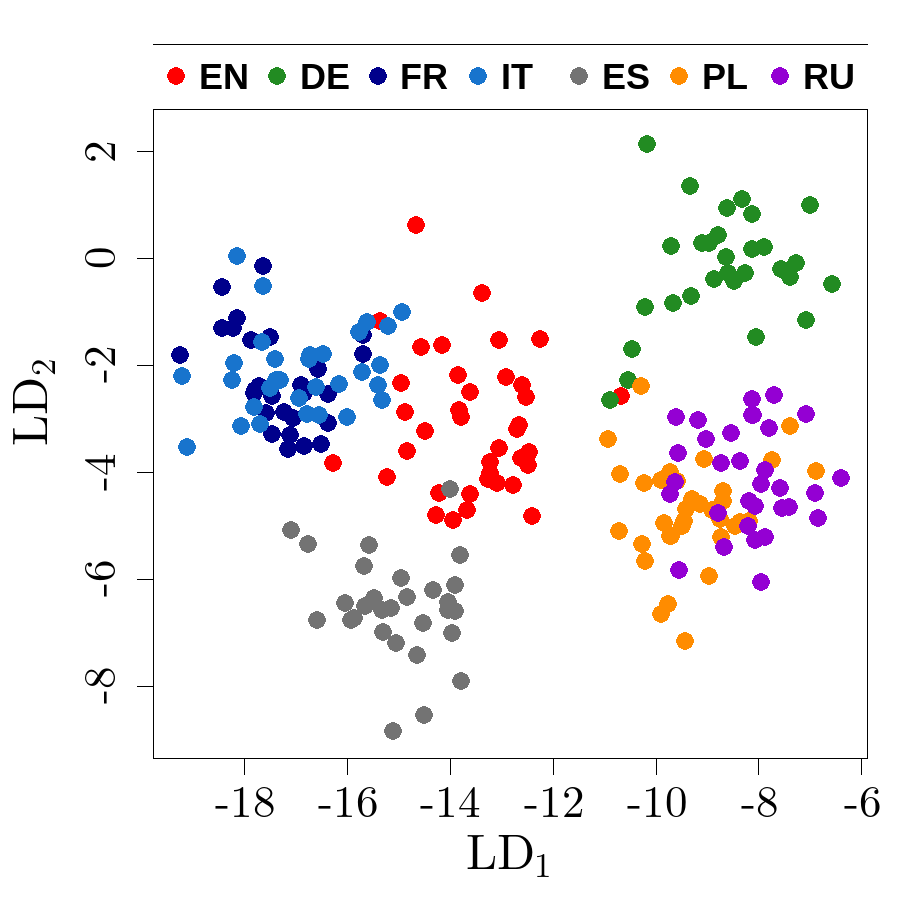}}
\subfloat[]{\includegraphics[width=0.475\textwidth]{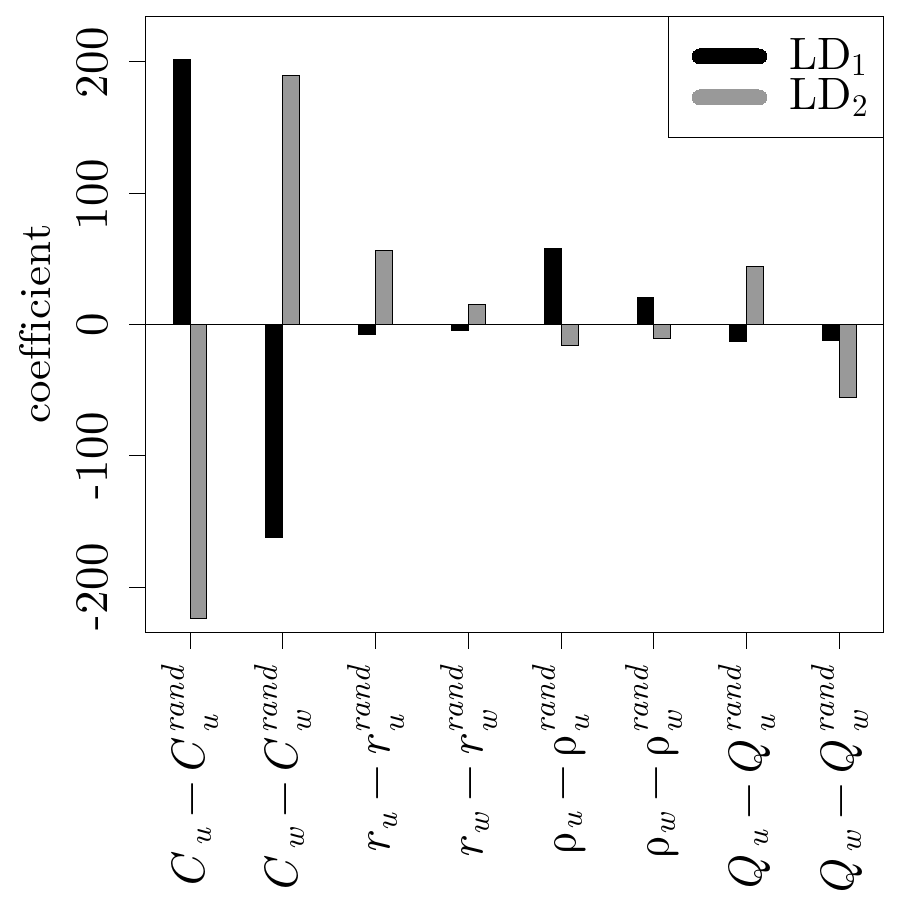}}
\vspace{-0.5em}
\caption{(a) Texts written in different languages (points represent individual texts and colors represent languages) embedded in 4-D space spanned by the selected parameters of the word-adjacency networks (see Fig.~\ref{fig_networks_global_characteristics_beeswarm}): the normalized global clustering coefficients in their unweighted and weighted versions ($C_{\rm u}^{\rm norm}$ and $C_{\rm w}^{\rm norm}$, respectively) and the normalized assortativity indices also in their unweighted and weighted versions ($r_{\rm u}^{\rm norm}$ and $r_{\rm w}^{\rm norm}$, respectively), projected on a 2-D subspace spanned by the 2 most prominent vectors ($\mathrm{LD_1}$ and $\mathrm{LD_2}$) of Linear Discriminant Analysis. (b) Coefficients of the linear combinations of the network parameters that lead to these spanning vectors. Absolute values of the coefficients measure the extent to which a particular characteristic allows one to separate texts in different languages. The clustering coefficient (both in its weighted and unweighted form) seems to be the most significant characteristic in this context.}
\label{fig_LDA_dataset_1}
\end{minipage}
\end{figure}

Differences between languages in the word-adjacency networks can also be illustrated by a dimensionality reduction technique, e.g., Linear Discriminant Analysis (LDA)~\cite{Duda2000}. Given a set of points in some space, each belonging to some predefined class and labeled by that class, LDA sequentially finds vectors that are orthogonal to each other such that the projection of the data points on the subspace spanned by these vectors results in the maximum possible separation between classes. The first vector maximizes the class separation, and each subsequent vector maximizes the separation under the condition that it is orthogonal to all the preceding vectors. Since the ability of each vector to discriminate between classes is weaker than for the preceding vectors, projecting the data onto the subspace spanned by the first few such vectors can be sufficient to detect the patterns of variability between classes. A projection of the dataset consisting of texts in 7 different Western languages onto a 2-dimensional space spanned by the most prominent vectors obtained by LDA in the space of four network characteristics: $C_{\rm u}^{\rm norm}$, $C_{\rm w}^{\rm norm}$, $Q_{\rm u}^{\rm norm}$, and $Q_{\rm w}^{\rm norm}$ is shown in Fig.~\ref{fig_LDA_dataset_1}. There can be distinguished the clusters of texts written in the same language yet there is still some overlap: it is particularly difficult to distinguish between the French and Italian texts and between the Polish and Russian ones. The other types of linguistic networks, for example the networks based on syntactic relationships between words, can also display different patterns of organization for different languages~\cite{Liu2010}.

\subsubsection{Word-adjacency networks and text authorship}

\begin{figure}
\centering
\begin{minipage}{\figurecustomwidth}
\centering
\subfloat[]{\includegraphics[width=0.49\textwidth]{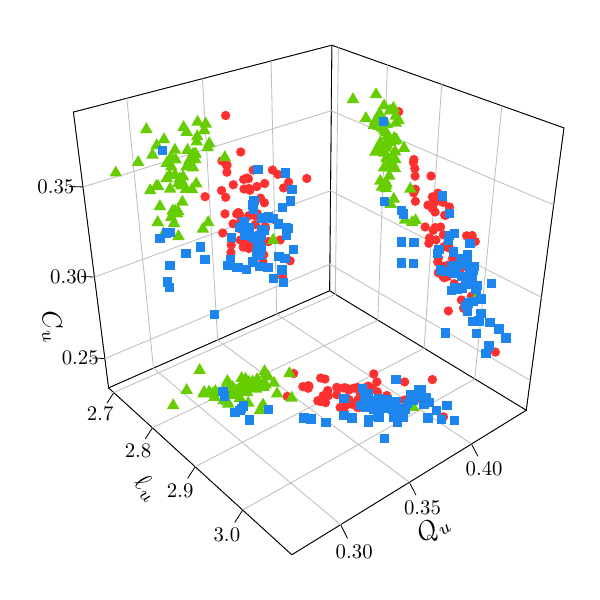}}
\hfill
\subfloat[]{\includegraphics[width=0.49\textwidth]{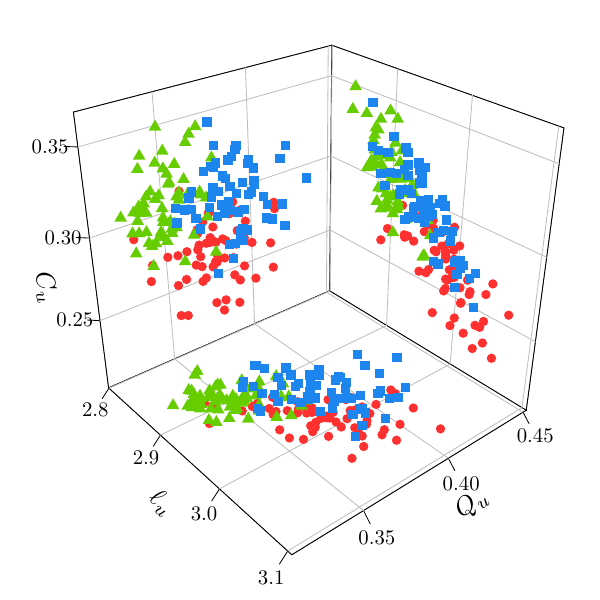}}
\hfill
\vspace{0.5cm}
\subfloat[]{\includegraphics[width=0.49\textwidth]{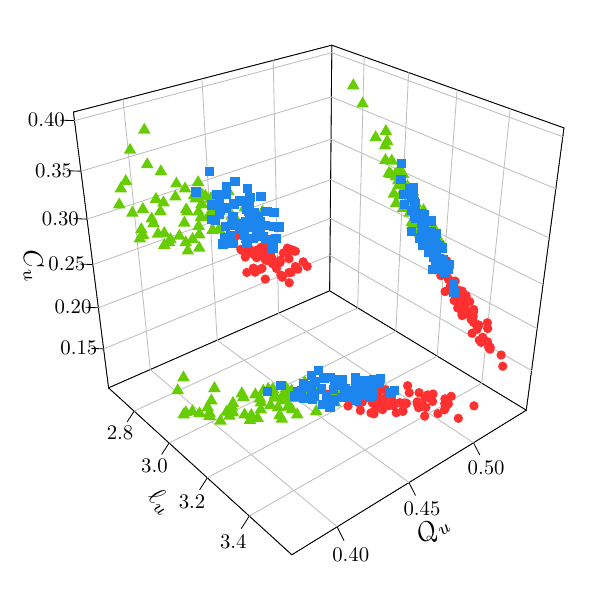}}
\hfill
\subfloat[]{\includegraphics[width=0.49\textwidth]{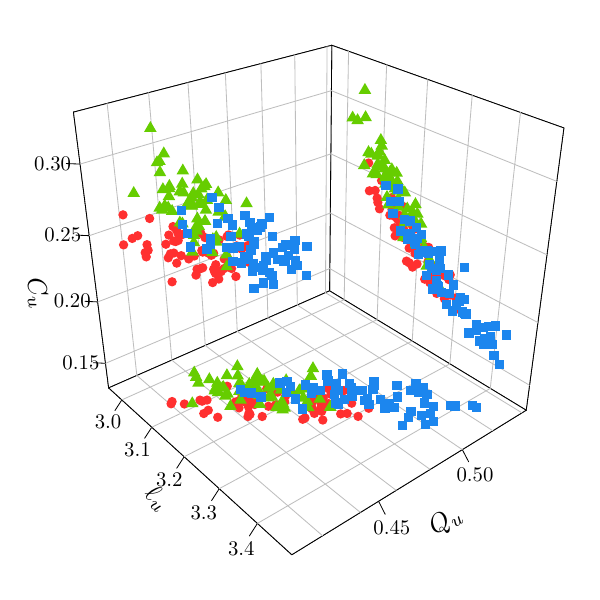}}
\hfill
\caption{Projections of the 3-D space of the parameters $\left(\ell_{\rm u},\rho_u,C_{\rm u}\right)$ of word-adjacency networks representing literary texts written in English (a),(b) or Polish (c),(d) onto 2-parameter planes. Each symbol pertains to a text sample of length of 5,000 words randomly chosen from works of different authors. The authors represented by symbols and colors (red circles, green triangles, and blue squares, respectively) are: (a) Charles Dickens, Daniel Defoe, and Mark Twain, (b) George Eliot, Jane Austen, and Joseph Conrad, (c) Władysław Reymont, Janusz Korczak, and Jan Lam, (d) Henryk Sienkiewicz, Józef Ignacy Kraszewski, and Stefan Żeromski. The characteristics were not normalized as all the samples had the same length~\cite{Stanisz2019}.}
\label{fig_char_triplets_in_3d}
\end{minipage}
\end{figure}

Word-adjacency networks can be applied in stylometry to investigate the authorship of texts. It is based on assumption that authors have individual styles of writing that can be reflected in the statistical properties of vocabulary and the specific relationship of words. Stanisz \textit{et al.} analyzed unweighted word-adjacency networks constructed from sample literary texts written in English and Polish and consisting of 5,000 words~\cite{Stanisz2019}. Three parameters were considered: the average shortest path length $\ell_{\rm u}$, global clustering coefficient $C_{\rm u}$, and modularity $Q_{\rm u}$. Although constructed from different books, the networks corresponding to texts written by the same author tended to be similar in terms of these parameters. Fig.~\ref{fig_char_triplets_in_3d} displays results for three authors representing each language. However, if the number of authors was larger, an overlap between the regions representing different authors would make the overall picture less evident.

\begin{figure}
\centering
\begin{minipage}{\figurecustomwidth}
\centering
\subfloat[]{\rotatebox[origin=c]{-90}{\includegraphics[height=0.49\textwidth]{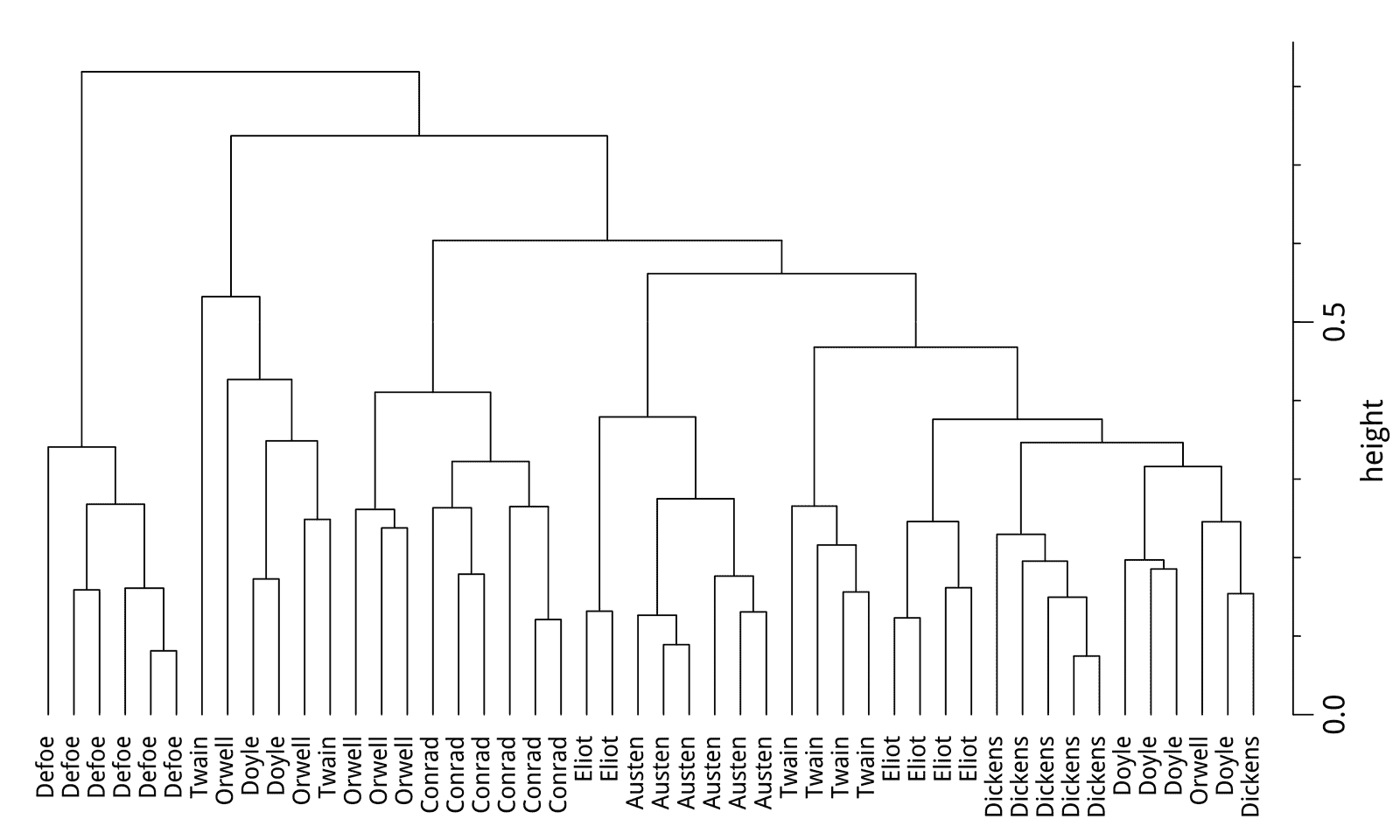}}}
\hfill
\subfloat[]{\rotatebox[origin=c]{-90}{\includegraphics[height=0.49\textwidth]{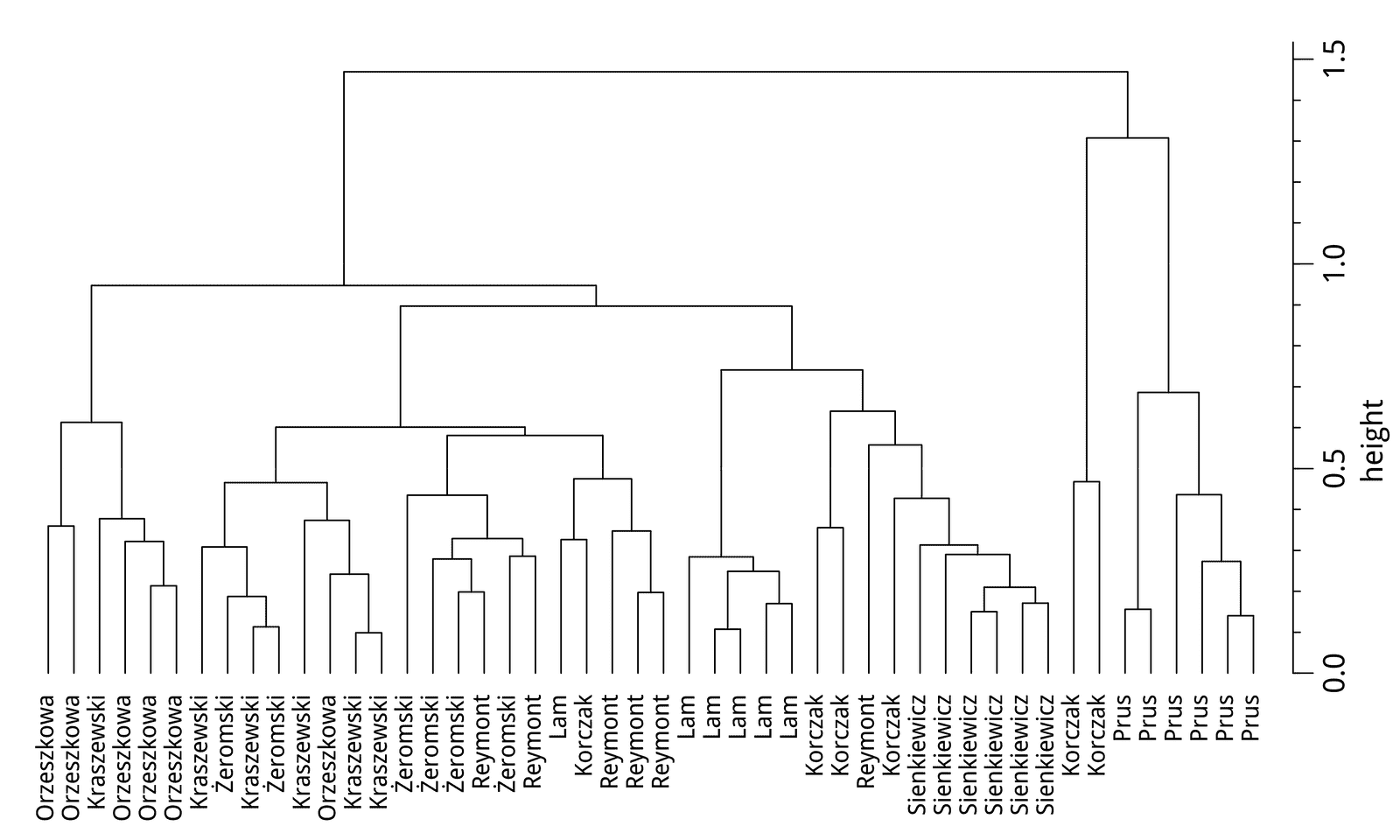}}}
\caption{Dendrograms of the hierarchical clustering for a set of (a) 48 English and (b) 48 Polish books carried out in the space of the local weighted clustering coefficients $C_{\rm w}^{\rm norm}(v)$ corresponding to the 12 most frequent words $v$ in each text (punctuation marks were included). Texts are labeled by the author surnames. (After~\cite{Stanisz2019}.)}
\label{fig_dendrogram_local_weighted_CC}
\end{minipage}
\end{figure}

The same authors studied two sets of books written in English or Polish (48 books per language) transformed to word-adjacency networks~\cite{Stanisz2019}. A hierarchical clustering was performed on the normalized local weighted clustering coefficients $C_{\rm w}^{\rm norm}(v)=C_{\rm w}(v)/C_{\rm w}^{\rm rand}(v)$ calculated for the 12 most frequent words (including punctuation marks) of each book. Their results showed that books written by a given author tend to group themselves together (Fig.~\ref{fig_dendrogram_local_weighted_CC}), which suggested that it was sufficient to analyze as few as a dozen the most frequent words to collect information about an author's individual writing style. This conclusion was further confirmed by a statistical classification carried out with algorithms of supervised machine learning on the same set of books. Indeed, the sharper the differences between the network structure for texts of different authors are, the easier it is to train a statistical classification algorithm to recognize authorship in the appropriate parameter space. By applying a decision tree bootstrap aggregating method~\cite{Sutton2005}, a classification in the space constructed from appropriately normalized vertex strengths and weighted clustering coefficients reached the accuracy of 86\% for English and 90\% for Polish, which is much more than the expected accuracy of a random classification for 8 indistinguishable authors (1/8=12.5\%). This proved usefulness of the word-adjacency network approach to authorship attribution. A parallel analysis of the same texts without punctuation marks (the word-adjacency networks that were analyzed did not contain the corresponding nodes) indicated that neglecting punctuation substantially decreases accuracy of the classification. This result supported the observation that specific patterns of punctuation usage constitute a non-negligible contribution to the writing style and they should not be omitted in stylometric analyses~\cite{Stanisz2019}.

\subsubsection{The word-adjacency network growth}

\begin{figure}[p]
\centering
\begin{minipage}{\figurecustomwidth}
\centering
\includegraphics[scale=0.17]{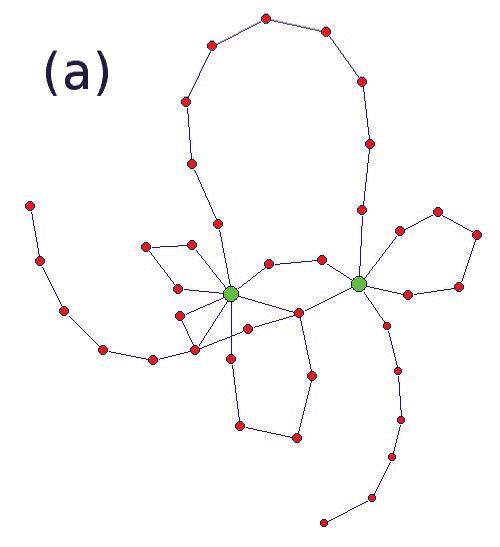}
\includegraphics[scale=0.17]{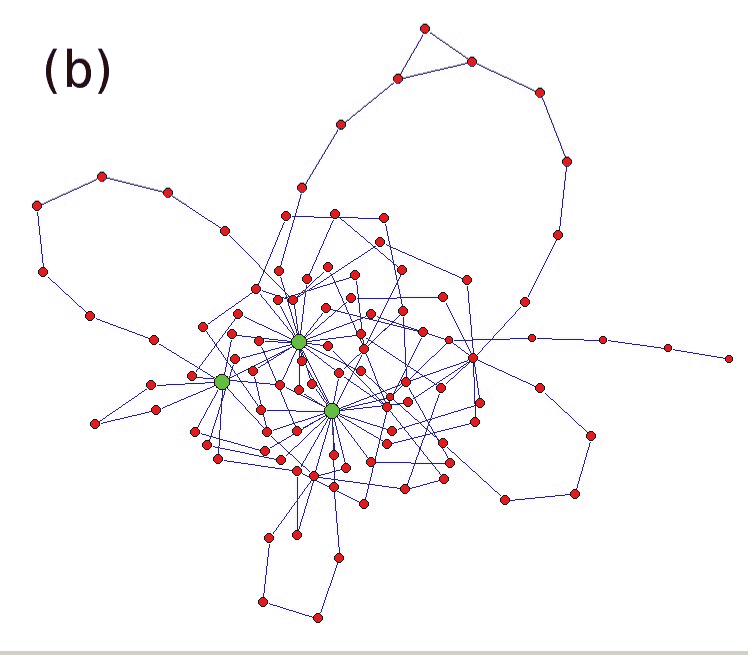}
\includegraphics[scale=0.17]{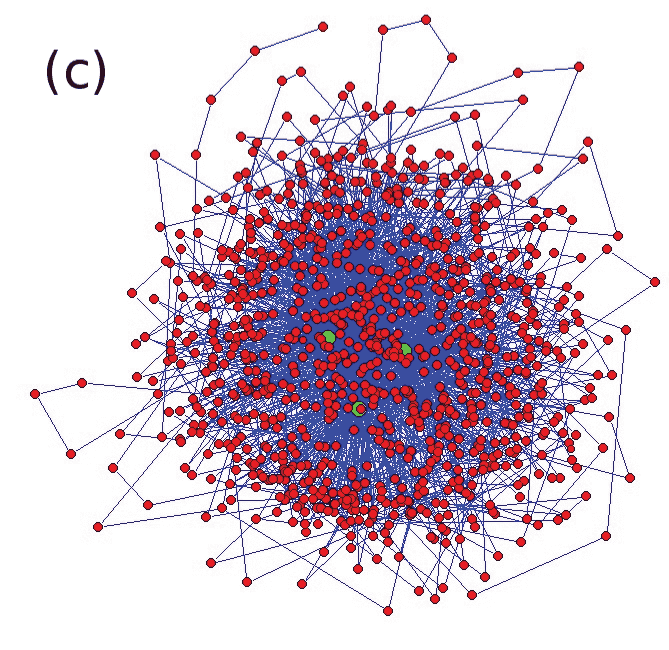}

\includegraphics[scale=0.17]{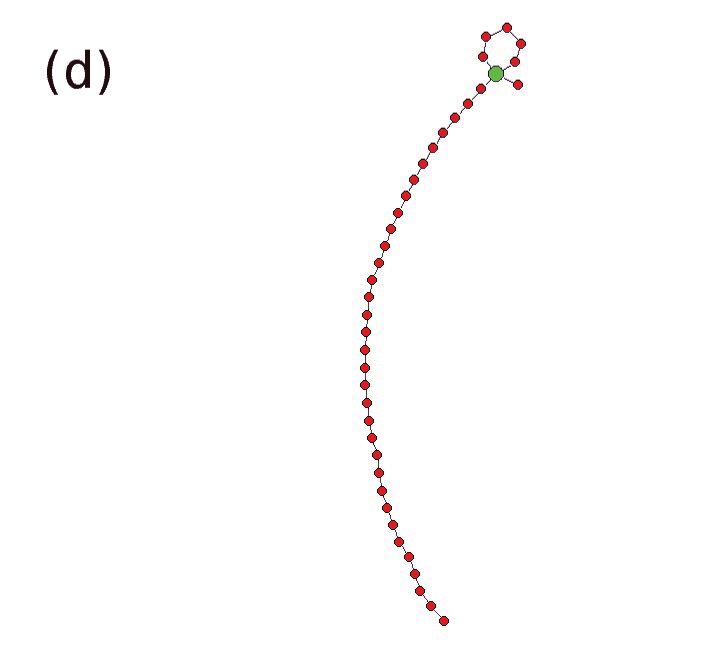}
\includegraphics[scale=0.17]{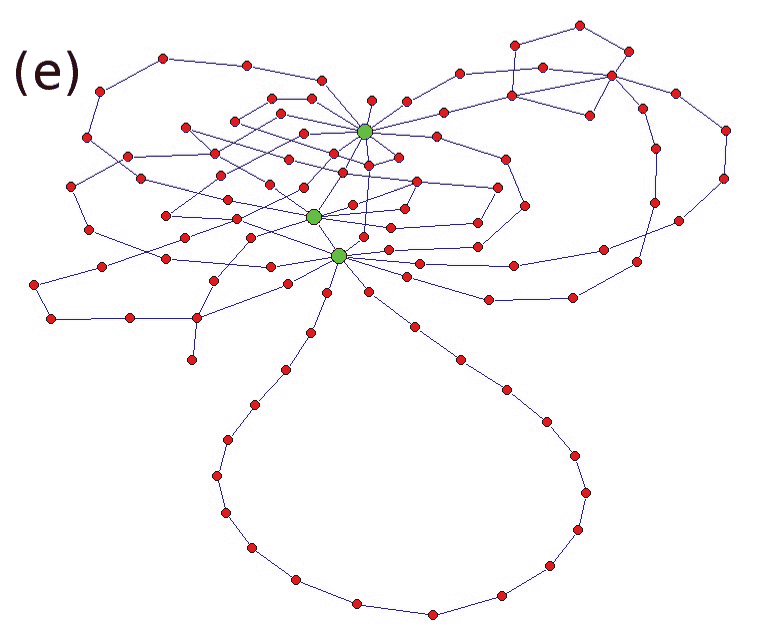}
\includegraphics[scale=0.17]{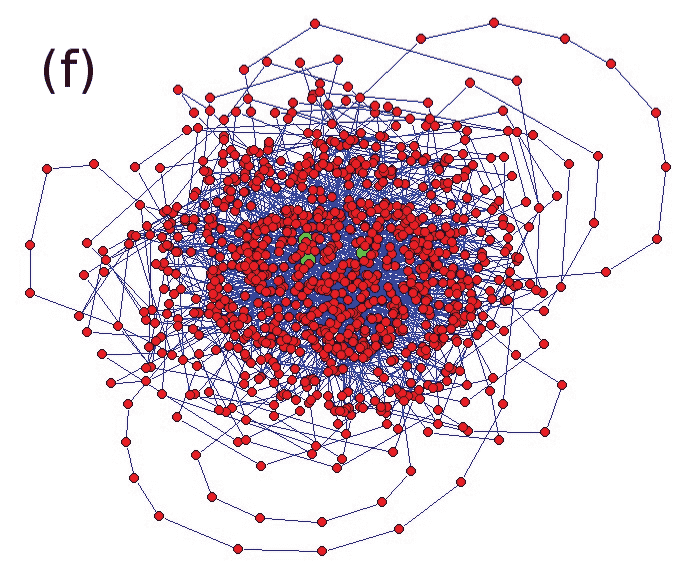}
\caption{Growing word-adjacency networks for sample literary texts: \textit{Ulysses} by J.~Joyce (English, top) and \textit{Lalka} by B.~Prus (Polish, bottom) in different development stages: (a),(d) an initial stage with few or even no repeated words ($N=40$), (b),(e) a stage in which hubs start to be distinguishable ($N=100$), and (c),(f) a stage in which old words are used more often than the new ones ($N=1000$).}
\label{fig::word-adjacency.network.growth.empirical}
\end{minipage}
\end{figure}

\begin{figure}[p]
\centering
\begin{minipage}{\figurecustomwidth}
\centering
\includegraphics[scale=0.6]{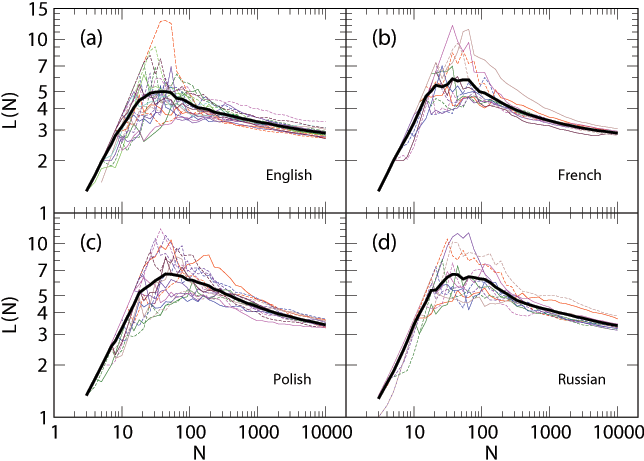}
\caption{Evolution of the average shortest path length $\ell(N)$ for growing word-adjacency networks created from opening pieces of sample literary texts representing different European languages: (a) English (23 texts), (b) French (14 texts), (c) Polish (17 texts), and (d) Russian (12 texts). Thin lines correspond to individual texts, while thick lines denote average over the texts written in the same language. (After~\cite{KuligA-2015a}).}
\label{fig::word-adjacency.network.empirical.aspl}
\end{minipage}
\end{figure}

Typically, if a text sample is studied, the related word-adjacency network is already mature and contains a large number of nodes and edges. It resembles taking a thermodynamic limit of large $N$ (number of nodes) without considering routes to such a limit. A dynamical approach to properties of such networks has to include studying of a process of their formation which is inherently related to a process of writing. Such an approach has seldom been reported in literature, however. A notable exception is the work~\cite{KuligA-2015a}, in which its authors noticed that there did not exist a network-growth model that would have been able to reproduce the observed properties of the word-adjacency networks as regards, for instance, the average shortest path length. Let us consider a network that is being created dynamically as a text sample is being written word by word. At first, only new words/nodes are added to the network since there has been no need to repeat any of the words yet. The network is a chain and the average shortest path length $\ell$ is growing with every added word (Fig.~\ref{fig::word-adjacency.network.growth.empirical}). At some moment, however, either grammar or meaning demands repeating of the already used words, which translates directly to a formation of loops and shortening of $\ell$. In the beginning, long loops dominate as relatively few words have been used repeatedly, but as the text length grows, the loops can also be shortened. The longer is the text, the more nodes contribute to the network, but also the more words have already occurred more than once. In accordance with the Heaps law (Eq.~(\ref{eq::Heaps_law})) fewer and fewer new words/nodes are added to the network and this inevitably condenses the network and further shortens $\ell$. This is why, for sufficiently large networks (i.e., for $N$ larger than a few dozens), the average shortest path length declines with $N$ (see Fig.~\ref{fig::word-adjacency.network.empirical.aspl} for sample empirical functions $\ell(N)$).

The above behaviour and the declining probability of adding new nodes distinguishes empirical word-adjacency networks from, for example, the Erd\"os-R\'enyi networks (for which $\ell(N) \sim \ln N$), the scale-free networks ($\ell(N) \sim \ln N$ for $\gamma > 3$, $\ell(N) \sim \ln N / \ln\ln N$ for $\gamma > 3$~\cite{CohenR-2003a}, and $\ell(N) \sim \ln\ln N$ or even $\lim_{N\to\infty} \ell(N) = 1/2 + 2/(3-\gamma)$ for $2 < \gamma < 3$~\cite{FronczakA-2004a}). There can be three stages of network maturity distinguished: an initial stage where $\ell(N)$ grows approximately linearly ($N \ge 20$, it does not depend on a sample and it is thus universal across the texts, authors, and languages), a saturation phase, in which the growth of $\ell(N)$ slows down and then stops completely reaching a maximum ($20 < N \ge 200$, it corresponds to a text piece from a line to a page, it is the most sample-specific, and thus it carries information about the individual styles), and finally a mature phase when $\ell(N)$ decreases ($N > 200$, it shows diminishing individual traits and, again, increasing universality, it therefore reflects the global statistical properties of language). In this mature stage, the empirical $\ell(N)$ can be approximated by~\cite{KuligA-2015a}:
\begin{equation}
\ell(N) \sim {\ln N \over \ln {c_0 \over \alpha+1} + \alpha \ln N},
\end{equation}
where $c_0 > 0$ and $\alpha > 0$ are free parameters.

\begin{figure}[t]
\centering
\begin{minipage}{\figurecustomwidth}
\centering
\includegraphics[scale=0.16]{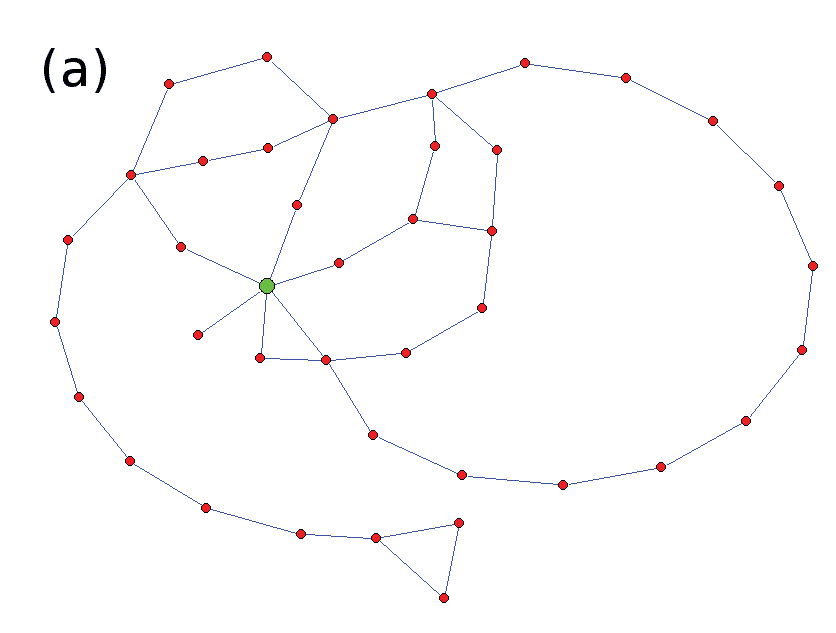}
\includegraphics[scale=0.16]{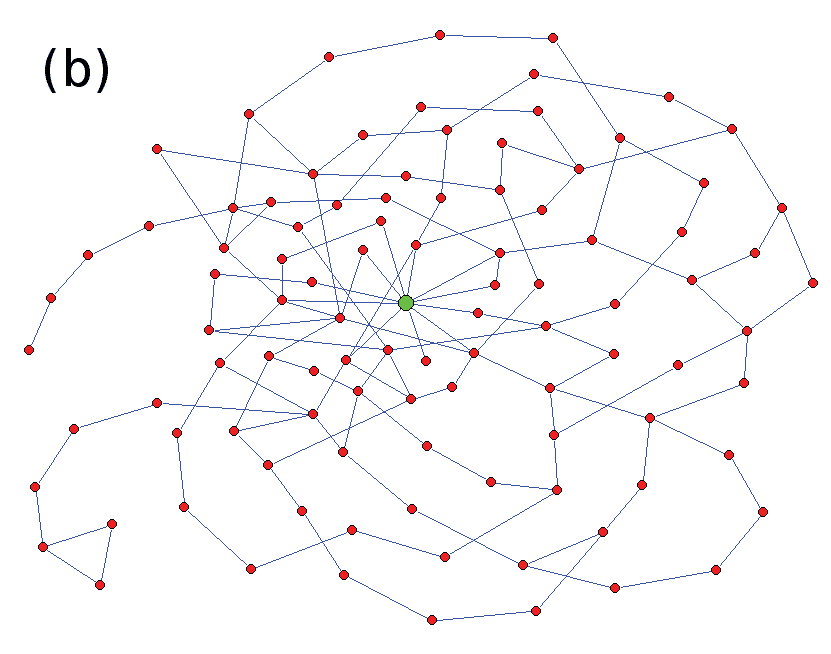}
\includegraphics[scale=0.16]{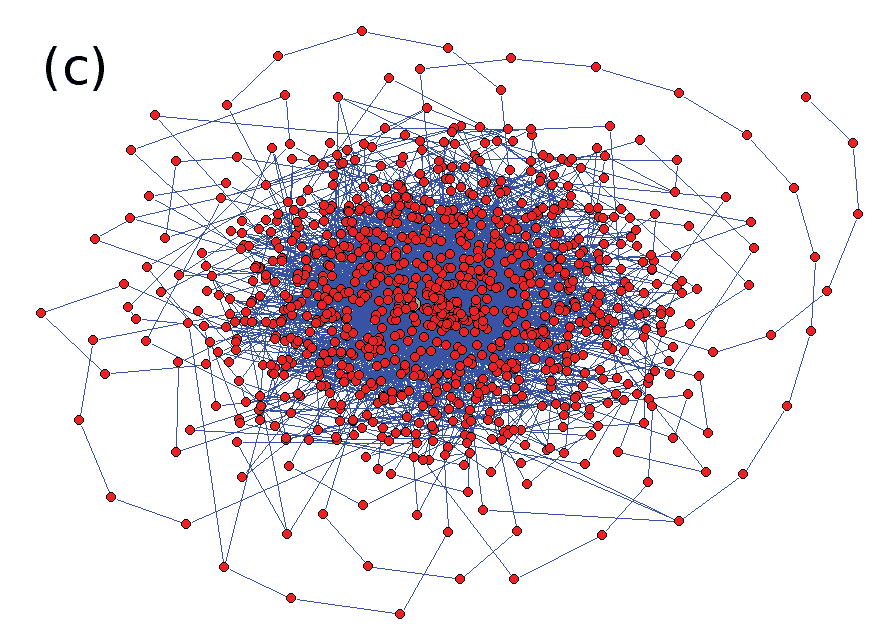}
\caption{Growth of a network constructed according to the hybrid model described in the text ($c_0 = 0.01$, $\alpha = 1.0$, $m=2$, $\mu=0.075$). Three snapshots of the network development are shown with the same values of $N$ as in Fig.~\ref{fig::word-adjacency.network.growth.empirical}: $N=40$ (a), 100 (b), and 1000 (c). (After~\cite{KuligA-2015a})}
\label{fig::hybrid.model.network.growth}
\end{minipage}
\end{figure}

\begin{figure}[t]
\centering
\begin{minipage}{\figurecustomwidth}
\centering
\includegraphics[scale=0.4]{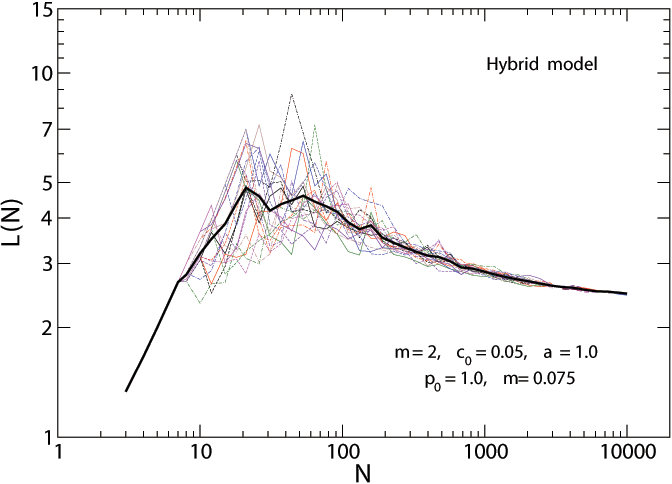}
\caption{Average shortest path length $\ell(N)$ for the networks constructed according to the hybrid model with $m=2$, $c_0=0.05$, $\alpha=1.0$, $p_0=1.0$, and $\mu=0.075$. Different realizations are represented by thin lines, while the average by thick one. (After~\cite{KuligA-2015a})}
\label{fig::hybrid.model.network.aspl}
\end{minipage}
\end{figure}

A generic model for the growing networks with the decreasing number of innovations is the Dorogovtsev-Mendes (DM) model of accelerated growth~\cite{DorogovtsevAA-2000a}. It is based on a modified BA model where the number of edges added to the network in each time step consists of a constant part $m$ and a continuous accelerating part $c(t)=c_0t^{\alpha}$ with $c_0>0$ and $\alpha > 0$. The exponent $\alpha$ is related to the Heaps exponent $\eta$ in Eq.~(\ref{eq::Heaps_law}) by the relation $\alpha=1/\eta-1$. For a realistic word-adjacency network it is required that $\alpha < 1$ (or $\alpha < 1 + \epsilon$, $\epsilon \ll 1$, if $c_0 \ll 1$) in order to avoid arriving at a fully connected state in a finite number of steps. However, the DM model does not work in the case of word-adjacency networks. A principal cause of that is the inability of the DM model to produce the long loops that are characteristic for the early stages of the empirical network growth. Therefore, Kulig \textit{et al.}~\cite{KuligA-2015a} introduced a hybrid preferential-attachment model that switches between the linear preferential attachment, responsible for the large-scale structure of the network, and the nonlinear preferential attachment with time-varying probability, which secures the existence of both the loops and the hubs in the early stage of the network growth. This is achieved by the amplification of the preferential attachment for small $t$:
\begin{equation}
\pi(k) \sim k^{\xi(t)}, \quad \xi(t)=c_1 t^{-\delta},
\end{equation}
where $c_1 > 0$ and $\delta > 0$. As the network grows, the attachment rule approaches the standard linear one, exactly as expected. Choosing the nonlinear regime occurs with a time-dependent probability:
\begin{equation}
p(t)=p_0 t^{-\mu},
\end{equation}
where $p_0\approx 1$, $\mu>0$, and $\mu \ll 1$, while choosing the DM regime with a probability $1-p(t)$. Fig.~\ref{fig::hybrid.model.network.growth} shows how the growing network constructed according to this hybrid model looks like and Fig.~\ref{fig::hybrid.model.network.aspl} shows the behaviour of $\ell(N)$ for a sample parameter value set. The agreement between these Figures and Figs.~\ref{fig::word-adjacency.network.growth.empirical} and~\ref{fig::word-adjacency.network.empirical.aspl} is substantial.

\subsection{Semantic networks and word-association networks}

\begin{figure}[ht]
\centering
\begin{minipage}{\figurecustomwidth}
\centering
\includegraphics[width=0.9\textwidth]{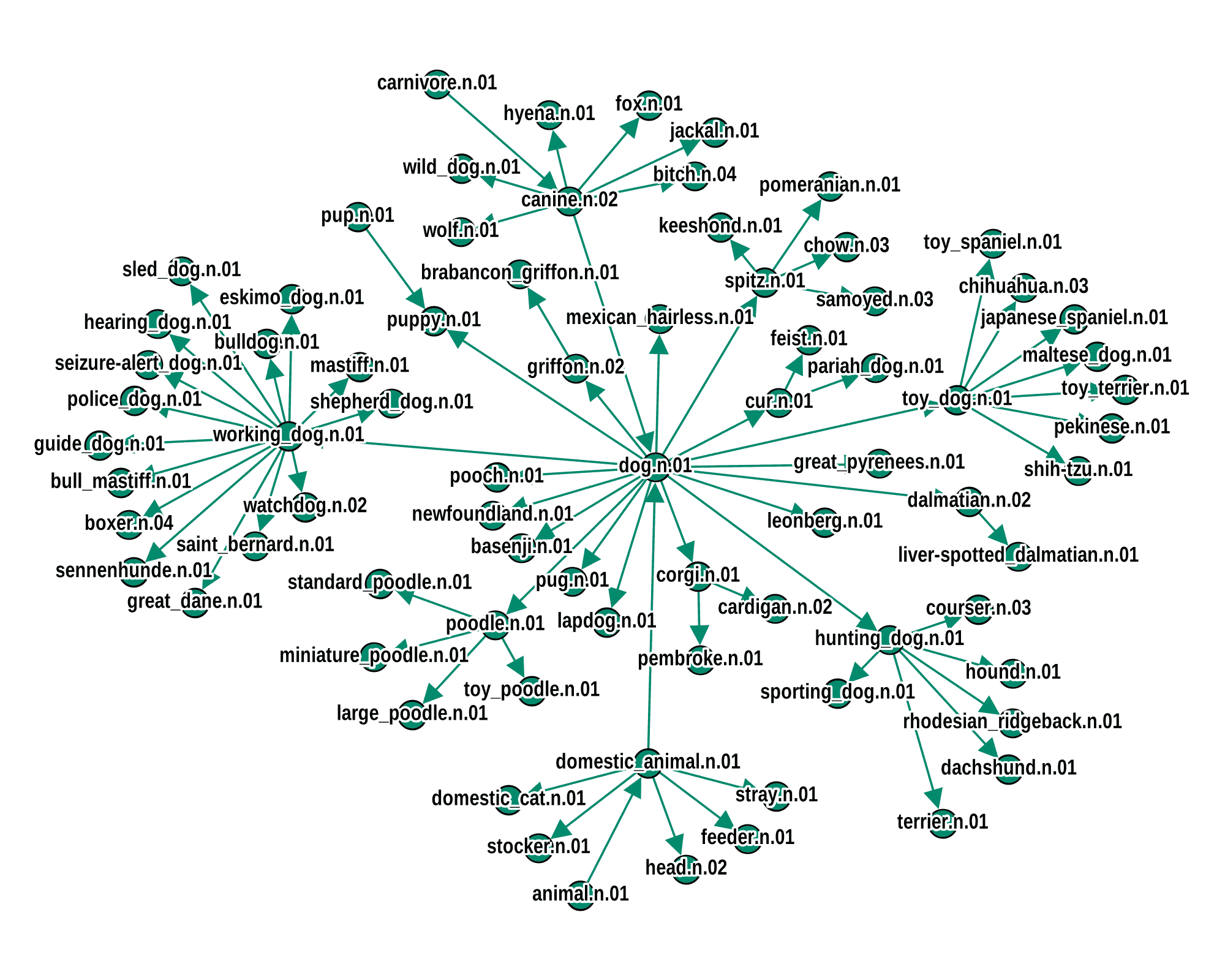}
\vspace{-0.5cm}
\caption{An excerpt from a network representing hypernym-hyponym relations between nouns in the WordNet database. Synsets are represented by nodes and hypernym-hyponym pairs are connected by unweighted directed edges from hypernyms to hyponyms. The presented subset of the original WordNet-based network consists of synsets that are at most 2 steps from the synset ``\textit{dog.n.01}'' (``n'' denotes nouns, the numbers are used to distinguish between synsets, which might otherwise be confused: for example, ``bank'' meaning a financial institution and ``bank'' meaning a sloping raised land).}
\label{fig_wordnet_dog.n.01_neighbourhood_of_order_2}
\end{minipage}
\end{figure}

A semantic network is a network in which nodes represent concepts and edges express semantic relationships between these concepts. Among the examples of semantic networks are networks representing the structure of certain linguistic databases like WordNet~\cite{Miller1995,Fellbaum1998}. WordNet is a lexical database consisting of words grouped into collections of synonyms, called synsets, which express certain concepts. Synsets can be connected with each other with various semantic relations, like hyponymy and hypernymy (if $Y$ is a subtype of $X$, then $Y$ is a hyponym of $X$ and $X$ is a hypernym of $Y$, e.g., \textit{plant-tree} is a hypernym-hyponym pair) or meronymy and holonymy (if $Y$ is a part of $X$, then $Y$ is a meronym of $X$ and $X$ is a holonym of $Y$, e.g., \textit{tree - leaf} is a holonym-meronym pair). An example of a network representing hypernymy-holonymy relations in an excerpt from WordNet database is shown in Fig.~\ref{fig_wordnet_dog.n.01_neighbourhood_of_order_2}.

Apart from having a number of applications in the field of automatic natural language processing~\cite{Fellbaum2015,Morato2004}, semantic networks are studied extensively in psycholinguistics -- a field that is focused on the cognitive mechanisms responsible for representing and processing language in human brain. For example WordNet, now serving as a lexical resource in a wide range of natural language processing solutions and language-related research, has been initially developed as a lexical database consistent with certain hypotheses regarding how semantic memory (the knowledge of words) is organized in human mind. Theories developed in 1960s and 1970s suggested that memory is organized in a hierarchical fashion, with concepts on deeper levels of hierarchy inheriting the properties assigned to relevant higher-level concepts~\cite{QuillianMR-1967a,CollinsAM-1969a,Fellbaum2010}. Although it has been recognized that some aspects of this view are oversimplified~\cite{Steyvers2005}, the idea of using network formalism to study the organization of words and concepts in human mind (often referred to as \textit{mental lexicon}) is more general than the mentioned theories and remains highly influential. An important class of networks in this context are word-association networks (or \textit{associative networks}). In this kind of network, nodes represent words and edges represent associations between words (often with weights representing strengths of individual associations). Word-association networks are constructed in experiments in which participants are presented words and are asked for writing down the first word that comes to their mind. Collecting data from many participants and for many different words allows to build a network with edge weights proportional to the number of participants giving a particular response. Significance of word associations has been investigated in a number of psycholinguistic experiments, involving tasks like word memorization or recognition (an example of word recognition task is deciding whether a given sequence of letters constitutes a word or not)~\cite{Deyne2014,Nelson2001}.

The fact that the characteristics of word-association networks allow one to make predictions regarding performance in tasks involving word processing and usage~\cite{Valba2021,DeDeyne2016} supports a claim that the structure of a word-association network can in some contexts be considered as a rough approximation of the lexicon structure in mind~\cite{Valba2021,DeDeyne2013}. This is why certain activities involving language processing, for example, a task consisting of finding a word that matches semantically to a set of given words, can be represented by a walk on a word-association network~\cite{Valba2021}. Therefore, studying the structure of word-association networks has a potential to give an insight into how language is organized and processed in human brain~\cite{Steyvers2005,DeDeyne2016,Deyne2008}.

Selected properties of word-association networks are presented here with the use of data coming from two datasets: \textit{University of South Florida Free Association Norms (USFFA)}~\cite{Nelson2004} and \textit{Edinburgh Associative Thesaurus (EAT)}~\cite{Kiss1973,Coltheart1981,Wilson1988}; both are available at~\cite{Batagelj2006} in a form allowing for their easy transformation into networks. The data in USFFA and EAT datasets was collected in experiments involving a large group of people and conducted according to the scheme mentioned above. These networks were preprocessed by removing nodes not representing ``typical'' words (e.g., sequences of digits) and transforming directed networks into undirected ones -- edge directions were ignored, while their weights were preserved (in cases where two oppositely directed edges were present between a pair of nodes, these edges were replaced with an undirected edge with weight equal to the sum of the weights of the original edges). Finally, the nodes with unit strengths were removed, because they represent words which appeared only once in the whole experiment. The so-preprocessed USFFA network contained 9958 nodes and 62491 edges and the EAT network contained 15184 nodes and 90236 edges.

\begin{figure}[p]
\centering
\begin{minipage}{\figurecustomwidth}
\centering
\subfloat[]{\includegraphics[width=0.49\textwidth]{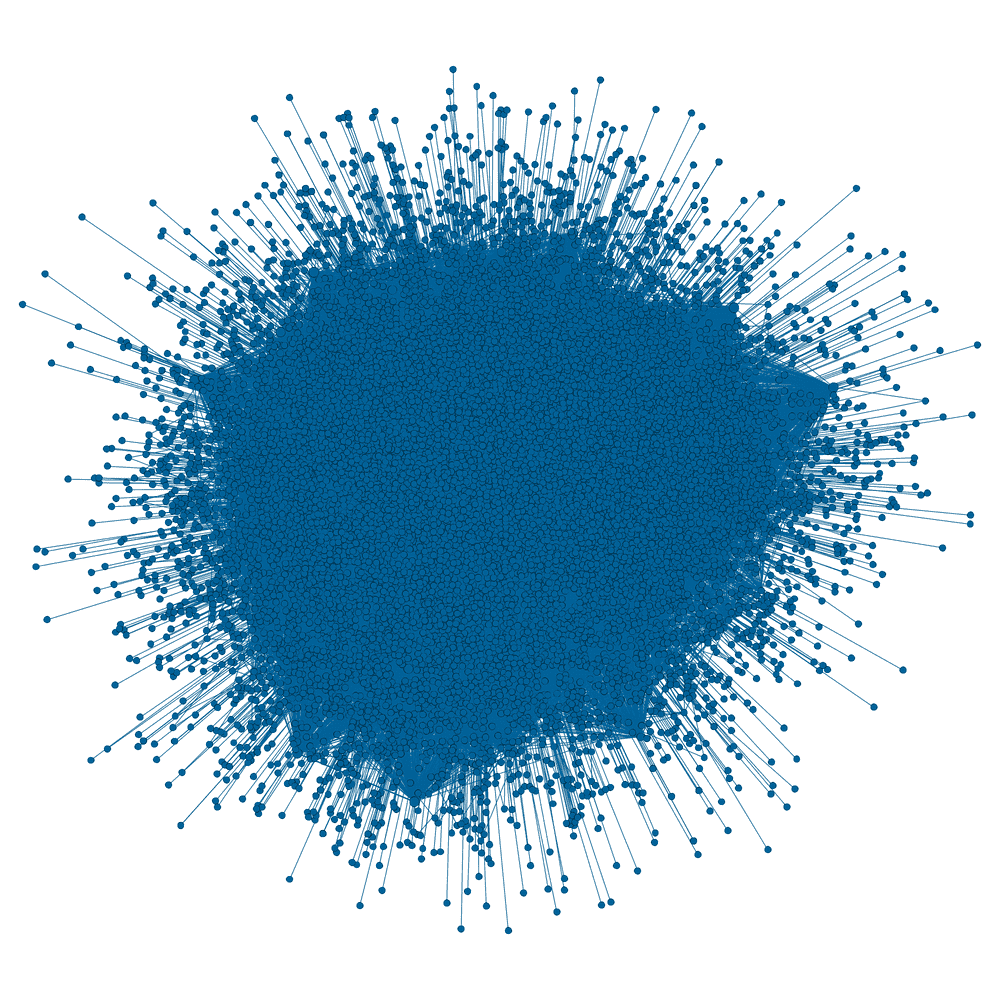}}
\hfill
\subfloat[]{\includegraphics[width=0.49\textwidth]{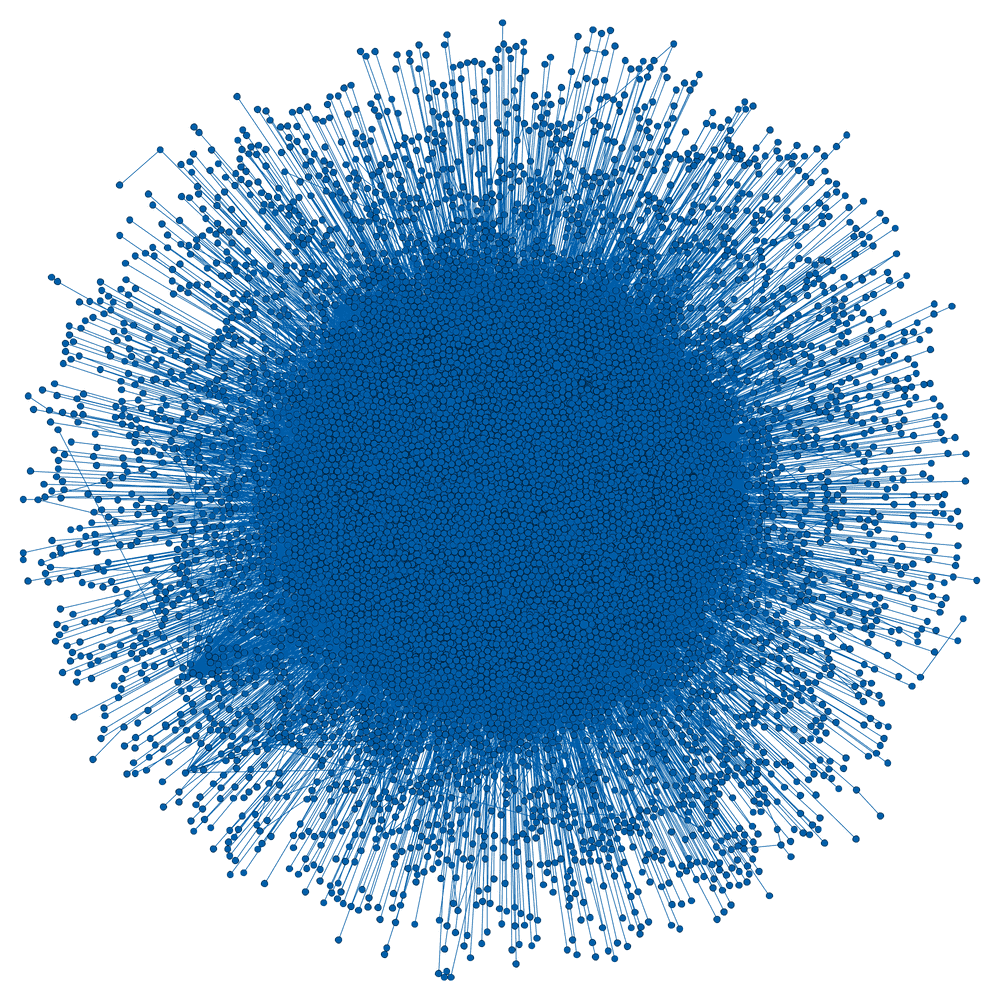}}
\vspace{1.0cm}
\subfloat[]{\includegraphics[width=0.49\textwidth]{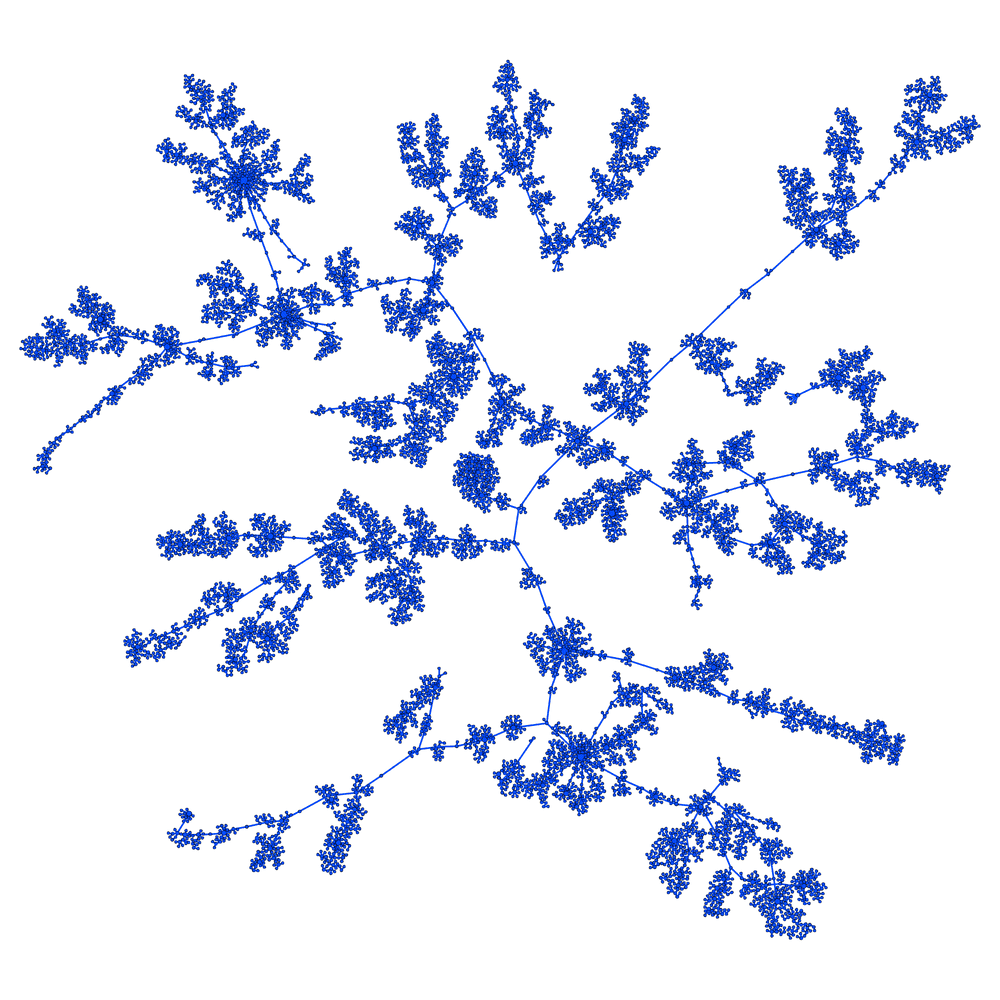}}
\hfill
\subfloat[]{\includegraphics[width=0.49\textwidth]{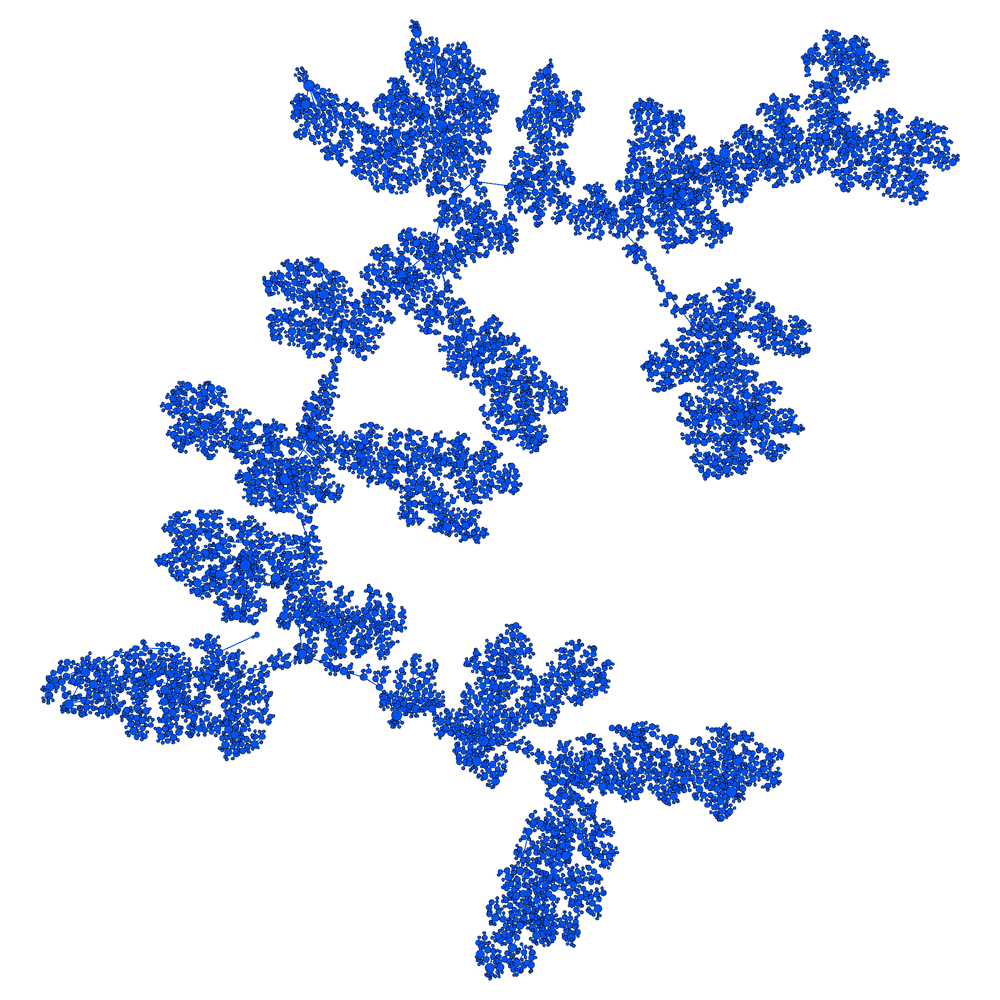}}
\caption{(a) A word-association network created from USFFA data set and (b) a sample realization of its randomized version. The corresponding minimum spanning trees are shown in (c) and (d), respectively.}
\label{fig_USFFA_network_and_randomization}
\end{minipage}
\end{figure}

\begin{figure}
\centering
\begin{minipage}{\figurecustomwidth}
\centering
\subfloat[]{\includegraphics[width=0.99\textwidth]{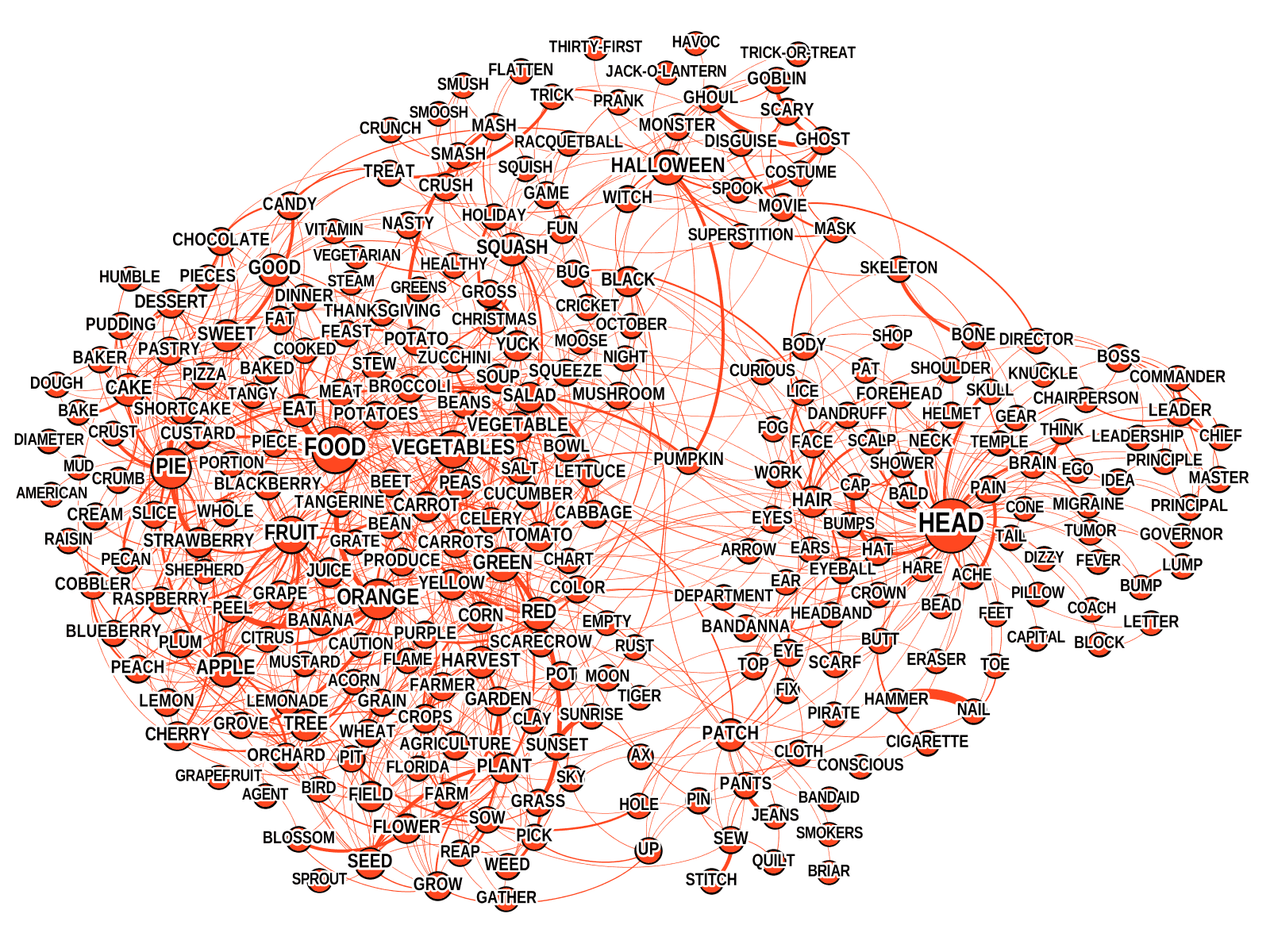}}
\hfill
\vspace{0.5cm}
\subfloat[]{\includegraphics[width=0.65\textwidth]{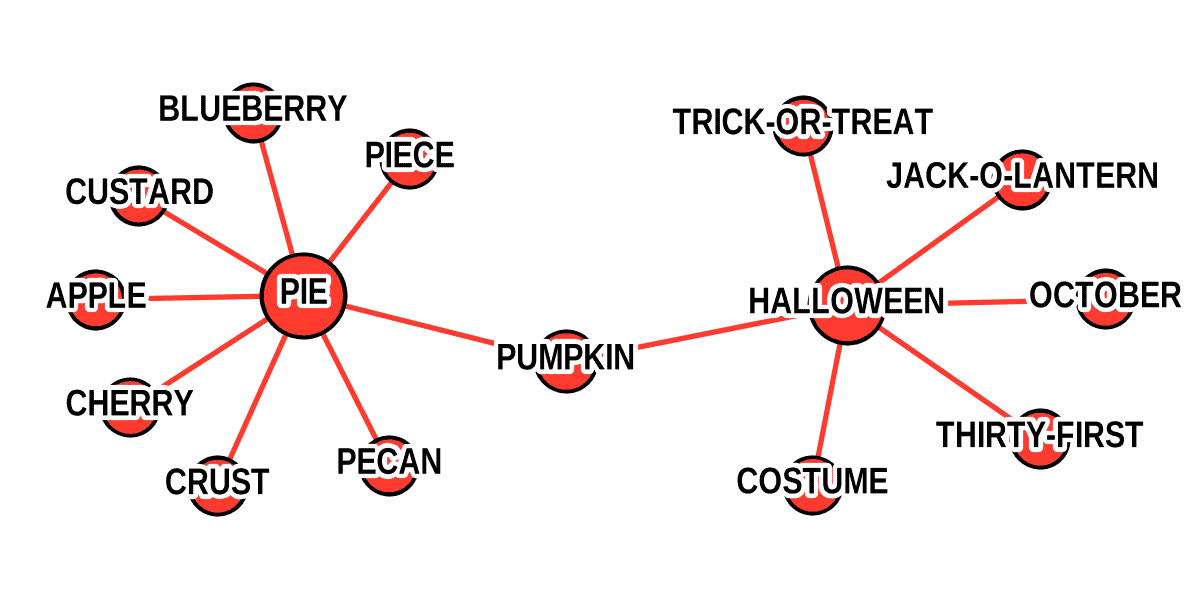}}
\caption{(a) A sample subnetwork of the USFFA network shown in Fig.~\ref{fig_USFFA_network_and_randomization} consisting of nodes that can be reached in at most 2 steps starting from the word ``pumpkin''. Node size and edge thickness represent node degrees and edge weights, respectively. (b) A subnetwork of the minimum spanning tree of the USFFA network, consisting of nodes selected by the same criterion as in the network shown in (a).}
\label{fig_USFFA_network_excerpt_2_step_from_pumpkin}
\end{minipage}
\end{figure}

The USFFA network is shown in Fig.~\ref{fig_USFFA_network_and_randomization}(a) together with its sample randomized version (b). The randomized networks serve as reference networks allowing one to decide whether the properties of the original networks are caused by a genuine network organization or they can entirely be attributed to the distributions of node degrees and edge weights. The randomization preserving these distributions was carried out in two steps: (1) after neglecting the edge weights, the network was shuffled according to the configuration model and (2) the original edge weights were randomly assigned to the edges of the randomized network. In addition, minimum spanning trees of the original and randomized USFFA networks were created with edge costs inversely proportional to the weights (Fig.~\ref{fig_USFFA_network_and_randomization}(c),(d)). The MSTs omit all the associations except for the strongest ones. Fig.~\ref{fig_USFFA_network_excerpt_2_step_from_pumpkin} displays a subnetwork of USFFA and a subnetwork of the corresponding MST, which illustrate how words are organized in a word-association network.

\begin{figure}
\centering
\begin{minipage}{\figurecustomwidth}
\centering
\subfloat[The USFFA network]{\includegraphics[width=0.49\textwidth]{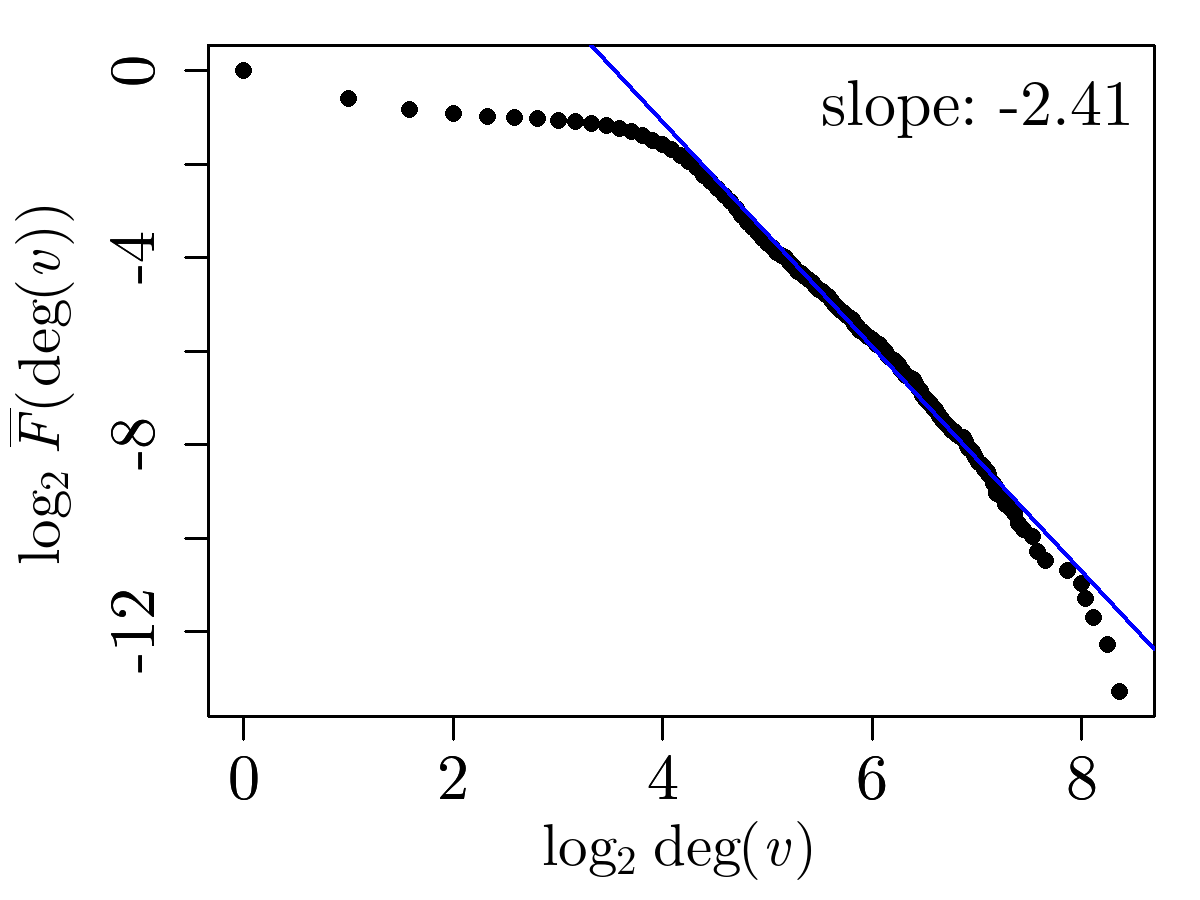}}
\hfill
\subfloat[The EAT network]{\includegraphics[width=0.49\textwidth]{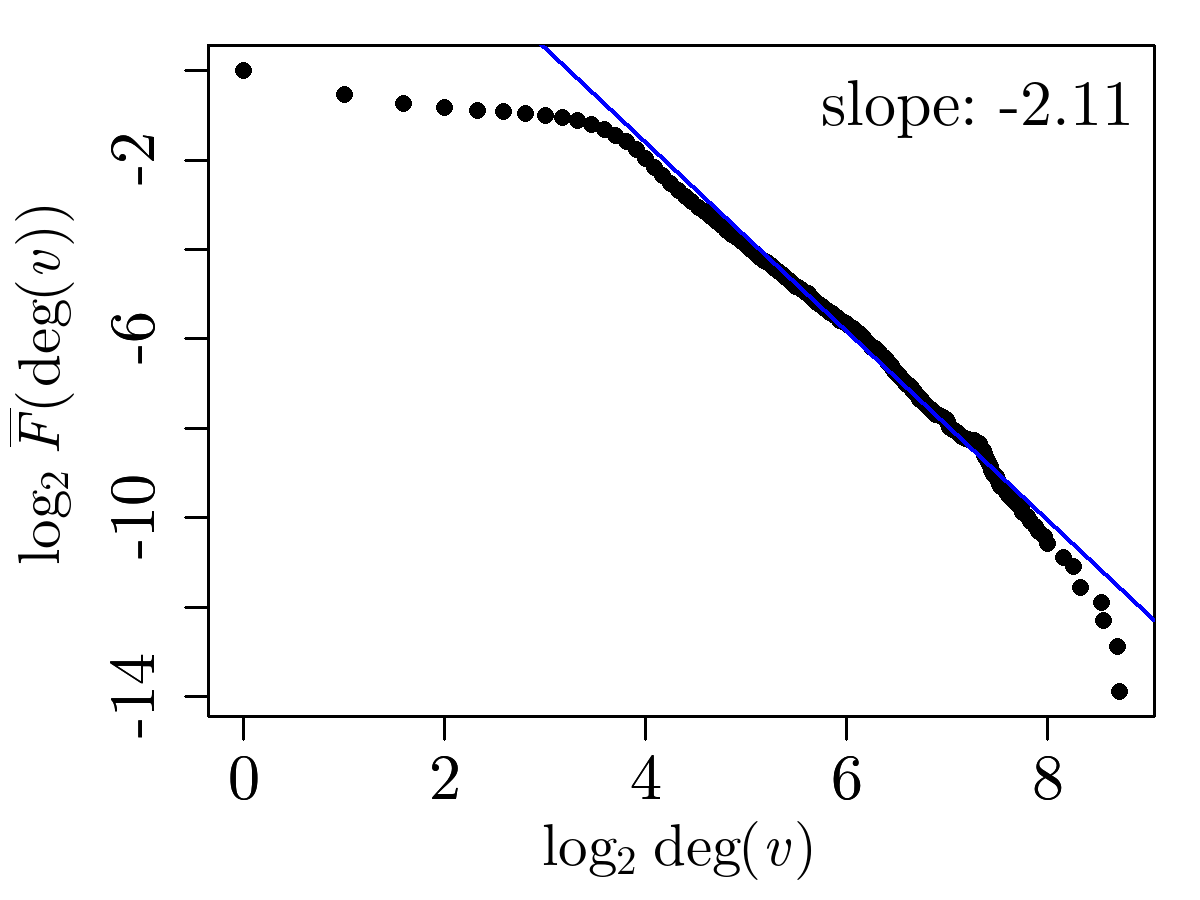}}
\hfill
\vspace{0.5cm}
\subfloat[The USFFA-based MST]{\includegraphics[width=0.49\textwidth]{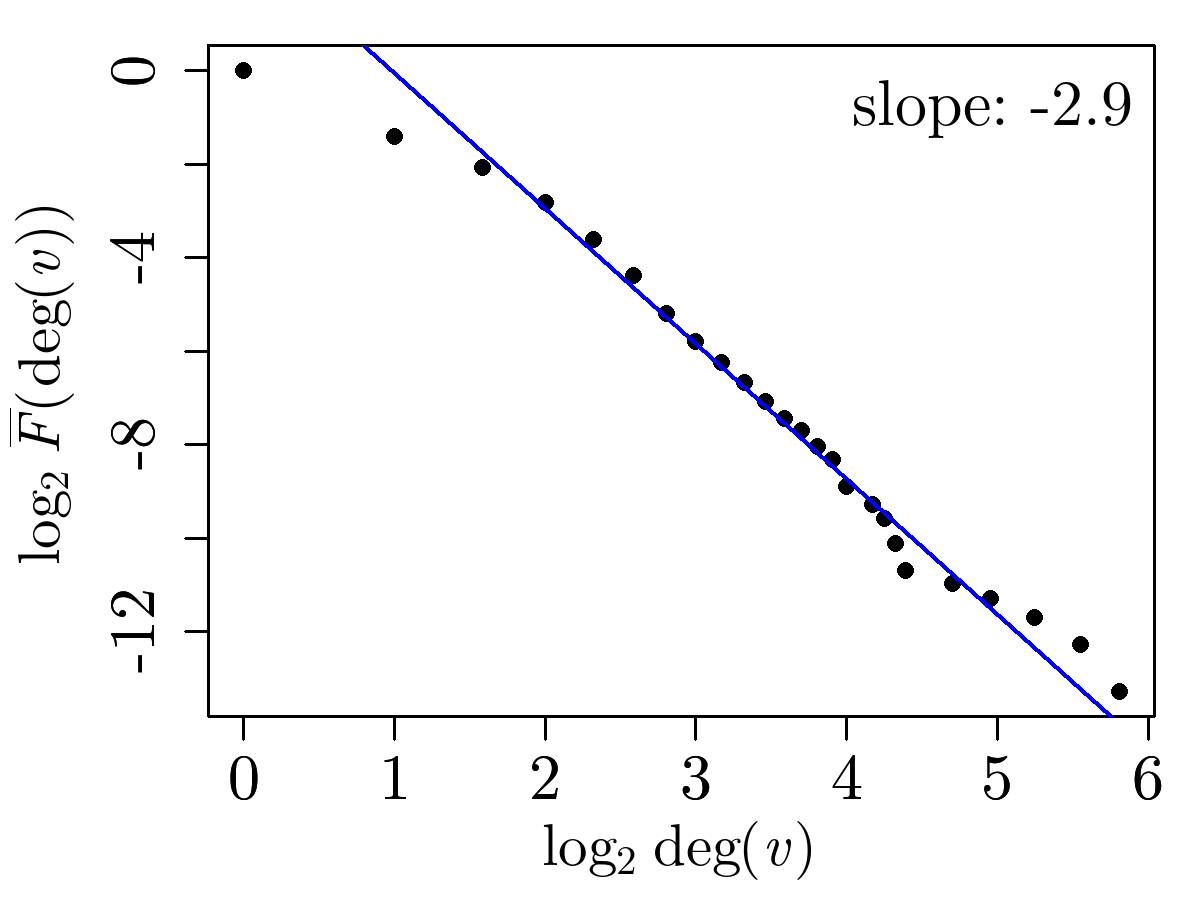}}
\hfill
\subfloat[The EAT-based MST]{\includegraphics[width=0.49\textwidth]{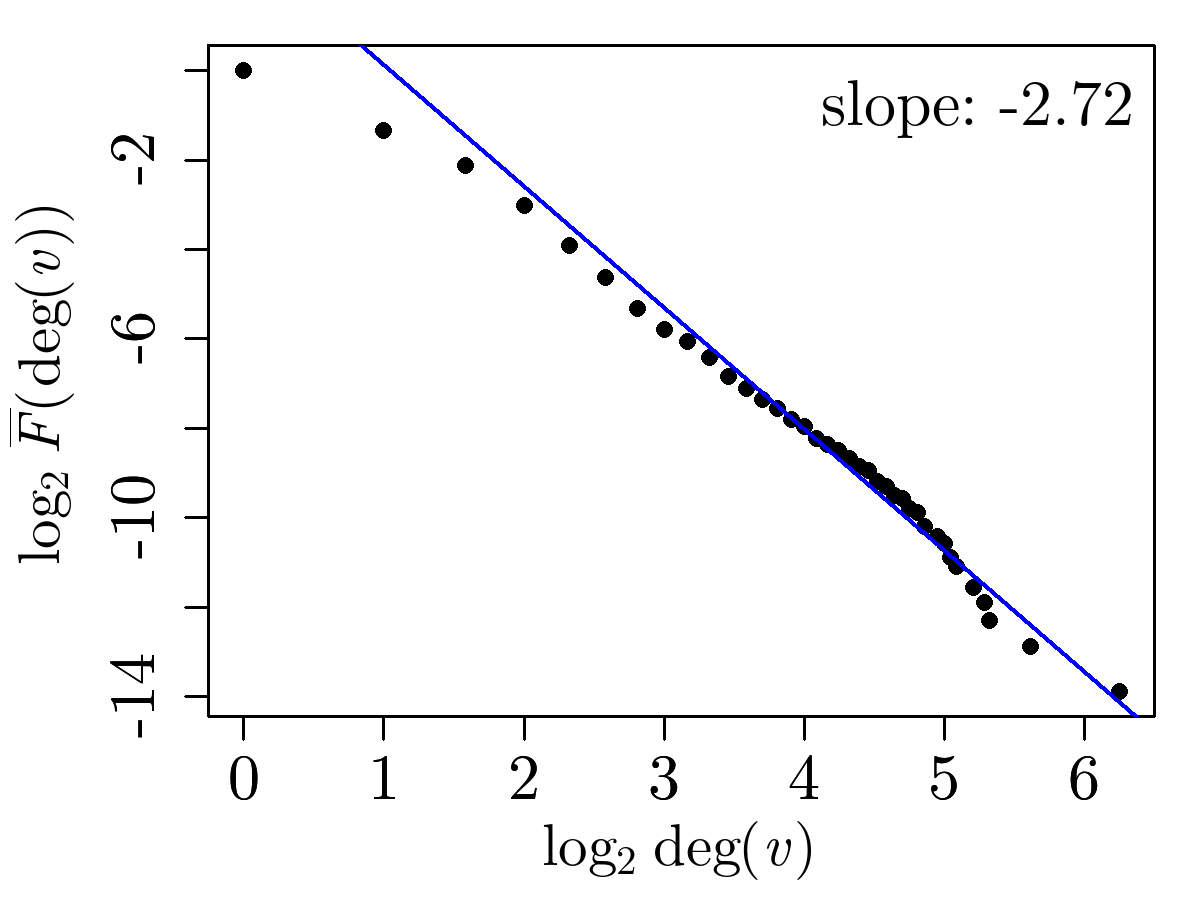}}
\hfill
\vspace{0.5cm}
\subfloat[The randomized USFFA-based MST]{\includegraphics[width=0.49\textwidth]{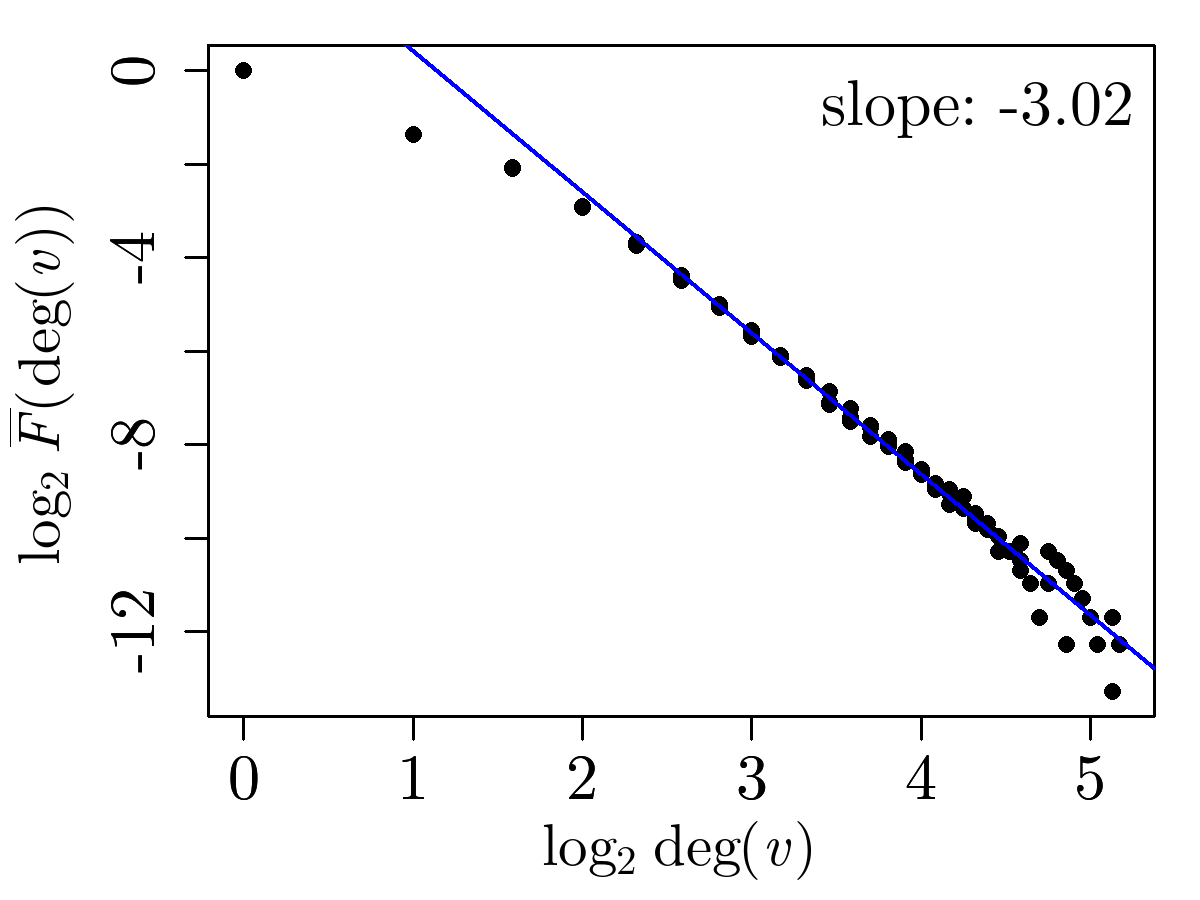}}
\hfill
\subfloat[The randomzied EAT-based MST]{\includegraphics[width=0.49\textwidth]{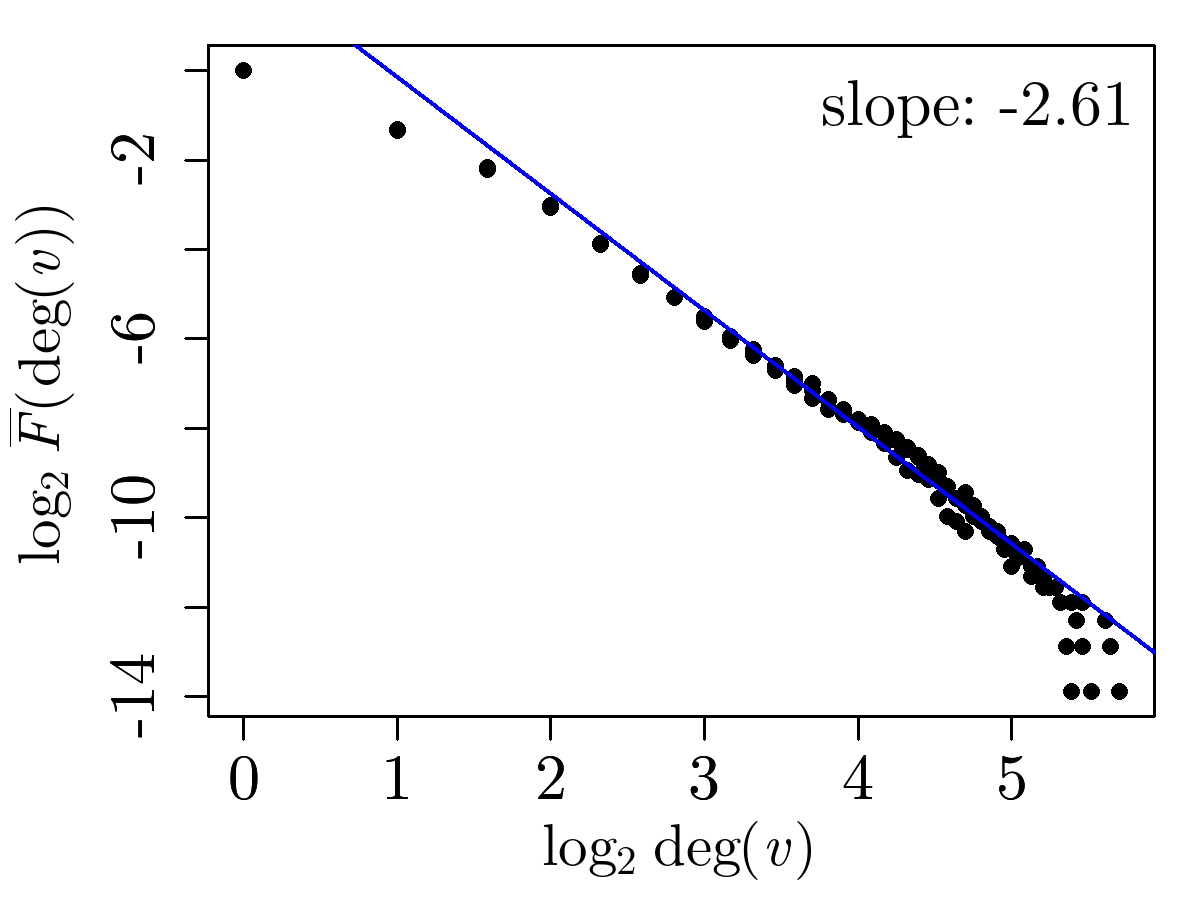}}
\hfill
\caption{Log-log plots of the survival functions of the node degree distributions $\bar{F}(\deg(v))$ for the USFFA network (a), the EAT network (b), the USFFA-based MST (c), the EAT-based MST (d), the randomized USFFA-based MST (e), and the randomized EAT-based MST (f). Slope indices of the blue lines are given in the top-right corner of each plot. In (e) and (f) superimposed results of three independent realizations of the randomization are shown.}
\label{fig_word_association_networks_degree_distributions}
\end{minipage}
\end{figure}

Fig.~\ref{fig_word_association_networks_degree_distributions} shows node degree distributions of the USFFA network, the EAT network, and their respective MSTs (the randomized networks have the same distributions). The distribution tails in the complete networks can be approximated by power laws. Distribution of node degrees in the MSTs can also be described by power laws in all the cases. From this perspective, USFFA and EAT are quite similar as regards the shape and power-law exponents of these distributions.

Selected global characteristics for the USFFA and EAT networks in their original and randomized versions are collected in Tab.~\ref{tab::associative_networks_characteristics}. While some results match the expectations (e.g., the global clustering coefficient $C_{\rm u}$ close to 0 in the randomized networks), the other ones provide more genuine information about the networks. The average shortest path length $\ell_{\rm u}$ in USFFA and EAT that is only slightly higher than in their randomized versions indicates that at least some ``shortcut'' edges connect the otherwise distant parts of the networks. They are responsible for keeping $\ell_{\rm u}$  relatively low, similar to the one observed in random networks. The relatively high values of modularity $Q_{\rm u}$ of the original networks (0.44 and 0.43) as compared to the randomized networks reflect the fact that words can be clustered in such a way that the associations among them are denser inside the clusters than outside of them. The presence of such clusters can also be related to the nonzero values of the clustering coefficient $C_{\rm u}$ in both USFFA and EAT. The negative values of assortativity coefficients $r_{\rm u}$ and $\rho_{\rm u}$ express a preference of edges to connect high-degree nodes with low-degree nodes. This is a common situation in networks that have many low-degree nodes and a certain number of high-degree hubs. However, in the USFFA and EAT networks, such an effect cannot be attributed solely to the degree distributions since both $r_{\rm u}$ and $\rho_{\rm u}$ are close to 0 in the randomized networks.

\begin{table}[t]
\centering
\begin{minipage}{\figurecustomwidth}
\centering
\caption{Selected global characteristics of the word-association networks constructed from the original and randomized USFFA and EAT data sets: the global clustering coefficient $C_{\rm u}$, the average shortest path length $\ell_{\rm u}$, modularity $Q_{\rm u}$, assortativity coefficient $r_{\rm u}$, and rank assortativity coefficient $\rho_{\rm u}$.}
{
\setlength{\tabcolsep}{8pt}
\renewcommand{\arraystretch}{1.6}
\setlength{\dashlinedash}{1pt}
\setlength{\dashlinegap}{1.5pt}
\setlength{\arrayrulewidth}{1pt}

\begin{tabular}{rrrrrr}
network & $C_{\rm u}$ & $\ell_{\rm u}$ & $Q_{\rm u}$ & $r_{\rm u}$ & $\rho_{\rm u}$ \\ 
\hline
\hline
EAT & 0.10 & 4.06 & 0.44 & -0.09 & -0.07 \\ 
EAT(rand) & 0.01 & 3.83 & 0.24 & -0.02 & -0.01 \\ 
USFFA & 0.12 & 3.95 & 0.43 & -0.08 & -0.07 \\ 
USFFA(rand) & 0.01 & 3.77 & 0.23 & -0.01 & -0.01 \\ 
\hline
\end{tabular}

}
\vspace{0.5\baselineskip}
\label{tab::associative_networks_characteristics}
\end{minipage}
\end{table}

\begin{figure}[!htb]
\centering
\begin{minipage}{\figurecustomwidth}
\centering
\subfloat[MST of the USFFA network]{\includegraphics[width=0.49\textwidth]{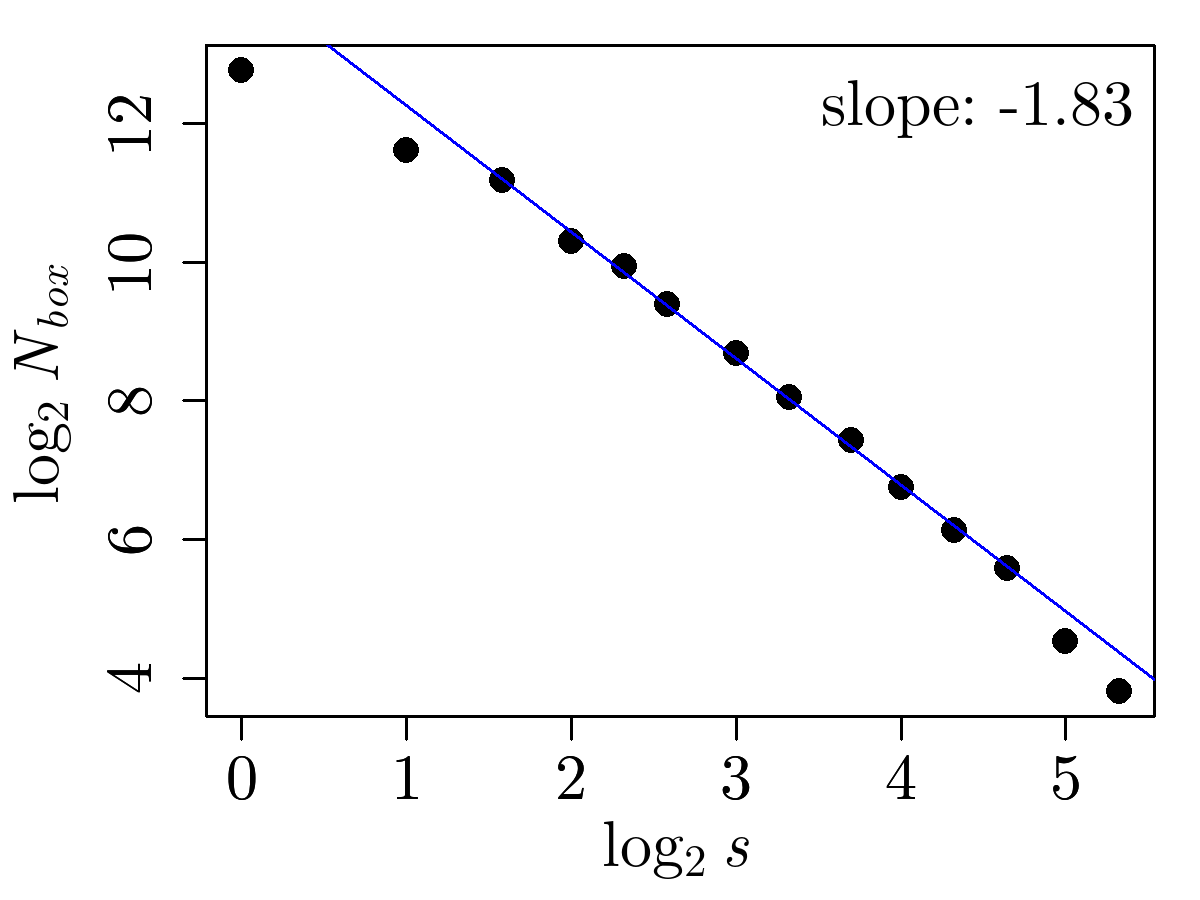}}
\hfill
\subfloat[MST of the EAT network]{\includegraphics[width=0.49\textwidth]{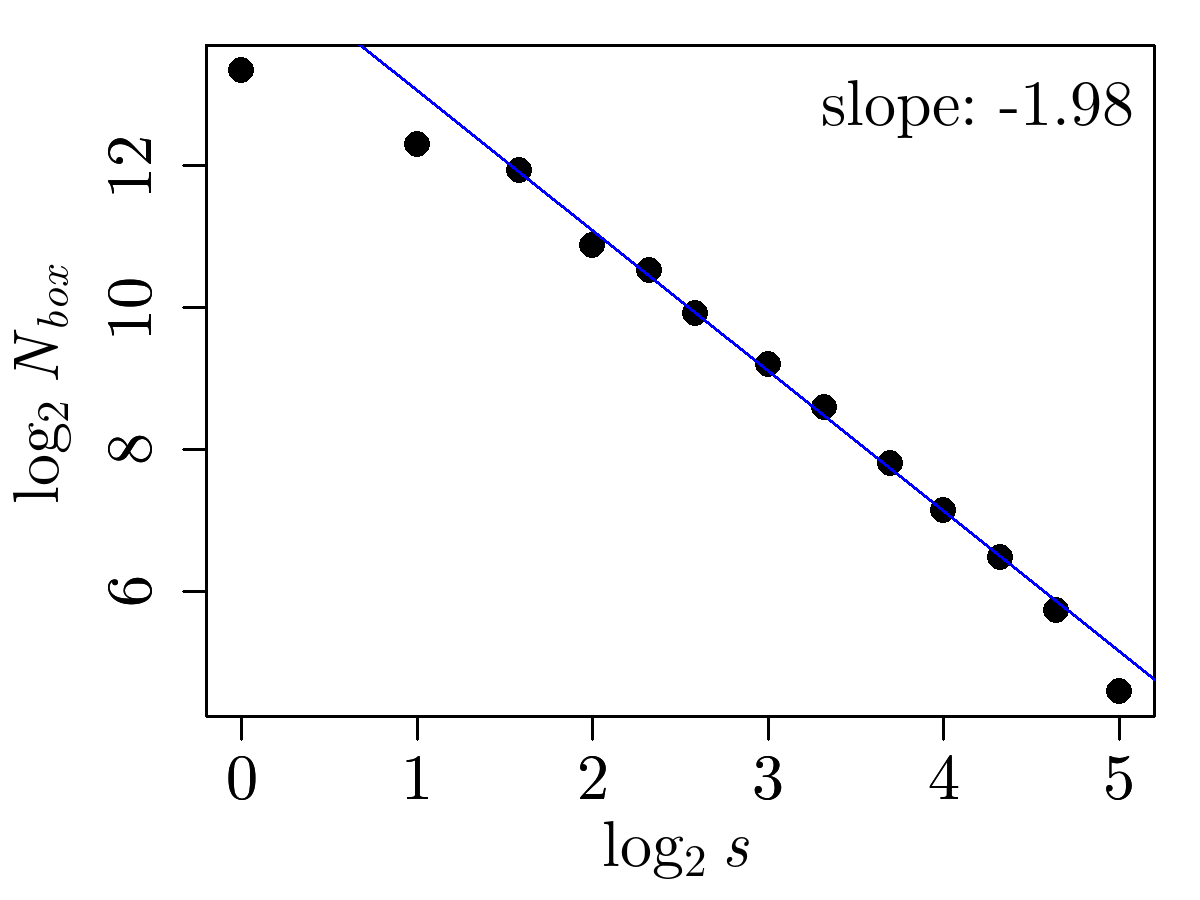}}
\hfill
\vspace{0.5cm}
\subfloat[MST of the randomized USFFA network]{\includegraphics[width=0.49\textwidth]{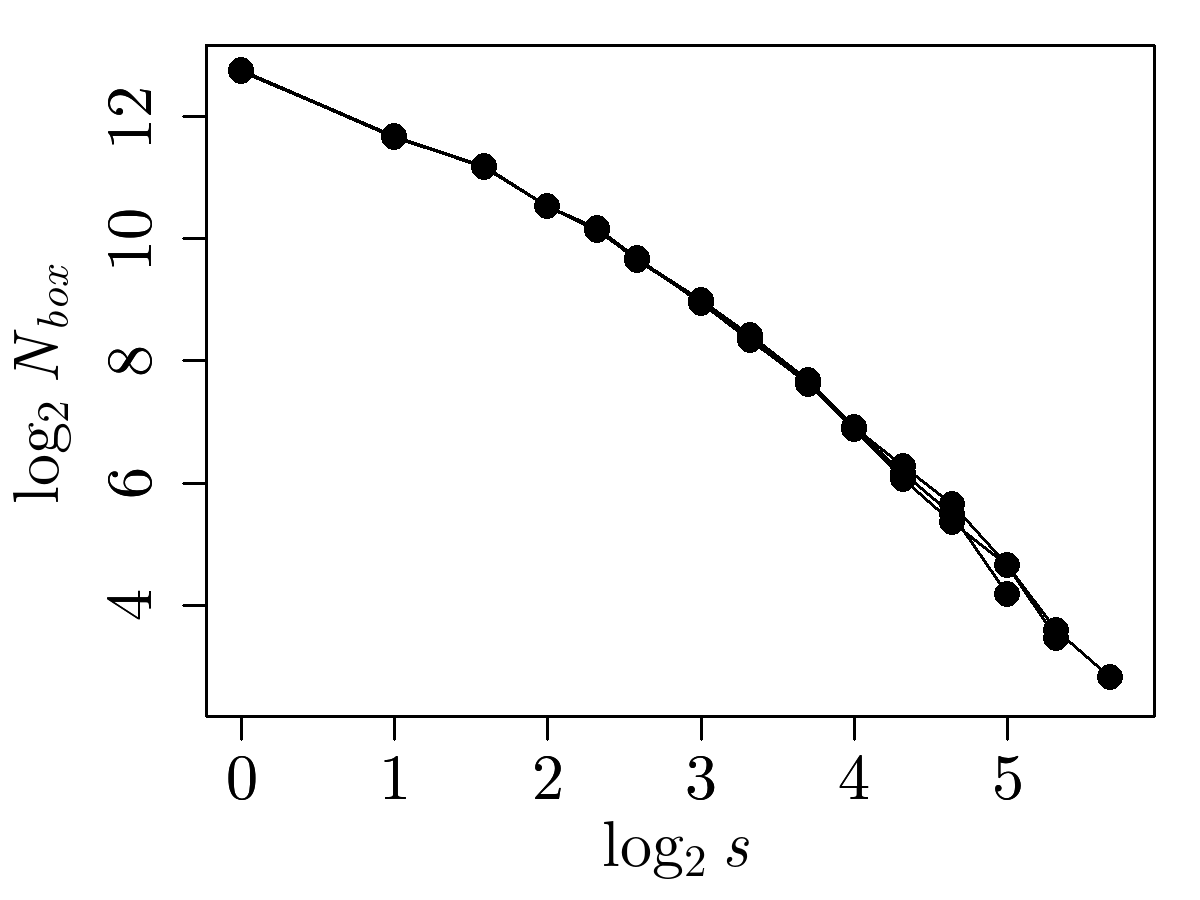}}
\hfill
\subfloat[MST of the randomized EAT network]{\includegraphics[width=0.49\textwidth]{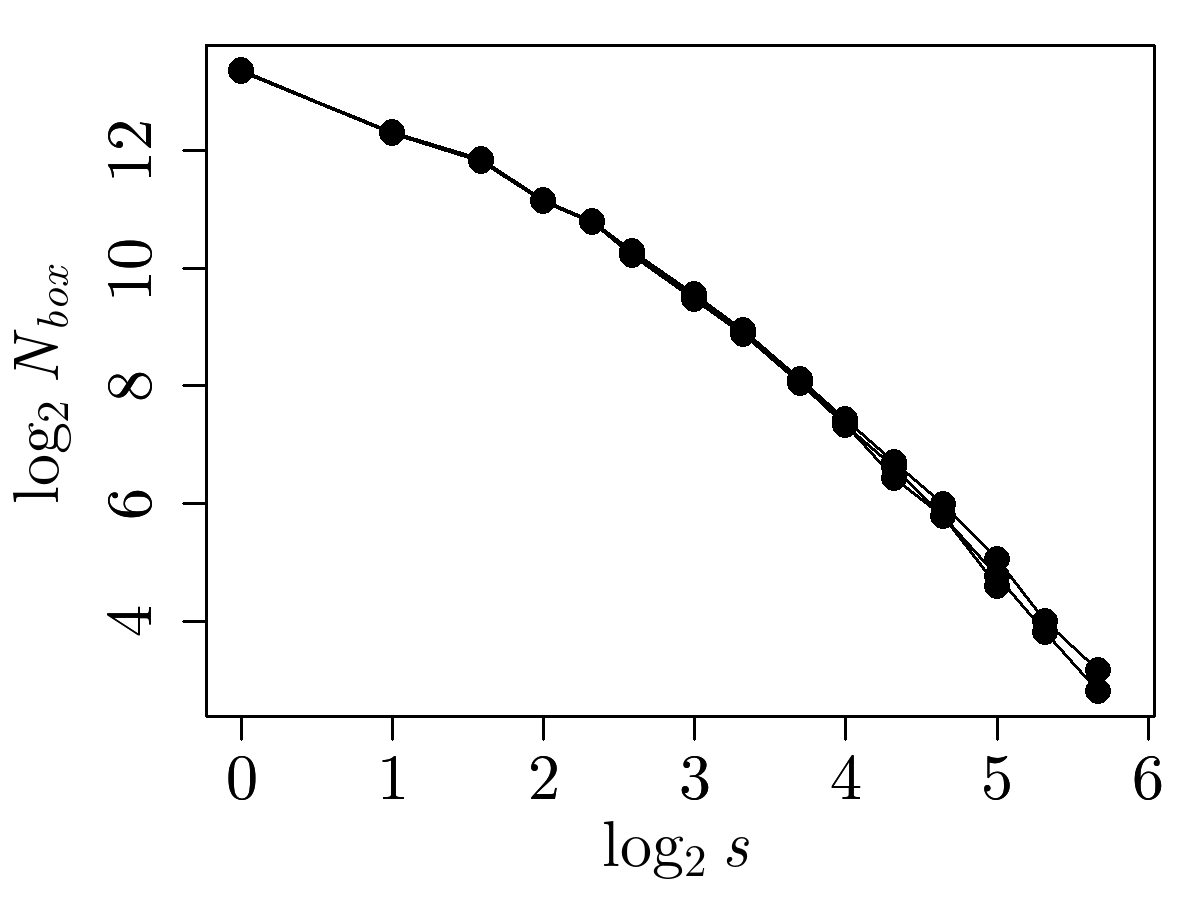}}
\hfill
\caption{Log-log plots of the number of boxes of size $s$ that are needed to cover (by using greedy coloring) the minimum spanning trees constructed from the USFFA (a) and EAT (b) networks and from their randomized versions (c) and (d). A scaling range of $N_{\rm box}(s)$ that indicates fractality can be identified in the MSTs for the original networks only. The slope indices of blue lines are given in the top-right corners. No scaling region can be observed in the randomized networks. The plots for the randomized networks are constructed from their multiple realizations.}
\label{fig::word_association_networks_box_counting}
\end{minipage}
\end{figure}

An interesting characteristic shared by USFFA and EAT networks is that their MSTs are organized in a hierarchical fashion and exhibit statistical self-similarity. In that regard, both the USFFA and EAT networks differ from their randomized counterparts: the MSTs for the randomized networks do not have fractal structure. The box-counting algorithm employed to the MSTs created from the USFFA and EAT networks gives the results shown in Fig.~\ref{fig::word_association_networks_box_counting}. The number of boxes $N_{\rm box}$ of size $s$ needed to cover each original network depends on box size like a power law: $N_{\rm box}(s)\sim s^{d_{\rm C}}$ and can be interpreted as the presence of fractality. Although this analysis might suffer from a relatively small size of the studied networks, it allows one to detect qualitative differences between the structure of the MSTs of the word-association networks and the structure of their randomizations. Such differences suggest that the ``MST-skeletons'' of the USFFA and EAT networks representing the strongest associations between words are organized into a genuine self-similar structure originating from the inner correlations, which would not be observed if associations were connecting words randomly. The estimated fractal dimensions of both networks have comparable values, $d_{\rm C}=1.83$ for the USFFA MST and $d_{\rm C}=1.98$ for the EAT MST. A~question arises if this result constitutes a general property of the word-association networks or they are characteristic for the considered data sets. However, a larger multilanguage set of data would be required to resolve this issue and make an association between fractality of word associations and human mind.

Steyvers and Tannenbaum studied empirical word-association networks created from different data sets including free association norms, WordNet, and Roget's Thesaurus~\cite{Steyvers2005}. The first data set was created in an association experiment of the type described above, while the third data set consisted of a bipartite (word vs. category) graph transformed into a simple network with edges connecting words representing the same semantic category. The analysis was focused on five network structural measures: sparsity (the number of existing edges with respect to the total number of possible edges), connectedness, the average short path lengths, clustering coefficient, and node degree distributions calculated for the actual and randomized networks. The results indicated that, although the word-adjacency networks were sparse with an average word being connected to less than 0.5\% of the words in a set, their largest components consisted of all the nodes if the networks were undirected and 96\% of nodes if the networks were directed. The average shortest path lengths for these networks were close to $\ln N$, while the global clustering coefficients were much larger than the random graph counterparts; both these results together situated the analyzed networks in the small-world network category. The node degree distributions for the undirected networks exhibited power-law tails with the power exponent $\gamma$ slightly larger than 3. For the directed version of the free association norms $\gamma$ the power exponent dropped to below 2 for the in-degree distribution, while for the out-degree distribution no power-law regime was found.

The same authors proposed a model for simulating network growth that was able to mimic the observed empirical characteristics. The model was based on the idea of preferential attachment but substantially modified with respect to the BA model. A new node is added to the network by differentiation, i.e., by inheriting a subset of the connections of an already existing node that was chosen with probability proportional to its degree. In addition, a parameter called \textit{utility}, which varied over time, was associated with each node and served as probability that this particular node would be selected for connection by the new node. In a directed version of the model, each new edge was attributed with its direction by a built-in mechanism. The preferential attachment allowed the model network to be organized into a scale-free form and to provide new nodes with a relatively significant initial degree. Owing to the differentiation mechanism, these new nodes were joining neighbourhoods of old hubs, which led to clustering and large global clustering coefficients. This model was thus able to reproduce the small-world property of the studied word-association networks. The authors of~\cite{Steyvers2005} argued that their model, especially its differentiation feature, was a plausible candidate for simulating how new meanings occur in a natural language. They also suggested that similar mechanism can be responsible for language development by an individual. They also claimed that an observed correlation between the age of word acquiring and the number of edges this word was connected by as well as a negative correlation between latency of word access and its significance in the association network could also be inferred from their model~\cite{Steyvers2005}.


\subsection{Other types of linguistic networks}

Networks built on different linguistic structures can also be applied to extracting meaningful information from texts. For example, to prepare a summary of a given text it requires to identify essential topics and condensate the related information that originally may be spread across the entire text. This may be done by a complex semantic analysis and employing of a language generator or, if simplicity is preferred over linguistic cohesion, it may suffice to identify the most informative sentences and use them unchanged in an extract. The latter approach was applied in~\cite{AntiqueiraL-2009a} where it was assumed that the principal carriers of information in a text are lemmatized nouns and these nouns were used to identify meaningful sentences. Networks were constructed from sentences considered to be network nodes and overlapping nouns considered to define edges with weights equal to the number of shared nouns. The authors proposed a method called ``complex network-based summarization'' that was based on ranking sentences according to a specific property characterizing individual nodes: degree, strength, the average shortest path length (i.e., a node's accessibility), and a modified matching index. Extraction of sentences was also based on the structures like $d$-rings, $k$-cores, communities, etc. A small corpus of 100 newspaper articles together with their human-written abstracts and machine-produced reference extracts based on these abstracts was considered. Results of the analysis that used the listed measures were then compared with the reference extract for each article by different correlation measures (see the source paper for details~\cite{AntiqueiraL-2009a}). Provided that the network-based extracts were preferred if they were strongly correlated with the reference extracts, it occurred that the results were the most satisfying for the extracting strategies based on node degrees and strengths, the average shortest path lengths, $d$-rings, and $k$-cores. However, a combined strategy that was choosing sentences that were jointly selected by other methods occurred to be the best one.

As word co-occurrence or sentence properties can be viewed as a microscale structure of texts, paragraph or section/chapter-level structure can be considered as a mesoscale structure. Such a structure was a subject of analysis in~\cite{eSouzaBC-2023a}. A set of 300 books of different genres written in English was considered there that were long enough to secure statistical validity of the analysis. Each paragraph was preprocessed in order to extract the most meaningful lemmatized tokens like subjects, predicates, and direct objects, which carry the information relevant for that particular piece of text. Paragraphs were then mapped into network nodes and token-similarity between them were mapped into edge weights. Accessibility of nodes and backbone symmetry of local structure in the so-constructed \textit{recurrence networks} for all the texts in the original and randomized (paragraph-level shuffled) form were investigated. Results showed that by using these metrics, it was relatively easy to discriminate between original and randomized texts as well as between different literary genres. It was also shown that such an approach offers better discriminatory power than methods based on word-adjacency networks and machine-learning-based doc2vec analysis~\cite{LeQ-2014a}.


\section{Summary}
\label{sect::summary and outlook}

Natural language is an extremely advanced and spectacularly efficient achievement of nature and it is for this reason alone that one can expect it to serve as a paradigm of a complex system. Furthermore, because the laws of nature can be so effectively formalized in the mathematical terms, one can also expect that certain characteristics of natural language can be described within the framework of the same mathematics. Indeed, as the present review documents various aspects of linguistic patterns and organization can be grasped using mathematical tools -- ranging from basic methods of statistics and time series analysis to fractal geometry and network theory -- designed to study systems exhibiting complexity. With the use of such tools a number of natural language characteristics can be described in a quantitative way.

The latest analysis of word frequency distributions in literary texts confirms and even strengthens the validity of a well-known statistical law of natural language -- the Zipf's law. At the same time studying word frequencies in more detail -- for example considering different parts of speech separately -- reveals the differences between these types in terms of their statistical properties. This fact indicates quite an intricate words' arrangement accompanying the linguistic Zipf's law. A particularly interesting result is obtained for word rank-frequency distributions approximated by Zipf-Mandelbrot law when punctuation marks are included into the analysis. It turns out that treating punctuation marks in the same way as words decreases the value of the constant $c$ responsible for the flattening of the rank-frequency distribution in the Zipf-Mandelbrot law. In other words, with the punctuation marks included, word frequency distribution is better approximated by a power law. This effect is present in all the Western languages studied. Its relative strength depends somewhat, however, on particular language. These facts provide an argument for including punctuation marks into statistical analysis of written language, especially from the standpoint of models attempting to explain power laws in word frequencies.

A valuable insight into the organization of natural language is obtained with the use of tools designed for time series analysis. Representing quantities like sentence lengths in the form of time series allows to reveal several signatures of complexity. They include the long-range correlations and fractal or even multifractal patterns. Multifractality is evidenced by a wide singularity spectrum often observed in texts using a narrative technique known as the stream of consciousness. An interesting perspective on certain aspects of the organization of written language is provided by analyzing the partition of a text determined by consecutive punctuation marks. A reason for which such a partition can be considered meaningful is the general purpose of punctuation in written language -- that is, splitting text into parts determined by grammatical or logical consistency. From a text segmented into pieces separated by punctuation marks one can construct a time series consisting of the lengths of consecutive intervals, measured by the number of words between consecutive punctuation marks. Such series also have properties indicating the presence of complex patterns of organization, like long-range correlations or mutlifractality. Typically, the strength of both of the mentioned effects for punctuation waiting times is weaker than in case of time series representing sentence lengths. A text having a wide singularity spectrum of sentence lengths usually has a singularity spectrum of punctuation waiting times of significantly smaller width.

The analysis of time series representing sentence lengths and punctuation waiting times emphasizes the significance of punctuation in written language. Since both types of series are determined by the arrangement of punctuation marks (either all of them or the ones being of a specific type), it can be stated that punctuation in general is responsible for organizing written language in a specific way which results in the presence of complex patterns. Pursuing further in that direction and investigating the probability distributions of the intervals between all the consecutive punctuation marks in texts points to the discrete Weibull distribution as an appropriate one. This allows to view the arrangement of punctuation marks in written language in terms of a process whose statistical properties can be quantitatively expressed by appropriate hazard functions. It is also interesting to notice in this connection that such a kind of distributions manifests itself in the survival analyses of expected duration of time until a particular event -- such as failure in mechanical system or death in biological organism -- occurs. This fact opens a broader perspective for understanding physiological aspects of the mechanism of natural language processing by humans. 

Studying linguistic networks designed to represent certain aspects of language structure also leads to a multitude of inspiring results. Word-adjacency networks -- the ones representing the co-occurrence of words in texts -- are a tool which allows to investigate a number of statistical properties of a given language sample and to express them in terms of quantities used to describe complex networks. Some basic characteristics of a text, which are usually studied with the use of a representation simpler than word-adjacency networks are incorporated into word-adjacency networks and can be easily retrieved. Word-adjacency networks allow to observe some effects which seem to be universal across languages, like the specific behaviour of the most frequent words and the words with moderate frequencies in terms of network's local characteristics, namely local clustering coefficients and average shortest path lengths. At the same time, word-adjacency networks are able to grasp certain differences between texts. The global characteristics like the clustering coefficient, assortativity and modularity, in several specific variants of networks constructed from texts in different languages have slightly different ranges of variability. This allows to observe that in the space of the mentioned characteristics, different languages tend to reside in somewhat different regions. 

The structure of a word-adjacency network can be characteristic not only to a particular language, but also to a specific style of writing. This gives an opportunity to use word-adjacency networks in stylometry. The effectiveness of stylometric analysis utilizing word-adjacency networks is demonstrated on an example of authorship attribution task. The analysis shows that networks constructed from texts of different authors differ in some of their structural properties. While in terms of networks' global properties this effect can be identified only to some degree, the local characteristics of selected words in word-adjacency networks allow to distinguish between texts of different authors with much better accuracy. Since local characteristics of a word-adjacency network describe certain statistical properties of word usage, and the considered words are the ones with the highest frequencies, it can be stated that structural differences between networks representing texts of different authors are a result of authors' individual patterns of using the most frequent words. A characteristic that seems to be particularly useful in grasping the information needed to recognize text authorship is the weighted variant of the clustering coefficient. Authorship attribution task performed in the sample set of books with the use of a general-purpose machine learning method -- decision tree ensemble -- is able to achieve accuracy of about 80-90\% when the clustering coefficients of only about 10-15 most frequent words and punctuation marks are taken into account. It is important to note that network-based approach to text classification can be combined with other methods to improve the quality of the results.

Applying other network representations of linguistic constructs, namely word-association networks designed to reveal associations between words in human mind, identifies them as also having complex organization. A~number of characteristics describing networks' basic statistical properties indicate that the two studied networks are similar to each other, although the data used to construct them comes from different, independent experiments. An interesting fact about these networks is that their minimum spanning trees, which might be considered sub-networks consisting of only the strongest associations, have a statistically self-similar, fractal organization. The estimated fractal dimensions of both minimum spanning trees have similar values. This raises a question whether such an observation is valid only for the discussed networks, or whether it represents a more general property of word-association networks. Investigating the generality of these results, as well as establishing their relationship with other findings regarding the organization of language in human mind and its interactions with the environment, in the light of the present review constitutes a desirable direction for future research within the science of complexity. 

\renewcommand\appendixname{}
\begin{appendices}

\renewcommand{\thesection}{}
\titleformat{\section}[block]{\normalsize\bfseries}{}{}{}

\phantomsection
\addcontentsline{toc}{section}{Appendix. Theoretical concepts related to natural language}
\section*{Appendix. Theoretical concepts related to natural language}

\renewcommand{\thesubsection}{A.\arabic{subsection}}

\subsection{Formal grammars}
\label{appendix::formal.grammars}

A formal grammar is one of the ways of specifying a formal language, providing a set of rules allowing to transform strings of symbols into other strings of symbols. Defining a formal grammar can be done in the following way. Let $\Sigma$ be a finite set, whose elements are called terminal symbols. Let $N$ be a finite set, whose elements are called nonterminal symbols, and which is disjoint with $\Sigma$. The distinction between terminal and nonterminal symbols is due to their role in strings -- terminal symbols are the symbols that are present in the ``final form'' of a string, while nonterminal symbols are the ones that occur at intermediate stages of string construction. Let $S$ be a distinguished symbol, belonging to $N$ and called the start symbol. Finally, let $P$ be a finite set of production rules. Each element of $P$ is a rule of the form: $\alpha \rightarrow \beta$, where $\alpha$ and $\beta$ are strings consisting of terminal and nonterminal symbols. $\beta$ is an arbitrary string (it can also be an empty string $\varepsilon$), while $\alpha$ must contain at least one nonterminal symbol. A formal grammar $G$ can be defined as a 4-tuple:
\begin{equation}
G = \left(\Sigma, N, S, P \right).
\end{equation}
A formal grammar can be interpreted as a string rewriting system, transforming strings into other strings. Starting from a string consisting solely of the start symbol, one can rewrite strings in such a way that each rewriting introduces modifications given by a selected production rule (the notation  $\alpha \rightarrow \beta$ represents replacing the substring $\alpha$ with the substring $\beta$). The set of all strings which contain only terminal symbols and can be constructed from the start symbol by applying some finite sequence of production rules is a formal language. Such a language is called a language generated by a given grammar.

As an example, let the following grammar be considered: $G  \!= \! \left(\Sigma, N, S, P \right)$, where
\begin{equation}
\begin{aligned}
& \Sigma = \left\lbrace a, b \right\rbrace, \\
& N = \left\lbrace A, B \right\rbrace, \\
& S = A, \\
& P = \left\lbrace \; A \rightarrow abBaa, \;\; B \rightarrow bBaa, \;\; B \rightarrow a \; \right\rbrace.
\end{aligned}
\label{eq_formal_grammar_example}
\end{equation}
Generating strings by this grammar is performed as follows. At the beginning, the string consists of one symbol, the start symbol $A$. The only rule that can be applied at this stage is the rule $A \rightarrow abBaa$, which rewrites $A$ into $abBaa$; therefore, the string becomes $abBaa$. Now, any of the rules $B \rightarrow bBaa$ and $B \rightarrow a$ can be applied. If the first one is chosen, then $B$ is replaced with $bBaa$ and the string becomes $abbBaaaa$. This new string also allows to apply both of the rules $B \rightarrow bBaa$ and $B \rightarrow a$. Using the first one again (one or more times) expands the string and allows for further expansion. Using the rule $B \rightarrow a$ at any stage removes $B$ from the string and inserts $a$. When this happens, no more operations on the string are possible. At this stage the string consists only of terminal symbols, and it can be considered a string belonging to a language generated by the grammar $G$. All strings in this language have a single $a$ as their first symbol, then $b$ is repeated $n$ times ($n=1,2,3,...$), and then $a$ is repeated $2n\!+\!1$ times. Therefore, all such strings are of the form $ab^na^{2n+1}$, where $n=1,2,3,...$.

Many important properties of formal grammars depend on the constraints imposed on their production rules. Such constraints have an impact on grammar generality. Formal grammars can be divided into types pertaining to that generality. A widely known classification of grammars is the Chomsky hierarchy~\cite{Chomsky1959,Levelt2008}, which distinguishes 4 types of grammars, labeled by numbers 0, 1, 2, and~3. Let $\alpha$, $\beta$ be arbitrary strings (possibly empty) of terminal and nonterminal symbols and let $\gamma$ be a nonempty string of terminal and nonterminal symbols. Let $A$ and $B$ be nonterminal symbols, and let $a$ denote a terminal symbol. The most general form of a production rule is:
\begin{equation}
\gamma \rightarrow \beta.
\end{equation}
A grammar which does not have any additional constraints imposed on its production rules, is a type-0 grammar (also called an unrestricted grammar). All languages that can be generated by such a grammar are called recursively enumerable languages. A grammar whose all production rules are of the form:
\begin{equation}
\alpha A \beta \rightarrow \alpha \gamma \beta
\end{equation}
is a type-1 grammar, also known under the name of a context-sensitive grammar. To make it possible for a context-sensitive grammar to generate empty strings, one additional rule is allowed: $S \rightarrow \varepsilon$, where $S$ is the start symbol, and $\varepsilon$ denotes an empty string. Each production rule of a context-sensitive grammar can be interpreted as a procedure transforming a single nonterminal symbol $A$ into a nonempty string, with a condition that such transformation may be dependent on the ``neighbourhood'' of $A$ (the context). Languages generated by a context-sensitive grammar are called context-sensitive languages.

A grammar which only has rules of the form:
\begin{equation}
A \rightarrow \alpha
\end{equation}
is called a type-2 grammar, or a context-free grammar. Languages generated by this type of grammar are context-free languages. Left-hand side of any production rule of a context-free grammar is a single nonterminal symbol. The name ``context-free'' reflects the fact that grammar's rules can be applied regardless of the context of a nonterminal symbol. Context-free grammars are a class of grammars particularly important in modeling and studying language. Their complexity is restricted enough to allow the construction of efficient parsing algorithms -- algorithms determining whether a given string of symbols belongs to the language generated by a given grammar, and, if so, finding the sequence of rules leading to the generation of this string. Yet they are general enough to be useful in studying natural language. An important example of their use is syntax analysis. Context-free grammars are also often the backbone of programming languages.

Type-3 grammars, known as regular grammars, can be divided into two groups: left-regular grammars and right-regular grammars. A left-regular grammar is a grammar having only the rules of one of the following types:
\begin{equation}
\begin{aligned}
& A \rightarrow a \\
& A \rightarrow Ba \\
& A \rightarrow \varepsilon
\end{aligned}
\end{equation}
where $\varepsilon$ denotes the empty string.
A right-regular grammar has only the rules of one of the following forms:
\begin{equation}
\begin{aligned}
& A \rightarrow a \\
& A \rightarrow aB \\
& A \rightarrow \varepsilon.
\end{aligned}
\end{equation}
Regular grammars generate regular languages. They are related to regular expressions -- special strings that represent certain patterns. A regular expression specifies a set of strings that match the given pattern. Each language that can be generated by a regular grammar can also be specified by a regular expression and vice versa. Regular expressions are a concise way of describing a set of strings sharing some properties; they have found application in various text  processing tools. Contemporary implementations of regular expressions (present, for example, in many programming languages) often extend their basic functionality and make them capable of specifying also the languages other than the ones generated by regular~grammars.

Chomsky hierarchy puts the types of grammars into order related to their generality -- consecutive types are more restricted than the  previous ones. If $L_i$ denotes the set of languages that can be generated by type-$i$ grammars, then:
\begin{equation}
L_3 \subset L_2 \subset L_1 \subset L_0.
\end{equation}
In other words, the set of context-sensitive languages is contained in the set of recursively enumerable languages, the set of context-free languages is contained in the set of context-sensitive languages, and the set of regular languages is contained in the set of context-free languages. All these inclusions are strict inclusions -- each $L_i$ contains languages that are not present in $L_{i+1}$.

\begin{figure}
\centering
\begin{minipage}{\figurecustomwidth}
\centering
\includegraphics[width=0.9\textwidth]{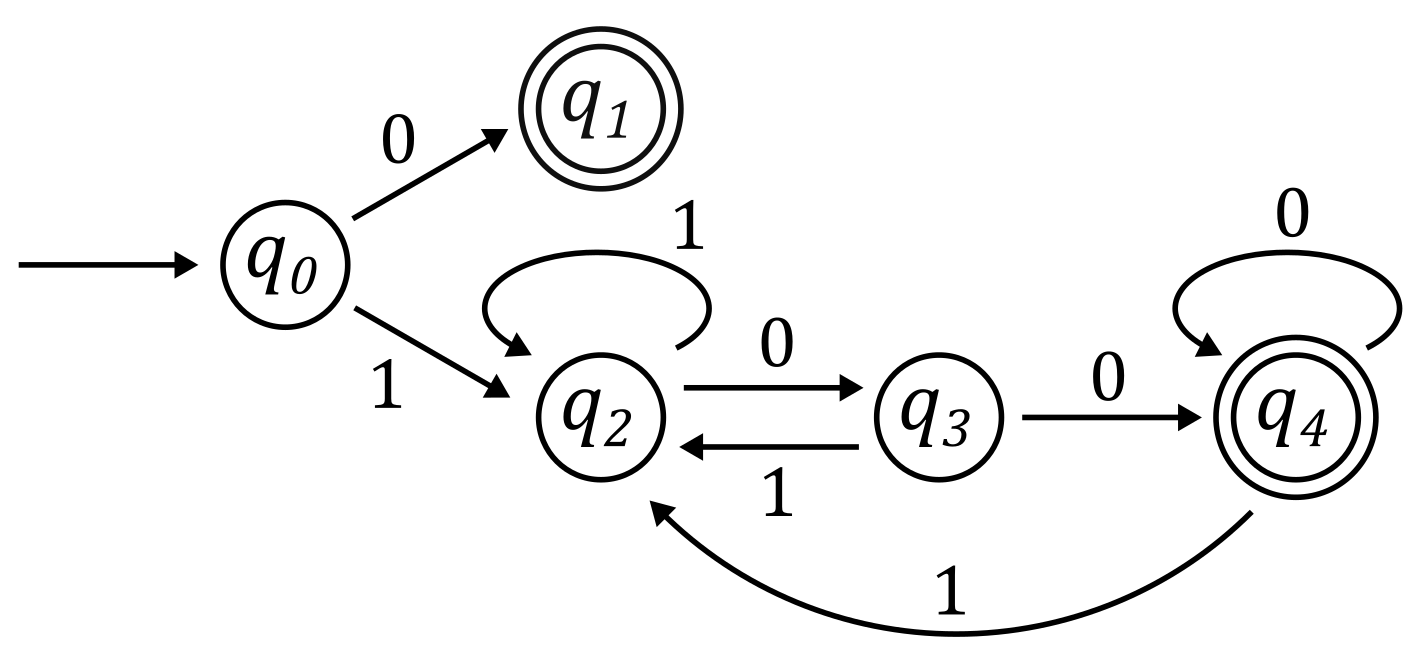}
\caption{An example of a state diagram of a (deterministic) finite-state automaton. The automaton reads strings consisting of zeros and ones and recognizes (accepts) a string if and only if the string is a binary representation of a natural number divisible by 4 (without any padding with leading zeros). States of the automaton are represented by circles (accepting states are marked by a double circle) and arrows correspond to transitions between states. The automaton switches its state to the one pointed by an arrow when the symbol read from the input matches the symbol labeling the arrow. The automaton utilizes the fact that numbers divisible by 4 have at least two zeros at the end of their binary representation. The accepting states are: $q_1$, which deals with the case when the whole input string is just one zero, and $q_4$, in which the automaton stays if it reads two or more zeros in a row. The string is recognized if its reading is finished and the automaton is in an accepting state. Any other situation (including, for example, the inability to move to another state from the state $q_0$ when reading from the input is not finished) leads to the rejection of the string. The~regular expression corresponding to the strings recognized by the presented automaton is ``0|1+[01]*00''.}
\label{fig_automaton_example}
\vspace{0.5cm}
\end{minipage}
\end{figure}

There is a close relationship between formal languages and automata theory.  For a given formal language, one can define an automaton (an abstract machine) capable of determining whether a given string belongs to that language. To do that, such an automaton (called an acceptor or a recognizer) starts from a starting state, reads the input string -- symbol after symbol -- in consecutive steps, and changes its internal state in each step. The combination of automaton's current state and the symbol being read determines the transition to the next state (if the automaton is deterministic) or the set of possible transitions (if the automaton is nondeterministic). If the input string is a sequence of symbols for which there exists a sequence of state transitions leading the automaton from the starting state to a predefined final state, then the automaton accepts the string. Otherwise, the string is rejected. The complexity of an acceptor depends on the type of language it is designed to recognize. Regular languages can be recognized by finite-state automata (an example of a finite-state automaton is presented in Fig.~\ref{fig_automaton_example}). Context-free languages can be recognized by nondeterministic pushdown automata (which can be thought of as nondeterministic finite-state automata with a stack capable of storing read symbols). Context-sensitive languages are recognized by linear bounded automata. The automaton needed to recognize a recursively enumerable language in a general case is a Turing machine. The more general the grammar, the more powerful automaton is required to recognize languages generated by that grammar.

\begin{figure}
\centering
\begin{minipage}{\figurecustomwidth}
\centering
\subfloat[]{\includegraphics[width=0.4\textwidth]{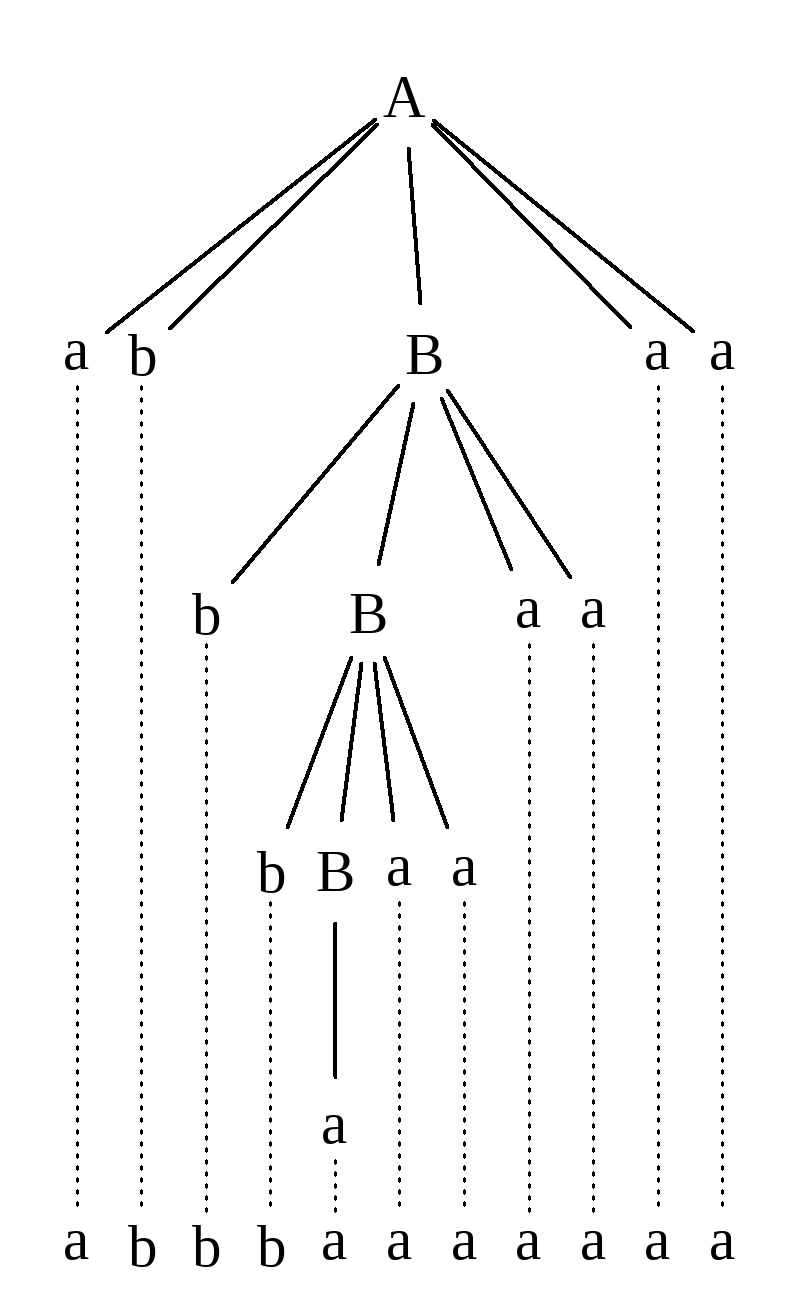}}
\hfill
\subfloat[]
{\includegraphics[width=0.55\textwidth]{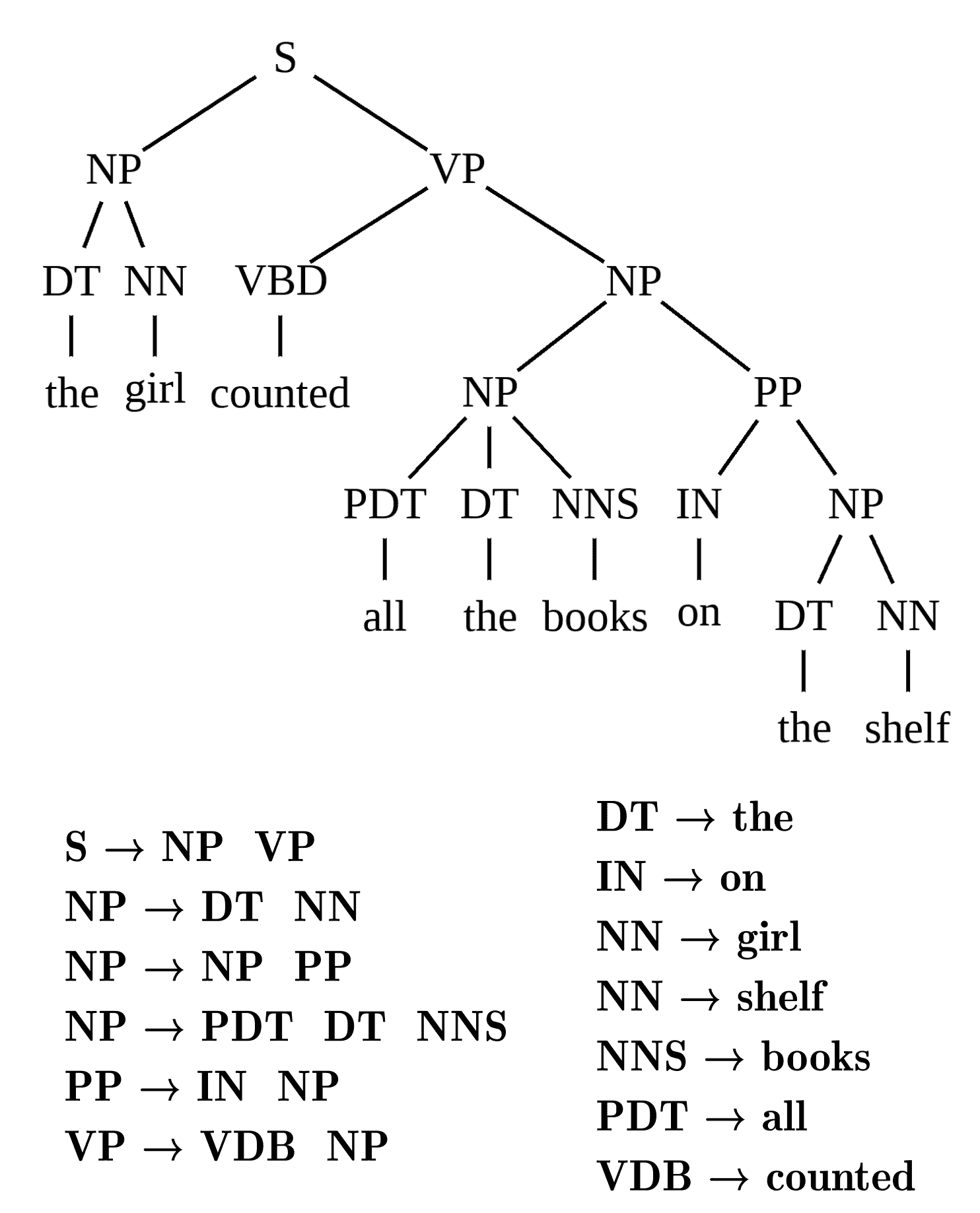}}
\caption{Examples of parse trees. The tree in (a) is the parse tree of the string $abbbaaaaaaa$, using the grammar defined in Eq.~(\ref{eq_formal_grammar_example}). The tree in (b) is the parse tree of the sentence \textit{``The girl counted all the books on the shelf.''}, using a context-free grammar modeling the syntax of English. The rules used to construct the tree are listed below the tree, while the whole grammar has many more rules. Abbreviations of syntactic categories are as follows: S - start symbol, NP - noun phrase, VP - verb phrase, DT - determiner, NN - noun, VBD - verb in past tense, PP - prepositional phrase, PDT - pre-determiner, NNS - noun in plural form, IN -preposition or subordinating conjunction.}
\label{fig_parse_trees}
\end{minipage}
\end{figure}

Formal grammars, automata, and related concepts are widely applied in both theoretical and practical scientific approach to natural language. An important example is syntactic analysis with the use of the so-called constituency grammars. Constituency grammars are grammars designed to express the syntactic structure by relations between constituents. A constituent is a word or a group of words that can be treated as a single unit in a larger grammatical construction. Constituents of the same type appear in similar syntactic environments, and can be treated as interchangeable -- to some extent -- form purely syntactic point of view. For example, in the sentence \textit{``The glass that was on the table fell on the floor''}, all of the phrases: \textit{the glass}, \textit{the table}, \textit{the floor}, as well as \textit{the glass that was on the table} can be treated as noun phrases -- constituents which perform the grammatical function of a noun. Swapping these phrases between one another could lead to a sentence which is semantically nonsensical, but syntactically correct. Constituency structure is hierarchical -- constituents might consist of other constituents, all the way down to individual words. Assigning a constituency structure to a given sentence is done by constructing a parse tree (also named a derivation tree) -- a graph whose structure corresponds to the relationships between constituents. An illustration of the concept of a parse tree is presented in Fig.~\ref{fig_parse_trees}.

Constituency grammars are used to describe the phrase structure of sentences, therefore they are also called phrase structure grammars. Another related name is generative grammar. Generative grammar is a broad term, which serves as a common name to multiple theories. What these theories have in common is the usage of formal grammars to model the grammar of natural language~\cite{Jurafsky2009}. It is worth noting that phrase structure grammars are only one of possible methods of the analysis of syntax. A line of inquiry often presented as opposing to constituency-based approach is concentrated on the so-called dependency grammars, which instead of relying on constituency relations, employ dependency relations -- binary relations between individual words, not groups of words. Each kind of approach has its advantages and disadvantages in particular situations, and both are important tools of syntactic analysis.

\subsection{Lindenmayer systems}
\label{appendix::lindenmayer.systems}

Lindenmayer systems, also known as L-systems, are string rewriting systems used in the analysis of growth processes involving branching. An L-system $G$ can be defined~as~\cite{Prusinkiewicz1990}:
\begin{equation}
G = \left(V, \omega, P \right),
\end{equation}
where $V$ is the set of symbols, $\omega$ is a nonempty string of symbols called the axiom, and $P$ is the set of production rules. The definition is very similar to the definition of a formal grammar, with the exception that the distinction between terminal and nonterminal symbols is not necessary and that the start symbol is replaced by the start string -- the axiom. L-systems are different from formal grammars in the way in which production rules are applied -- rules of formal grammars are applied sequentially (one rule at a time), while L-systems apply their rules in parallel -- at each iteration the rewriting is performed in all possible places in the string (in all places where a production rule can be applied). Like formal grammars, L-systems can be divided into types, according to the properties of their production rules. In that sense, an important class of L-systems are context-free L-systems, with all production rules of the form:
\begin{equation}
A \rightarrow \alpha,
\end{equation}
where $A$ is a single symbol from $V$ and $\alpha$ is a string of symbols from $V$. If an L-system is designed to model the growth or the development of a certain object or system, then the assumption that the L-system is context-free can be related to the assumption that individual parts of the modeled object develop independently, without interactions between each other. If for each symbol in $V$ there is exactly one rule which has that symbol on its left-hand side, then the system is deterministic. If any symbol appears as a left-hand side in more than one production rule (which means that it can be rewritten in multiple ways), then the system is called a stochastic L-system. In such a system, each time when multiple rules can be applied to a given symbol, the rule is chosen randomly from the set of possible rules; each rule in that set has some probability of being chosen.

How an L-system works can be illustrated by the following example. Let $G = \left(V, \omega, P \right),$ be an L-system where
\begin{equation}
\begin{aligned}
& V = \left\lbrace A, B \right\rbrace, \\
& \omega = A, \\
& P = \left\lbrace A \rightarrow ABA , \; B \rightarrow BBB \right\rbrace.
\end{aligned}
\label{eq_Lsystem_Cantor_set}
\end{equation}
String production in such an L-system proceeds as follows. At the beginning, the string contains one symbol, $A$. In the first iteration, the rule $A \rightarrow ABA$ produces the string $ABA$. In the second iteration, each $A$ in the string is replaced by $ABA$, and each $B$ (here only one) is replaced by $BBB$. Hence, the string $ABABBBABA$ is obtained. The process is then repeated in consecutive iterations, up to the point when a predefined number of iterations is reached.

\begin{figure}
\captionsetup[subfigure]{width=0.9\textwidth}
\centering
\begin{minipage}{\figurecustomwidth}
\centering
\subfloat[Generating an image of the Cantor set. Symbol set: $V = \left\lbrace A, B \right\rbrace$, axiom: $\omega = A$, production rule set: $P = \left\lbrace A \rightarrow ABA , \; B \rightarrow BBB \right\rbrace$ (this is the L-system defined in Eq.~(\ref{eq_Lsystem_Cantor_set})). Commands assigned to symbols: $A$ - move forward a fixed distance and draw a line, $B$ - move forward a fixed distance without drawing.]{\includegraphics[width=0.75\textwidth]{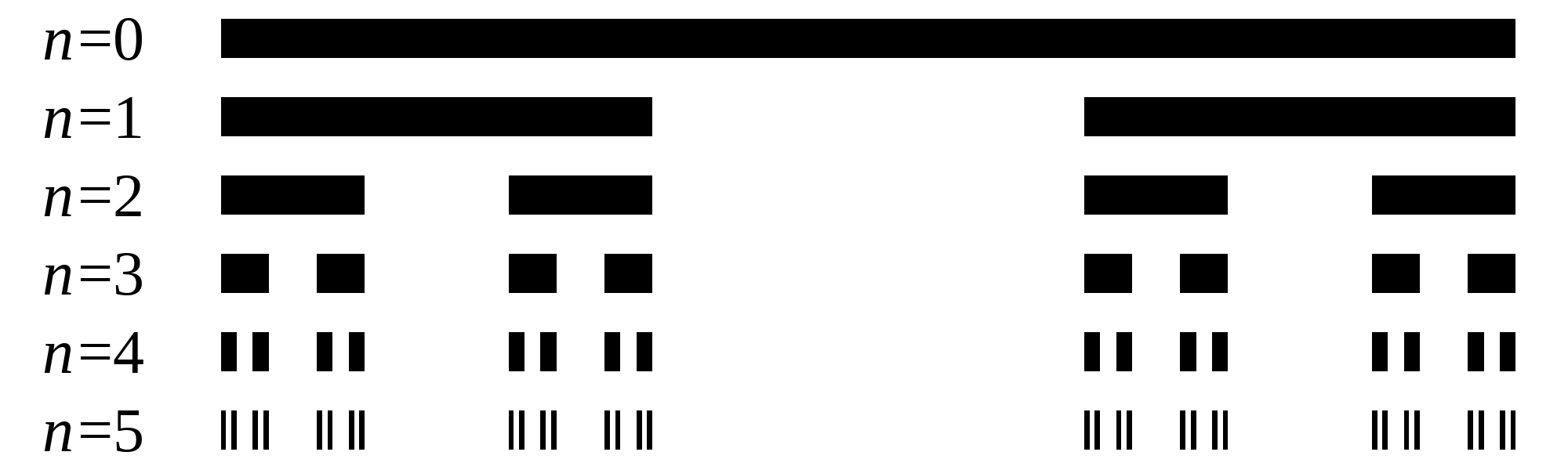}}
\vspace{3mm}
\subfloat[Generating the Koch curve. Symbol set: $V = \left\lbrace F, L, R \right\rbrace$, axiom: $\omega = F$, production rule set: $P = \left\lbrace F \rightarrow FLFRRFLF \right\rbrace$. Commands assigned to symbols: $F$ - move forward a fixed distance and draw a line, $L/R$ - turn 60 degrees left/right.]{\includegraphics[width=0.75\textwidth]{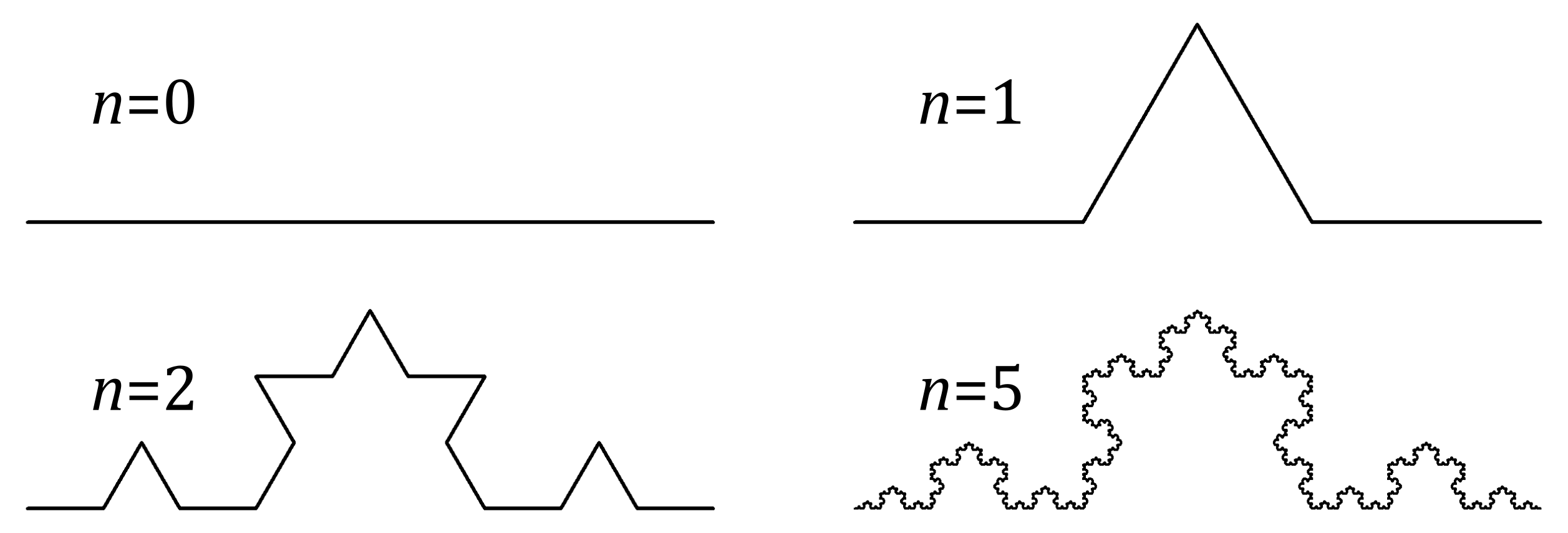}}
\vspace{6mm}
\subfloat[Generating a binary tree. Symbol set: $V = \left\lbrace F, G, L, R, <, > \right\rbrace$, axiom: $\omega = G$, production rule set: $P = \left\lbrace G \rightarrow \textit{F<LG><RG>}, \, F \rightarrow FF \right\rbrace$. Commands assigned to symbols: $F$ or $G$ - move forward a fixed distance and draw a line, $L/R$ - turn 45 degrees left/right, $<\!/\!>$ - push/pop current position and angle onto/from the stack (a LIFO queue allowing to save the state of the plotting device and restore it later).]{\includegraphics[width=0.75\textwidth]{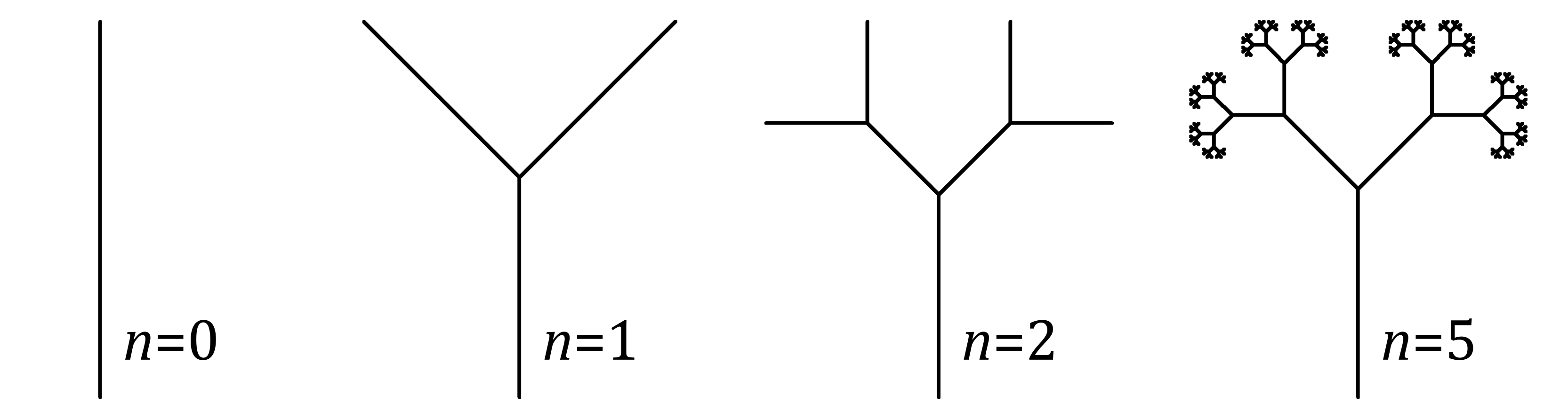}}
\vspace{6mm}
\subfloat[Generating a tree. Symbol set: $V = \left\lbrace F, G, L, R, <, > \right\rbrace$, axiom: $\omega = G$, production rule set: $P = \lbrace G \rightarrow \textit{FF<LGRGRG><RRGLGLG>}$, \, $ F \rightarrow FF \rbrace$. Commands assigned to symbols: $F$ or $G$ - move forward a fixed distance and draw a line, $L/R$ - turn 20 degrees left/right, $<\!/\!>$ - push/pop current position and angle onto/from the stack (a LIFO queue allowing to save the state of the plotting device and restore it later).]{\includegraphics[width=0.75\textwidth]{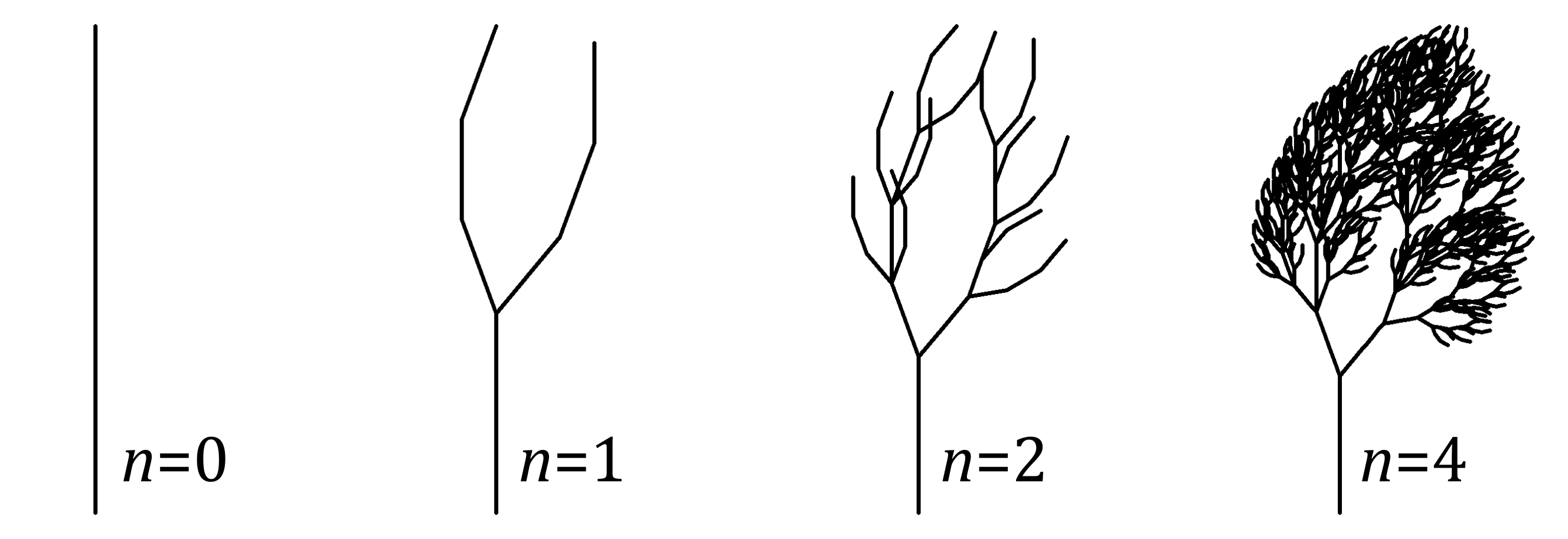}}
\caption{Examples of images generated with the use of L-systems and turtle graphics. Each subfigure (a)-(d) presents several images; each image corresponds to a given number $n$ of string generation iterations ($n=0$ corresponds to the starting string -- the axiom~$\omega$). Images within each subfigure are rescaled, so that they all have the same size. Symbol set $V$, axiom $\omega$ and production rule set $P$ of each used L-system are given along with the plotting device operations assigned to each symbol.}
\label{fig_Lsystems}
\end{minipage}
\end{figure}

Strings generated by L-systems can be represented graphically, using turtle graphics -- a method of creating graphics in which an imaginary object (called the turtle) moves around the drawing area according to a sequence of commands, and the trail left by this object is the desired output. If each symbol in a string generated by an L-system is treated as a command to a drawing device, then a graphical representation of this string can be plotted. In the context of computer graphics, a noteworthy example of L-systems' application is creating models of plants. Examples of images created with the use of L-systems are presented in Fig.~\ref{fig_Lsystems}.

\end{appendices}

\clearpage

\footnotesize

\bibliographystyle{elsarticle-num}

\phantomsection
\addcontentsline{toc}{section}{References}
\setlength{\bibsep}{0.18\baselineskip}
\bibliography{bibliography_dir/bibliography_bibfile}

\end{document}